\documentclass[11pt,a4paper,twoside,openright]{report}

\usepackage[utf8]{inputenc}
\usepackage[english]{babel}
\usepackage[T1]{fontenc}
\usepackage{bm}
\usepackage{amsmath}
\usepackage{amsfonts}
\usepackage{amsthm}
\usepackage{amssymb}
\usepackage{pifont}
\usepackage{graphicx}
\usepackage[colorlinks=true,allcolors=blue]{hyperref}
\usepackage{wasysym}
\usepackage{cite}
\usepackage[font=footnotesize]{caption}
\usepackage[makeroom]{cancel}
\usepackage{xcolor}
\usepackage{mathtools}
\usepackage{xfrac}
\usepackage{afterpage}

\usepackage[a4paper,top=3cm,bottom=4cm,left=3cm,right=3cm]{geometry}

\usepackage{fancyhdr}
\usepackage{titlesec, blindtext, color}
\titleformat{\chapter}[display]{\huge \bfseries}{\thechapter}{0pt}{\titlerule \vspace{10pt} \huge\bfseries}[\vspace{10pt} \titlerule]

\newcommand{\tr}{\mathrm{tr}}
\newcommand{\wh}{\widehat}
\newcommand{\di}{{\rm d}}
\newcommand{\de}{\partial}

\newcommand{\xT}{{\bf x}_{\rm T}}
\newcommand{\kT}{{\bf k}_{\rm T}}
\newcommand{\pT}{{\bf p}_{\rm T}}
\newcommand{\mT}{m_{\rm T}}
\newcommand\blankpage{%
    \null
    \thispagestyle{empty}%
    \addtocounter{page}{0}%
    \newpage}
\newcommand{\ed}{\rho}
\newcommand{\pl}{p^{\rm L}}
\newcommand{\pt}{p^{\rm T}}

\title{}
\author{}
\date{}

\begin{document}

\afterpage{\blankpage}
\begin{figure}
	\begin{center}
		\includegraphics[height=3cm]{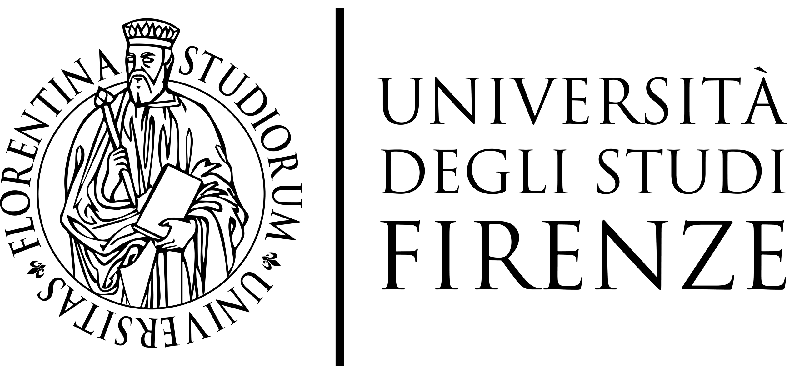}
		\label{fig:logo}
	\end{center}
\end{figure}
\begin{center}
	\begin{large}
		\textsc{Dipartimento di Fisica e Astronomia\\
		\vspace{1cm}
		Dottorato di Ricerca in Fisica e Astronomia\\
		Indirizzo Fisica -- XXXIII Ciclo\\
		Settore Scientifico Disciplinare FIS/02}\\
		\vfill
		\begin{huge}
			\textsc{\bfseries Entropy Current in\\
			Relativistic Quantum\\
			Statistical Mechanics\\}
		\end{huge}
		\vspace{2cm}
		\begin{LARGE}
		    \textit{Davide Rindori}
		\end{LARGE}
 		\vfill
	\end{large}
\end{center}
\textbf{Tutor}
\hfill
\textbf{Coordinatore}\\
Prof. Francesco Becattini
\hfill
Prof. Raffaello D'Alessandro\\
\vspace{1.5cm}
\begin{center}
    \begin{large}
        \textsc{2017 -- 2020}
    \end{large}
\end{center}
\thispagestyle{empty}
\newpage


\afterpage{\blankpage}
\vspace*{5cm}
\begin{Large}
    \begin{center}
        \textit{To my parents}
    \end{center}
\end{Large}
\vfill
\thispagestyle{empty}


\chapter*{Acknowledgements}
This PhD thesis is the culmination of a long journey I was able to walk not only on my own two feet, but also thanks to the help and support of many people that I would like to thank here.
I apologise in advance for those I may accidentally forget to mention.

First of all, I thank my supervisor Prof.\ Francesco Becattini.
He definitely is a leading expert in his field who puts true passion in his work conveying it to people around him.
Curiously, the first time we ever met was at the PhD interview at Florence University, Italy, where he was my interviewer.
I never attended any of his courses before.
He must have had some good feeling about me during the interview, because after I won the scholarship, he reached out for me with a project proposal for my PhD that I ended up taking.
I am grateful he did that back then, otherwise things may have gone quite differently by now.

I also thank his other current and former students and my PhD fellows Dr.\ Eduardo Grossi, Dr.\ Matteo Buzzegoli and Andrea Palermo.
Although we worked on different projects, we have all been involved in the same physics area, so it has been very important for me to have them for discussions, comparisons and advice.

A fundamental part of my PhD took place at the Institut f\"ur Theoretische Physik of the Goethe-Universit\"at in Frankfurt am Main, Germany.
This has been an experience of working as much as of life that I will always carry in my memory.
I cannot thank enough Prof.\ Dr.\ Dirk H.\ Rischke for giving me this opportunity by hosting me and finding time to work with me, but also for his constant help, his kindness, always making me feel part of the team and appreciating my work.
His human side has been as valuable for me as his knowledge of physics.
Vielen Dank f\"ur alles, ich werde es nicht vergessen.

A person without whom my work in Germany would have never been possible is Dr.\ Leonardo Tinti.
I am grateful to him for his guidance and for sharing his deep expertise with me, as well as for never complaining for all the times I knocked at his office door with some dumb question, for picking me up when I first arrived late at night in Frankfurt am Main and taking me to my room at the guesthouse.
His mentoring and friendship have really made the difference during my stay.

This visiting has been my first time away from my hometown, family and friends for such a long time.
I am grateful for the time spent with the Italian guys, especially Dr.\ Enrico Speranza, Dr.\ Alessandro Sciarra and Dr.\ Lucia Oliva.
Having lunch together at the canteen while talking, often comparing life in Italy and in Germany, having dinner out at a traditional restaurant, sipping some gl\"uwein at the Christmas markets and meeting at the gym may have seemed small gestures, but they helped me settle in and made me feel part of the group.
I wish you guys all the best and hope to see you again some day.

Being in the same boat with someone every day for three years creates a bond.
Thus, a big thank goes to my office mates in Florence: Lorenzo Maffi and Alessio Caddeo.
We sure shared some physics together, but nowhere near as all the laughs, the coffee breaks, the curses for calculations that did not add up and for the typical overcomplicated and confusing Italian bureaucracy, the celebrations for the first published paper, some personal confidence.
Thank you for going through the good times and the bad together.

I also want to thank some colleagues and dear friends: Antonio De Cristofaro, Emanuele Viviani and Dr.\ Simone Blasi.
After having been undergraduate students together, we went our more or less separate ways, but we never stopped being in touch.
I cannot even tell half of the fun and the crazy moments we have had through the years.
I am certainly glad that physics shaped my mind and taught me things, but I am also grateful that it introduced me to friends like them that made the process enjoyable and worth doing.
Our friendship is of great value for me.

A very special thank goes to my girlfriend Silvia Tammaro.
Life has not been gentle with her, and I am well aware I put her through a lot when my visiting in Germany took place and I had to actually leave.
I cannot thank her enough for realizing how important that experience was for me and finding the strength to endure it together.
I do not usually seek for comfort or help, but ever since she has been by my side, she has constantly provided me with support without me having to ask.
Thank you for being with me and giving me the stability to walk along this path.

My final and deepest thank goes to my parents Daniela Fontanelli and Roberto Rindori.
Sometimes I may not be good at saying thank you out loud, so I am glad I here have the occasion to do that in writing.
I am grateful to them for believing in me, even more than I myself have done at times, constantly providing me with help in any possible way and unconditioned love to support my education, my passions, my happiness, my life.
I want them to know how much I appreciate all of this, how much I respect it, that I never took it for granted, and that I will always strive to make the best use I can of their efforts.
They deserve credit for anything good I will ever do in life.


\tableofcontents
\newpage


\chapter*{Notation}

In this work, we will use the the natural unit system in which $\hbar=c=k_{\rm B}\equiv 1$.
If necessary, the natural constants will be restored with an explicit notification.

We adopt the ``mostly minus'' convention for the metric tensor, so the Minkowski metric will be $g_{\mu \nu}={\rm diag}(1,-1,-1,-1)$.
Therefore, timelike and spacelike vector fields will have positive and negative magnitude respectively.

Tensors will be denoted by their components, e.g.\ $P^{\mu}$, where Greek indices will take on values $\mu=0,1,2,3$.
Vectors in the 3-dimensional space will be indicated either with a bold letter, e.g.\ ${\bf p}$, or with their components, e.g.\ $p_i$, where Latin indices will take on values $i=1,2,3$.

The square modulus of a vector field $g_{\mu \nu}\beta^{\mu}\beta^{\nu}=\beta^{\mu}\beta_{\mu}$ will be denoted as $\beta^2$, hence the modulus $\sqrt{\beta^2}$ or $|\beta^{\mu}|$.
The square modulus of a 3-dimensional vector ${\bf p}\cdot {\bf p}$ will either be indicated as $|{\bf p}|^2$ or simply ${\bf p}^2$, hence the modulus $|{\bf p}|$ unless otherwise specified.

The Einstein convention of index summation whenever the same index is repeated in a subscript and a superscript will be adopted, e.g.\ $\alpha^{\mu}\beta_{\mu}=\sum_{\mu=0}^3\alpha^{\mu}\beta_{\mu}$.
The scalar product of two 3-dimensional vectors will be denoted by the symbol ``$\cdot$'', e.g.\ ${\bf k}\cdot {\bf p}=\sum_{i=1}^3k_ip_i$.

The notations $a_{(\mu}b_{\nu)}$ and $a_{[\mu}b_{\nu]}$ will stand for symmetrization and antisymmetrization of indices respectively, that is $a_{(\mu}b_{\nu)}=\frac{1}{2}(a_{\mu}b_{\nu}+a_{\nu}b_{\mu})$ and $a_{[\mu}b_{\nu]}=\frac{1}{2}(a_{\mu}b_{\nu}-a_{\nu}b_{\mu})$ respectively.

For the Levi-Civita symbol we use the convention $\epsilon^{0123}=1$, while $\nabla_{\mu}$ and $\de_{\mu}=\de/\de x^{\mu}$ will indicate the covariant and the standard derivative respectively.

Quantum perators in Hilbert spaces will be denoted by a wide hat, e.g.\ $\wh{O}$, whereas a regular hat may be used for vector fields of unit magnitude, e.g.\ $\hat{e}^{\mu}$ such that $\hat{e}^{\mu}\hat{e}_{\mu}=\pm 1$.

The commutator of two operators will be denoted as $[\wh{A},\wh{B}]$, while $:{}:$ will stand for normal ordering with respect to creation and annihilation operators.

Sansserif letters will be used to indicate elements of a group, e.g.\ ${\sf R}\in {\rm SO}(3)$.

The symbol $i$ will denote the imaginary unit, while the symbol ``${\rm e}$'' the Euler's number.

The complex and the Hermitean conjugates will be indicated by ${}^*$ and ${}^{\dagger}$ respectively.
The real and the imaginary parts of a complex number $z$ will be denoted by ${\rm Re}(z)$ and ${\rm Im}(z)$ respectively.

The symbols ``$\tr$'', ``${\rm det}$'' and ``$\log $'' will stand for trace, determinant and natural logarithm respectively.

Other notation may be introduced as needed.

\chapter{Introduction}
As our intuition would guess, hydrodynamics is indeed a theory describing the evolution of fluids, but the definition of a fluid is way more general than one would probably expect.
Roughly speaking, in modern physics a fluid is any system comprised of a large number of degrees of freedom whose characteristic scale of the microscopic interactions is significantly smaller than its macroscopic size.
Thus, although microscopically consisting of a substrate of discrete particles, such systems can be regarded macroscopically as a continuum, called a fluid, whose dynamics can be studied collectively.
In this sense, hydrodynamics establishes a connection between the microscopic and the macroscopic properties of a system.
If the velocities of the fluid's constituents are close to the speed of light, or if so is the velocity of the fluid as a whole, or if the macroscopic gravitational field is strong, the fluid, whence the hydrodynamic theory, becomes relativistic.
Much endeavour has been put through the years to build and refine a consistent and predictive theory of relativistic hydrodynamics, capable of describing successfully not only the simplest case of so-called ideal fluids, but also including dissipations so that transport and entropy-producing phenomena could be accounted for.
Today, the result of this effort is used to describe with remarkable precision what we observe, for instance, when colliding two heavy ions moving at near the speed of light, when studying racing sailing boats, or even when explaining the most catastrophic events from the remote corners of the universe \cite{landau2013fluid, Eckart:1940, Israel:1976tn, Israel:1979wp, Hiscock:1983zz, Danielewicz:1984ww, Rischke:1998fq, Muronga:2003ta, Romatschke:2009im, rezzolla2013relativistic}.

The fundamental nature underlying the collective classical picture of all fluids is essentially quantum field theoretical.
Quantum effects can emerge due to the acceleration or rotation of a fluid, but in most cases they are dominated by thermal effects: very low temperatures are usually required for them to be comparable.
A well-known example is indeed the famous Unruh effect, a relativistic quantum phenomenon arising at low temperatures due to acceleration, which was discovered more than forty years ago and still is a vibrant subject of investigation \cite{Unruh:1976db}.
In short, the Unruh effect states that uniformly accelerated observers in Minkowski spacetime associate a thermal bath of particles to the no-particle state of inertial observers, that is the Minkowski vacuum, and that its temperature is proportional to the magnitude of the acceleration.
However, in order to have an Unruh temperature of order $1\,{\rm K}$, an acceleration of order $10^{20}\,{\rm m/s^2}$ is needed; therefore, for typical values of the acceleration, the Unruh effect is quite small \cite{Crispino:2007eb}.
Nevertheless, sometimes matter can show up in conditions so extreme that quantum and thermal effects are of similar order even at high temperatures.
This is certainly the case of the quark-gluon plasma formed in heavy-ion collisions, which has been found experimentally to be a nearly ideal relativistic quantum fluid at local thermodynamic equilibrium \cite{Heinz:2000bk, Becattini:2007sr, Teaney:2009qa, Heinz:2013th}.
In order for it to be produced, the heavy ions have to be strongly accelerated to a velocity approaching the speed of light, and since most of the collisions are non-central, the quark-gluon plasma ends up spinning incredibly fast.
Therefore, despite the temperature being very high, the quantum effects due to rotation and acceleration are in fact not negligible with respect to the thermal ones.
This rotational structure can lead to a spin polarization of the hadrons that eventually emerge from the plasma, providing experimental access to flow substructure in great detail.
These measurements are in broad agreement with hydrodynamic and transport-based calculations and reveal that the quark-gluon plasma is the most vortical fluid ever observed \cite{STAR:2017ckg, Adam:2018ivw, Becattini:2020ngo}.

Thus, with the development of the hydrodynamic theory of multiparticle production, as well as relativistic astrophysics and cosmology, at some point it became necessary to go beyond the framework of the phenomenological linear relativistic hydrodynamics.
The aim was then to derive systematically the hydrodynamic equations, the transport relations and the transport coefficients without relying in the general case on the assumption that the state of the system is close to thermodynamic equilibrium.
Moreover, the need of describing quantum phenomena in relativistic fluids started to convey a great deal of attention into the quantum statistical foundations of relativistic hydrodynamics, and it still is to date.
In the late 70's, a theoretical setup was proposed by D.\ N.\ Zubarev to address these and other issues.
His approach to relativistic quantum statistical mechanics is based on a definition of a density operator at local thermodynamic equilibrium which is fully covariant and built in terms of the quantum operators of the conserved currents by means of the maximum entropy principle \cite{Zubarev:1979, Vanweert1982133, zubarev1996statistical, zubarev1997statistical}.
Global thermodynamic equilibrium and non-equilibrium states can be obtained from it, and the classical quantities entering the hydrodynamic equations are understood as thermal expectation values of the corresponding quantum operators calculated with that density operator.
As such, it provides us with a formulation of relativistic hydrodynamics which has a clear geometrical interpretation and is capable of dealing, in principle, with full non-equilibrium states, without having to be close to thermodynamic equilibrium.
Furthermore, the underlying microscopic theory is in general taken to be a Quantum Field Theory: this is important especially for strongly coupled fluids such as the quark-gluon plasma, for which the description in terms of a relativistic kinetic theory breaks down \cite{Adams:2012th, CasalderreySolana:2011us}.
After being first put forward, this approach has somehow gone unnoticed for quite some time, but then it was rediscovered and recently reworked in a more modern language \cite{Hosoya:1983id, Becattini:2011ev, Becattini:2012tc, Becattini:2012pp, Becattini:2014yxa, Hayata:2015lga, Becattini:2019dxo}.

The Zubarev formulation of the quantum statistical foundations of relativistic hydrodynamics is being used to tackle a variety of problems, such as the derivation and use of the Kubo formulae \cite{Hosoya:1983id, Huang:2011dc}, the apparent thermodynamic inequivalence of the quantum energy-momentum and spin tensors \cite{Becattini:2011ev, Becattini:2012pp}, the study of relativistic quantum fluids with rotation and acceleration \cite{Becattini:2014yxa, Becattini:2017ljh}, polarization, vorticity and chiral effects in the quark-gluon plasma \cite{Becattini:2020ngo, Buzzegoli:2020ycf} and many more.
Another such interesting topic is certainly the entropy production occurring in quantum irreversible processes.
The entropy is indeed one of the most fundamental and fascinating concepts in relativistic hydrodynamics and physics in general.
The second law of thermodynamics constrains it to never decrease in time, in particular it stays constant for systems at thermodynamic equilibrium and increases for those out of equilibrium because of irreversible processes, that are the dissipations.
An important property of the entropy is the extensivity, meaning that it depends on the volume of the system.
Consequently, the second law of thermodynamics as above reported is a global statement, in the sense that it concerns the system in its entirety.
This is fine in a non-relativistic theory, but from a relativistic standpoint the meaningful statements ought to be local.
In order to embed the extensivity and the second law of thermodynamics in a covariant language, the existence of a vector field, called the \textit{entropy current}, is postulated such that its integral on some 3-dimensional spacelike hypersurface gives the entropy.
With this definition, the local version of the second law of thermodynamics is simply the fact that the divergence of the entropy current, referred to as the entropy production rate, ought to be non-negative, namely vanishing at thermodynamic equilibrium and being positive otherwise.
Thus, the form of the entropy current entails the constitutive equations of the conserved currents in terms of the derivatives of the intensive thermodynamic parameters, making it a quantity one would most likely want to know exactly.

In the very Israel-Stewart formulation of relativistic hydrodynamics, however, the form of the entropy current is obtained in a more or less phenomenological fashion by writing down all the possible terms with given properties, but it certainly it is not derived \cite{Israel:1976tn}.
Also, from a quantum statistical standpoint, an equation for the entropy production rate at local thermodynamic equilibrium is obtained in the framework of the Zubarev approach, but it provides us with information on the divergence of the entropy current, not the entropy current itself \cite{Zubarev:1979, Becattini:2019dxo}.
Over the last decade, there has been a large number of studies where the structure of the entropy current was involved, and there have also been attempts to reformulate relativistic hydrodynamics without one \cite{Loganayagam:2008is, Bhattacharya:2012zx, Chattopadhyay:2014lya, Banerjee:2014ita, Glorioso:2017fpd, Haehl:2015uoc, Haehl:2018uqv, Jensen:2018hhx, Hattori:2019lfp, Jensen:2012jh}.
In most of these studies, the structure of the entropy current is postulated based on some classical form of the laws of thermodynamics supplemented by methods to include the dissipations, but technically speaking it is not derived.
In the language of relativistic quantum statistical mechanics, the reason for this apparently insufficient definition is that the entropy current, unlike the energy-momentum tensor and the charged currents, is not the thermal expectation value of an operator built with quantum fields.
Yet another reason for this indeterminacy is the fact that while the entropy has a precise definition in terms of the density operator given by the von Neumann formula, the entropy current, which should be a more fundamental quantity in a relativistic framework, has not.
\newline

In this work, we will show that it is indeed possible to provide a rigorous definition of the entropy current at local thermodynamic equilibrium in relativistic quantum statistical mechanics and thereby to derive its form.
It should be emphasized that this will hold at local thermodynamic equilibrium at most, since no such equation is currently known for 
relativistic quantum fluids fully out of thermodynamic equilibrium.
Our method for 
the entropy current will be based on the expression of the local thermodynamic equilibrium density operator put forward by Zubarev.
The key step to show that an entropy current exists at local thermodynamic equilibrium will be the proof that, in a relativistic theory, the logarithm of the partition function is extensive, meaning that it can be expressed as the integral on a 3-dimensional spacelike hypersurface of a vector field called the thermodynamic potential current.
We will provide, to the best of our knowledge, the first general proof of this usually tacitly understood hypothesis under very general conditions \cite{Becattini:2019poj}.

After the general method, we will present two specific non-trivial instances of calculation of the entropy current.
The first system we will consider is a relativistic quantum fluid at global thermodynamic equilibrium with acceleration field of constant magnitude in Minkowski spacetime.
From a phenomenological perspective, this is particularly interesting in the context of heavy-ion collisions, for the quark-gluon plasma is a fluid at local thermodynamic equilibrium with large values of acceleration and vorticity, local thermodynamic equilibrium being approximated by global thermodynamic equilibrium at first order in the thermodynamic fields.
From a more theoretical standpoint, an acceleration involves particular quantum field theoretical effects at low temperature such as the Unruh effect, which will in fact emerge, and the entropy current will indeed contain a quantum correction due to the acceleration.
Interestingly enough, the entropy obtained by integration of this entropy current will coincide with the entanglement entropy between two subspaces of Minkowski spacetime \cite{Becattini:2019poj}.

The second system we will consider is a relativistic quantum fluid with longitudinal boost invariance, also known in literature as the Bjorken model.
The phenomenological interest towards this system comes again from the physics of the quark-gluon plasma, for J.\ D.\ Bjorken first proposed it as approximately realized in the central-rapidity region in heavy-ion collisions \cite{Bjorken:1982qr}.
Modern hydrodynamic models can transcend it by including transverse expansion and, eventually, by breaking boost invariance itself; however, they still follow its general principles \cite{Florkowski:2010zz}.
On the theoretical side, boost-invariant fluids are inherently out of thermodynamic equilibrium, thus notoriously hard to study.
Nevertheless, boost invariance will provide us with special constraints allowing for the analytic calculation of the thermal expectation value at local thermodynamic equilibrium, making it the first ever case of an exactly solvable system in that configuration.
The entropy current thereby obtained will include no apparent quantum correction.
A full non-equilibrium analysis will also be carried out in some asymptotic limit, making this system a benchmark for the vast and mostly unexplored subject that is non-equilibrium thermal field theory \cite{Rindori:2021quq}.
\newline

This work is organized as follows.
We will provide in Chapter \ref{chapter:relhydro} a quick review on relativistic hydrodynamics, starting by defining what fluids and a hydrodynamic theory are in Section \ref{sec:relhydro_fluids}, over to ideal and dissipative hydrodynamics in Sections \ref{sec:relhydro_ideal_hydro} and \ref{sec:relhydro_dissipative_hydro} respectively, touching only the points most essential for this work.
Special emphasis will be put into highlighting the importance of the entropy current in this theory.

Chapter \ref{chapter:zubarev} will start with a survey of quantum statistical mechanics in Section \ref{sec:zubarev:quant_stat_mech} followed by the core of the Chapter in Section \ref{sec:zubarev_zub_approach}, that is the Zubarev approach to relativistic quantum statistical mechanics.
As a comprehensive and pedagogical review on this formalism seems to lack in literature, we will seize this opportunity to put forward one ourselves to the best of our ability.
Section \ref{sec:zubarev_entropy_current_method} will contain the detailed derivation of our method for the entropy current, therefore it will be of particular relevance for this work.

The case of a relativistic quantum fluid at global thermodynamic equilibrium with acceleration will be analyzed in Chapter \ref{chapter:gteacceleration}, starting with a presentation of the density operator describing this particular configuration in Section \ref{sec:gte_gte_with_acceleration}.
The thermal expectation value of the energy-momentum tensor will be carried out in Section \ref{sec:gte_tev_unruh} exploring renormalization and the relation with the Unruh effect.
In Section \ref{sec:gte_entropy_current_in_RRW} we will calculate the entropy current with our method, followed by a discussion of the result.

In Chapter \ref{chapter:boost}, we will study a relativistic quantum fluid out of thermodynamic equilibrium with boost invariance, starting with thorough introduction on the boost-invariant density operator in Section \ref{sec:boost_boost_invariance}.
The thermal expectation value of the energy-momentum tensor at local thermodynamic equilibrium will be worked out exactly and renormalization explored in Section \ref{sec:boost_lte_analysis}, followed by the calculation of the entropy current.
Finally, in Section \ref{sec:boost_ne_analysis} we will carry out an analysis of the energy-momentum tensor fully out of thermodynamic equilibrium in some asymptotic limit, examining again renormalization.

We will conclude with a brief summary of our results in Chapter \ref{chapter:conclusions}.

\chapter{Relativistic Hydrodynamics}
\label{chapter:relhydro}
In this Chapter, we will provide a very quick and rather qualitative review of the vast topic that is relativistic hydrodynamics, with the aim of introducing those concepts that will be important for the rest of this work.
We will start by defining what a fluid and a hydrodynamic theory are, then moving over to the simplest instance of a hydrodynamic theory, namely ideal hydrodynamics.
Next, we will extend it by including dissipations and present the relativistic Navier-Stokes theory, which, although reproducing the correct non-relativistic limit and fulfilling the laws of thermodynamics, will also have deal-breaking features.
As a solution, we will briefly introduce the Israel-Stewart theory.
Finally, we will conclude with a short summary.

As for the references, in this work we will loosely follow the modern textbook by L.~Rezzolla and O.~Zanotti \cite{rezzolla2013relativistic}, but we also ought to mention the classical textbook by L.\ D.\ Landau and E.\ M.\ Lifshitz \cite{landau2013fluid} and the famous work by W.\ Israel and J.\ M.\ Stewart \cite{Israel:1976tn, Israel:1979wp}.


\section{Fluids}
\label{sec:relhydro_fluids}

When we stumble upon the word ``hydro'', our thoughts instinctively go to water and liquids in general.
This is certainly not wrong, as hydrodynamics has indeed to do with the motion of fluids, but in modern Physics the concept of fluid is far more general than that.

In order to get some insight on this point, let us consider a system of $N$ particles interacting through some coupling and look at the ratio ${\cal R}\equiv \lambda_{\rm DB}/l$, where $\lambda_{\rm DB}$ is the de Broglie wavelength associated to each particle and $l$ is the characteristic distance between particles.
If ${\cal R}\gtrsim 1$ there is a considerable overlap of the wavefunctions of the particles, thus the quantum mechanical nature of the system is important and the evolution will be governed by the Schr\"odinger equation for the $N$-particle wavefunction.
On the other hand, if ${\cal R}\ll 1$ the particles are so far away from each other that quantum interference can be neglected, so there will be $N$ single-particle wavefunctions each evolving according to the Schr\"odinger equation, moving like classical particles.
This is but the Ehrenfest's theorem, stating that the evolution of the expectation value of each observable coincides with the evolution described by classical mechanics.
Of course, if $N$ is very large, solving $N$ Schr\"odinger equations does not seem practically feasible, a statistical approach would then be necessary.

However, there is also a regime where another kind of description is possible.
Suppose that the number of particles $N$ is very large, and that the system extends over a length scale $L$ so much larger than $l$ (and of course $\lambda_{\rm DB}$) that the single-particle dynamics cannot be followed, that is ${\cal R}\lll 1$.
This separation between the microscopic and the macroscopic scales is quantified by the Knudsen number $K_{\rm N}\equiv l/L$, which should be $K_{\rm N}\ll 1$ in the above configuration.
In this case, the system can be conveniently approximated as a continuum whose dynamics can be studied collectively.
This continuum is called a \textit{fluid}, a system whose large-scale properties can be described effectively without having to worry about the features that the constituent elements have at much smaller length-scales.
A fluid is usually thought as divided in so-called \textit{fluid elements}, namely ``cells'' large enough to contain a great number of particles, but still small compared to the macroscopic scale $L$ in order to guarantee homogeneity within them.
As a consequence, in each fluid element the particles have the same average velocity and are at thermodynamic equilibrium.
The properties of neighbouring fluid elements, which can be different and even discontinuous, represent then the global properties of the fluid.
The collective dynamics of the fluid is called \textit{fluid dynamics} or \textit{hydrodynamics}: its aim is the study of the evolution of the fluid by solving its equations of motion, the \textit{hydrodynamic equations}.

Thus, hydrodynamics is an effective classical theory describing the evolution of a system in terms of few quantities, although the underlying microscopic theory might be quantum mechanical and possibly complicated.
Moreover, this definition of a fluid appears to be quite general: it certainly includes the liquids and gases we intuitively expect, but actually many other different kinds of systems as well.
These are some of the reasons why hydrodynamics finds applications in a vast number of phenomena ranging from meteorology over to heavy-ion collisions, relativistic astrophysics and cosmology.

As hydrodynamics establishes a connection between the microscopic and the macroscopic properties of a system, when does it become \textit{relativistic}?
This can occur essentially in three different contexts, which are not mutually exclusive in general.
The first one is when the velocities of the constituent particles within a fluid element are close to the speed of light, or, equivalently, when their Lorentz factor is significantly larger than one.
The second context applies instead when the Lorentz factor of the macroscopic motion of the fluid is significantly larger than one, quite independently of the microscopic properties of the fluid.
The third and final one emerges whenever the macroscopic gravitational field is strong enough to require a description in terms of General Relativity.
In this latter case, no assumption is made about the velocity of the fluid, which can even be at rest as, for instance, in a stationary relativistic star.

As we mentioned, a relativistic fluid is thought as divided in many fluid elements.
What we can calculate in those elements are averages of an underlying particle substrate.
The distribution function to calculate those averages usually comes from a relativistic kinetic theory, a model describing the non-equilibrium dynamics of a fluid whose microscopic degrees of freedom can be regarded as a diluted system of weakly interacting particles.
The averages thereby obtained enter the hydrodynamic equations, so in this sense relativistic hydrodynamics can be thought as a coarse graining of an underlying relativistic kinetic theory.
This approach works well when the mean free path of the particles is much larger than their thermal wavelength, however, it breaks down for strongly interacting systems, which require a full quantum field theoretical description in general.
This does not mean that a hydrodynamic theory cannot be built for such systems, but only that it should be done without necessarily relying on a relativistic kinetic theory.
The problem of calculating the averages entering the hydrodynamic equations in the case of an underlying Quantum Field Theory will be the subject of next Chapter.
In the present one, we assume they have been somehow obtained and we focus on their hydrodynamic evolution.

The first problem we come across when we forget about a particle substrate is the definition of a hydrodynamic velocity field $u^{\mu}$, which, in a kinetic theory, would intuitively be the average velocity of the particles.
Thus, the idea is to consider as the fundamental quantity the energy-momentum tensor $T^{\mu \nu}$, which always exists, and define a velocity field from it alone.
It is known that if $T^{\mu \nu}u_{\mu}u_{\nu}\ge 0$ for any timelike vector field $u^{\mu}$, called the \textit{weak energy condition}, then there exists a unique timelike  vector field ${u_{\rm L}}^{\mu}$ with magnitude ${u_{\rm L}}^{\mu}{u_{\rm L}}_{\mu}=1$ such that $T^{\mu \nu}{u_{\rm L}}_{\mu}=\lambda {u_{\rm L}}^{\nu}$ for some $\lambda \ge 0$.
In other words, if the energy density measured by any observer is non-negative, the energy-momentum tensor has a unique timelike eigenvector.
The reference frame comoving with ${u_{\rm L}}^{\mu}$ is called the \textit{local rest frame}, and it is such that $T^{00}=\lambda$ and $T^{0i}=0$ for some $T^{ij}$.
So, one way to proceed is to take ${u_{\rm L}}^{\mu}$ as the hydrodynamic velocity field of the fluid.
However, this frame choice, called the \textit{Landau frame} (hence the subscript ${\rm L}$), is not the only possible one.
In fact, the system could also possess a charged current $j^{\mu}$, for instance an electromagnetic or baryonic current, as it often happens in astrophysics, so we might as well define the hydrodynamic velocity field as the direction of the charged current ${u_{\rm E}}^{\mu}\equiv j^{\mu}/\sqrt{j^2}$, where by $j^2$ we mean $j^2\equiv j^{\mu}j_{\mu}$.
This frame choice, called the \textit{Eckart frame} (hence the subscript ${\rm E}$), is in some sense the most straightforward generalization of the velocity field of non-relativistic hydrodynamics.
In fact, if $j^{\mu}$ is taken as the baryonic current, ${u_{\rm E}}^{\mu}$ coincides with the average particles' velocity of non-relativistic hydrodynamics.
Anyway, whatever the frame we choose, the philosophy is that the fundamental quantities are the energy-momentum tensor and the charged currents, which always exist, and the velocity field is built from them.
This is somehow the other way round of what is usually done in non-relativistic hydrodynamics, where all the quantities are built starting from the velocity field.

Once the hydrodynamic velocity field is given, the other kinematic quantities can be worked out.
The acceleration vector field, for instance, is defined as the derivative of the velocity field along the flow
\begin{equation}\label{eqrelhydro23}
    A^{\mu}\equiv \frac{\di u^{\mu}}{\di \tau}=u^{\nu}\nabla_{\nu}u^{\mu},
\end{equation}
where $\tau$ is the fluid proper time, namely the proper time of an observer comoving with the fluid, and $\nabla_{\mu}$ the covariant derivative.
Of course, if the spacetime is flat, the covariant derivative is simply replaced by the standard derivative $\de_{\mu}$.
The hydrodynamic velocity is timelike, has unit magnitude by definition
\begin{equation}\label{eqrelhydro01}
    u^2\equiv u^{\mu}u_{\mu}=1
\end{equation}
and it is orthogonal to the acceleration field, which is spacelike
\begin{equation}
    u^{\mu}A_{\mu}=0.
\end{equation}
For future purposes, it is also convenient to define the \textit{projection tensor} $\Delta_{\mu \nu}$, which projects tensors on the hypersurface orthogonal to the velocity field, as
\begin{equation}
    \Delta_{\mu \nu}\equiv g_{\mu \nu}-u_{\mu}u_{\nu},
\end{equation}
where $g_{\mu \nu}$ is the metric tensor.
Sometimes it is also useful to decompose the covariant derivative along the velocity field and orthogonally to it, namely
\begin{equation}
    \nabla_{\mu}\equiv u_{\mu}D+D_{\mu},
\end{equation}
where
\begin{equation}
    D\equiv u^{\mu}\nabla_{\mu}=\frac{\di}{\di \tau},\qquad
    D_{\mu}\equiv {\Delta^{\nu}}_{\mu}\nabla_{\nu}
\end{equation}
with $D$ often called the \textit{convective derivative}.


\section{Ideal hydrodynamics}
\label{sec:relhydro_ideal_hydro}

In non-relativistic hydrodynamics, a \textit{perfect} or \textit{ideal fluid} is a fluid such that, at thermodynamic equilibrium, taken any section, all the inner forces are orthogonal to that section.
This could be rephrased by saying that the stress tensor $T^{ij}$ has vanishing shear stress, which in turn means that $T^{ij}$ is diagonal with all equal elements on the diagonal, i.e.\ $T^{ij}=-p\delta^{ij}$ for some eigenvalue $p$ with $\delta^{ij}$ the Kronecker delta.

Of course, in a relativistic theory we should consider the full energy-momentum tensor instead of the stress tensor alone.
Thus, in order to generalize this definition to the relativistic case, we could exploit the aforementioned known result stating that, if the energy-momentum tensor fulfills the weak energy condition, in its local rest frame its components are $T^{00}\ge 0$, $T^{0i}=0$ and some $T^{ij}$.
Therefore, we might think of defining an ideal relativistic fluid as a system whose energy-momentum tensor at thermodynamic equilibrium is diagonal and isotropic in its local rest frame, namely
\begin{equation}
    T^{\mu \nu}=(\rho+p)u^{\mu}u^{\nu}-pg^{\mu \nu}=
    \rho u^{\mu}u^{\nu}-p\Delta^{\mu \nu}.
\end{equation}
Here, $\rho$ and $p$ have the meaning of equilibrium energy density and equilibrium pressure respectively, while $g^{\mu \nu}$, being in the local rest frame, is simply the Minkowskian metric tensor.
Likewise, the charged current $j^{\mu}$ reads
\begin{equation}
    j^{\mu}=nu^{\mu},
\end{equation}
where $n$ is the equilibrium charge density.
\textit{Ideal hydrodynamics} corresponds to adopting the thermodynamic equilibrium forms of the energy-momentum tensor and charged current also out of equilibrium, promoting the energy density, pressure, charge density and velocity field to slowly-varying functions of the spacetime point, that is
\begin{equation}
    T^{\mu \nu}(x)=\left[\rho(x)+p(x)\right]u^{\mu}(x)u^{\nu}(x)-p(x)g^{\mu \nu}(x),
\end{equation}
\begin{equation}
    j^{\mu}(x)=n(x)u^{\mu}(x).
\end{equation}
The above expressions were obtained in the local rest frame, but being fully covariant they generalize to any frame, so $g^{\mu \nu}$ is in fact a generic metric tensor.
In the following, the dependence on the spacetime point will be omitted for ease of notation, unless otherwise specified.
Note that, under this hypothesis, the Landau frame coincides with the Eckart frame, namely ${u_{\rm L}}^{\mu}={u_{\rm E}}^{\mu}\equiv u^{\mu}$, so in this sense there is no ambiguity in the frame choice in ideal hydrodynamics.
This feature will not be preserved in non-ideal hydrodynamics.

The hydrodynamic equations are simply the conservation laws of the energy-momentum tensor and the charged current
\begin{equation}\label{eqrelhydro02}
    \nabla_{\mu}T^{\mu \nu}=0,
\end{equation}
\begin{equation}\label{eqrelhydro03}
    \nabla_{\mu}j^{\mu}=0.
\end{equation}
They amount to a set of 5 equations in 6 unknowns: the energy density $\rho$, the pressure $p$, the charge density $n$ and the three independent components of the velocity field $u^{\mu}$ (recall that $u^{\mu}$ satisfies the normalization condition \eqref{eqrelhydro01}).
Thus, in order to solve the system, we need an extra relation.
Technically, any relation between two unknowns is fine, for instance between the energy density and the pressure.
However, we can as well assume that, although globally out of equilibrium in general, for sufficiently small fluid elements the system is at thermodynamic equilibrium locally in space and time.
In this hypothesis, we can define a temperature and a chemical potential for each fluid element, so the thermodynamic quantities $\rho$, $p$ and $n$ can be related to them by an \textit{equation of state} reducing the number of unknowns to 5.
The achievement of this configuration, called \textit{local thermodynamic equilibrium}, is therefore equivalent to assuming the existence of the temperature and chemical potential for each fluid element, $T=T(x)$ and $\mu=\mu(x)$, which, from a relativistic standpoint, are defined as those measured in the local rest frame of the fluid element.
Once $T$, $\mu$ and $u^{\mu}$ are given on some three-dimensional spacelike hypersurface, one can try to solve the ideal hydrodynamic equations.

The four energy-momentum conservation equations are usually decomposed into one energy conservation equation and three momentum conservation equations, obtained by projecting \eqref{eqrelhydro02} along the velocity field and orthogonally to it respectively.
Together with the charged current conservation, they read
\begin{subequations}
    \begin{align}
        u_{\nu}\nabla_{\mu}T^{\mu \nu}=&\nabla_{\mu}(\rho u^{\mu})+p\nabla_{\mu}u^{\mu}=0,\label{eqrelhydro04}\\
        {\Delta^{\lambda}}_{\nu}\nabla_{\mu}T^{\mu \nu}=&(\rho+p)A^{\lambda}-{\Delta^{\lambda}}_{\nu}\nabla^{\nu}p=0,\label{eqrelhydro05}
    \end{align}
\end{subequations}
\begin{equation}\label{eqrelhydro07}
    \nabla_{\mu}j^{\mu}=u^{\mu}\nabla_{\mu}n+n\nabla_{\mu}u^{\mu}=0.
\end{equation}
In the non-relativistic limit, equations \eqref{eqrelhydro04} and \eqref{eqrelhydro05} reproduce the well-known continuity and Euler equations respectively
\begin{subequations}
    \begin{align}
        \nabla_{\mu}(\rho u^{\mu})+p\nabla_{\mu}u^{\mu}=0&\xrightarrow[\nabla_{\mu}\to \de_{\mu}]{p\ll \rho c^2}\de_t \rho+\nabla \cdot (\rho {\bf v})=0\\
        (\rho+p)A^{\lambda}-{\Delta^{\lambda}}_{\nu}\nabla^{\nu}p=0&\xrightarrow[\nabla_{\mu}\to \de_{\mu}]{p\ll \rho c^2}\rho {\bf a}+\nabla p=0
    \end{align}
\end{subequations}
where $\rho$ is now the mass density, ${\bf a}$ is the spatial component of the acceleration field $A^{\mu}$ and $\nabla$ is the gradient with respect to ${\bf x}$.

The next step is to consider the thermodynamics of ideal relativistic fluids.
The entropy $S$ of a system is constrained to never decrease in time by the second law of thermodynamics,
\begin{equation}\label{eqrelhydro18}
    \frac{\di S}{\di \tau}\ge 0.
\end{equation}
In particular, the responsible for entropy production are irreversible processes, while reversible ones do not produce any entropy.
It is also known that entropy is an extensive quantity, meaning that it depends on the volume of the system.
Now, as expressed in \eqref{eqrelhydro18}, the second law of thermodynamics is a global statement, meaning that it concerns the system in its entirety.
This is fine in a non-relativistic theory, but from a relativistic standpoint the meaningful statements ought to be local.
In order to embed the extensivity and the second law of thermodynamics in a covariant language, it is postulated that there exists a vector field $s^{\mu}$, called the \textit{entropy current}, such that its integral on some 3-dimensional spacelike hypersurface $\Sigma$ gives the entropy
\begin{equation}\label{eqrelhydro19}
    S\equiv \int_{\Sigma}\di \Sigma_{\mu}\,s^{\mu}.
\end{equation}
Here, $\di \Sigma_{\mu}\equiv \di \Sigma \,n_{\mu}$ with $\di \Sigma$ and $n^{\mu}$ the measure on $\Sigma$ and the timelike vector field of magnitude $n^{\mu}n_{\mu}=1$ orthogonal to $\Sigma$ respectively.
Equation \eqref{eqrelhydro19} is in fact the extensivity property, as it expresses the entropy as a volume integral.
With this definition of the entropy current, the local version of the second law of thermodynamics for any spacelike hypersurface $\Sigma$ is
\begin{equation}
    \nabla_{\mu}s^{\mu}\ge 0.
\end{equation}
Now the question is what the expression of the entropy current looks like and what constraints stem from the second law of thermodynamics.

As we mentioned, in ideal hydrodynamics we assume that the energy density, the pressure and the charge density have, point by point, the same type of dependence on the temperature and chemical potential, often referred to as the \textit{thermodynamic fields}, as they do at thermodynamic equilibrium.
Thus, the equations of state read
\begin{subequations}
    \begin{align}
        \rho=&\rho(x)=\rho_{\rm eq}(T(x),\mu(x)),\\
        p=&p(x)=p_{\rm eq}(T(x),\mu(x)),\\
        n=&n(x)=n_{\rm eq}(T(x),\mu(x)).
    \end{align}
\end{subequations}
This means that we can use our knowledge of equilibrium thermodynamics to study non-equilibrium thermodynamics.
For instance, at thermodynamic equilibrium we know that
\begin{equation}\label{eqrelhydro06}
    Ts=\rho+p-\mu n,
\end{equation}
where $s$ is the entropy density, namely the entropy per unit volume.
In particular, given the above equations of state, equation \eqref{eqrelhydro06} becomes in fact a definition of the entropy density out of equilibrium.
Moreover, the Gibbs-Duhem equation must hold
\begin{equation}\label{eqrelhydro09}
    T\,\di s=\di \rho-\mu \,\di n,
\end{equation}
which combined with the differentiation of \eqref{eqrelhydro06} gives
\begin{equation}
    \di p=s\,\di T+n\,\di \mu,
\end{equation}
hence
\begin{equation}
    s=\left(\frac{\de p}{\de T}\right)_{\mu},\qquad
    n=\left(\frac{\de p}{\de \mu}\right)_T,
\end{equation}
the symbols meaning that when deriving with respect to a thermodynamic field, the other one is kept fixed.
Thus
\begin{equation}
    \nabla_{\mu}p=\left(\frac{\de p}{\de T}\right)_{\mu}\nabla_{\mu}T+\left(\frac{\de p}{\de \mu}\right)_T\nabla_{\mu}\mu=s\nabla_{\mu}T+n\nabla_{\mu}\mu,
\end{equation}
and projecting along the velocity field
\begin{equation}
    u^{\mu}\nabla_{\mu}p=su^{\mu}\nabla_{\mu}T+nu^{\mu}\nabla_{\mu}\mu.
\end{equation}
Together with the derivative of \eqref{eqrelhydro06} along the flow, this implies
\begin{equation}\label{eqrelhydro10}
    Tu^{\mu}\nabla_{\mu}s=u^{\mu}\nabla_{\mu}\rho-\mu u^{\mu}\nabla_{\mu}n.
\end{equation}
The first term at right-hand side can be worked out by using the equation of motion \eqref{eqrelhydro04}, while for the second one we use \eqref{eqrelhydro07}
\begin{equation}
    u^{\mu}\nabla_{\mu}\rho=-(\rho+p)\nabla_{\mu}u^{\mu},\qquad
    u^{\mu}\nabla_{\mu}n=-n\nabla_{\mu}u^{\mu},
\end{equation}
hence
\begin{equation}
    Tu^{\mu}\nabla_{\mu}s=-(\rho+p-\mu n)\nabla_{\mu}u^{\mu}=-Ts\nabla_{\mu}u^{\mu},
\end{equation}
where in the last equality we used \eqref{eqrelhydro06}.
This is in fact a conservation equation
\begin{equation}\label{eqrelhydro08}
    \nabla_{\mu}s^{\mu}=0,
\end{equation}
where we defined the entropy current of an ideal relativistic fluid as
\begin{equation}
    s^{\mu}\equiv su^{\mu}.
\end{equation}
Equation \eqref{eqrelhydro08} tells us an important fact: an ideal fluid does not produce entropy, its entropy is constant in time.
As we will see, it will not be the same in general for non-ideal fluids.
In fact, the entropy current of a generic fluid will have more components other than the ideal $su^{\mu}$, those causing the entropy production rate $\nabla_{\mu}s^{\mu}$ to be greater than zero are called \textit{dissipations}.


\section{Dissipative hydrodynamics}
\label{sec:relhydro_dissipative_hydro}

Ideal hydrodynamics is obtained under the simplest hypothesis that, out of equilibrium, the energy-momentum tensor and the charged current have the same tensor structure as they do at thermodynamic equilibrium.
Indeed this is a strong assumption, which clearly becomes a poor approximation when the microscopic time scales are comparable to the macroscopic ones, namely when the condition of local thermodynamic equilibrium breaks down.
When this occurs, the ideal fluid description has to be extended by including dissipative terms,
that is by considering \textit{dissipative} or \textit{non-ideal fluids}.

The first problem we encounter when dealing with non-ideal fluids is the definition of the velocity field.
As we mentioned, for an ideal fluid the Landau and the Eckart frames coincide, making the choice of the velocity field unambiguous.
For a non-ideal fluid, this is known to no longer hold true, so one has to make a frame choice.
Both the frames have their own advantages.
For instance, the Eckart frame is more intuitive, being a quite straightforward generalization of the non-relativistic case and the continuity equation taking on a simple form, however it is not well-defined for systems with no net charge.
On the other hand, the Landau frame can be convenient as it simplifies the expression of the energy-momentum tensor, the price to pay is that the definition of the velocity field is implicit.
For more details on the frame choice in dissipative hydrodynamics, see \cite{Tsumura:2007ji}.

Whatever the frame we choose, in order to capture the effects of dissipations we have to go beyond the assumption of ideal hydrodynamics and allow the energy-momentum tensor and the charged current to contain other terms in addition to the ideal ones.
With this in mind, we can write in general
\begin{equation}
    T^{\mu \nu}=\rho u^{\mu}u^{\nu}-p\Delta^{\mu \nu}+\delta T^{\mu \nu}
\end{equation}
\begin{equation}
    j^{\mu}=nu^{\mu}+\delta j^{\mu \nu}
\end{equation}
where $\delta T^{\mu \nu}$ and $\delta j^{\mu}$ are the dissipative terms.
As such, they must vanish at thermodynamic equilibrium and depend on the derivatives of the thermodynamic fields.
We demand the energy and charge densities to be left unchanged by the dissipative terms, so we enforce the so-called \textit{Landau matching conditions}
\begin{equation}
    \delta T^{\mu \nu}u_{\mu}u_{\nu}=0,\qquad
    \delta j^{\mu}u_{\mu}=0.
\end{equation}
These allow us to write down the following irreducible tensor decompositions
\begin{equation}\label{eqrelhydro11}
    T^{\mu \nu}=\rho u^{\mu}u^{\nu}-(p+\Pi)\Delta^{\mu \nu}+q^{\mu}u^{\nu}+q^{\nu}u^{\mu}+\Pi^{\mu \nu}
\end{equation}
\begin{equation}\label{eqrelhydro12}
    j^{\mu}=nu^{\mu}+v^{\mu}.
\end{equation}
Here, $q^{\mu}$, $v^{\mu}$ and $\Pi^{\mu \nu}$ are tensor fields orthogonal to $u^{\mu}$, $\rho$ and $n$ are still the energy and charge densities while $p$ is in fact the equilibrium component of the pressure $p=p_{\rm eq}$, and being $p+\Pi$ the total pressure, $\Pi$ represents the \textit{non-equilibrium pressure}.
In formulae
\begin{subequations}
    \begin{align}
        \rho=&T^{\mu \nu}u_{\mu}u_{\nu}\\
        p+\Pi=&\frac{1}{3}T^{\mu \nu}\Delta_{\mu \nu}\\
         q^{\mu}=&T^{\alpha \beta}{\Delta^{\mu}}_{\alpha}u_{\beta}\\
         \Pi^{\mu \nu}=&\left[\frac{1}{2}\left({\Delta^{\mu}}_{\alpha}{\Delta^{\nu}}_{\beta}+{\Delta^{\nu}}_{\alpha}{\Delta^{\mu}}_{\beta}\right)-\frac{1}{3}\Delta^{\mu \nu}\Delta_{\alpha \beta}\right]T^{\alpha \beta}\\
         n=&j^{\mu}u_{\mu}\\
         v^{\mu}=&j^{\nu}{\Delta^{\mu}}_{\nu}.
    \end{align}
\end{subequations}

Being the dissipations irreversible processes, a non-ideal fluid will produce entropy, but in the limit of dissipations going to zero, the result of ideal hydrodynamics of vanishing entropy production rate ought to be recovered.
As a consequence, we expect the entropy current $s^{\mu}$ to have two components: one is the ideal term $su^{\mu}$ along the velocity field, while the other is orthogonal to that and depends on the dissipations.
Thermodynamic equilibrium is characterized by the absence of transport phenomena, namely by having $\Pi$, $q^{\mu}$ and $\Pi^{\mu \nu}$ all equal to zero, therefore we might as well say that, not surprisingly, the thermodynamic equilibrium state corresponds to a vanishing entropy production rate.
In the following, we will see how two approaches to dissipative hydrodynamics differ in prescribing the dissipative component of the entropy current while maintaining the same properties for the thermodynamic equilibrium state.


\subsection{Navier-Stokes theory}

This formulation of dissipative hydrodynamics was first proposed by Eckart \cite{Eckart:1940} and then slightly modified by Landau and Lifshitz \cite{landau2013fluid}.
It represents the simplest relativistic generalization of Navier-Stokes and Fourier equations of non-relativistic dissipative hydrodynamics, and it is usually referred to as \textit{classical irreversible thermodynamics}.
The fundamental feature of this theory is that the dissipative terms of the entropy current depend linearly on the dissipations $\Pi$, $q^{\mu}$ and $\Pi^{\mu \nu}$, therefore the entropy current contains up to the first derivatives of the thermodynamic parameters.
For this reason, classical irreversible thermodynamics is called a \textit{first-order theory}.

If we limit ourselves to first derivatives, meaning that deviations from thermodynamic equilibrium are small, and we still identify $s=s^{\mu}u_{\mu}$ with the entropy density, the thermodynamic relations \eqref{eqrelhydro06} and \eqref{eqrelhydro09} should keep holding.
Note that the total pressure is $p+\Pi$, so $p$ in \eqref{eqrelhydro06} must now be replaced with $p+\Pi$.
We have shown already that with \eqref{eqrelhydro06} and \eqref{eqrelhydro09} satisfied at each point, \eqref{eqrelhydro10} is obtained.
The first term at right-hand side is worked out from the projection of the equations of motion $\nabla_{\mu}T^{\mu \nu}=0$ along the flow, in particular we note that
\begin{equation}
    u_{\nu}\nabla_{\mu}T^{\mu \nu}=0
    \qquad \Rightarrow \qquad
    \nabla_{\mu}(T^{\mu \nu}u_{\nu})-T^{\mu \nu}\nabla_{\mu}u_{\nu}=0.
\end{equation}
By using the expression \eqref{eqrelhydro11} of the energy-momentum tensor, we have
\begin{equation}
    u^{\mu}\nabla_{\mu}\rho=-\rho \nabla_{\mu}u^{\mu}-\nabla_{\mu}q^{\mu}-(p+\Pi)\nabla_{\mu}u^{\mu}+q^{\mu}A_{\mu}+\Pi^{\mu \nu}\nabla_{\mu}u_{\nu}.
\end{equation}
For the second term, we consider the equation of motion $\nabla_{\mu}j^{\mu}=0$.
Using \eqref{eqrelhydro12} we find
\begin{equation}
    u^{\mu}\nabla_{\mu}n=-n\nabla_{\mu}u^{\mu}-\nabla_{\mu}v^{\mu}.
\end{equation}
Plugging these two equations into \eqref{eqrelhydro10}, we get
\begin{equation}
    \nabla_{\mu}(su^{\mu})=-\frac{1}{T}\nabla_{\mu}q^{\mu}+\frac{q^{\mu}}{T}A_{\mu}+\frac{\Pi^{\mu \nu}}{T}\nabla_{\mu}u_{\nu}+\frac{\mu}{T}\nabla_{\mu}v^{\mu}.
\end{equation}
Finally, by defining $\zeta \equiv \mu/T$, we have
\begin{equation}\label{eqrelhydro13}
    \nabla_{\mu}\left(su^{\mu}+\frac{q^{\mu}}{T}-\zeta v^{\mu}\right)=-\frac{q^{\mu}}{T^2}\nabla_{\mu}T+\frac{q^{\mu}}{T}A_{\mu}+\frac{\Pi^{\mu \nu}}{T}\nabla_{\mu}u_{\nu}-v^{\mu}\nabla_{\mu}\zeta.
\end{equation}
The non-relativistic limit of this equation is the known equation for the entropy production rate of non-relativistic dissipative hydrodynamics, provided that $q^{\mu}$ and $v^{\mu}$ are interpreted as the \textit{heat flow} and the \textit{current flow} respectively.
Moreover, $\Pi^{\mu \nu}$ is called the \textit{viscous stress tensor}.
Thus, we are persuaded to identifying the entropy current of dissipative hydrodynamics with
\begin{equation}\label{eqrelhydro16}
    s^{\mu}=su^{\mu}+\frac{q^{\mu}}{T}-\zeta v^{\mu}.
\end{equation}

So far, we do not have any information on the expression of the dissipations.
This is obtained by enforcing the second law of thermodynamics, namely by demanding \eqref{eqrelhydro13} to be non-negative.
Being the dissipative terms are independent from each other, we will have a constraint for each of them.
The following expressions of the dissipations are such that the second law is fulfilled for any value of the temperature and velocity field
\begin{subequations}
    \begin{align}
        q^{\mu}\equiv&\kappa \Delta^{\mu \nu}(\nabla_{\nu}T-TA_{\nu})\label{eqrelhydro15a}\\
        \Pi^{\mu \nu}\equiv&2\eta \sigma^{\mu \nu}\label{eqrelhydro15b}\\
        \Pi \equiv&\xi \nabla_{\mu}u^{\mu}\label{eqrelhydro15c}\\
        v^{\mu}\equiv&{\cal D}\nabla_{\mu}\zeta \label{eqrelhydro15d}
    \end{align}
\end{subequations}
where $\sigma^{\mu \nu}$ is the \textit{shear tensor} 
\begin{equation}\label{eqrelhydro21}
    \begin{split}
        \sigma_{\mu \nu}\equiv &
        \frac{1}{2}\left({\Delta^{\alpha}}_{\mu}{\Delta^{\beta}} _{\nu}+{\Delta^{\alpha}}_{\nu}{\Delta^{\beta}}_{\mu}-\frac{2}{3}\Delta_{\mu \nu}\Delta^{\alpha \beta}\right)\nabla_{\alpha}u_{\beta}=\\
        =&\nabla_{(\mu}u_{\nu)}-A_{(\mu}u_{\nu)}-\frac{1}{3}\Delta_{\mu \nu}\nabla_{\lambda}u^{\lambda},
    \end{split}
\end{equation}
in fact, the entropy production rate
\begin{equation}
    \nabla_{\mu}s^{\mu}=-\frac{q^{\mu}q_{\mu}}{\kappa T^2}+\frac{\Pi^2}{\xi T}+\frac{\Pi^{\mu \nu}\Pi_{\mu \nu}}{2\eta T}+\frac{v^{\mu}v_{\mu}}{\cal D}
\end{equation}
is non-negative provided that
\begin{equation}
    \kappa \ge 0,\qquad \xi \ge 0,\qquad \eta \ge 0,\qquad {\cal D}\ge 0.
\end{equation}
The parameters $\kappa$, $\xi$, $\eta$ and ${\cal D}$ are called the \textit{transport coefficients}; in particular, $\kappa$ is the \textit{thermal conductivity}, $\xi$ the \textit{bulk viscosity}\footnote{The bulk viscosity is usually indicated as $\zeta$, but throughout this work and papers relevant for it, $\zeta$ always stands for the ratio between chemical potential and temperature $\zeta=\mu/T$. We choose to ``sacrify'' the traditional nomenclature because we will often deal with $\mu/T$, but never again with the bulk viscosity.}, $\eta$ the \textit{shear viscosity} and ${\cal D}$ the \textit{diffusion coefficient}.
Equations \eqref{eqrelhydro15a}--\eqref{eqrelhydro15d} expressing the transport coefficients in terms of the derivatives of the thermodynamic parameters as a consequence of the second law of thermodynamics are called the \textit{constitutive equations}.
Together with the equations of motion \eqref{eqrelhydro02} and \eqref{eqrelhydro03}, they close the set of dissipative hydrodynamics equations.
We will not go into the calculations for they are quite long, but it can be shown that this set of equations reduces to the well-known Navier-Stokes ones in the non-relativistic limit.

The constitutive equations can be used to further characterize the properties of the thermodynamic equilibrium state.
At thermodynamic equilibrium the dissipations must vanish for any value of the transport coefficients, so equations \eqref{eqrelhydro15a}--\eqref{eqrelhydro15d} imply
\begin{subequations}
    \begin{align}
        \nabla_{\mu}T-TA_{\mu}=&0\label{eqrelhydro20a}\\
        \sigma^{\mu \nu}=&0\label{eqrelhydro20b}\\
        \nabla_{\mu}u^{\mu}=&0\label{eqrelhydro20c}\\
        \nabla_{\mu}\zeta=&0\label{eqrelhydro20d}.
    \end{align}
\end{subequations}
By plugging the definition \eqref{eqrelhydro21} of the shear tensor into the condition \eqref{eqrelhydro20b}, combining it with \eqref{eqrelhydro20c} and dividing everything by the temperature, we obtain
\begin{equation}
    \begin{split}
        0=&
        \frac{\nabla_{\mu}u_{\nu}}{T}+\frac{\nabla_{\nu}u_{\mu}}{T}-\frac{A_{\mu}u_{\nu}}{T}-\frac{A_{\nu}u_{\mu}}{T}=\\
        =&\nabla_{\mu}\frac{u_{\nu}}{T}-\frac{u_{\nu}}{T^2}\nabla_{\mu}T+\nabla_{\nu}\frac{u_{\mu}}{T}-\frac{u_{\mu}}{T^2}\nabla_{\nu}T-\frac{A_{\mu}u_{\nu}}{T}-\frac{A_{\nu}u_{\mu}}{T}.
    \end{split}
\end{equation}
Finally, by using \eqref{eqrelhydro20a} we are simply left with
\begin{equation}\label{eqrelhydro22}
    \nabla_{\mu}\frac{u_{\nu}}{T}+\nabla_{\nu}\frac{u_{\mu}}{T}=0.
\end{equation}
Equations \eqref{eqrelhydro22} and \eqref{eqrelhydro20d} represent the conditions that the thermodynamic fields must fulfill in order to have thermodynamic equilibrium, namely $u^{\mu}/T$ must be a Killing vector field and $\zeta=\mu/T$ must be constant.
This statement is of fundamental importance, and will be discussed more thoroughly in Subsection \ref{sec:LE_to_GE_and_NE} in the next Chapter.
As it pertains the thermodynamic equilibrium state, it will hold true even in the framework of extended irreversible thermodynamics that we are soon going to introduce.
\newline

Although classical irreversible thermodynamics includes the dissipations in such a way to ensure the second law of thermodynamics and it reproduces the known equations in the non-relativistic limit, it also has some undesirable features.
In particular, it is unstable under perturbations and it allows for the propagation of information at infinite speed, thus breaking causality \cite{Hiscock:1983zz, Hiscock:1985zz, Hiscock:1987zz}.
Non-causality, in particular, is due to the algebraic nature of the constitutive equations \eqref{eqrelhydro15a}--\eqref{eqrelhydro15d}, which makes \textit{parabolic} equations emerge and, as a consequence, the thermodynamic fluxes react instantaneously to the gradients of the thermodynamic fields.
Actually, this feature is present in the non-relativistic Navier-Stokes theory as well, where the Fourier law for the heat flux leads to a parabolic diffusion equation for the temperature.
This is not a conceptual problem \textit{per se} in a non-relativistic theory, though a big one in a relativistic context.
A recent discussion on this subject can be found in \cite{Hoult:2020eho}, where an interpretation and a solution different from the traditional one is presented.

In order to convince ourselves of the causality breaking, we can study sound waves propagation.
For the sake of simplicity, let us consider a fluid initially at thermodynamic equilibrium and at rest in a flat spacetime, and suppose that shear viscosity is the only dissipation present \cite{Romatschke:2009im}.
To make this even simpler, we take a perturbation of the velocity field along the $x$ direction, independent of $y$ and $z$
\begin{equation}
    u^{\mu}={u_0}^{\mu}+\delta u^{\mu}(t,x),
\end{equation}
where the initial velocity field is ${u_0}^{\mu}=(1,{\bf 0})$, being the system initially at rest.
The density and pressure will change in response to this perturbation
\begin{equation}
    \rho=\rho_0+\delta \rho(t,x),\qquad
    p=p_0+\delta p(t,x),
\end{equation}
where $\rho_0$ and $p_0$ are their initial values.
In this setup we have the hydrodynamic equation
\begin{equation}
    {\Delta^{\alpha}}_{\nu}\nabla_{\mu}T^{\mu \nu}=0
    \qquad \longrightarrow \qquad
    (\rho+p)A^{\alpha}-{\Delta^{\alpha}}_{\nu}\de^{\nu}p+{\Delta^{\alpha}}_{\nu}\de_{\mu}\Pi^{\mu \nu}=0
\end{equation}
and after some algebra we obtain the following diffusive equation at first order in the perturbations for the $y$ component $\delta u^y$
\begin{equation}
    (\rho_0+p_0)\frac{\de \delta u^y(t,x)}{\de t}-\eta \frac{\de^2 \delta u^y(t,x)}{\de x^2}=0.
\end{equation}
This is in fact a parabolic equation.
As is customary in perturbative analysis, we assume that the perturbation has a simple harmonic behaviour of the type
\begin{equation}
    \delta u^y(t,x)\propto {\rm e}^{-\omega t+ikx},
\end{equation}
which leads to the quadratic dispersion relation
\begin{equation}
    \omega=\frac{\eta}{\rho_0+p_0}k^2.
\end{equation}
We can now estimate the velocity of propagation of the mode with wavenumber $k$
\begin{equation}
    \frac{\di \omega}{\di k}=\frac{2\eta}{\rho_0+p_0}k.
\end{equation}
Clearly, there exists a critical value $k_{\rm crit}$ for the wavenumber such that this velocity equals the speed of light and exceeds it for greater values, thus breaking causality.
While this can be acceptable within a non-relativistic description, it is clearly unsatisfactory within a relativistic theory in which all signals must be contained in the light cone.

In summary: the ultimate reason for the non-causality of the Navier-Stokes theory is the parabolicity of the equations, which, in fact, turns out to be a general feature of first-order theories.
In order to avoid it and restore causality, we need \textit{hyperbolic} instead of parabolic equations.
The traditional solution to this problem is the replacement of a first-order theory with a second-order one.


\subsection{Israel-Stewart theory}

Several approaches have been developed to overcome the causality breaking and instability of classical irreversible thermodynamics.
The basic idea is to extend the space of variables by considering the dissipations as conserved variables of an ideal fluid, which allows to restore causality under some conditions.
The resulting conservation equations describe the evolution of these new ``extended dissipations'', and turn out to be hyperbolic.
The price to pay is that the theory now mathematically more complicated, as it involves a greater number of variables and parameters, and the physical meaning is not always crystal clear.
This class of theories generally goes under the name of \textit{extended irreversible thermodynamics} because of this extension of the space of variables.
A non-relativistic extended theory was first proposed by M\"uller \cite{Muller:1967zza} and later generalised to the relativistic framework by Israel \cite{Israel:1976tn} and Stewart \cite{Israel:1979wp}.
Here we just touch upon the main qualitative features without going into technical calculations.

Supported by evidence from relativistic kinetic theory, Israel and Stewart argued that the entropy current in equation \eqref{eqrelhydro16} is not the most general one, but is in fact the first-order truncation of a more elaborate expression that also contains terms quadratic in the dissipation, which is therefore of second-order in the derivatives of the thermodynamic fields.
In the Eckart frame, for simplicity, it reads
\begin{equation}
    s^{\mu}=su^{\mu}+\frac{q^{\mu}}{T}-\left(\beta_0\Pi^2-\beta_1q^{\nu}q_{\nu}+\beta_2\Pi^{\alpha \beta}\Pi_{\alpha \beta}\right)\frac{u^{\mu}}{2T}-\alpha_0\frac{\Pi q^{\mu}}{T}+\alpha_1\frac{\Pi^{\mu \nu}q_{\nu}}{T}.
\end{equation}
Here, the thermodynamic coefficients $\alpha_0$ and $\alpha_1$ control the couplings between viscosity and heat fluxes, while $\beta_0$, $\beta_1$ and $\beta_2$ govern the scalar, vector and tensor contributions to the entropy density respectively.
As we can see, both the component of the entropy current along the velocity field and orthogonal to it are modified with respect to their counterparts in the Navier-Stokes theory.

As in the Navier-Stokes theory, the simplest way to enforce the second law of thermodynamics is to impose linear relations between the dissipative fluxes and the corresponding ``extended thermodynamic forces''.
In this case, however, being the dissipations regarded as dynamic variables, the extended forces contain derivatives of their respective fluxes along the velocity field.
The resulting constitutive equations replacing \eqref{eqrelhydro15a}--\eqref{eqrelhydro15d}, which are in fact the equations of motion of the dissipations, read \cite{Maartens:1996vi}
\begin{subequations}
    \begin{align}
        \tau_{\Pi}D\Pi+\Pi=&-\xi \nabla_{\mu}u^{\mu}-\frac{1}{2}T\xi \nabla_{\mu}\left(\frac{\tau_{\Pi}}{\xi T}u^{\mu}\right)\Pi+\tau_0{\Delta^{\nu}}_{\mu}\nabla_{\nu}q^{\mu}\label{eqrelhydro17a}\\
        \tau_qDq_{\mu}+q_{\mu}=&\kappa \left({\Delta^{\nu}}_{\mu}\nabla_{\nu}T-TA_{\mu}\right)+\frac{1}{2}T^2\kappa \nabla_{\nu}\left(\frac{\tau_q}{\kappa T^2}u^{\nu}\right)q_{\mu}\nonumber \\
        &-\tau_0{\Delta^{\lambda}}_{\nu}{\Pi^{\nu}}_{\mu}-\tau_2{\Delta^{\nu}}_{\mu}\nabla_{\nu}\Pi \label{eqrelhydro17b} \\
        \tau_{\pi}D\Pi_{\mu \nu}+\Pi_{\mu \nu}=&\eta \sigma_{\mu \nu}-\frac{1}{2}T\eta \nabla_{\lambda}\left(\frac{\tau_{\pi}}{\eta T}u^{\lambda}\right)\Pi_{\mu \nu}\nonumber \\
        &+\frac{1}{2}\tau_1\left(\Delta_{\mu \alpha}\Delta_{\nu \beta}+\Delta_{\nu \alpha}\Delta_{\mu \beta}-\frac{2}{3}\Delta_{\mu \nu}\Delta_{\alpha \beta}\right)\nabla^{\alpha}q^{\beta}.\label{eqrelhydro17c}
    \end{align}
\end{subequations}
Here, $\tau_{\Pi}$, $\tau_q$, $\tau_{\pi}$, $\tau_0$, $\tau_1$ and $\tau_2$ are new transport coefficients related to $\xi$, $q^{\mu}$ and $\Pi^{\mu \nu}$ by
\begin{equation}
    \begin{split}
        \tau_{\Pi}\equiv &\xi \beta_0,\qquad
        \tau_q\equiv \kappa T^2\beta_1,\qquad
        \tau_{\pi}\equiv \eta \beta_2,\\
        \tau_0\equiv &\xi \alpha_0,\qquad
        \tau_1\equiv \kappa T^2\alpha_1,\qquad
        \tau_2\equiv 2\eta \alpha_1.
    \end{split}
\end{equation}
It can be shown that they measure the time scales over which the system evolves to a new equilibrium, so they can be regarded as relaxation times.
They are not known \textit{a priori}, and must be worked out from the underlying microscopic theory.
Finally, note that the Navier-Stokes theory is simply recovered in the limit where the new terms are negligible.

The constitutive equations \eqref{eqrelhydro17a}--\eqref{eqrelhydro17c} are sometimes referred to as the \textit{Israel-Stewart equations} of extended irreversible thermodynamics.
They are 9 first-order-in-time partial differential equations which, together with the 5 energy-momentum and charge conservation equations \eqref{eqrelhydro02} and \eqref{eqrelhydro03}, form a set of 14 equations in the 14 unknowns represented by the independent components of the variables $\{ \rho,n,u^{\mu},\Pi,q^{\mu},\Pi^{\mu \nu}\}$.

Clearly, this theory is mathematically more involved than ideal hydrodynamics or the Navier-Stokes theory: more equations in more variables come into play, more input from the underlying microscopic theory is needed and sometimes the physical meaning can be clouded by technicalities.
However, the feature that the dissipations relax with specific time scales instead of instantaneously reach their asymptotic values is the key to obtain a hyperbolic set of equations, thus a causal and stable theory of relativistic hydrodynamics \cite{Israel:1979wp}.
In fact, by repeating a perturbative analysis similar to the one previously discussed, one can indeed show that the resulting equations are hyperbolic in nature. 
It is not straightforward to see whether hyperbolicity holds in any condition or not, but we are not going into that.

Over the years, a great variety of predictions of the Israel-Stewart theory have successfully matched with experiments, both in the relativistic and non-relativistic regimes.
Most of the recent interest came in fact from heavy-ion collisions, where the need of a relativistic theory of non-ideal fluids has become crucial to explain the spectra of particles produced in ultrarelativistic nucleus-nucleus collisions.


\section{Summary and outlook}

Before moving to the next Chapter, we briefly wrap up the content of the present one.
Hydrodynamics is an effective classical theory describing the evolution of fluids, systems whose microscopic and macroscopic characteristics are significantly separated.
Although their fundamental nature can be quantum mechanical and possibly complicated, whenever that scale separation occurs their evolution can be effectively described in terms of few classical degrees of freedom.
The simplest instance is that of a perfect or ideal fluid, namely ideal hydrodynamics.
This relies on the assumption of local thermodynamic equilibrium, a configuration wherein the energy-momentum tensor and possible charged currents, which are the fundamental ingredients upon which hydrodynamics is built, although out of thermodynamic equilibrium preserve the same structure they have at thermodynamic equilibrium, promoting the energy density, pressure and velocity field to slowly varying functions of space and time.
Thus, it is possible to define the thermodynamic fields for each fluid element and do thermodynamics.
The key quantity allowing the enforcement of the second law of thermodynamics is the entropy current, a vector field that integrated on some three-dimensional spacelike hypersurface gives the entropy of the system.
It is shown that ideal fluids do not produce any entropy.
On the contrary, entropy is produced by non-perfect or non-ideal fluids, systems where dissipations are present, giving rise to irreversible transport phenomena, whose track is kept in the entropy current, that in turn make the entropy increase.
The most straightforward theory of dissipative hydrodynamics is the Navier-Stokes one: although reproducing the correct non-relativistic limit and fulfilling the laws of thermodynamics, it turns out to be unstable and to break causality.
The ultimate cause for that is the parabolic nature of the equations.
The solution to this problem proposed by Israel and Stewart is the extension of the dissipations from fixed parameters to dynamical variables.
The resulting entropy current captures more terms, which are of second order in the derivatives of the thermodynamic fields and which in turn give rise to hyperbolic equations that restore stability and causality.
The price to pay is that the theory is mathematically more involved and the physical interpretation can be clouded by technicalities, but the predictions of this theory has proven to match many experimental results so far.

As mentioned, the fundamental ingredients of relativistic hydrodynamics are the conserved currents, namely the energy-momentum tensor and some possible charged currents.
In order to enter the hydrodynamic equations, they must be worked out as averages in the underlying microscopic theory.
This could be, for instance, a relativistic kinetic theory, that is a model describing the non-equilibrium dynamics of a fluid whose microscopic degrees of freedom can be regarded as a diluted system of weakly interacting particles.
The relativistic kinetic approach works well when the mean free path of the particles is much larger than their thermal wavelength, but it breaks down for strongly interacting fluids requiring a full quantum field theoretical description in general.
Thus, when the underlying microscopic theory is a Quantum Field Theory, those averages are in fact thermal expectation values calculated either with a generating functional or a density operator depending on the state of the system.
Throughout the rest of this work, we will use a method to calculate thermal expectation values based on a density operator put forward by Zubarev in the late 70's, which will be the subject of the next Chapter.






\chapter{Relativistic Quantum Statistical Mechanics}
\label{chapter:zubarev}
The fundamental quantities entering the hydrodynamic equations are the conserved currents, namely the energy-momentum tensor $T^{\mu \nu}$ and some possible charged current $j^{\mu}$.
These are regarded as averages calculated from the underlying microscopic theory, which in this work is taken to be a Quantum Field Theory since not all systems can be described by means of a relativistic kinetic theory, for instance strongly interacting ones such as the quark-gluon plasma formed in heavy-ion collisions \cite{Adams:2012th, CasalderreySolana:2011us}.
From a quantum statistical perspective, these averages are renormalized thermal expectation values of quantum operators calculated with some density operator depending on the state of the system
\begin{equation}\label{zubeq29}
    T^{\mu \nu}=\tr(\wh{\rho}\wh{T}^{\mu \nu})_{\rm ren},\qquad
    j^{\mu}=\tr(\wh{\rho}\wh{j}^{\mu})_{\rm ren},
\end{equation}
where $\wh{\rho}$ is the density operator, $\wh{T}^{\mu \nu}$ and $\wh{j}^{\mu}$ are quantum operators and ``ren'' means that the expectation value is somehow renormalized.
Thus, the hydrodynamic equations read
\begin{equation}
    \nabla_{\mu}T^{\mu \nu}=\nabla_{\mu}\tr(\wh{\rho}\wh{T}^{\mu \nu})_{\rm ren}=\tr(\wh{\rho}\nabla_{\mu}\wh{T}^{\mu \nu})_{\rm ren}=0
\end{equation}
\begin{equation}
    \nabla_{\mu}j^{\mu}=\nabla_{\mu}\tr(\wh{\rho}\wh{j}^{\mu})_{\rm ren}=\tr(\wh{\rho}\nabla_{\mu}\wh{j}^{\mu})_{\rm ren}=0,
\end{equation}
making explicit, in this form, that the conservation laws of the thermal expectation values descend from the more fundamental conservation laws of the corresponding quantum operators.
Here $\nabla_{\mu}\wh{\rho}=0$ was exploited, $\wh{\rho}$ being a stationary state in the Heisenberg picture.
The Heisenberg picture is usually preferred in a relativistic theory for it is manifestly covariant, so it will be adopted in this work too.
Equations \eqref{zubeq29} make it clear how quantum effects can arise in relativistic hydrodynamics.
There, the quantum operators and the renormalization procedure depend on the details of the underlying Quantum Field Theory, examples will be shown in Chapters \ref{chapter:gteacceleration} and \ref{chapter:boost}.
Here, we will address instead the issue of seeking the appropriate density operator and, directly from its expression, derive a general method for the calculation of the entropy current at local thermodynamic equilibrium.

We will start with a quick review on quantum statistical mechanics by recalling what a density operator is and how it is determined according to the maximum entropy principle.
Moving towards a covariant formulation, the core of this Chapter will be the Zubarev approach to relativistic quantum statistical mechanics.
Since a comprehensive and pedagogical review on this powerful formalism seems not to exist in literature, which is part of the reason why it has long gone unnoticed unfairly in our opinion, we would like to seize the opportunity to put forward one ourselves to the best of our ability.
A special note will concern the derivation of a general formula for the entropy current at local thermodynamic equilibrium: this will be the fundamental result of this work, and will be applied in later Chapters to different kinds of systems.
Finally, we will conclude with a short summary.

The review on quantum statistical mechanics will be loosely based on the textbook by R.\ Balian \cite{balian2006microphysics}.
As for the Zubarev approach, we sure have to mention the original work by D.\ N.\ Zubarev \textit{et al.} \cite{Zubarev:1979} followed closely by that of Ch.\ G.\ van Weert \cite{Vanweert1982133}, but also the more recent works by F.\ Becattini \textit{et al.} \cite{Becattini:2014yxa} and T.\ Hayata \textit{et al.} \cite{Hayata:2015lga}.
For a reworking of the formalism in a more modern language, on which part of this Chapter is based, we suggest \cite{Becattini:2019dxo} by F.\ Becattini \textit{et al.}.
Finally, the method for the entropy current at local thermodynamic equilibrium was put forward in \cite{Becattini:2019poj} by F. Becattini and D.\ R..


\section{Quantum statistical mechanics}
\label{sec:zubarev:quant_stat_mech}

The aim of statistical mechanics is the description of systems which are known solely by the specification of some global variables.
Typically these systems are macroscopic, in the sense that they involve a great number of degrees of freedom, and in quantum mechanical language, the given macroscopic data are expectation values of operators.
At the microscopic level, however, there are in general several possible states producing the same values for the macroscopic variables, therefore the 
state is not completely known: all we can give is the probability that the system is described by one state or another.
Therefore, such a system is represented by the set $\{|\psi_i\rangle,p_i\}$ of the possible states $|\psi_i\rangle$ compatible with the given macroscopic data each with its own probability of occurrence $p_i\in [0,1]$ such that $\sum_ip_i=1$.
The set $\{|\psi_i\rangle,p_i\}$ is called a \textit{statistical ensemble}, and each state $|\psi_i\rangle$ is an element of that ensemble.
Thus, such a system embodies two probabilities of different origin: the one is quantum mechanical and due to the $|\psi_i\rangle$, the other is classical and ascribed to our incomplete knowledge of the state represented by the $p_i$.
If the system is exactly in one of the $|\psi_i\rangle$ it is said to be in a \textit{pure state}, otherwise it is in a \textit{mixed state}.
For a pure state, all the $p_i$ vanish except one, which is equal to 1.
On the contrary, for a maximally mixed state, all the $|\psi_i\rangle$ have the same probability, which is $p_i=1/N$ with $N$ the dimension of the Hilbert space the $|\psi_i\rangle$ belong to.
For a system described by the statistical ensamble $\{|\psi_i\rangle,p_i\}$, the expectation value of an operator $\wh{O}$ is
\begin{equation}\label{zubeq19}
    \langle \wh{O}\rangle=\sum_ip_i\langle \psi_i|\wh{O}|\psi_i\rangle=\tr(\wh{\rho}\wh{O}),
\end{equation}
where $\wh{\rho}$ such that $\tr(\wh{\rho})=1$ is called the \textit{density operator} and is defined as
\begin{equation}\label{zubeq08}
    \wh{\rho}\equiv \sum_ip_i|\psi_i\rangle \langle \psi_i|.
\end{equation}
The density operator embodies in a compact form the statistical nature of the macroscopic system and allows us to calculate expectation values of operators, therefore it is the natural object used in quantum statistical mechanics to describe a statistical ensemble.

Our knowledge of the system is more or less complete: clearly our information is maximum when we can make predictions with full certainty, and it is larger when the system is in a pure state than in a mixed one.
Moreover, the system is better known when the number of possible 
states is small or when the probability for one of them is close to unity than when there is a large number of possible 
states with all approximately the same probability.
The idea is then to identify the missing information with a quantitative measure for the degree of disorder existing in a system whose preparation has a random nature.
This quantity is the \textit{von Neumann entropy}
\begin{equation}\label{zubeq01}
    S\equiv -\tr(\wh{\rho}\log \wh{\rho}),
\end{equation}
which is, in fact, maximum for a maximally mixed state and minimum and equal to zero for a pure state, as desired.
The von Neumann entropy satisfies a number of other properties, but to our intents and purposes it suffices to say that it is a good measure of our missing information on a system.
Recall that the thermodynamic entropy is the von Neumann entropy multiplied by the Boltzmann constant, which in this work is set to $k_{\rm B}=1$, so we regard them as the same quantity which we indicate as $S$ and simply refer to as ``entropy''.

In order to actually be able to make predictions on a system prepared in some given way, we must know how to assign it a density operator representing the physical configuration that we want to describe.
If we know nothing about the system, the answer is simple: we must assume that all the possible 
states are equally probable, any other choice would arbitrarily introduce an order for which there are no reasons of believing that it exists.
This is but the maximally mixed state, namely the one that maximizes the 
entropy.
If the system is partially known, meaning that we are given some data, the same kind of philosophy applies: among all the possible 
states, we must select those compatible with the given data and assign them equal probabilities.
In other words, we must look for the maximum of the 
entropy while reproducing the known data.
We assume therefore that the best guess for $\wh{\rho}$ is provided by the following prescription: amongst all the density operators compatible with the available data, we must represent the system by the one which has the largest value of the 
entropy.
This is called the \textit{maximum entropy principle}.
The states determined accordingly are said to be at \textit{thermodynamic equilibrium}, all other states are called of \textit{non-equilibrium}.
Mathematically, the available information is represented by a set of expectation values $\{\langle \wh{O}_i\rangle\}$ of a set of observables $\{\wh{O}_i\}$.
The density operator must reproduce the data, therefore it must fulfill that constraints
\begin{equation}
    \tr(\wh{\rho}\wh{O}_i)=\langle \wh{O}_i\rangle,\qquad \forall i.
\end{equation}
For each constraint, we introduce a Lagrange multiplier $\lambda_i$ and maximize the functional expression
\begin{equation}\label{zubeq02}
    -\tr(\wh{\rho}\log \wh{\rho})-\sum_i\lambda_i\left(\tr(\wh{\rho}\wh{O}_i)-\langle \wh{O}_i\rangle \right)
\end{equation}
with respect to $\wh{\rho}$.
The solution has the well-known form
\begin{equation}\label{zubeq03}
    \wh{\rho}=\frac{1}{Z}\exp \left[-\sum_i\lambda_i\wh{O}_i\right],\qquad
    Z=\tr \left(\exp \left[-\sum_i\lambda_i\wh{O}_i\right]\right),
\end{equation}
where $Z$ is the \textit{partition function} enforcing the normalization $\tr(\wh{\rho})=1$.
Depending on which set of observables is chosen, a different statistical ensamble is obtained.

We conclude this review with a remark on the time evolution.
In the Schr\"odinger picture, the evolution of the 
states $|\psi_i\rangle$ is governed by the Schr\"odinger equation, whose solution is a unitary transformation of $|\psi_i\rangle$
\begin{equation}
    i\frac{\di}{\di t}|\psi_i\rangle=\wh{H}|\psi_i\rangle
    \qquad \Rightarrow \qquad
    |\psi_i(t)\rangle=\wh{U}(t)|\psi_i(0)\rangle,\qquad
    \wh{U}(t)={\rm e}^{-i\wh{H}t}
\end{equation}
with $\wh{H}$ the Hamiltonian.
As a consequence, the evolution of the density operator is governed by the von Neumann equation, which is similar to the Heisenberg equation but with opposite sign.
Its solution is again a unitary transformation of $\wh{\rho}$
\begin{equation}\label{zubeq14}
    \frac{\di \wh{\rho}}{\di t}=-i[\wh{H},\wh{\rho}]
    \qquad \Rightarrow \qquad
    \wh{\rho}(t)=\wh{U}(t)\wh{\rho}(0)\wh{U}^{\dagger}(t).
\end{equation}
A known property of the 
entropy, which can be checked by using the definition \eqref{zubeq01}, is the invariance under unitary transformations of the density operator, therefore it is in fact stationary in the Schr\"odinger picture.
This is more obvious in the Heisenberg picture, where the 
states $|\psi_i\rangle$ are constant in time, and so is the density operator.
We must conclude that, regardless of the picture, the 
entropy is always stationary in quantum statistical mechanics.
However, it is well-known that, in classical physics, irreversible processes that produce entropy can definitely occur, as we discussed also in Chapter \ref{chapter:relhydro}.
From a kinetic theory perspective, for instance, if a system is initially out of thermodynamic equilibrium, the collisions between particles make the entropy increase until it reaches a maximum corresponding to a state of thermodynamic equilibrium.
Thus, in analogy with the classical theory, is there a way to describe irreversible processes where the entropy can actually increase in the quantum theory?
This question leads us to the next Section.


\section{The Zubarev approach}
\label{sec:zubarev_zub_approach}

The aim of D.\ N.\ Zubarev was actually way broader than just the description of quantum processes with entropy production.
With the development of relativistic astrophysics, cosmology, and the hydrodynamic theory of multiparticle production, in the late 70's it became necessary to go beyond the framework of the phenomenological linear relativistic hydrodynamics.
Thus, his purpose back then was rather to derive systematically the equations of non-linear quantum relativistic hydrodynamics, the transport relations and the transport coefficients without relying in the general case on the assumption that the state of the system is close to thermodynamic equilibrium.
Dealing with full non-equilibrium and including quantum effects, his formulation appears to be 
completely general
and also well suited to describe, among other things, the entropy production of quantum processes.

\subsection{Local thermodynamic equilibrium density operator}

The core of the Zubarev formulation of the quantum statistical foundations of relativistic hydrodynamics is the so-called \textit{local thermodynamic equilibrium density operator}.
This is defined, according to the maximum entropy principle, as the operator $\wh{\rho}_{\rm LE}$ that maximizes the functional expression $-\tr(\wh{\rho}_{\rm LE}\log \wh{\rho}_{\rm LE})$ with constraints of given energy, momentum and possible charge densities.
This definition does not assume any underlying kinetic theory, and it also is completely unambiguous in the non-relativistic theory as it only demands for the densities to vary significantly over distances much larger than the characteristic microscopic scale \cite{balian2006microphysics, Becattini:2014yxa}.
Its relativistic extension, however, requires a little more effort due to the stronger frame-dependence of the densities.

The first step towards a covariant expression of $\wh{\rho}_{\rm LE}$ is the foliation of the spacetime with a 1-parameter family of 3-dimensional spacelike hypersurfaces $\{ \Sigma(\tau)\}$.
Indeed the timelike unit vector field $n^{\mu}$ normal to the hypersurfaces defines worldlines of observers, but $\tau$ does not in general coincide with the proper time of comoving observers.
By the Frobenius theorem, in order for the foliation to be well-defined, the normal unit vector must fulfill the so-called \textit{vorticity-free condition}
\begin{equation}\label{zubeq07}
    \epsilon_{\mu \nu \rho \sigma}n^{\mu}(\partial^{\nu}n^{\rho}-\partial^{\rho}n^{\nu})=0,
\end{equation}
with $\epsilon_{\mu \nu \rho \sigma}$ the Levi-Civita symbol.
Such a foliation, also known as the Arnowitt-Deser-Misner (ADM) foliation, will hereafter be assumed to exist for the geometry at hand.
Whatever the expression of $\wh{\rho}_{\rm LE}$ is, the renormalized energy-momentum and charge densities on a hypersurface $\Sigma(\tau)$ of the foliation calculated with $\wh{\rho}_{\rm LE}$ are respectively
$n_{\mu}\tr(\wh{\rho}_{\rm LE}\wh{T}^{\mu \nu})_{\rm ren}$ and $n_{\mu}\tr(\wh{\rho}_{\rm LE}\wh{j}^{\mu})_{\rm ren}$.
These are constrained to equal the actual values, that is
\begin{equation}\label{zubeq04}
    n_{\mu}\tr(\wh{\rho}_{\rm LE}\wh{T}^{\mu \nu})_{\rm ren}=n_{\mu}T^{\mu \nu},\qquad
    n_{\mu}\tr(\wh{\rho}_{\rm LE}\wh{j}^{\mu})_{\rm ren}=n_{\mu}j^{\mu}.
\end{equation}
In the following, these will be referred to as the \textit{constraint equations}.
The quantities at right-hand side are the actual ones, finite, however they are known or defined.
In other words, they are what we called the given data in the previous Section.
The plain expectation values of $\wh{T}^{\mu \nu}$ and $\wh{j}^{\mu}$ are divergent in general, therefore it is important to emphasize that the quantities at left-hand side must be suitably renormalized in order to match the actual ones.
The renormalization procedure depends on the details of the underlying Quantum Field Theory.
Throughout the rest of this work, we will be concerned with the simplest case of free field theories, therefore renormalization is most readily established by subtraction of the vacuum expectation value
\begin{subequations}
    \begin{align}
        \tr(\wh{\rho}_{\rm LE}\wh{T}^{\mu \nu})_{\rm ren}=&
        \tr(\wh{\rho}_{\rm LE}\wh{T}^{\mu \nu})-\langle 0|\wh{T}^{\mu \nu}|0\rangle=
        \tr(\wh{\rho}_{\rm LE}:\wh{T}^{\mu \nu}:)=\langle :\wh{T}^{\mu \nu}:\rangle_{\rm LE}\\
        \tr(\wh{\rho}_{\rm LE}\wh{j}^{\mu})_{\rm ren}=&
        \tr(\wh{\rho}_{\rm LE}\wh{j}^{\mu})-\langle 0|\wh{j}^{\mu}|0\rangle=
        \tr(\wh{\rho}_{\rm LE}:\wh{j}^{\mu}:)=\langle :\wh{j}^{\mu}:\rangle_{\rm LE},
    \end{align}
\end{subequations}
which is tantamount to normally ordering the creation and annihilation operators because the conserved currents are quadratic in the fields.
In the last step, we have used the notation $\langle \wh{O}\rangle_{\rm LE}\equiv \tr(\wh{\rho}_{\rm LE}\wh{O})$ for the expectation value of an operator $\wh{O}$ calculated with $\wh{\rho}_{\rm LE}$, much in the same was as \eqref{zubeq19}.
Thus, in analogy with the non-relativistic case \eqref{zubeq02}, we introduce a set of Lagrange multipliers and maximize following functional expression with respect to $\wh{\rho}_{\rm LE}$
\begin{equation}
    -\tr(\wh{\rho}_{\rm LE}\log \wh{\rho}_{\rm LE})+\int_{\Sigma(\tau)}\di \Sigma_{\mu}\left[
    \left(\tr(\wh{\rho}_{\rm LE}\wh{T}^{\mu \nu})_{\rm ren}-T^{\mu \nu}\right)\beta_{\nu}-\zeta \left(\tr(\wh{\rho}_{\rm LE}\wh{j}^{\mu})_{\rm ren}-j^{\mu}\right)\right],
\end{equation}
where $\di \Sigma_{\mu}\equiv \di \Sigma \,n_{\mu}$ with $\di \Sigma$ the measure on $\Sigma(\tau)$.
Once again in analogy with \eqref{zubeq03}, the solution of the variational problem is
\begin{equation}\label{zubeq05}
    \wh{\rho}_{\rm LE}=\frac{1}{Z_{\rm LE}}\exp \left[-\int_{\Sigma(\tau)}\di \Sigma_{\mu}\left(\wh{T}^{\mu \nu}\beta_{\nu}-\zeta \wh{j}^{\mu}\right)\right]
\end{equation}
\begin{equation}\label{zubeq06}
    Z_{\rm LE}=\tr \left(\exp \left[-\int_{\Sigma(\tau)}\di \Sigma_{\mu}\left(\wh{T}^{\mu \nu}\beta_{\nu}-\zeta \wh{j}^{\mu}\right)\right]\right)
\end{equation}
where the partition function $Z_{\rm LE}$ ensures $\tr(\wh{\rho}_{\rm LE})=1$.
It is important to notice that it can be kept in this simple form, without subtraction of the vacuum expectation value, because the latter is a non-operator term which would appear also in the partition function, hence cancelling out in the ratio.
In formulae
\begin{equation}\label{zubeq24}
    \begin{split}
        &\frac{\exp \left[-\int_{\Sigma(\tau)}\di \Sigma_{\mu}\left(:\wh{T}^{\mu \nu}:\beta_{\nu}-\zeta :\wh{j}^{\mu}:\right)\right]}{\tr \left(\exp \left[-\int_{\Sigma(\tau)}\di \Sigma_{\mu}\left(:\wh{T}^{\mu \nu}:\beta_{\nu}-\zeta :\wh{j}^{\mu}:\right)\right]\right)}\\
        &=\frac{\exp \left[-\int_{\Sigma(\tau)}\di \Sigma_{\mu}\left((\wh{T}^{\mu \nu}-\langle 0|\wh{T}^{\mu \nu}|0\rangle)\beta_{\nu}-\zeta (\wh{j}^{\mu}-\langle 0|\wh{j}^{\mu}|0\rangle)\right)\right]}{\tr \left(\exp \left[-\int_{\Sigma(\tau)}\di \Sigma_{\mu}\left((\wh{T}^{\mu \nu}-\langle 0|\wh{T}^{\mu \nu}|0\rangle)\beta_{\nu}-\zeta (\wh{j}^{\mu}-\langle 0|\wh{j}^{\mu}|0\rangle)\right)\right]\right)}\\
        &=\frac{\exp \left[-\int_{\Sigma(\tau)}\di \Sigma_{\mu}\left(\wh{T}^{\mu \nu}\beta_{\nu}-\zeta \wh{j}^{\mu}\right)\right]}{\tr \left(\exp \left[-\int_{\Sigma(\tau)}\di \Sigma_{\mu}\left(\wh{T}^{\mu \nu}\beta_{\nu}-\zeta \wh{j}^{\mu}\right)\right]\right)}
        =\wh{\rho}_{\rm LE}.
    \end{split}
\end{equation}
In other words, $\wh{\rho}_{\rm LE}$ is invariant under addition or subtraction of a non-operator term such as a vacuum expectation value.
This property will come in handy later on.

Equation \eqref{zubeq05} is the fully covariant expression of the local thermodynamic equilibrium density operator, and it holds in a general curved spacetime.
It was first obtained by Zubarev \cite{Zubarev:1979} and van Weert \cite{Vanweert1982133}, and has been recently reworked \cite{Becattini:2014yxa, Hayata:2015lga}.


In order to enforce the constraint equations \eqref{zubeq04}, we introduced the Lagrange multipliers $\beta^{\mu}$ and $\zeta$: their physical meaning is that of thermodynamic fields, and they are determined as solutions of the constraint equations themselves with $\wh{\rho}_{\rm LE}$ given by \eqref{zubeq05}.
Now, it is clear that the local thermodynamic equilibrium density operator depends first on the chosen foliation and then on the particular hypersurface $\Sigma(\tau)$, whence on the vector field $n^{\mu}$.
In turn, the thermal expectation values at left-and side in the constraint equations also depend on $\Sigma(\tau)$.
Thus, explicitly, the constraint equations read
\begin{equation}
    n_{\mu}\tr(\wh{\rho}_{\rm LE}[\beta^{\mu},\zeta,n^{\mu}]\wh{T}^{\mu \nu})_{\rm ren}=n_{\mu}T^{\mu \nu},\qquad
    n_{\mu}\tr(\wh{\rho}_{\rm LE}[\beta^{\mu},\zeta,n^{\mu}]\wh{j}^{\mu})_{\rm ren}=n_{\mu}j^{\mu}
\end{equation}
where the dependence of $\wh{\rho}_{\rm LE}$ on $\beta^{\mu}$, $\zeta$ and $n^{\mu}$ can be functional.
For any given spacetime foliation, these are 5 equations in 5 unknowns, and can be solved in principle to determine the thermodynamic fields.
In the general case this is a non-trivial task, however a great simplification is achieved if $n^{\mu}$ is taken to be the direction of $\beta^{\mu}$, that is $n^{\mu}=\beta^{\mu}/\sqrt{\beta^2}$, which is possible as $\beta^{\mu}$ is known to be timelike, provided it satisfies the vorticity-free condition \eqref{zubeq07}.
Nevertheless, in \cite{Becattini:2014yxa} it is argued that a definition of $n^{\mu}$ based on $\beta^{\mu}$ can be put forward also in the vorticous case, at least in Minkowski spacetime.
With this choice, the number of independent variables is reduced, and the above equations become
\begin{equation}
    \beta_{\mu}\tr(\wh{\rho}_{\rm LE}[\beta^{\mu},\zeta]\wh{T}^{\mu \nu})_{\rm ren}=\beta_{\mu}T^{\mu \nu},\qquad
    \beta_{\mu}\tr(\wh{\rho}_{\rm LE}[\beta^{\mu},\zeta]\wh{j}^{\mu})_{\rm ren}=\beta_{\mu}j^{\mu}.
\end{equation}
Thus, although being true that there is an overall ambiguity due to the choice of the hypersurface $\Sigma(\tau)$, this can in fact be chosen consistently by using $\beta^{\mu}$ itself.
This choice is known as the $\beta$-\textit{frame}, or the \textit{thermodynamic frame}, represented by the velocity field defined as
\begin{equation}\label{zubeq11}
    u^{\mu}\equiv \frac{\beta^{\mu}}{\sqrt{\beta^2}},\qquad
    T\equiv \frac{1}{\sqrt{\beta^2}}.
\end{equation}
The Lorentz scalar $T$ defines a \textit{proper temperature}, namely the one measured by an ideal thermometer comoving with the system, and is in general different from the one marked by a fixed ideal thermometer who sees the system passing by, that is $T_{\rm fix}=1/\beta^0$.
By virtue of this interpretation, $\beta^{\mu}$ is called the \textit{four-temperature} vector field.
Finally, $\zeta$, sometimes called the \textit{fugacity}, is the ratio between the chemical potential $\mu$ associated to the charged current $\wh{j}^{\mu}$ and the proper temperature, $\zeta \equiv \mu/T$, as we saw in Chapter \ref{chapter:relhydro}.

Much more could actually be said about the $\beta$-frame and its properties, but, to all intents and purposes of this work, we would like the takeaway to be just the following.
First of all, it is true that the local thermodynamic equilibrium density operator depends on the choice of the foliation and then on the spacelike hypersurface onto which the densities are given.
Actually, this is not completely new, as in the usual thermal field theory it is just as well necessary to assign initial values on a Cauchy hypersurface.
However, the four-temperature can in fact help fixing this ambiguity.
And second, while in relativistic hydrodynamics the field $\beta^{\mu}$ is defined starting from the velocity field and the temperature, here we showed that the other way round is equally possible, namely we regarded the four-temperature as a fundamental quantity from which the velocity field and proper temperature were derived \cite{Van:2013sma}.
For a thorougher discussion of the $\beta$-frame and the reasons why it should be regarded as a privileged frame, see \cite{Becattini:2014yxa}.

\subsection{From local equilibrium to global and non-equilibrium}
\label{sec:LE_to_GE_and_NE}

The key observation we now want to point out is that the local thermodynamic equilibrium density operator \eqref{zubeq05} is not stationary in general.
The reason is that the quantum operators $\wh{T}^{\mu \nu}$ and $\wh{j}^{\mu}$ appearing in it are built from the quantum fields, which are $\tau$-dependent, hence so is $\wh{\rho}_{\rm LE}$ through the choice of the hypersurface $\Sigma(\tau)$.
This implies that $\wh{\rho}_{\rm LE}$ cannot be a state of the system, for it is not in the Heisenberg picture.
Recall that it was worked out starting from the quantum operators instead of the microscopic states, therefore it cannot be expressed in general in the form \eqref{zubeq08} as a state would.
The actual, stationary state in the Heisenberg picture $\wh{\rho}$ remains unknown, however $\wh{\rho}_{\rm LE}$ contains information about it in the thermodynamic fields $\beta^{\mu}$ and $\zeta$.
In fact, at the right-hand side of the constraint equations \eqref{zubeq04}, which define the thermodynamic fields, there are the actual densities determined as renormalized expectation values calculated with the actual state $\wh{\rho}$.
The non-stationarity of $\wh{\rho}_{\rm LE}$ is a property so important that is convenient emphasizing by making it explicit, therefore from now on we will write $\wh{\rho}_{\rm LE}(\tau)$.

If the local thermodynamic equilibrium density operator is not even a state of the system, why did we put so much effort in defining it?
Are we missing anything?
The answer, which is an amendment of Zubarev's original idea, is overly simple: if at some initial $\tau_0$ the system is known to be at local thermodynamic equilibrium, the actual state is $\wh{\rho}=\wh{\rho}_{\rm LE}(\tau_0)$, and it remains such at any $\tau$ by virtue of stationarity.
In other words, the actual state is $\wh{\rho}$ at any time, but if $\wh{\rho}_{\rm LE}(\tau)$ is designed in such a way that at some $\tau_0$ we have $\wh{\rho}_{\rm LE}(\tau_0)=\wh{\rho}$, then the system is said to be at local thermodynamic equilibrium at $\tau_0$.
That is to say that the state still depends on the choice of the foliation, but the only hypersurface it really depends on is the one where the initial data, namely the energy-momentum and charge densities, is given, that is $\Sigma(\tau_0)$.
Note that this hypothesis does not discriminate between thermodynamic equilibrium and non-equilibrium, the actual state we are referring to can be either.
Note also that, although $\wh{\rho}$ might be equal to $\wh{\rho}_{\rm LE}(\tau_0)$ at some $\tau_0$, their derivatives are different: the derivatives of $\wh{\rho}$ vanish identically because of stationarity, while those of $\wh{\rho}_{\rm LE}(\tau_0)$ do not in principle.


Then, how do we check if a system is at local thermodynamic equilibrium at a given $\tau_0$?
The local thermodynamic equilibrium density operator is determined by fixing the expectation values of energy-momentum and charge densities equal to the actual ones.
The expectation value is a first moment.
We could consider, for instance, the second moments, namely the variances, and check if they agree with the measured values within the experimental errors at $\tau_0$.
If they do, we still do not know what the actual state of system is, but it is reasonable to assume it to be $\wh{\rho}_{\rm LE}(\tau_0)$, which becomes more and more legitimate as one checks moments of higher and higher order.
This procedure can be far from easy to do in practice, but there is no conceptual fallacy in principle.

\begin{figure}
    \begin{center}
	    \includegraphics[width=10cm]{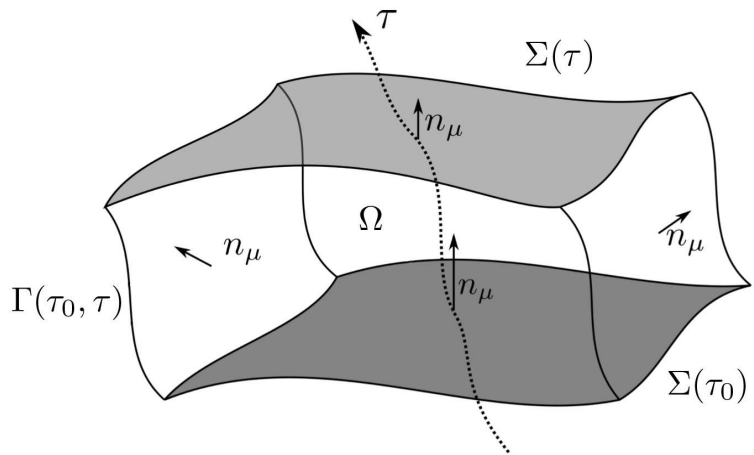}
	    \caption{Here it is shown the spacetime volume $\Omega$ enclosed between the two spacelike hypersurfaces of the foliation $\Sigma(\tau_0)$ and $\Sigma(\tau)$, where $\tau_0$ is the time where the system is at local thermodynamic equilibrium.
	    The timelike boundary at infinity joining them is indicated as $\Gamma(\tau_0,\tau)$, and $n^{\mu}$ is the timelike unit vector field orthogonal to the foliation.}
	    \label{fig01}
	\end{center}
\end{figure}

If at $\tau_0$ the system is at local thermodynamic equilibrium in the above sense, we can obtain the relation between $\wh{\rho}$ and $\wh{\rho}_{\rm LE}(\tau)$ by means of Gauss' theorem.
Let us consider a volume $\Omega$ of spacetime enclosed by two spacelike hypersurfaces of the foliation: one $\Sigma(\tau_0)$ at the initial $\tau_0$ and the other $\Sigma(\tau)$ at the ``present'' $\tau$.
Let also $\Gamma(\tau_0,\tau)$ be the timelike boundary at infinity joining $\Sigma(\tau_0)$ and $\Sigma(\tau)$.
This is configuration is shown in Figure \ref{fig01}.
Assuming that the quantum fields the conserved currents are built with satisfy boundary conditions such that the flux through $\Gamma(\tau_0,\tau)$ vanishes \cite{Becattini:2012tc}, by Gauss' theorem we have
\begin{equation}
    \int_{\Omega}{\rm d}\Omega \left(\wh{T}^{\mu \nu}\nabla_{\mu}\beta_{\nu}-\wh{j}^{\mu}\nabla_{\mu}\zeta \right)=\int_{\Sigma(\tau)}{\rm d}\Sigma_{\mu}\left(\wh{T}^{\mu \nu}\beta_{\nu}-\zeta \wh{j}^{\mu}\right)-\int_{\Sigma(\tau_0)}{\rm d}\Sigma_{\mu}\left(\wh{T}^{\mu \nu}\beta_{\nu}-\zeta \wh{j}^{\mu}\right),
\end{equation}
where the conservation of the energy-momentum tensor and charged current operators were used.
Thus, the actual state is
\begin{equation}\label{zubeq09}
    \begin{split}
        \wh{\rho}=&\wh{\rho}_{\rm LE}(\tau_0)=
        \frac{1}{Z_{\rm LE}(\tau_0)}\exp \left[-\int_{\Sigma(\tau_0)}{\rm d}\Sigma_{\mu}\left(\wh{T}^{\mu \nu}\beta_{\nu}-\zeta \wh{j}^{\mu}\right)\right]\\
        =&
        \frac{1}{Z_{\rm LE}(\tau_0)}\exp \left[-\int_{\Sigma(\tau)}{\rm d}\Sigma_{\mu}\left(\wh{T}^{\mu \nu}\beta_{\nu}-\zeta \wh{j}^{\mu}\right)+\int_{\Omega}{\rm d}\Omega \left(\wh{T}^{\mu \nu}\nabla_{\mu}\beta_{\nu}-\wh{j}^{\mu}\nabla_{\mu}\zeta \right)\right],
    \end{split}
\end{equation}
where in the first line it is understood that not only the hypersurface, but the argument of the integral too are evaluated at $\tau_0$.
Note that the first integral in the second line is the one appearing in $\wh{\rho}_{\rm LE}(\tau)$.
Although hardly useful at a practical level, this expression is exact and represents the actual, stationary state of the system, whether equilibrium or non-equilibrium.
However, at thermodynamic equilibrium we can take it a step further.

At thermodynamic equilibrium, the entropy is maximized with constrained values of energy-momentum and charge densities at any $\tau$.
Since $\wh{\rho}_{\rm LE}(\tau)$ is also obtained by maximizing the entropy, thermodynamic equilibrium is realized if $\wh{\rho}_{\rm LE}(\tau)=\wh{\rho}_{\rm LE}(\tau')$ for any $\tau$ and $\tau'$.
This implies that the volume term in \eqref{zubeq09} must vanish, which happens for any volume $\Omega$ if the argument of the integral vanishes
\begin{equation}
    \wh{T}^{\mu \nu}\nabla_{\mu}\beta_{\nu}-\wh{j}^{\mu}\nabla_{\mu}\zeta=\frac{1}{2}\wh{T}^{\mu \nu}\left(\nabla_{\mu}\beta_{\nu}+\nabla_{\nu}\beta_{\mu}\right)-\wh{j}^{\mu}\nabla_{\mu}\zeta=0,
\end{equation}
where we used the symmetry of the energy-momentum tensor operator.
Being the thermodynamic fields $\beta^{\mu}$ and $\zeta$ independent from each other, the condition for thermodynamic equilibrium is
\begin{equation}\label{zubeq10}
    \nabla_{\mu}\beta_{\nu}+\nabla_{\nu}\beta_{\mu}=0,\qquad 
    \nabla_{\mu}\zeta=0
\end{equation}
namely $\zeta$ must be constant and $\beta^{\mu}$ a Killing vector \cite{Chrobok:2006rr}.
Taking into account for the definition \eqref{zubeq11} of the four-temperature, this is exactly the same result as \eqref{eqrelhydro22}.
However, the latter was obtained in the context of the Navier-Stokes theory, which is a first-order theory, while \eqref{zubeq10} was obtained in the Zubarev theory, which is not truncated to any order in the derivatives of the thermodynamic fields.
This particular stationary configuration is called a \textit{global thermodynamic equilibrium} state
\begin{equation}\label{zubeq28}
    \wh{\rho}=\frac{1}{Z}\exp \left[-\int_{\Sigma}{\rm d}\Sigma_{\mu}\left(\wh{T}^{\mu \nu}\beta_{\nu}-\zeta \wh{j}^{\mu}\right)\right],
\end{equation}
\begin{equation}
    Z=\tr \left(\exp \left[-\int_{\Sigma}{\rm d}\Sigma_{\mu}\left(\wh{T}^{\mu \nu}\beta_{\nu}-\zeta \wh{j}^{\mu}\right)\right]\right).
\end{equation}
The $\tau$-dependence is omitted, as in fact there is none being $\wh{\rho}$ a stationary state in the Heisenberg picture.
The above expression is formally analogous to the one of the local thermodynamic equilibrium density operator \eqref{zubeq05}, with the important difference that it is an actual state and the thermodynamic fields fulfill the conditions \eqref{zubeq10}.
This means that we have much more information on a system when it is at global thermodynamic equilibrium than out of equilibrium, for in the former case the thermodynamic fields are bounded to be a Killing vector and a constant, whereas in the latter they can be whatever as long as they are solutions of the constraint equations.
This is one of the reasons that make systems at local thermodynamic equilibrium harder to study, as we will have the chance to convince ourselves in the next Chapters.
However, not all is lost if local thermodynamic equilibrium proves to be too hard to deal with.
We will not go into details here, but in \cite{Becattini:2014yxa} it is shown, at least in Minkowski spacetime, that in the $\beta$-frame $\wh{\rho}_{\rm LE}(\tau)$ can be expanded in the derivatives of the thermodynamic fields by using linear response theory around a configuration of global thermodynamic equilibrium.
The idea is that, if at local thermodynamic equilibrium the derivatives of the thermodynamic fields are small compared to their global thermodynamic equilibrium values, the so-called \textit{hydrodynamic limit}, the main contribution to thermal expectation values comes from global thermodynamic equilibrium.
This is what makes global thermodynamic equilibrium interesting even for systems at local thermodynamic equilibrium, as we shall see in the next Chapter.

We conclude this part with the following remark.
Suppose that the actual state $\wh{\rho}=\wh{\rho}_{\rm LE}(\tau_0)$ has some symmetry, namely it commutes with some unitary representation of subgroup ${\rm G}$ of transformations of the proper orthochronous Poincar\'e group ${\rm IO}(1,3)^{\uparrow}_+$.
Let it be $\wh{U}(g)$ in the Hilbert space.
Then we have
\begin{equation}
    \begin{split}
        \wh{U}(g)\wh{\rho}\wh{U}^{-1}(g)=&
        \frac{1}{Z}\exp \left[-\int_{\Sigma(\tau_0)}\di \Sigma_{\mu}(x)\left(\wh{U}(g)\wh{T}^{\mu \nu}(x)\wh{U}^{-1}(g)\beta_{\nu}(x)\right.\right.\\
        &\left.\left.-\zeta(x)\wh{U}(g)\wh{j}^{\mu}(x)\wh{U}^{-1}(g)\right)\right]\\
        =&\frac{1}{Z}\exp \left[-\int_{\Sigma(\tau_0)}\di \Sigma_{\mu}(x)\left({D(g^{-1})^{\mu}}_{\rho}{D(g^{-1})^{\nu}}_{\sigma}\wh{T}^{\rho \sigma}(g(x))\beta_{\nu}(x)\right.\right.\\
        &\left.\left.-\zeta(x){D(g^{-1})^{\mu}}_{\rho}\wh{j}^{\rho}(g(x))\right)\right].
    \end{split}
\end{equation}
By setting $y\equiv g(x)$ and using the fact that $\di \Sigma_{\mu}(x)={D(g)^{\nu}}_{\mu}\di \Sigma_{\nu}(x)$ we get
\begin{equation}
    \begin{split}
        \wh{U}(g)\wh{\rho}\wh{U}^{-1}(g)=&
        \frac{1}{Z}\exp \left[-\int_{g(\Sigma(\tau_0))}\di \Sigma_{\rho}(y)\left(\wh{T}^{\mu \nu}(y){D(g^{-1})^{\nu}}_{\sigma}\beta_{\nu}(g^{-1}(y))\right.\right.\\
        &\left.\left.-\zeta(g^{-1}(y))\wh{j}^{\rho}(y)\right)\right],
    \end{split}
\end{equation}
so, if the hypersurface is invariant under the transformation $g$ and if
\begin{equation}\label{zubeq12}
    {D(g^{-1})^{\nu}}_{\sigma}\beta_{\nu}(g^{-1}(y))=\beta_{\sigma}(y),\qquad
    \zeta(g^{-1}(y))=\zeta(y),
\end{equation}
then the density operator is invariant under the unitary transformation, $\wh{U}(g)\wh{\rho}\wh{U}^{-1}(g)=\wh{\rho}$.
Equations \eqref{zubeq12} specify the symmetry conditions on the thermodynamic fields in order for $\wh{\rho}$ to be invariant.
An invariance of $\wh{\rho}$ has consequences on the thermal expectation values of operators.
For instance, for the energy-momentum tensor
\begin{equation}\label{zubeq13}
    \begin{split}
        T^{\mu \nu}=&
        \tr(\wh{\rho}\wh{T}^{\mu \nu}(x))=
        \tr(\wh{\rho}\wh{U}^{-1}(g)\wh{T}^{\mu \nu}(x)\wh{U}(g))\\
        =&{D(g)^{\mu}}_{\rho}{D(g)^{\nu}}_{\sigma}\tr(\wh{\rho}\wh{T}^{\rho \sigma}(g^{-1}(x)))=
        {D(g)^{\mu}}_{\rho}{D(g)^{\nu}}_{\sigma}T^{\rho \sigma}(g^{-1}(x)).
    \end{split}
\end{equation}
If we consider a 1-parameter subgroup of transformations $g_{\phi}$, for instance the rotations around some axis, then, by equations \eqref{zubeq12} and \eqref{zubeq13}, the Lie derivative of the field under consideration along the vector field $X(x)=\di g_{\phi}/\di \phi$ vanishes
\begin{equation}
    {\cal L}_X(\beta)^{\mu}=0,\qquad
    {\cal L}_X(T)^{\mu \nu}=0.
\end{equation}
Now the following question arises: if ${\rm G}$ is a symmetry subgroup for the actual state, is it so also for the local thermodynamic equilibrium density operator at any time?
In formulae, does the following implication hold at any $\tau$?
\begin{equation}
    \wh{U}(g)\wh{\rho}\wh{U}^{-1}(g)=\wh{\rho}
    \qquad \Rightarrow \qquad
    \wh{U}(g)\wh{\rho}_{\rm LE}(\tau)\wh{U}^{-1}(g)=\wh{\rho}_{\rm LE}(\tau).
\end{equation}
It can be shown that if ${\rm G}$ transforms $\Sigma(\tau_0)$ into itself and the thermodynamic fields fulfill \eqref{zubeq12}, then the answer is yes.
Recall that, by definition, $\wh{\rho}_{\rm LE}(\tau)$ is determined by maximizing the entropy while satisfying the constraint equations.
Clearly, both the entropy and the constraint equations are invariant under unitary transformations of $\wh{\rho}_{\rm LE}(\tau)$, therefore, if $\wh{\rho}_{\rm LE}(\tau)$ is a solution of the variational problem, so is $\wh{U}(g)\wh{\rho}_{\rm LE}(\tau)\wh{U}^{-1}(g)$.
This implies that either $\wh{U}(g)\wh{\rho}_{\rm LE}(\tau)\wh{U}^{-1}(g)$ coincides with $\wh{\rho}_{\rm LE}(\tau)$, or it is an entirely different solution.
In either case, it is possible to generate a symmetric solution under the subgroup ${\rm G}$ by taking a particular solution $\wh{\rho}_{\rm LE}^{(0)}(\tau)$ and summing over all the elements of ${\rm G}$
\begin{equation}
    \wh{\rho}_{\rm LE}(\tau)=\frac{1}{M({\rm G})}\sum_{g\in {\rm G}}\wh{U}(g)\wh{\rho}_{\rm LE}^{(0)}(\tau)\wh{U}^{-1}(g),
\end{equation}
where $M({\rm G})$ is the volume of ${\rm G}$.
Thus, a sufficient condition for $\wh{\rho}_{\rm LE}(\tau)$ to be symmetric under ${\rm G}$ at any $\tau$, is that the thermodynamic fields fulfill \eqref{zubeq12} at any $\tau$.
This will be important especially in Chapter \ref{chapter:boost}.

\subsection{Entropy production rate}
\label{sec:entr_prod_rate}

We conclude this part with an argument linking back to the end of the previous Section, namely the entropy production of irreversible quantum processes.
We will follow the steps of \cite{Becattini:2019dxo} with a few important adjustments concerning the renormalization.

As it should be clear by now, the local thermodynamic equilibrium density operator is not a state of the system for it is not stationary.
However, it still is an operator, so we can take traces of it and do all the operations we do on operators.
Thus, if we pretend it is a state and we treat it as such, the corresponding entropy is in fact $\tau$-dependent
\begin{equation}\label{zubeq15}
    S(\tau)=-\tr(\wh{\rho}_{\rm LE}(\tau)\log \wh{\rho}_{\rm LE}(\tau)).
\end{equation}
That is because $\wh{\rho}_{\rm LE}(\tau)$ is built from the quantum operators as in \eqref{zubeq05}, not the microscopic states as in \eqref{zubeq08}, therefore its $\tau$-dependence is governed by $\Sigma(\tau)$ instead of the unitary transformation \eqref{zubeq14}.
In other words, the invariance of the entropy under unitary transformations of the state will not prevent it from being $\tau$-dependent this time, because, as $\wh{\rho}_{\rm LE}(\tau)$ is not a state, its time evolution is not controlled by a unitary transformation but by $\Sigma(\tau)$ instead.
Thus, up to more convincing proposals, equation \eqref{zubeq15} is conventionally taken as the definition of a non-stationary entropy, despite $\wh{\rho}_{\rm LE}(\tau)$ not being a state.

The next step is finding an expression for the entropy production rate, that is $\di S(\tau)/\di \tau$.
Recall that, by definition, the entropy \eqref{zubeq15} can be expressed in the form of \eqref{eqrelhydro19} as
\begin{equation}
    S(\tau)=\int_{\Sigma(\tau)}\di \Sigma_{\mu}s^{\mu}.
\end{equation}
Let us emphasize that out of equilibrium, since $\nabla_{\mu}s^{\mu}>0$, the entropy is a frame-dependent quantity as it varies with the integration hypersurface $\Sigma(\tau)$.
The derivative of the entropy with respect to $\tau$ can be computed by exploiting a general formula for the variation of an integral between two infinitesimally close hypersurfaces, which we do not prove here
\begin{equation}\label{zubeq16}
    \frac{\di S(\tau)}{\di \tau}=
    \int_{\Sigma(\tau)}\di \Sigma_{\nu}U^{\nu}\nabla_{\mu}s^{\mu}+
    \frac{1}{2}\int_{\de \Sigma(\tau)}\di \tilde{S}_{\mu \nu}(s^{\mu}U^{\nu}-s^{\nu}U^{\mu}),
\end{equation}
where $\de \Sigma(\tau)$ is the 2-dimensional boundary of $\Sigma(\tau)$, $\di \tilde{S}_{\mu \nu}=-\frac{1}{2}\epsilon_{\mu \nu \rho \sigma}\di S^{\rho \sigma}$ is the dual of the measure $\di S^{\rho \sigma}\equiv \di x^{\rho}\wedge \di x^{\sigma}$ and $U^{\mu}\equiv \di x^{\mu}/\di \tau$ \cite{Becattini:2019dxo}.
Assuming that the boundary term vanishes, we are left with
\begin{equation}\label{zubeq18}
    \frac{\di S(\tau)}{\di \tau}=
    \int_{\Sigma(\tau)}\di \Sigma_{\nu}U^{\nu}\nabla_{\mu}s^{\mu}.
\end{equation}
Now we want to repeat the same steps with \eqref{zubeq15} and compare with the above result.
As we are going to exploit the invariance of $\wh{\rho}_{\rm LE}(\tau)$ under the subtraction of a vacuum expectation value, let us first define out of convenience the following notation
\begin{equation}
    :Z_{\rm LE}(\tau):\,\equiv \tr \left(\exp \left[-\int_{\Sigma}\di \Sigma_{\mu}\left(:\wh{T}^{\mu \nu}:\beta_{\nu}-\zeta :\wh{j}^{\mu}:\right)\right]\right).
\end{equation}
By using the expression \eqref{zubeq05} of $\wh{\rho}_{\rm LE}(\tau)$ and the above mentioned invariance, we find
\begin{equation}
    \begin{split}
        S(\tau)=&
        -\tr \left(\wh{\rho}_{\rm LE}(\tau)\log \frac{\exp \left[-\int_{\Sigma(\tau)}\di \Sigma_{\mu}\left(\wh{T}^{\mu \nu}\beta_{\nu}-\zeta \wh{j}^{\mu}\right)\right]}{Z_{\rm LE}(\tau)}\right)\\
        =&-\tr \left(\wh{\rho}_{\rm LE}(\tau)\log \frac{\exp \left[-\int_{\Sigma(\tau)}\di \Sigma_{\mu}\left(:\wh{T}^{\mu \nu}:\beta_{\nu}-\zeta :\wh{j}^{\mu}:\right)\right]}{:Z_{\rm LE}(\tau):}\right)\\
        =&\log :Z_{\rm LE}(\tau):+\tr \left(\wh{\rho}_{\rm LE}(\tau)\int_{\Sigma(\tau)}\di \Sigma_{\mu}\left(:\wh{T}^{\mu \nu}:\beta_{\nu}-\zeta :\wh{j}^{\mu}:\right)\right)\\
        =&\log :Z_{\rm LE}(\tau):+\int_{\Sigma(\tau)}\di \Sigma_{\mu}\left(\langle :\wh{T}^{\mu \nu}:\rangle_{\rm LE}\beta_{\nu}-\zeta \langle :\wh{j}^{\mu}:\rangle_{\rm LE}\right)\\
        =&\log :Z_{\rm LE}(\tau):+\int_{\Sigma(\tau)}\di \Sigma_{\mu}\left(T^{\mu \nu}\beta_{\nu}-\zeta j^{\mu}\right).
    \end{split}
\end{equation}
In the last step, we used the definition $\di \Sigma_{\mu}=\di \Sigma \,n_{\mu}$ and the constraint equations \eqref{zubeq04} to replace the renormalized energy-momentum and charge densities at local thermodynamic equilibrium with the actual ones.
By deriving with respect to $\tau$, using the formula \eqref{zubeq16} and assuming once again that the boundary term vanishes, we obtain
\begin{equation}\label{zubeq17}
    \frac{\di S(\tau)}{\di \tau}=\frac{\di \log :Z_{\rm LE}(\tau):}{\di \tau}+\int_{\Sigma(\tau)}\di \Sigma_{\lambda}U^{\lambda}\left(T^{\mu \nu}\nabla_{\mu}\beta_{\nu}-j^{\mu}\nabla_{\mu}\zeta \right),
\end{equation}
where the conservation law of the energy-momentum tensor and charged current, namely the hydrodynamic equations \eqref{eqrelhydro02} and \eqref{eqrelhydro03}, were also used.
The derivative of the logarithm of the partition function requires a little more effort.
Let us start by defining the operator $\wh{\Upsilon}(\tau)$, which will appear again later on, as
\begin{equation}\label{zubeq23}
    Z_{\rm LE}(\tau)=\tr \left({\rm e}^{-\wh{\Upsilon}(\tau)}\right),\qquad
    \wh{\Upsilon}(\tau)\equiv \int_{\Sigma(\tau)}\di \Sigma_{\mu}\left(\wh{T}^{\mu \nu}\beta_{\nu}-\zeta \wh{j}^{\mu}\right),
\end{equation}
hence the derivative
\begin{equation}
    \begin{split}
        \frac{\di \log :Z_{\rm LE}(\tau):}{\di \tau}=&
        \frac{1}{:Z_{\rm LE}(\tau):}\frac{\di :Z_{\rm LE}(\tau):}{\di \tau}=
        \frac{1}{:Z_{\rm LE}(\tau):}\frac{\di}{\di \tau}\tr \left({\rm e}^{-:\wh{\Upsilon}(\tau):}\right)\\
        =&-\frac{1}{:Z_{\rm LE}(\tau):}\tr \left({\rm e}^{-:\wh{\Upsilon}(\tau):}\frac{:\wh{\Upsilon}(\tau):}{\di \tau}\right)=
        -\left\langle \frac{\di :\wh{\Upsilon}(\tau):}{\di \tau}\right\rangle_{\rm LE}.
    \end{split}
\end{equation}
We use again the formula \eqref{zubeq16} assuming that the boundary term vanishes, and with the conservation laws of the energy-momentum tensor and charged current operators we get
\begin{equation}
    \frac{\di :\wh{\Upsilon}(\tau):}{\di \tau}=
    \int_{\Sigma(\tau)}\di \Sigma_{\lambda}\left(:\wh{T}^{\mu \nu}:\nabla_{\mu}\beta_{\nu}-:\wh{j}^{\mu}:\nabla_{\mu}\zeta \right),
\end{equation}
hence
\begin{equation}
    \frac{\di \log :Z_{\rm LE}(\tau):}{\di \tau}=-\int_{\Sigma(\tau)}\di \Sigma_{\lambda}U^{\lambda}\left(\langle :\wh{T}^{\mu \nu}:\rangle_{\rm LE}\nabla_{\mu}\beta_{\nu}-\langle :\wh{j}^{\mu}:\rangle_{\rm LE}\nabla_{\mu}\zeta \right),
\end{equation}
and plugging into \eqref{zubeq17} we finally obtain
\begin{equation}
    \frac{\di S(\tau)}{\di \tau}=
    \int_{\Sigma(\tau)}\di \Sigma_{\lambda}U^{\lambda}\left[\left(T^{\mu \nu}-\langle :\wh{T}^{\mu \nu}:\rangle_{\rm LE}\right)\nabla_{\mu}\beta_{\nu}-\left(j^{\mu}-\langle :\wh{j}^{\mu}:\rangle_{\rm LE}\right)\nabla_{\mu}\zeta \right].
\end{equation}
Comparing with \eqref{zubeq18} and taking into account that the equation should hold at any $\tau$, we finally obtain the following result for the entropy production rate
\begin{equation}\label{zubeq30}
    \nabla_{\mu}s^{\mu}=\left(T^{\mu \nu}-\langle :\wh{T}^{\mu \nu}:\rangle_{\rm LE}\right)\nabla_{\mu}\beta_{\nu}-\left(j^{\mu}-\langle :\wh{j}^{\mu}:\rangle_{\rm LE}\right)\nabla_{\mu}\zeta,
\end{equation}
which was first found in \cite{Zubarev:1979} and later reworked in \cite{Becattini:2019dxo}.
As we can see, the entropy production rate depends on the deviation of the actual values of the conserved currents from those at local thermodynamic equilibrium and, as expected, on the derivatives of the thermodynamic fields, which essentially are the dissipations.
By the second law of thermodynamics, the above quantity is bounded to always be non-negative.
At global thermodynamic equilibrium the thermodynamic fields fulfill the conditions \eqref{zubeq10}, so the entropy production rate vanishes as expected.
It is important to emphasize that equation \eqref{zubeq30} represents the entropy production rate at local thermodynamic equilibrium, which is the underlying assumption of relativistic hydrodynamics and contains global thermodynamic equilibrium as a special case.
However, no analogous formula is known for the entropy production rate or relativistic quantum fluids fully out of thermodynamic equilibrium.

Let us seize the opportunity of having mentioned the divergencelessness of the entropy current at global thermodynamic equilibrium to make the following geometrical remark.
Being $\nabla_{\mu}s^{\mu}=0$, the entropy current can be expressed as the Hodge dual of an exact 3-form, which, if the domain is topologically contractible, can in turn be written as the exterior derivative of a 2-form.
This eventually amounts to state that the original entropy current can be cast into the divergence of an antisymmetric tensor field
\begin{equation}
    \nabla_{\mu}s^{\mu}=0\qquad \Rightarrow \qquad
    s^{\mu}=\nabla_{\nu}\varsigma^{\mu \nu},\qquad
    \varsigma^{\mu \nu}=-\varsigma^{\nu \mu}.
\end{equation}
We call this $\varsigma^{\mu \nu}$ a \textit{potential} of the entropy current.
Whence, by using Stoke's theorem
\begin{equation}\label{zubeq31}
    S=\int_{\Sigma}\di \Sigma_{\mu}s^{\mu}=
    \int_{\Sigma}\di \Sigma_{\mu}\nabla_{\nu}\varsigma^{\mu \nu}=
    \frac{1}{2}\int_{\de \Sigma}\di \tilde{S}_{\mu \nu}\varsigma^{\mu \nu}=
    -\frac{1}{4}\int_{\de \Sigma}\di S^{\rho \sigma}\sqrt{|g|}\epsilon_{\mu \nu \rho \sigma}\varsigma^{\mu \nu},
\end{equation}
where $\de \Sigma$ is the 2-dimensional boundary of $\Sigma$, $\di S^{\rho \sigma}=\di x^{\rho}\wedge \di x^{\sigma}$ is the measure on $\de \Sigma$, and $\di \tilde{S}_{\mu \nu}=-\frac{1}{2}\sqrt{|g|}\epsilon_{\mu \nu \rho \sigma}\di S^{\rho \sigma}$ is its dual.
In other words, the entropy can be expressed not only as a 3-dimensional integral on $\Sigma$, but also as a 2-dimensional integral on the boundary of $\Sigma$.
This fact is known in literature \cite{Wald:1993nt, Eling:2012xa, Majhi:2011ws, Majhi:2012tf} and has been reworked in \cite{Becattini:2019poj} including a practical example of calculation, which we will present in Subsection \ref{subsec:gte_entropy}.

Although elegantly concise and quite straightforward in its meaning, equation \eqref{zubeq30} can only provide us information on the divergence of the entropy current, not the entropy current itself.
In fact, in principle, there are infinitely many different entropy currents sharing the same divergence.
But as we saw in the last Chapter, the structure of the entropy current is crucial in defining the nature of the relativistic hydrodynamic theory, so it would be far more meaningful to know the expression of the entropy current instead of its divergence.
In the next Section we will put forward a general method to derive the entropy current directly in terms of the conserved currents.


\section{The entropy current method}
\label{sec:zubarev_entropy_current_method}

In this Section we retrace the steps of \cite{Becattini:2019poj}, where we determined, to the best of our knowledge, the first proof of existence of the entropy current at local thermodynamic equilibrium and put forward a general method to calculate it.
The crucial point will be the fact that the logarithm of the partition function is also extensive, meaning that it can be expressed as an integral of a vector field on a hypersurface of the foliation, which is usually assumed without proof.

Let us start our journey by noticing that the entropy straight out of the local thermodynamic equilibrium density operator \eqref{zubeq05} reads
\begin{equation}
    S(\tau)=\log Z_{\rm LE}(\tau)+\int_{\Sigma(\tau)}\di \Sigma_{\mu}\left(\langle \wh{T}^{\mu \nu}\rangle_{\rm LE}\beta_{\nu}-\zeta \langle \wh{j}^{\mu}\rangle_{\rm LE}\right).
\end{equation}
By definition of the entropy current \eqref{eqrelhydro19}, this is also equal to
\begin{equation}
    S(\tau)=\int_{\Sigma(\tau)}\di \Sigma_{\mu}s^{\mu}.
\end{equation}
We readily understand that, in order for the entropy to be extensive and so for the entropy current to exist, the logarithm of the partition function must also be extensive, namely there must be a vector field $\phi^{\mu}$, called the \textit{thermodynamic potential current}, such that
\begin{equation}
    \log Z_{\rm LE}(\tau)\equiv \int_{\Sigma(\tau)}\di \Sigma_{\mu}\phi^{\mu}.
\end{equation}
This way, we have the following expression for the entropy current
\begin{equation}
    s^{\mu}=\phi^{\mu}+\langle \wh{T}^{\mu \nu}\rangle_{\rm LE}\beta_{\nu}-\zeta \langle \wh{j}^{\mu}\rangle_{\rm LE}.
\end{equation}
The existence of the thermodynamic potential current and, in turn, of the entropy current is usually assumed.
This was made clear in Chapter \ref{chapter:relhydro}, when the entropy current was postulated in order to express the second law of thermodynamics in a local covariant fashion so that it could be enforced by relativistic hydrodynamics.
In this Section we provide the first general proof of this typically understood hypothesis and we put forward a method to calculate the thermodynamic potential current, whence the entropy current, at local thermodynamic equilibrium.

The first step is a modification of the local thermodynamic equilibrium density operator by the introduction of a dimensionless parameter $\lambda$ in the following way
\begin{equation}\label{zubeq21}
    \wh{\rho}_{\rm LE}(\tau,\lambda)\equiv
    \frac{1}{Z_{\rm LE}(\tau,\lambda)}\exp \left[-\lambda \int_{\Sigma(\tau)}\di \Sigma_{\mu}\left(\wh{T}^{\mu \nu}\beta_{\nu}-\zeta \wh{j}^{\mu}\right)\right],
\end{equation}
\begin{equation}\label{zubeq22}
    Z_{\rm LE}(\tau,\lambda)\equiv
    \tr \left(\exp \left[-\lambda \int_{\Sigma(\tau)}\di \Sigma_{\mu}\left(\wh{T}^{\mu \nu}\beta_{\nu}-\zeta \wh{j}^{\mu}\right)\right]\right).
\end{equation}
This is purposely designed in such a way to recover the original density operator \eqref{zubeq05} and partition function \eqref{zubeq06} for $\lambda=1$.
By taking the derivative with respect to $\lambda$ of the logarithm of this new $\lambda$-dependent partition function, we obtain
\begin{equation}\label{zubeq20}
    \frac{\de \log Z_{\rm LE}(\tau,\lambda)}{\de \lambda}=
    -\int_{\Sigma(\tau)}\di \Sigma_{\mu}\left(\langle \wh{T}^{\mu \nu}\rangle_{\rm LE}(\lambda)\beta_{\nu}-\zeta \langle \wh{j}^{\mu}\rangle_{\rm LE}(\lambda)\right),
\end{equation}
where by the symbol $\langle \wh{O}\rangle_{\rm LE}(\lambda)$ we mean the expectation value of $\wh{O}$ calculated with $\wh{\rho}_{\rm LE}(\tau,\lambda)$, that is $\langle \wh{O}\rangle_{\rm LE}(\lambda)\equiv \tr(\wh{\rho}_{\rm LE}(\tau,\lambda)\wh{O})$.
Note that the $\lambda$-dependence affects only the expectation values and not the thermodynamic fields, for it stems from the density operator with whom the expectation values are calculated.
If the above result looks confusing at a glance, let us just point out that it is but the same as the following known formula, which is obtained from \eqref{zubeq03} and can be easily checked
\begin{equation}
    Z=\tr \left(\exp \left[-\sum_i\lambda_i\wh{O}_i\right]\right)
    \qquad \Rightarrow \qquad
    \frac{\de \log Z}{\de \lambda_k}=-\langle \wh{O}_k\rangle.
\end{equation}
Let us now integrate both sides of \eqref{zubeq20} with respect to $\lambda$ from some $\lambda_0$ to $\lambda=1$
\begin{equation}
    \log Z_{\rm LE}(\tau)-\log Z_{\rm LE}(\tau,\lambda_0)=-\int_{\lambda_0}^1\di \lambda \int_{\Sigma(\tau)}\di \Sigma_{\mu}\left(\langle \wh{T}^{\mu \nu}\rangle_{\rm LE}(\lambda)\beta_{\nu}-\zeta \langle \wh{j}^{\mu}\rangle_{\rm LE}(\lambda)\right),
\end{equation}
where we used the feature that the $\lambda$-modification is so that $\log Z_{\rm LE}(\tau,\lambda=1)=\log Z_{\rm LE}(\tau)$.
If we can exchange the order of the $\lambda$-integration and the $\Sigma(\tau)$ one, we have
\begin{equation}
    \log Z_{\rm LE}(\tau)-\log Z_{\rm LE}(\tau,\lambda_0)=\int_{\Sigma(\tau)}\di \Sigma_{\mu} \int_1^{\lambda_0}\di \lambda \left(\langle \wh{T}^{\mu \nu}\rangle_{\rm LE}(\lambda)\beta_{\nu}-\zeta \langle \wh{j}^{\mu}\rangle_{\rm LE}(\lambda)\right).
\end{equation}
Thus, if there exists a particular $\lambda_0$ such that $\log Z_{\rm LE}(\tau,\lambda_0)=0$, we have both the proof of the extensivity of the logarithm of the partition function and a formula for the thermodynamic potential current
\begin{equation}
    \log Z_{\rm LE}(\tau)=\int_{\Sigma(\tau)}\di \Sigma_{\mu}\phi^{\mu},\qquad
    \phi^{\mu}=\int_1^{\lambda_0}\di \lambda \left(\langle \wh{T}^{\mu \nu}\rangle_{\rm LE}(\lambda)\beta_{\nu}-\zeta \langle \wh{j}^{\mu}\rangle_{\rm LE}(\lambda)\right).
\end{equation}

The key quantity for the determination of such a special $\lambda_0$ is the operator $\wh{\Upsilon}(\tau)$ already defined in \eqref{zubeq23}, in particular its spectrum.
Suppose that $\wh{\Upsilon}(\tau)$ is bounded from below, namely there exists a lowest eigenvalue $\Upsilon_0(\tau)$ and a corresponding eigenvector $|0_{\Upsilon}(\tau)\rangle$, and also let the latter be non-degenerate.
In the following, we will omit the $\tau$-dependence in the operator $\wh{\Upsilon}(\tau)$, in the eigenvalues $\Upsilon_i(\tau)$ and in the eigenvector $|0_{\Upsilon}(\tau)\rangle$ in order to ease the notation and highlight the $\lambda$-dependence instead, but it should be kept in mind that they are all $\tau$-dependent in principle.
In this hypothesis, the eigenvalues can be arranged in ascending order as $\Upsilon_0<\Upsilon_1<\Upsilon_2<\cdots$ and the partition function can be written as
\begin{equation}\label{zubeq25}
    Z_{\rm LE}(\tau,\lambda)={\rm e}^{-\lambda \Upsilon_0}\left(1+{\rm e}^{-\lambda(\Upsilon_1-\Upsilon_0)}+{\rm e}^{\lambda(\Upsilon_2-\Upsilon_0)}+\cdots \right).
\end{equation}
Then, if $\Upsilon_0=0$ and we let $\lambda \to +\infty$, we notice that
\begin{equation}
    \lim_{\lambda \to +\infty}Z_{\rm LE}(\tau,\lambda)=1
    \qquad \Rightarrow \qquad \lim_{\lambda \to +\infty}\log Z_{\rm LE}(\tau,\lambda)=0,
\end{equation}
therefore $\lambda_0=+\infty$ would be our candidate; however, we might expect $\Upsilon_0$ to be different from zero in general.
So, in order for these wheels to spin the way we want, we exploit the above mentioned property \eqref{zubeq24} of the local thermodynamic equilibrium density operator, namely the invariance under addition or subtraction of non-operator terms in the argument of the exponential function, provided the same is done in the partition function.
Recall that the entropy depends only on the density operator, therefore this is also an invariance of the entropy.
Thus, with the replacement
\begin{equation}
    \wh{\rho}_{\rm LE}(\tau,\lambda)=\frac{{\rm e}^{-\lambda \wh{\Upsilon}(\tau)}}{Z_{\rm LE}(\tau,\lambda)}
    \qquad \mapsto \qquad
    \wh{\rho}'_{\rm LE}(\tau,\lambda)\equiv \frac{{\rm e}^{-\lambda \wh{\Upsilon}'(\tau)}}{Z'_{\rm LE}(\tau,\lambda)}
\end{equation}
where
\begin{equation}
    \begin{split}
        \wh{\Upsilon}'\equiv&
        \wh{\Upsilon}-\Upsilon_0
        =\wh{\Upsilon}-\langle 0_{\Upsilon}|\wh{\Upsilon}|0_{\Upsilon}\rangle \\
        =&\int_{\Sigma(\tau)}\di \Sigma_{\mu}\left[\left(\wh{T}^{\mu \nu}-\langle 0_{\Upsilon}|\wh{T}^{\mu \nu}|0_{\Upsilon}\rangle \right)\beta_{\nu}-\zeta \left(\wh{j}^{\mu}-\langle 0_{\Upsilon}|\wh{j}^{\mu}|0_{\Upsilon}\rangle\right)\right],
    \end{split}
\end{equation}
the entropy is in fact unchanged.
In other words, if this is useful to find an expression for the entropy current, we are free to make such a tweak and rest assured that the entropy stays the same.
The rationale behind this transformation is that the new operator $\wh{\Upsilon}'$ has a vanishing lowest eigenvalue, $\Upsilon'_0=\Upsilon_0-\Upsilon_0=0$, therefore the new partition function
\begin{equation}
    Z'_{\rm LE}(\tau,\lambda)=\tr \left(\exp \left[-\lambda \int_{\Sigma(\tau)}\di \Sigma_{\mu}\left[\left(\wh{T}^{\mu \nu}-\langle 0_{\Upsilon}|\wh{T}^{\mu \nu}|0_{\Upsilon}\rangle \right)\beta_{\nu}-\zeta \left(\wh{j}^{\mu}-\langle 0_{\Upsilon}|\wh{j}^{\mu}|0_{\Upsilon}\rangle\right)\right]\right]\right),
\end{equation}
which, like \eqref{zubeq25} and taking into account that $\Upsilon'_i=\Upsilon_i-\Upsilon_0$, can also be written as
\begin{equation}
    Z'_{\rm LE}(\tau,\lambda)=1+{\rm e}^{-\lambda(\Upsilon_1-\Upsilon_0)}+{\rm e}^{-\lambda(\Upsilon_2-\Upsilon_0)}+\cdots,
\end{equation}
does have a vanishing logarithm for $\lambda \to +\infty$, therefore $\lambda_0=+\infty$ is indeed the sought solution.
So, by repeating the above steps with $Z_{\rm LE}(\tau,\lambda)$ replaced by $Z'_{\rm LE}(\tau,\lambda)$, the thermodynamic potential current and the entropy current read respectively
\begin{equation}\label{zubeq26}
    \phi^{\mu}=\int_1^{+\infty}\di \lambda \left[\left(\langle \wh{T}^{\mu \nu}\rangle_{\rm LE}-\langle 0_{\Upsilon}|\wh{T}^{\mu \nu}|0_{\Upsilon}\rangle\right)(\lambda)\beta_{\nu}-\zeta \left(\langle \wh{j}^{\mu}\rangle_{\rm LE}-\langle 0_{\Upsilon}|\wh{j}^{\mu}|0_{\Upsilon}\rangle \right)(\lambda)\right],
\end{equation}
\begin{equation}\label{zubeq27}
    s^{\mu}=\phi^{\mu}+\left(\langle \wh{T}^{\mu \nu}\rangle_{\rm LE}-\langle 0_{\Upsilon}|\wh{T}^{\mu \nu}|0_{\Upsilon}\rangle \right)\beta_{\nu}-\zeta \left(\langle \wh{j}^{\mu}\rangle_{\rm LE}-\langle 0_{\Upsilon}|\wh{j}^{\mu}|0_{\Upsilon}\rangle \right).
\end{equation}

In summary, a sufficient condition for the existence of the thermodynamic potential current and the entropy current at local thermodynamic equilibrium is that the operator $\wh{\Upsilon}$ must be bounded from below with non-degenerate lowest eigenvalue.
In this case, the two ingredients necessary to calculate the entropy current are:
\begin{enumerate}
    \item the expectation values of the conserved currents at local thermodynamic equilibrium, i.e.\ $\langle \wh{T}^{\mu \nu}\rangle_{\rm LE}$ and $\langle \wh{j}^{\mu}\rangle_{\rm LE}$, and
    \item the eigenvector $|0_{\Upsilon}\rangle$ corresponding to the lowest eigenvalue $\Upsilon_0$ of the operator $\wh{\Upsilon}$.
\end{enumerate}
The quantum operators of the conserved currents $\wh{T}^{\mu \nu}$ and $\wh{j}^{\mu}$ are built by using the quantum field expansion, which in turn is obtained by solving the quantum field equations of motion of the Quantum Field Theory at hand, so the preliminary step of the method is to do that first.
Then, the algorithm for the calculation is the following:
\begin{enumerate}
    \item Take the expectation values at local thermodynamic equilibrium $\langle \wh{T}^{\mu \nu}\rangle_{\rm LE}$ and $\langle \wh{j}^{\mu}\rangle_{\rm LE}$ and subtract from them the expectation values in $|0_{\Upsilon}\rangle$, namely $\langle 0_{\Upsilon}|\wh{T}^{\mu \nu}|0_{\Upsilon}\rangle$ and $\langle 0_{\Upsilon}|\wh{j}^{\mu}|0_{\Upsilon}\rangle$.
    This must be done using $\widehat{\rho}_{\rm LE}(\tau,\lambda)$, so the result will be $\lambda$-dependent.
    \item Multiply by the corresponding thermodynamic fields, which are not $\lambda$-dependent, and integrate with respect to $\lambda$ from $\lambda=1$ to $\lambda=+\infty$ in order to obtain the thermodynamic potential current $\phi^{\mu}$ according to \eqref{zubeq26}.
    \item Plug the result into \eqref{zubeq27} and obtain the entropy current $s^{\mu}$.
\end{enumerate}

Let us emphasize once again that the $\tau$-dependence have been omitted from $\wh{\Upsilon}$, $\Upsilon_i$ and $|0_{\Upsilon}\rangle$ in order to ease the notation, but they are all $\tau$-dependent quantities in principle.
Note also that, although the operator $\wh{\Upsilon}$ can be shown to reduce to the Hamiltonian in the special case of global thermodynamic equilibrium in Minkowski spacetime with $\zeta=0$ and $\beta^{\mu}=(1/T)(1,\mathbf{0})$, it is not so in general.
Thus, $|0_{\Upsilon}\rangle$ is not the vacuum state of the theory, and the expectation values in \eqref{zubeq26} and \eqref{zubeq27} do not coincide in general with the physically normally ordered ones.
If we wish not to be misled, we should therefore regard $|0_{\Upsilon}\rangle$ simply as the lowest eigenvector of some operator instead of some vacuum state.

Moreover, we stress once more that the result \eqref{zubeq27} is the entropy current at local thermodynamic equilibrium, which is the underlying assumption of relativistic hydrodynamics and includes global thermodynamic equilibrium as a special case.
However, no analogous expression is known for the entropy current for relativistic quantum fluids fully out of thermodynamic equilibrium.

Note that the expression for the entropy current we just derived fulfills the feature of being divergenceless at global thermodynamic equilibrium.
First of all, at global thermodynamic equilibrium we have $\nabla_{\mu}\langle \wh{O}\rangle=\langle \nabla_{\mu}\wh{O}\rangle$, because $\nabla_{\mu}\wh{\rho}=0$ thanks to stationarity.
Same thing for $\nabla_{\mu}\langle 0_{\Upsilon}|\wh{O}|0_{\Upsilon}\rangle=\langle 0_{\Upsilon}|\nabla_{\mu}\wh{O}|0_{\Upsilon}\rangle$, as the non-degenerate state $|0_{\Upsilon}\rangle$ is stationary at global thermodynamic equilibrium.
Thus, for the thermodynamic potential current we have
\begin{equation}
    \begin{split}
        \nabla_{\mu}\phi^{\mu}=&
        \nabla_{\mu}\int_1^{+\infty}\di \lambda \left[\left(\langle \wh{T}^{\mu \nu}\rangle_{\rm LE}-\langle 0_{\Upsilon}|\wh{T}^{\mu \nu}|0_{\Upsilon}\rangle\right)(\lambda)\beta_{\nu}-\zeta \left(\langle \wh{j}^{\mu}\rangle_{\rm LE}-\langle 0_{\Upsilon}|\wh{j}^{\mu}|0_{\Upsilon}\rangle \right)(\lambda)\right]\\
        =&\int_1^{+\infty}\di \lambda \left[\left(\langle \nabla_{\mu}\wh{T}^{\mu \nu}\rangle-\langle 0_{\Upsilon}|\nabla_{\mu}\wh{T}^{\mu \nu}|0_{\Upsilon}\rangle \right)(\lambda)\beta_{\nu}-
        \zeta \left(\langle \nabla_{\mu}\wh{j}^{\mu}\rangle-\langle 0_{\Upsilon}|\nabla_{\mu}\wh{j}^{\mu}|0_{\Upsilon}\rangle \right)(\lambda)\right.\\
        &+\left.\left(\langle \wh{T}^{\mu \nu}\rangle-\langle 0_{\Upsilon}|\wh{T}^{\mu \nu}|0_{\Upsilon}\rangle \right)(\lambda)\nabla_{\mu}\beta_{\nu}-
        \nabla_{\mu}\zeta \left(\langle \wh{j}^{\mu}\rangle-\langle 0_{\Upsilon}|\wh{j}^{\mu}|0_{\Upsilon}\rangle \right)\right]=0.
    \end{split}
\end{equation}
The second line vanishes because of the conservation laws of the conserved currents quantum operators, and the last line vanishes because of the geometric conditions \eqref{zubeq10} for global thermodynamic equilibrium.
For the exact same reasons, we have
\begin{equation}
    \begin{split}
        \nabla_{\mu}s^{\mu}=&
        \nabla_{\mu}\phi^{\mu}+
        \nabla_{\mu}\left(\langle \wh{T}^{\mu \nu}\rangle-\langle 0_{\Upsilon}|\wh{T}^{\mu \nu}|0_{\Upsilon}\rangle \right)\beta_{\nu}-
        \zeta \nabla_{\mu}\left(\langle \wh{j}^{\mu}\rangle-\langle 0_{\Upsilon}|\wh{j}^{\mu}|0_{\Upsilon}\rangle \right)\\
        &+\left(\langle \wh{T}^{\mu \nu}\rangle-\langle 0_{\Upsilon}|\wh{T}^{\mu \nu}|0_{\Upsilon}\rangle \right)\nabla_{\mu}\beta_{\nu}-
        \nabla_{\mu}\zeta \left(\langle \wh{j}^{\mu}\rangle-\langle 0_{\Upsilon}|\wh{j}^{\mu}|0_{\Upsilon}\rangle \right)=0.
    \end{split}
\end{equation}


\section{Summary and outlook}

Over the years, relativistic hydrodynamics has proven to be a powerful and versatile theory for describing, among others, astrophysical, cosmological and nuclear phenomena.
At some point in time, it became necessary to both go beyond the usual first or second order theory and also implement effects of quantum origin in the hydrodynamic framework.
An example of the latter is the entropy production of irreversible quantum processes.
It is in this atmosphere that the study of the quantum statistical foundations of relativistic hydrodynamics started to take hold, and it still is to date.

To these purposes, in the late 70's D.\ N.\ Zubarev put forward an approach to relativistic quantum statistical mechanics based on a covariant expression of the density operator at local thermodynamic equilibrium, $\wh{\rho}_{\rm LE}(\tau)$, built according to the maximum entropy principle in terms of the quantum operators of the conserved currents.
After a quick review of quantum statistical mechanics, we started this Chapter by introducing this very object and presenting its most important and interesting features.
A fundamental one is that this operator is not stationary, therefore it cannot be a state of the system in the Heisenberg picture.
Notwithstanding, if at some $\tau_0$ the system is known to be at local thermodynamic equilibrium, the actual stationary state of the system $\wh{\rho}$ is in fact $\wh{\rho}_{\rm LE}(\tau_0)$, whether at or out of thermodynamic equilibrium.
We also took this a step further by showing that stationarity at global thermodynamic equilibrium translates into geometrical conditions on the thermodynamic fields, specifically the timelike four-temperature $\beta^{\mu}$ must be a Killing vector field and the fugacity $\zeta$ must be constant.
These are a prerogative of global thermodynamic equilibrium: at local thermodynamic equilibrium and at full non-equilibrium, the thermodynamic fields can be whatever, as long as they solve the constraint equations, what makes this kind of systems much harder to study.

Concerning the problem of entropy production, which has indeed an important place in this work, we showed that an equation for the entropy production rate at local thermodynamic equilibrium can be derived directly from the expression of the density operator.
However, as elegant and clear as this result may be, it does not provide us in fact with any information on the very structure of the entropy current, which, from a relativistic hydrodynamic standpoint, would be desirable to know.
In this respect, in the last Section we showed that a sufficient condition for the existence of the entropy current at local thermodynamic equilibrium is that the operator $\wh{\Upsilon}(\tau)$ should be bounded from below with non-degenerate lowest eingevalue.
In this case, we proved that the logarithm of the partition function is extensive and, a consequence, that an entropy current exists at local thermodynamic equilibrium, making it the first general proof of this usually tacitly understood hypothesis.
With our method, we derived an expression for the entropy current at local thermodynamic equilibrium based on two main ingredients: the expectation values of the conserved currents at local thermodynamic equilibrium and the eigenvector corresponding to the lowest eigenvalue of $\wh{\Upsilon}(\tau)$.

In Chapters \ref{chapter:gteacceleration} and \ref{chapter:boost}, we will use our method and the Zubarev approach in general to study two different kinds of systems.
As we shall see, aside from their entropy currents, they will be interesting in their own right, both theoretically and phenomenologically.


\chapter{Relativistic Quantum Fluid at Equilibrium with Acceleration}
\label{chapter:gteacceleration}
In this Chapter, we will apply our method for the entropy current to the case of a relativistic quantum fluid at global thermodynamic equilibrium with acceleration, which is a non-trivial state of global thermodynamic equilibrium in Minkowski spacetime.
From a phenomenological perspective, this system is particularly interesting in the context of heavy-ion collisions, for the quark-gluon plasma is a fluid at local thermodynamic equilibrium with strong values of acceleration and vorticity, local thermodynamic equilibrium being approximated by global thermodynamic equilibrium at first order in the thermodynamic fields.
From a more theoretical standpoint, it is known that an acceleration involves particular relativistic quantum effects at low temperature, such as the Unruh effect, which, although dating back to the 70's, is still a vibrant subject of investigation.

We will start by presenting the possible global thermodynamic equilibria in Minkowski spacetime and then focus on the one with acceleration.
As we shall see, in order to make thermodynamics out of the four-temperature field, we will have to restrict ourselves to a subspace of the whole Minkowski spacetime, namely the right Rindler wedge.
The Unruh effect will emerge naturally from thermal expectation values there, in particular the Unruh temperature will be an absolute lower bound for the temperature in a sense that will be made clear later on.
With the thermal expectation value of the energy-momentum tensor, we will be able to use our method to calculate the entropy current, and we will see that it receives a quantum correction due to the acceleration field.
The entropy obtained by integration of this entropy current will possess properties in full agreement with previous literature and, interestingly enough, it will coincide with the entanglement entropy of the right Rindler wedge with the left Rindler wedge.

The general discussion on global thermodynamic equilibrium with acceleration, part of the thermal expectation values and the Unruh effect will be based on \cite{Becattini:2017ljh} by F.\ Becattini.
For more details on the Unruh effect, we suggest the reading of W.\ G.\ Unruh's original paper \cite{Unruh:1976db} and the modern review by L.\ C.\ B.\ Crispino \textit{et al.} \cite{Crispino:2007eb}.
As for the rest of the thermal expectation values, the calculation of the entropy current and the total entropy, we will follow \cite{Becattini:2019poj} by F.\ Becattini and D.\ R..


\section{Global thermodynamic equilibrium with acceleration}
\label{sec:gte_gte_with_acceleration}

The first step in order to make any thermal field theory is the determination of the density operator representing the state at hand, which will then allow for the calculation of thermal expectation values.
In this Section, we will present the state of global thermodynamic equilibrium with acceleration, a non-trivial instance of global thermodynamic equilibrium in Minkowski spacetime.


\subsection{Global thermodynamic equilibrium in Minkowski spacetime}

As we showed in Chapter \ref{chapter:zubarev}, in order for a state to be of global thermodynamic equilibrium, the thermodynamic fields must fulfill the geometric conditions \eqref{zubeq10}, that is the four-temperature $\beta^{\mu}$ must be a (timelike) Killing vector field and the fugacity $\zeta$ must be a constant.
The existence of global timelike Killing vector fields, whence of global thermodynamic equilibrium states, is a prerogative of so-called \textit{stationary} spacetimes, of which Minkowski is an example.
The solution of the Killing equation \eqref{zubeq10} in Minkowski spacetime has the form
\begin{equation}\label{gteq05}
    \beta_{\mu}=b_{\mu}+\varpi_{\mu \nu}x^{\nu},
\end{equation}
where $b_{\mu}$ is a constant vector field and $\varpi_{\mu \nu}$ a constant tensor given by the exterior derivative of the four-temperature,
\begin{equation}\label{gteq06}
    \varpi_{\mu \nu}=-\frac{1}{2}\left(\partial_{\mu}\beta_{\nu}-\partial_{\nu}\beta_{\mu}\right),
\end{equation}
called the \textit{thermal vorticity} \cite{Becattini:2012tc}.
This can be further decomposed, in general, in a way similar to how the electromagnetic tensor is decomposed into the comoving electric and magnetic fields \cite{Buzzegoli:2017cqy}
\begin{equation}\label{gteq07}
    \varpi_{\mu \nu}=\alpha_{\mu}u_{\nu}-\alpha_{\nu}u_{\mu}+\epsilon_{\mu \nu \rho \sigma}w^{\rho}u^{\sigma},
\end{equation}
where $\alpha^{\mu}$ and $w^{\mu}$ are two spacelike vector fields defined as
\begin{equation}
    \alpha_{\mu}\equiv \varpi_{\mu \nu}u^{\nu},\qquad
    w_{\mu}\equiv -\frac{1}{2}\epsilon_{\mu \nu \rho \sigma}\varpi^{\nu \rho}u^{\sigma}.
\end{equation}
By expressing them in terms of the four-temperature and the proper temperature, one can show that their physical meaning is that of acceleration and vorticity fields respectively both divided by the proper temperature
\begin{equation}\label{gteq24}
    \alpha_{\mu}=\frac{A_{\mu}}{T},\qquad w_{\mu}=\frac{\omega_{\mu}}{T},
\end{equation}
where the acceleration field was defined in \eqref{eqrelhydro23} while the \textit{vorticity} field is defined as $\omega_{\mu}\equiv -\frac{1}{2}\sqrt{\beta^2}\epsilon_{\mu \nu \rho \sigma}\de^{\rho}u^{\nu}u^{\sigma}$.
Thus, we understand that the thermal vorticity accounts, in some sense, for thermal effects due to the acceleration and rotation of the fluid.
To get an idea of their order of magnitude, it is convenient to go to the local rest frame and express the above vector fields in terms of the local acceleration ${\bf a}$ and local angular velocity $\boldsymbol{\omega}$, that is $\alpha^{\mu}=(0,{\bf a}/T)$ and $w^{\mu}=(0,\boldsymbol{\omega}/T)$.
In order for the decomposition \eqref{gteq07} to have the correct dimensions, it is realized that the appropriate combination of natural constants is
\begin{equation}\label{gteq25}
    |\alpha^{\mu}|=\frac{\hbar}{ck_{\rm B}}\frac{|{\bf a}|}{T},\qquad
    |w^{\mu}|=\frac{\hbar}{k_{\rm B}}\frac{|\boldsymbol{\omega}|}{T},
\end{equation}
so they are terms of relativistic and/or quantum origin possibly appreciable at low temperature.
For typical systems on Earth, they are in fact negligible both because of the suppression coming from the natural constants and the room temperature being large with respect to the magnitude of the acceleration and rotation.
However, although possessing a very high temperature ($k_{\rm B}T\simeq 200\,{\rm MeV}$), the quark-gluon plasma produced in heavy-ion collisions has such large values of acceleration ($\hbar |{\bf a}|/c\simeq 10\,{\rm MeV}$) and vorticity ($h|\boldsymbol{\omega}|\simeq 12\,{\rm MeV}$) that these terms are in fact meaningful ($|\alpha^{\mu}|\simeq 0.05$, $|w^{\mu}|\simeq 0.06$), enough so to generate observable effects such as the polarization of $\Lambda$ hyperions \cite{STAR:2017ckg, Adam:2018ivw}.

With equations \eqref{gteq05} and \eqref{gteq06}, the density operator at global thermodynamic equilibrium \eqref{zubeq28} readily becomes
\begin{equation}
    \begin{split}
            \wh{\rho}=&
            \frac{1}{Z}\exp \left[-\int_{\Sigma}\di \Sigma_{\mu}\left(\wh{T}^{\mu \nu}\left(b_{\nu}+\varpi_{\nu \lambda}x^{\lambda}\right)-\zeta \wh{j}^{\mu}\right)\right]\\
            =&\frac{1}{Z}\exp \left[-b_{\mu}\int_{\Sigma}\di \Sigma_{\mu}\wh{T}^{\mu \nu}+\frac{1}{2}\varpi_{\nu \lambda}\int_{\Sigma}\di \Sigma_{\mu}\left(x^{\nu}\wh{T}^{\mu \lambda}-x^{\lambda}\wh{T}^{\mu \nu}\right)+\zeta \int_{\Sigma}\di \Sigma_{\mu}\wh{j}^{\mu}\right],
    \end{split}
\end{equation}
where $b_{\mu}$, $\varpi_{\mu \nu}$ and $\zeta$ are taken out of the integral as they are constant.
Here we recognize the generators of the Poincaré group
\begin{equation}\label{gteq43}
    \wh{P}^{\mu}=\int_{\Sigma}\di \Sigma_{\mu}\wh{T}^{\mu \nu},\qquad
    \wh{J}^{\nu \lambda}=\int_{\Sigma}\di \Sigma_{\mu}\left(x^{\nu}\wh{T}^{\mu \lambda}-x^{\lambda}\wh{T}^{\mu \nu}\right)
\end{equation}
the former being the four-momentum operator and the latter the generators of Lorentz transformations, and also the conserved charge associated to the current $\wh{j}^{\mu}$
\begin{equation}
    \wh{Q}=\int_{\Sigma}\di \Sigma_{\mu}\wh{j}^{\mu},
\end{equation}
therefore we find
\begin{equation}\label{gteq16}
    \wh{\rho}=\frac{1}{Z}\exp \left[-b_{\mu}\wh{P}^{\mu}+\frac{1}{2}\varpi_{\mu \nu}\wh{J}^{\mu \nu}+\zeta \wh{Q}\right].
\end{equation}
In other words, aside from a constant fugacity associated to the conserved charge ascribed to internal symmetries, the most general expression of the density operator at global thermodynamic equilibrium in Minkowski spacetime is given by a linear combination of the generators of the Poincaré group with ten constant coefficients.
Depending on how these coefficients are specified, different kinds of global thermodynamic equilibrium are obtained.

In the simplest case of a vanishing thermal vorticity $\varpi_{\mu \nu}=0$, the four-temperature is just $\beta_{\mu}=b_{\mu}$, and the following familiar density operator is recovered
\begin{equation}\label{gteq11}
    \wh{\rho}=\frac{1}{Z}\exp \left[-b_{\mu}\wh{P}^{\mu}+\zeta \wh{Q}\right]=\frac{1}{Z}\exp \left[-\frac{\wh{H}}{T_0}-\zeta \wh{Q}\right],
\end{equation}
where the last expression is obtained by setting $\beta_{\mu}=b_{\mu}=(1/T_0)(1,{\bf 0})$ with $T_0$ a constant thermodynamic parameter with the dimensions of a temperature and $\wh{H}$ the Hamiltonian.
This state is called \textit{homogeneous global thermodynamic equilibrium}.

As for a non-vanishing thermal vorticity, there are two possible cases.
The first one is that of a thermal vorticity with only a transverse (or ``magnetic'') component
\begin{equation}
    b_{\mu}=\frac{1}{T_0},\qquad
    \varpi_{\mu \nu}=\frac{\omega}{T_0}\left(g_{1\mu}g_{2\nu}-g_{1\nu}g_{2\mu}\right),
\end{equation}
where $\omega$ is a constant thermodynamic parameter with the dimensions of an angular velocity.
The four-temperature then reads
\begin{equation}
    \beta_{\mu}=\frac{1}{T_0}(1,\omega y,-\omega x,0),
\end{equation}
hence the density operator
\begin{equation}\label{gteq03}
    \wh{\rho}=\frac{1}{Z}\exp \left[-\frac{\wh{H}}{T_0}+\frac{\omega}{T_0}\wh{J}_z+\zeta \wh{Q}\right],
\end{equation}
with $\wh{J}_z$ the angular momentum operator along $z$.
As it is known, this expression represents a fluid at thermodynamic equilibrium rigidly rotating around the $z$ direction with constant angular velocity $\omega$ \cite{landau2013statistical}, therefore it is called \textit{global thermodynamic equilibrium with rotation}.
This was used for instance in \cite{Vilenkin:1980zv, Becattini:2011ev} to study quantum effects in rotating relativistic matter and reworked more thoroughly in \cite{Becattini:2014yxa}.

The last and less known case is that of a thermal vorticity with only a longitudinal (or ``electric'') component
\begin{equation}
    b_{\mu}=\frac{1}{T_0}(1,{\bf 0}),\qquad
    \varpi_{\mu \nu}=\frac{a}{T_0}\left(g_{0\nu}g_{3\mu}-g_{0\mu}g_{3\nu}\right),
\end{equation}
where $a$ is a thermodynamic parameter with the dimensions of an acceleration.
The four-temperature then reads
\begin{equation}\label{gteq02}
    \beta_{\mu}=\frac{1}{T_0}\left(1+az,0,0,-at\right),
\end{equation}
hence the density operator
\begin{equation}\label{gteq01}
    \wh{\rho}=\frac{1}{Z}\exp \left[-\frac{\wh H}{T_0}+\frac{a}{T_0}\wh{K}_z+\zeta \wh{Q}\right],
\end{equation}
with $\wh{K}_z$ the boost operator along $z$.
As we shall shortly see, this expression represents a fluid at thermodynamic equilibrium with acceleration field of constant magnitude along the flow, therefore it is called \textit{global thermodynamic equilibrium with acceleration}.
This was studied in detail in \cite{Becattini:2017ljh}, where it was shown that the Unruh effect is indeed recovered, and partly also in \cite{Becattini:2019poj}, where the calculation of the thermal expectation value of the energy-momentum tensor was completed and the entropy current calculated.

Equations \eqref{gteq03} and \eqref{gteq01} are the only two independent global thermodynamic equilibria in Minkowski spacetime with non-vanishing thermal vorticity, all other cases are combinations thereof.
For a discussion on the general case with both rotation and acceleration, see \cite{Korsbakken:2004bv}.

We conclude this part with the following remark.
As we mentioned in Chapter \ref{chapter:zubarev}, the symmetries of the density operator constrain the form of the thermal expectation values calculated with it.
In general, by looking at equation \eqref{gteq16}, the thermal expectation value of any local operator $\wh{O}$ can depend on $b_{\mu}$, $\varpi_{\mu \nu}$, $x_{\mu}$ and $g_{\mu \nu}$.
Concerning the dependence on $x_{\mu}$, let us indicate the translation operator as $\wh{\sf T}(x)\equiv \exp[-ix_{\mu}\wh{P}^{\mu}]$, hence
\begin{equation}\label{gteq17}
    \begin{split}
        \langle \wh{O}(x)\rangle=&
        \tr \left(\wh{\rho}\wh{\sf T}(x)\wh{O}(0)\wh{\sf T}^{-1}(x)\right)=
        \tr \left(\wh{\sf T}^{-1}(x)\wh{\rho}\wh{\sf T}(x)\wh{O}(0)\right)\\
        =&\frac{1}{Z}\tr \left(\exp \left[b_{\mu}\wh{P}^{\mu}+\frac{1}{2}\varpi_{\mu \nu}\wh{\sf T}^{-1}(x)\wh{J}^{\mu \nu}\wh{\sf T}(x)\right]\wh{O}(0)\right)\\
        =&\frac{1}{Z}\tr \left(\exp \left[-\beta_{\mu}(x)\wh{P}^{\mu}+\frac{1}{2}\varpi_{\mu \nu}\wh{J}^{\mu \nu}\right]\wh{O}(0)\right)=
        \langle \wh{O}(0)\rangle_{\beta_{\mu}(x)},
    \end{split}
\end{equation}
where the known relation of the Poincaré algebra was used
\begin{equation}
    \wh{\sf T}^{-1}(x)\wh{J}_{\mu \nu}\wh{\sf T}(x)=\wh{J}_{\mu \nu}-x_{\nu}\wh{P}_{\mu}+x_{\mu}\wh{P}_{\nu}.
\end{equation}
Equation \eqref{gteq17} implies that the dependence on the spacetime point of the thermal expectation value of any quantum operator cannot be arbitrary, but it can be only through the four-temperature.
This property will come in handy later on.


\subsection{Global thermodynamic equilibrium with acceleration}

Let us now focus on the configuration of global thermodynamic equilibrium with acceleration.
From here on, we will set $\zeta=0$ for simplicity without loss of generality, namely we will forget about the charged current and consider only the energy-momentum tensor.
The expression of the density operator is then
\begin{equation}\label{gteq04}
    \wh{\rho}=\frac{1}{Z}\exp \left[-\frac{\wh{H}}{T_0}+\frac{a}{T_0}\wh{K}_z\right].
\end{equation}
From the general discussion in the previous Chapter, we recall that in the argument of the exponential function there appear the quantum operators corresponding to the conserved charges.
One particular feature of \eqref{gteq04} is that the Hamiltonian $\wh{H}$ and the boost operator $\wh{K}_z$ do not commute with each other, and yet they are both conserved.
This occurs because the boost operator is explicitly time-dependent
\begin{equation}
    \wh{K}_z=\wh{J}_{30}=t\wh{P}_z-\int \di^3{\rm x}\,z\wh{T}^{00},
\end{equation}
so its Heisenberg equation reads
\begin{equation}
    i\frac{\di \wh{K}_z}{\di t}=[\wh{K}_z,\wh{H}]+i\frac{\de \wh{K}_z}{\de t}=-i\wh{P}_z+i\wh{P}_z=0.
\end{equation}
Although possible at a mathematical level, the question arises if this state can be physically realized.
As we mentioned in the last Chapter, the global thermodynamic equilibrium density operator is in fact an approximation of the local thermodynamic equilibrium one at first order in the derivatives of the thermodynamic fields \cite{Becattini:2014yxa}.
This makes \eqref{gteq04} a reasonable approximation for describing fluids at local thermodynamic equilibrium with non-vanishing acceleration (and vorticity), such as the quark-gluon plasma, in the hydrodynamic limit.

In order to work out the kinematic quantities, it is convenient to shift the origin of the $z$ coordinate by defining $z'\equiv z+1/a$, so that the four-temperature in \eqref{gteq02} simply reads
\begin{equation}\label{gteq08}
    \beta^{\mu}=\frac{a}{T_0}(z',0,0,t).
\end{equation}
Then, according to \eqref{zubeq11}, the velocity field and proper temperature defined by its direction and inverse magnitude are respectively given by
\begin{equation}\label{gteq09}
    u^{\mu}=\frac{1}{\sqrt{{z'}^2-t^2}}(z',0,0,t),\qquad
    T=\frac{T_0}{a\sqrt{{z'}^2-t^2}}.
\end{equation}
It is easily realized that the flow lines, namely the curves with $u^{\mu}$ as tangent vector field, are hyperbolae of constant values of ${z'}^2-t^2$, and the proper temperature is constant along them.
This should be no surprise, being in fact a general feature of Killing vectors to have constant magnitude along their integral curves.
Now, by using the definition \eqref{eqrelhydro23} of the acceleration field, we readily obtain the expression
\begin{equation}\label{gteq37}
    A^{\mu}=\frac{1}{{z'}^2-t^2}(t,0,0,z'),
\end{equation}
which is clearly orthogonal to the four-temperature and whose magnitude is constant along the flow lines.
In other words, the density operator \eqref{gteq04} represents a fluid at global thermodynamic equilibrium with acceleration field of constant magnitude along the flow, hence the name ``global thermodynamic equilibrium with acceleration''.
Furthermore, referring to the general decomposition \eqref{gteq07} of the thermal vorticity, in our specific case at hand the vorticity field is identically null, so we are actually left with
\begin{equation}\label{gteq18}
    \varpi_{\mu \nu}=\alpha_{\mu}u_{\nu}-\alpha_{\nu}u_{\mu},
\end{equation}
where $\alpha^{\mu}$ has constant magnitude in the whole spacetime
\begin{equation}\label{gteq23}
    \alpha^2=\frac{A^2}{T^2}=-\frac{a^2}{T_0^2}.
\end{equation}

\begin{figure}
    \begin{center}
	    \includegraphics[width=0.745\textwidth]{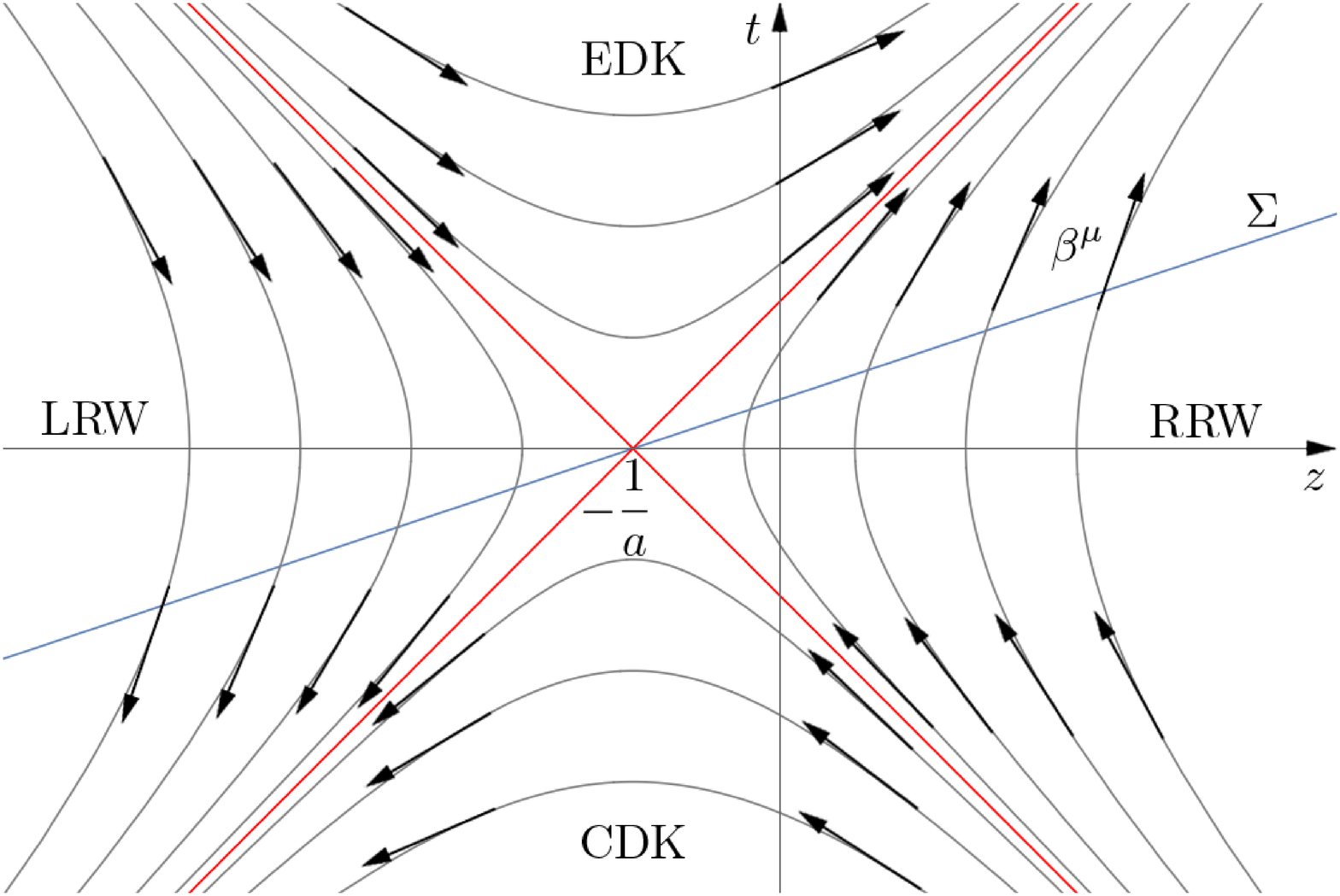}
	    \caption{2-dimensional $(t,z)$ section of Minkowski spacetime.
	    The hyperbolae are the flow lines of the four-temperature, the latter being the tangent vector field.
	    The red lines crossing at $z'=0$, i.e.\ $z=-1/a$, are the bifurcated Killing horizon of $\beta^{\mu}$ splitting the spacetime into four different subspaces: the right Rindler wedge (RRW), the left Rindler wedge (LRW), the expanding degenerate Kasner universe (EDK) and the contracting degenerate Kasner universe (CDK).
	    The four-temperature is globally timelike and future-oriented only in the right Rindler wedge.
	    The blue line through $z'=0$ and contained in the two Rindler wedges is a hyperplane $\Sigma$ orthogonal to $\beta^{\mu}$: such hyperplanes can be used to foliate the subspaces where the four-temperature is timelike.}
	    \label{fig:wedges}
	\end{center}
\end{figure}

As for the thermodynamics, the key point is that the four-temperature \eqref{gteq08} is not globally timelike for it has a bifurcated Killing horizon at $|z'|=t$ splitting Minkowski spacetime into four different subspaces as shown in Figure \ref{fig:wedges}:
\begin{itemize}
    \item $|t|<z'$ is called the \textit{Right Rindler Wedge} (RRW),
    \item $|t|<-z'$ is called the \textit{Left Rindler Wedge} (LRW),
    \item $t>|z'|$ is called the \textit{Expanding Degenerate Kasner Universe} (EDK),
    \item $t<-|z'|$ is called the \textit{Contracting Degenerate Kasner Universe} (EDK).
\end{itemize}
In particular, in order to make hydrodynamics and thermodynamics out of $\beta^{\mu}$, we have to figure out in which subspace it is both timelike and future-oriented globally, so that the interpretation of $u^{\mu}$ and $T$ as a velocity field and a proper temperature respectively makes sense.
By looking at Figure \ref{fig:wedges}, we can tell that the subspace we are looking for is the right Rindler wedge, so that is where we must restrict.
An observer moving along a constant-acceleration path, that is along the flow, is called a \textit{Rindler observer}.
Note that the expression of the proper temperature in equation \eqref{gteq09} is but an instance of Tolman's law \cite{Buchholz:2015fqa}, but it is most naturally obtained in the Zubarev approach by demanding the four-temperature to be a Killing vector field.

\subsection{Factorization of the density operator}

In \cite{Becattini:2017ljh}, it is shown that the density operator \eqref{gteq04} has an important factorization property.
Rather than \eqref{gteq04}, it is convenient to consider the general expression \eqref{zubeq28}, which still holds of course, and rewrite it by defining $\wh{\Pi}$ as
\begin{equation}\label{gteq10}
    \wh{\rho}=\frac{1}{Z}\exp \left[-\frac{\wh{\Pi}}{T_0}\right],\qquad
    \wh{\Pi}\equiv T_0\int_{\Sigma}\di \Sigma_{\mu}\wh{T}^{\mu \nu}\beta_{\nu}=\wh{H}-a\wh{K}_z.
\end{equation}
At global thermodynamic equilibrium, this is independent on the choice of the spacelike hypersurface $\Sigma$ of the foliation.
To define the foliation, we can exploit the fact that the four-temperature \eqref{gteq08} fulfills the vorticity-free condition \eqref{zubeq07}, that is
\begin{equation}
    \epsilon^{\mu \nu \rho \sigma}\beta_{\sigma}\de_{\nu}\beta_{\rho}=0.
\end{equation}
In particular, the family of 3-dimensional spacelike hypersurfaces orthogonal to it is given by the hyperplanes through $z'=0$ and entirely contained in the right and left Rindler wedges, as shown in Figure \ref{fig:wedges}, so let $\Sigma$ be any of them.
Thus, let us decompose the hyperplane as $\Sigma=\Sigma_{\rm R}\cup \Sigma_{\rm L}$ with $\Sigma_{\rm R}$ and $\Sigma_{\rm L}$ entirely contained in the right and left Rindler wedge respectively, namely $\Sigma_{\rm R}$ being the $z'>0$ part of $\Sigma$ and $\Sigma_{\rm L}$ the $z'<0$ part.
This way, $\wh{\Pi}$ is given by the sum of two terms, each depending only on the field degrees of freedom in one of the two Rindler wedges
\begin{equation}
    \wh{\Pi}=\wh{\Pi}_{\rm R}-\wh{\Pi}_{\rm L}
\end{equation}
\begin{equation}
    \wh{\Pi}_{\rm R}\equiv T_0\int_{\Sigma_{\rm R}}\di \Sigma_{\mu}\wh{T}^{\mu \nu}\beta_{\nu},\qquad
    \wh{\Pi}_{\rm L}\equiv -T_0\int_{\Sigma_{\rm L}}\di \Sigma_{\mu}\wh{T}^{\mu \nu}\beta_{\nu}.
\end{equation}
The minus sign in the definition of $\wh{\Pi}_{\rm L}$ is due to the fact that $n^{\mu}$ orthogonal to $\Sigma_{\rm L}$ has opposite orientation with respect to $\beta^{\mu}$ in the left Rindler wedge.
Now the key point is the following.
The only contribution to the commutator $[\wh{\Pi}_{\rm R},\wh{\Pi}_{\rm L}]$ stems from $z'=0$, which is the only point where the quantum field operators in the energy-momentum tensor have non-vanishing commutators.
However, the four-temperature altogether vanishes there, as we can convince ourselves by noticing that the proper temperature \eqref{gteq09} diverges there, or simply just by looking at Figure \ref{fig:wedges}, therefore the commutator is in fact null.
This implies that the density operator is factorized as
\begin{equation}
    \wh{\rho}=
    \frac{1}{Z}\exp \left[-\frac{\wh{\Pi}_{\rm R}-\wh{\Pi}_{\rm L}}{T_0}\right]=
    \frac{1}{Z}\exp \left[-\frac{\wh{\Pi}_{\rm R}}{T_0}\right]\exp \left[\frac{\wh{\Pi}_{\rm L}}{T_0}\right].
\end{equation}
Formally, since $\wh{\Pi}_{\rm R}$ and $\wh{\Pi}_{\rm L}$ involve only field degrees of freedom each in its corresponding Rindler wedge, in the Hilbert space of the field states we actually have $\wh{\Pi}_{\rm R}\equiv \wh{\Pi}_{\rm R}\otimes \wh{I}_{\rm L}$ and $\wh{\Pi}_{\rm L}\equiv \wh{I}_{\rm R}\otimes \wh{\Pi}_{\rm L}$, with $\wh{I}_{\rm R}$ and $\wh{I}_{\rm L}$ the identities in the right and left Rindler wedge respectively.
In turn, this implies that the partition function is factorized as well
\begin{equation}
    Z=\tr \left(\exp \left[-\frac{\wh{\Pi}_{\rm R}-\wh{\Pi}_{\rm L}}{T_0}\right]\right)=
    \tr_{\rm R}\left(\exp \left[-\frac{\wh{\Pi}_{\rm R}}{T_0}\right]\right)
    \tr_{\rm L}\left(\exp \left[\frac{\wh{\Pi}_{\rm L}}{T_0}\right]\right)\equiv Z_{\rm R}Z_{\rm L},
\end{equation}
where by $\tr_{\rm R}$ and $\tr_{\rm L}$ we mean that the trace is calculated only on the Hilbert space spanned by the field degrees of freedom in the right and left Rindler wedge respectively, which is called a \textit{partial trace}.
In summary, the density operator becomes factorized into two density operators commuting with each other
\begin{equation}\label{gteq41}
    \wh{\rho}=\wh{\rho}_{\rm R}\otimes \wh{\rho}_{\rm L},\qquad
    [\wh{\rho}_{\rm R},\wh{\rho}_{\rm L}]=0
\end{equation}
each involving the field degrees of freedom in either of the Rindler wedges
\begin{equation}\label{gteq12}
    \wh{\rho}_{\rm R}=
    \frac{1}{Z_{\rm R}}\exp \left[-\frac{\wh{\Pi}_{\rm R}}{T_0}\right]=
    \frac{1}{Z_{\rm R}}\exp \left[-\int_{\Sigma_{\rm R}}\di \Sigma_{\mu}\wh{T}^{\mu \nu}\beta_{\nu}\right],
\end{equation}
\begin{equation}
    \wh{\rho}_{\rm L}=
    \frac{1}{Z_{\rm L}}\exp \left[\frac{\wh{\Pi}_{\rm L}}{T_0}\right]=
    \frac{1}{Z_{\rm L}}\exp \left[-\int_{\Sigma_{\rm L}}\di \Sigma_{\mu}\wh{T}^{\mu \nu}\beta_{\nu}\right].
\end{equation}
They are obtained by taking the partial trace of the whole density operator on the opposite wedge, so, referring to the terminology of Quantum Information Theory, they are called \textit{reduced density operators}
\begin{equation}\label{gteq42}
    \wh{\rho}_{\rm R}=\tr_{\rm L}(\wh{\rho}),\qquad
    \wh{\rho}_{\rm L}=\tr_{\rm R}(\wh{\rho}).
\end{equation}
Thus, the two Rindler wedges are completely disconnected.
As a consequence, the thermal expectation value of an operator $\wh{O}(x)$ with $x$ in the right Rindler wedge, for instance, will only depend on the reduced density operator $\wh{\rho}_{\rm R}$, regardless of the field states in the left Rindler wedge
\begin{equation}
    \langle \wh{O}(x)\rangle=\tr(\wh{\rho}\wh{O}(x))=\tr_{\rm R}(\wh{\rho}_{\rm R}\wh{O}(x)).
\end{equation}
Equation \eqref{gteq10} defining the operator $\wh{\Pi}$ is reminiscent of the homogeneous global thermodynamic equilibrium density operator \eqref{gteq11}.
Indeed, by calculating the commutator with the quantum field operator, one can show that $\wh{\Pi}$ is the generator of translations along the flow in Minkowski spacetime \cite{Korsbakken:2004bv}, same for $\wh{\Pi}_{\rm R}$ and $\wh{\Pi}_{\rm L}$ in their respective Rindler wedges \cite{Becattini:2017ljh}.


\section{Thermal expectation values and the Unruh effect}
\label{sec:gte_tev_unruh}

With the density operator, we are now in a position to calculate thermal expectation values of quantum operators.
In particular, we are interested in that of the energy-momentum tensor, for it is both of physical concern in its own right and also necessary for the calculation of the entropy current.
As already mentioned, the quantum operators are built with the quantum field operators, and as such they depend on the Quantum Field Theory underlying the hydrodynamic theory.
In this work, we will consider the simple case of a free real scalar field in the right Rindler wedge.
As we shall see, the Unruh effect will emerge from thermal expectation values in a natural way.

\subsection{Free scalar field theory in the right Rindler wedge}

The problem of a free real scalar field in the right Rindler wedge is indeed a well-known one in Physics.
Here, we just present a short summary highlighting the characteristics most salient for this work, for more details we refer to the review \cite{Crispino:2007eb}.

In order to build the energy-momentum tensor operator and calculate its thermal expectation value, we first need the expression of the quantum field operator, which is obtained as a solution of its equation of motion.
The Lagrangian density $\wh{\cal L}$ of a free real scalar field $\wh{\psi}$ of mass $m$ in Minkowski spacetime is
\begin{equation}\label{eq19}
    \wh{\cal L}=\frac{1}{2}\left(g^{\mu \nu}\partial_{\mu}\wh{\psi} \partial_{\nu}\wh{\psi}-m^2\wh{\psi}^2\right)
\end{equation}
and the corresponding equation of motion is the Klein-Gordon equation
\begin{equation}
    \left(\Box+m^2\right)\wh{\psi}=0.
\end{equation}
Provided initial data, the solutions of Klein-Gordon equation can be uniquely determined in Minkowski spacetime, for it is globally hyperbolic with constant-time hypersurfaces as Cauchy surfaces.
Given any two solutions $\phi_1$ and $\phi_2$, we can build the \textit{Klein-Gordon inner product} to define orthogonality and normalization of the solutions
\begin{equation}
    (\phi_1,\phi_2)_{\rm KG}\equiv i\int_{\Sigma}{\rm d}\Sigma_{\mu}\left(\phi_1^*\de^{\mu}\phi_2-\phi_2\de^{\mu}\phi_1^*\right),
\end{equation}
with $\Sigma$ a spacelike hypersurface with future-oriented unit orthogonal vector field.
This is independent of $\Sigma$, and also independent of time if $\Sigma$ is a constant-time hypersurface.
Let $\{u_i\}$ be a set of solutions orthonormalized according to the above inner product, that is
\begin{equation}
    (u_i,u_j)_{\rm KG}=\delta_{ij}=-(u_i^*,u_j^*)_{\rm KG},\qquad (u_i^*,u_j)_{\rm KG}=0=(u_i,u_j^*)_{\rm KG}.
\end{equation}
The general solution of the Klein-Gordon equation can then be expanded as
\begin{equation}
    \wh{\psi}=\sum_i\left(u_i\wh{a}_i+u_i^*\wh{a}_i^{\dagger}\right),
\end{equation}
where, $\wh{a}_i^{\dagger}$ and $\wh{a}_i$ are the creation and annihilation operators respectively.
Using the orthonormality of the solutions $u_i$, one can show that
\begin{equation}
    \wh{a}_i=(u_i,\wh{\psi})_{\rm KG},\qquad \wh{a}_i^{\dagger}=-(u_i^*,\wh{\psi})_{\rm KG},
\end{equation}
which, together with the canonical commutation relations, imply as expected
\begin{equation}
    [\wh{a}_i,\wh{a}_j^{\dagger}]=\delta_{ij},\qquad [\wh{a}_i,\wh{a}_j]=[\wh{a}_i^{\dagger},\wh{a}_j^{\dagger}]=0.
\end{equation}
The vacuum state is defined by requiring it to be annihilated by all the annihilation operators, and the Fock space is built by applying the creation operators to it.
The annihilation operators, in turn, are determined by the solutions $u_i$, so it is actually their choice that determines the vacuum.
Such choice is not unique, and so is the vacuum in principle, however Minkowski spacetime is \textit{static}.
In a static spacetime it is natural for the solutions $u_i$ to have a time-dependence of the form ${\rm e}^{-i\omega_it}$, where $\omega_i$ are positive constants interpreted as the energies of the particles with respect to the future-oriented Killing vector $\partial_t$.
For this reason, the solutions $u_i$ are called \textit{positive-frequency modes}, and the $u_i^*$ \textit{negative-frequency modes}.
In Minkowski spacetime, the quantum field expansion usually reads
\begin{equation}\label{gteq27}
    \wh{\psi}(x)=\int_{-\infty}^{+\infty}\di^3{\rm p}\left(v_{\bf p}\wh{a}_{\bf p}+v^*_{\bf p}\wh{a}^{\dagger}_{\bf p}\right),
\end{equation}
where the positive-frequency modes are the plane waves
\begin{equation}\label{gteq49}
    v_{\bf p}=\frac{1}{\sqrt{(2\pi)^32\omega_{\bf p}}}{\rm e}^{-i(\omega_{\bf p} t-{\bf p}\cdot {\bf x})}.
\end{equation}
If a spacetime is both static and globally hyperbolic, the choice of positive-frequency modes with time-dependence of the form ${\rm e}^{-i\omega_it}$ leads to a natural vacuum state that preserves time-translation symmetry, called the \textit{static vacuum}.
The static vacuum in Minkowski spacetime is called the \textit{Minkowski vacuum}, indicated as $|0_{\rm M}\rangle$, that is the state annihilated by all the annihilation operators $\wh{a}_{\bf p}$.

\begin{figure}
    \begin{center}
	    \includegraphics[width=0.75\textwidth]{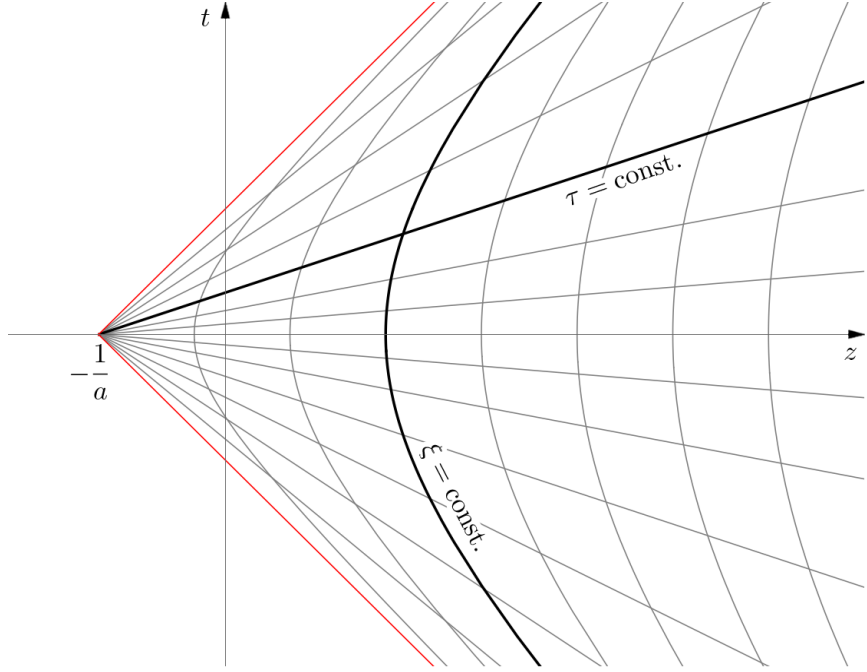}
	    \caption{2-dimensional section of the right Rindler wedge in the $(t,z)$ plane, spanned by the Rindler coordinates $(\tau,\xi)$.
	    The straight lines through $z=-1/a$, i.e.\ $z'=0$, are $\tau={\rm const.}$ hypersurfaces, while the hyperbolae are $\xi={\rm const.}$ hypersurfaces.}
	    \label{fig:rindlercoord}
	\end{center}
\end{figure}

The right Rindler wedge is also a static and globally hyperbolic spacetime, and it possesses a timelike future-oriented Killing vector given by the boost generator $z'\partial_t+t\partial_{z'}$ playing the role of time-translations generator, so we can solve the Klein-Gordon equation therein as well.
To do so, it is convenient to introduce a set of hyperbolic coordinates called the \textit{Rindler coordinates} $(\tau,\xT,\xi)$, where the ``transverse coordinates'' $\xT \equiv (x,y)$ are the same as Minkowski's, and $(\tau,\xi)$ are related to $(t,z')$ by
\begin{equation}
    \tau \equiv \frac{1}{2a}\log \left(\frac{z'+t}{z'-t}\right),\qquad \xi \equiv \frac{1}{2a}\log \left[a^2\left({z'}^2-t^2\right)\right]
\end{equation}
whose inverse read
\begin{equation}
    t=\frac{{\rm e}^{a\xi}}{a}\sinh(a\tau),\qquad
    z'=\frac{{\rm e}^{a\xi}}{a}\cosh(a\tau)
\end{equation}
spanning indeed the right Rindler wedge, as shown in Figure \ref{fig:rindlercoord}.
Plugging them into the Klein-Gordon equation, the positive-frequency modes are obtained
\begin{equation}\label{gteq30}
    u_{\omega,\kT}(\tau,\xT,\xi)\equiv \sqrt{\frac{1}{4\pi^4a}\sinh \left(\frac{\pi \omega}{a}\right)}{\rm K}_{i\frac{\omega}{a}}\left(\frac{\mT{\rm e}^{a\xi}}{a}\right){\rm e}^{-i(\omega \tau-\kT \cdot \xT)},
\end{equation}
where ${\rm K}_{i\frac{\omega}{a}}$ are the modified Bessel functions, $\mT$ is the \textit{transverse mass}
\begin{equation}
    \mT=\sqrt{\kT^2+m^2},
\end{equation}
$\kT \equiv (k_x,k_y)$ is the ``transverse momentum'' and $\omega \ge 0$ is the frequency.
Together with the negative-frequency modes $u_{\omega,\kT}^*$, they are orthonormalized according to the Klein-Gordon inner product.
Thus, the field can be expanded as
\begin{equation}\label{gteq13}
    \wh{\psi}(\tau,\xT,\xi)=\int_0^{+\infty}\di \omega \int_{-\infty}^{+\infty}\di^2{\rm k_T}\left(u_{\omega,\kT}\wh{a}^{\rm R}_{\omega,\kT}+u^*_{\omega,\kT}\wh{a}^{\rm R\dagger}_{\omega,\kT}\right),
\end{equation}
where $\wh{a}^{\rm R\dagger}_{\omega,\kT}$ and $\wh{a}^{\rm R}_{\omega,\kT}$ are the creation and annihilation operators respectively, given by
\begin{equation}
    \wh{a}^{\rm R}_{\omega,\kT}=(u_{\omega,\kT},\wh{\psi})_{\rm KG},\qquad \wh{a}^{\rm R\dagger}_{\omega,\kT}=-(u^*_{\omega,\kT},\wh{\psi})_{\rm KG}.
\end{equation}
With the canonical commutation relations, the above equations imply the usual algebra
\begin{equation}\label{gteq15}
    \begin{split}
        [\wh{a}^{\rm R}_{\omega,\kT},\wh{a}^{\rm R\dagger}_{\omega',\kT'}]=&\delta(\omega-\omega')\,\delta^2(\kT-\kT'),\\
        [\wh{a}^{\rm R}_{\omega,\kT},\wh{a}^{\rm R}_{\omega',\kT'}]=&0=[\wh{a}^{\rm R\dagger}_{\omega,\kT},\wh{a}^{\rm R\dagger}_{\omega',\kT'}].
    \end{split}
\end{equation}
%
In the left Rindler wedge, the coordinates $t\equiv a^{-1}{\rm e}^{a\bar{\xi}}\sinh(a\bar{\tau})$ and $z'\equiv -a^{-1}{\rm e}^{a\bar{\xi}}\cosh(a\bar{\tau})$ are introduced in order to solve the Klein-Gordon equation.
An expansion analogous to \eqref{gteq13} holds there, with the important difference that the role of the positive- and negative-frequency modes is interchanged as a consequence of the fact that the boost generator, acting as a time-translations generator, is past-oriented
\begin{equation}\label{gteq26}
    \wh{\psi}(\bar{\tau},\xT,\bar{\xi})=\int_0^{+\infty}\di \omega \int_{-\infty}^{+\infty}\di^2{\rm k_T}\left(u_{\omega,\kT}^*\wh{a}^{\rm L}_{\omega,\kT}+u_{\omega,\kT}\wh{a}^{\rm L\dagger}_{\omega,\kT}\right),
\end{equation}
where $u_{\omega,\kT}$ has the same expression as \eqref{gteq30} with $(\tau,\xi)$ replaced by $(\bar{\tau},\bar{\xi})$.
%
The static vacuum in the Rindler wedges, called the \textit{Rindler vacuum} and indicated as $|0_{\rm R}\rangle$, is the state annihilated by all $\wh{a}^{\rm R}_{\omega,\kT}$ and $\wh{a}^{\rm L}_{\omega,\kT}$, namely $\wh{a}^{\rm R}_{\omega,\kT}|0_{\rm R}\rangle=0=\wh{a}^{\rm L}_{\omega,\kT}|0_{\rm R}\rangle$.

As for the energy-momentum tensor, we consider the canonical one
\begin{equation}\label{gteq19}
    \wh{T}^{\mu \nu}=\frac{1}{2}\left(\de^{\mu}\wh{\psi}\de^{\nu}\wh{\psi}+\de^{\nu}\wh{\psi}\de^{\mu}\wh{\psi}\right)-g^{\mu \nu}\wh{\cal L},
\end{equation}
but this is not the only possible choice in principle.
Plugging the quantum field expansion into it, and then the result into \eqref{gteq12}, the following expressions of the reduced density operators in the Rindler wedges are obtained
\begin{equation}\label{gteq28}
    \wh{\rho}_{\rm R}=\frac{1}{Z_{\rm R}}\exp \left[-\frac{\wh{\Pi}_{\rm R}}{T_0}\right],\qquad
    \wh{\Pi}_{\rm R}=\int_0^{+\infty}\di \omega \int_{-\infty}^{+\infty}\di^2{\rm k_T}\,\omega \wh{a}^{\rm R\dagger}_{\omega,\kT}\wh{a}^{\rm R}_{\omega,\kT}
\end{equation}
\begin{equation}\label{gteq29}
    \wh{\rho}_{\rm L}=\frac{1}{Z_{\rm L}}\exp \left[\frac{\wh{\Pi}_{\rm L}}{T_0}\right],\qquad
    \wh{\Pi}_{\rm L}=\int_0^{+\infty}\di \omega \int_{-\infty}^{+\infty}\di^2{\rm k_T}\,\omega \wh{a}^{\rm L\dagger}_{\omega,\kT}\wh{a}^{\rm L}_{\omega,\kT}.
\end{equation}
They are diagonal in the Rindler creation and annihilation operators, so they have the same vacuum as the quantum field \eqref{gteq13} and \eqref{gteq26}, the Rindler vacuum $|0_{\rm R}\rangle$.
Moreover, their being diagonal is very convenient for it allows thermal expectation values to be calculated by using standard methods.


\subsection{Thermal expectation values}

Operators of physical interest, such as the energy-momentum tensor, are quadratic in the quantum field, therefore their thermal expectation values are given in terms of thermal expectation values of products of creation and annihilation operators.
These can be calculated in the Rindler wedges by using standard methods, thanks to the reduced density operators \eqref{gteq28} and \eqref{gteq29} being diagonal.
In particular, in the right Rindler wedge, which is the subspace of our interest, they are \cite{Becattini:2017ljh}
\begin{subequations}
    \begin{align}
        \langle \wh{a}^{\rm R\dagger}_{\omega,\kT}\wh{a}^{\rm R}_{\omega',\kT'}\rangle=&n_{\rm B}\,\delta(\omega-\omega')\,\delta^2(\kT-\kT')\label{gteq14a}\\
        \langle \wh{a}^{\rm R}_{\omega,\kT}\wh{a}^{\rm R\dagger}_{\omega',\kT'}\rangle=&\left(n_{\rm B}+1\right)\delta(\omega-\omega')\,\delta^2(\kT-\kT')\label{gteq14b}\\
        \langle \wh{a}^{\rm R}_{\omega,\kT}\wh{a}^{\rm R}_{\omega',\kT'}\rangle=&0=\langle \wh{a}^{\rm R\dagger}_{\omega,\kT}\wh{a}^{\rm R\dagger}_{\omega',\kT'}\rangle \label{gteq14c},
    \end{align}
\end{subequations}
where $n_{\rm B}$ is the Bose-Einstein distribution
\begin{equation}
    n_{\rm B}=\frac{1}{{\rm e}^{\omega/T_0}-1}.
\end{equation}
We prove the first one as an example.
For the sake of simplicity, let us temporarily introduce $\beta_0\equiv 1/T_0$ as a new parameter unrelated to the 0-component of $\beta_{\mu}$, and define
\begin{equation}\label{gteq50}
    \wh{a}^{\rm R\dagger}_{\omega,\kT}(\beta_0)\equiv {\rm e}^{-\beta_0\wh{\Pi}_{\rm R}}\wh{a}^{\rm R\dagger}_{\omega,\kT}{\rm e}^{\beta_0\wh{\Pi}_{\rm R}}.
\end{equation}
The derivative with respect to $\beta_0$ reads
\begin{equation}
    \begin{split}
        \frac{\de}{\de \beta_0}\wh{a}^{\rm R\dagger}_{\omega,\kT}(\beta_0)=&
        -[\wh{\Pi}_{\rm R},\wh{a}^{\rm R\dagger}_{\omega,\kT}(\beta_0)]=
        -{\rm e}^{-\beta_0\wh{\Pi}_{\rm R}}[\wh{\Pi}_{\rm R},\wh{a}^{\rm R\dagger}_{\omega,\kT}]{\rm e}^{\beta_0\wh{\Pi}_{\rm R}}\\
        =&-\omega {\rm e}^{-\beta_0\wh{\Pi}_{\rm R}}\wh{a}^{\rm R\dagger}_{\omega,\kT}{\rm e}^{\beta_0\wh{\Pi}_{\rm R}}=
        -\omega \wh{a}^{\rm R\dagger}_{\omega,\kT}(\beta_0),
    \end{split}
\end{equation}
whose solution is
\begin{equation}
    \wh{a}^{\rm R\dagger}_{\omega,\kT}(\beta_0)={\rm e}^{-\beta_0\omega }\wh{a}^{\rm R\dagger}_{\omega,\kT}.
\end{equation}
By exploiting the cyclic property of the trace, we have
\begin{equation}
    \begin{split}
        \langle \wh{a}^{\rm R\dagger}_{\omega,\kT}\wh{a}^{\rm R}_{\omega',\kT'}\rangle=&
        \tr_{\rm R}\left(\wh{\rho}_{\rm R}\wh{a}^{\rm R\dagger}_{\omega,\kT}\wh{a}^{\rm R}_{\omega',\kT'}\right)=
        \tr_{\rm R}\left(\frac{{\rm e}^{-\beta_0\wh{\Pi}_{\rm R}}}{Z_{\rm R}}\wh{a}^{\rm R\dagger}_{\omega,\kT}\wh{a}^{\rm R}_{\omega',\kT'}\right)\\
        =&\frac{1}{Z_{\rm R}}\tr_{\rm R}\left(\wh{a}^{\rm R}_{\omega',\kT'}{\rm e}^{-\beta_0\wh{\Pi}_{\rm R}}\wh{a}^{\rm R\dagger}_{\omega,\kT}{\rm e}^{\beta_0\wh{\Pi}_{\rm R}}{\rm e}^{-\beta_0\wh{\Pi}_{\rm R}}\right)\\
        =&\tr_{\rm R}\left(\frac{{\rm e}^{-\beta_0\wh{\Pi}_{\rm R}}}{Z_{\rm R}}\wh{a}^{\rm R}_{\omega',\kT'}\wh{a}^{\rm R\dagger}_{\omega,\kT}(\beta_0)\right)=
        {\rm e}^{-\beta_0\omega}\tr_{\rm R}\left(\wh{\rho}_{\rm R}\wh{a}^{\rm R}_{\omega',\kT'}\wh{a}^{\rm R\dagger}_{\omega,\kT}\right)\\
        =&{\rm e}^{-\beta_0\omega}\langle \wh{a}^{\rm R\dagger}_{\omega,\kT}\wh{a}^{\rm R}_{\omega',\kT'}+[\wh{a}^{\rm R}_{\omega',\kT'},\wh{a}^{\rm R\dagger}_{\omega,\kT}]\rangle \\
        =&{\rm e}^{-\beta_0\omega}\left(\langle \wh{a}^{\rm R\dagger}_{\omega,\kT}\wh{a}^{\rm R}_{\omega',\kT'}\rangle+\delta(\omega-\omega')\,\delta^2(\kT-\kT')\right),
    \end{split}
\end{equation}
thus
\begin{equation}\label{gteq51}
    \langle \wh{a}^{\rm R\dagger}_{\omega,\kT}\wh{a}^{\rm R}_{\omega',\kT'}\rangle \left(1-{\rm e}^{-\beta_0\omega}\right)=\delta(\omega-\omega')\,\delta^2(\kT-\kT'),
\end{equation}
hence the result \eqref{gteq14a}.
Equations \eqref{gteq14b} and \eqref{gteq14c} are proven in the same way.

The $+1$ term in \eqref{gteq14b} stems from the commutation relations \eqref{gteq15} and gives rise to divergencies, therefore some renormalization will be needed at some point.
More on this will be said shortly, for now we just want to point out that, in the Rindler vacuum, the above thermal expectation values have the same form evaluated at $T_0=0$.
This means that renormalization in the Rindler vacuum is but the subtraction of the $T_0=0$ contribution, which is tantamount to canceling the $+1$ term.

The structure of thermal expectation values depends on the symmetries of the density operator and by the quantities at disposal.
At global thermodynamic equilibrium, these are $b_{\mu}$, $\varpi_{\mu \nu}$, $x_{\mu}$ and $g_{\mu \nu}$, but as shown in \eqref{gteq17}, the dependence on $x_{\mu}$ can only be through $\beta_{\mu}$, which also includes $b_{\mu}$.
As for the thermal vorticity $\varpi_{\mu \nu}$, in our case of global thermodynamic equilibrium with acceleration, this is decomposed in terms of $u_{\mu}$ and $\alpha_{\mu}$ as in \eqref{gteq18}, and the dependence on $u_{\mu}$ is the same as on $\beta_{\mu}$.
The only non-vanishing component of the first derivative of $\beta_{\mu}$ is $\varpi_{\mu \nu}$, which is constant, therefore derivatives of higher order are identically null.
In summary, the most general structure of the thermal expectation value of the energy-momentum tensor is
\begin{equation}
    \langle \wh{T}^{\mu \nu}\rangle=F_1\beta^{\mu}\beta^{\nu}+F_2g^{\mu \nu}+F_3\alpha^{\mu}\alpha^{\nu}+F_4(\alpha^{\mu}\beta^{\nu}+\alpha^{\nu}\beta^{\mu}).
\end{equation}
The only Lorentz scalar we can build with $\beta^{\mu}$ and $\alpha^{\mu}$ are $\beta^2=1/T^2$ and $\alpha^2=-a^2/T_0^2$, being $\beta^{\mu}$ and $\alpha^{\mu}$ mutually orthogonal.
Thus, the scalar functions $F_1$ and $F_2$ will depend on $\beta^2$ and $\alpha^2$, while $F_3$ and $F_4$ are expected to be directly proportional to $\alpha^2$ only as the ideal form of $\langle \wh{T}^{\mu \nu}\rangle$ ought to be recovered for vanishing acceleration field.
On the other hand, in our case of free real scalar field theory, the Hamiltonian $\wh{H}$ in \eqref{gteq04} is even under time-reversal, thus so is the density operator.
In formulae, $\wh{\Theta}\wh{\rho}\wh{\Theta}^{-1}=\wh{\rho}$, where $\wh{\Theta}$ is the time-reversal operator.
Care must be taken that $\wh{\Theta}$ depends on the hypersurface of simultaneity with respect to which time is reflected, so the above equation and the following ones should be understood with respect to the $t=0$ hypersurface.
On the other hand, the momentum $\wh{T}^{0i}$ is odd under time-reversal, i.e.\ $\wh{\Theta}\wh{T}^{0i}\wh{\Theta}^{-1}=-\wh{T}^{0i}$.
Together with the evenness of $\wh{\rho}$ and the cyclicity of the trace, this implies that $\langle \wh{T}^{0i}\rangle=0$; conversely, $\langle \wh{T}^{00}\rangle$ and $\langle \wh{T}^{ij}\rangle$ do not vanish.
Now, in the above equation $\alpha^i$ and $\beta^0$ do not vanish at $t=0$, which means that the coupling between $\alpha^{\mu}$ and $\beta^{\mu}$ breaks time-reversal symmetry, therefore $F_4$ must be identically zero, and we are left with
\begin{equation}
    \langle \wh{T}^{\mu \nu}\rangle=F_1\beta^{\mu}\beta^{\nu}+F_2g^{\mu \nu}+F_3\alpha^{\mu}\alpha^{\nu}.
\end{equation}
The functions $F_i$ can be renamed using a more familiar notation in order to make their physical meaning more apparent
\begin{equation}\label{gteq45}
    \langle \wh{T}^{\mu \nu}\rangle=\rho u^{\mu}u^{\nu}-p\Delta^{\mu \nu}+kA^{\mu}A^{\nu}.
\end{equation}
Here, $\rho$ and $p$ reconstruct the ideal form, so they are the energy density and the pressure respectively, while $k$ is an anisotropic pressure term.
Interestingly enough, although being a non-ideal term, $k$ is not technically a dissipation, for entropy is not produced at global thermodynamic equilibrium.
Explicitly, they read
\begin{equation}
    \rho=\langle \wh{T}^{\mu \nu}\rangle u_{\mu}u_{\nu},
\end{equation}
\begin{equation}
    p=\frac{1}{2}\langle \wh{T}^{\mu \nu}\rangle \left(\frac{A_{\mu}A_{\nu}}{A^2}-\Delta_{\mu \nu}\right),
\end{equation}
\begin{equation}
    k=\frac{1}{2A^2}\langle \wh{T}^{\mu \nu}\rangle \left(3\frac{A_{\mu}A_{\nu}}{A^2}-\Delta_{\mu \nu}\right).
\end{equation}
Their expressions depend on the specific Quantum Field Theory underlying the hydrodynamic theory, so we now calculate them in our case of free real scalar field theory in the right Rindler wedge.
We take the energy density $\rho$ as an example, $p$ and $k$ are worked out following analogous steps.

As for the energy density, note that we can either take the thermal expectation value $\langle \wh{T}^{\mu \nu}\rangle$ first and then project onto $u_{\mu}u_{\mu}$ or vieceversa, so let us first project onto $u_{\mu}u_{\nu}$ and then take the thermal expectation value $\langle \wh{T}^{\mu \nu}u_{\mu}u_{\nu}\rangle$.
Using the canonical energy-momentum tensor \eqref{gteq19}, we see that the energy density is given by the difference of two terms
\begin{equation}\label{gteq20}
    \rho=\langle (u^{\mu}\de_{\mu}\wh{\psi})^2\rangle-\langle \wh{\cal L}\rangle.
\end{equation}
Let us focus on the first term.
In Rindler coordinates, the convective derivative is simply $u^{\mu}\de_{\mu}={\rm e}^{-a\xi}\de_{\tau}$, so with the quantum field expansion \eqref{gteq13} and the equations \eqref{gteq14a}--\eqref{gteq14c} we come to
\begin{equation}\label{gteq35}
    \langle (u^{\mu}\de_{\mu}\wh{\psi})^2\rangle=
    \int_0^{+\infty}\di \omega \int_{-\infty}^{+\infty}\di^2{\rm k_T}\,\omega^2|u_{\omega,\kT}|^2(2n_{\rm B}+1).
\end{equation}
The $+1$ term stemming from the commutation relations \eqref{gteq15} gives rise to divergencies and needs renormalization.
More on the renormalization scheme will be said shortly, for the time being let us renormalize with respect to the Rindler vacuum $|0_{\rm R}\rangle$, thus subtract the $T_0=0$ contribution, which is tantamount to cancel the $+1$ term.
Then
\begin{equation}
    \begin{split}
        \langle (u^{\mu}\de_{\mu}\wh{\psi})^2\rangle&-\langle 0_{\rm R}|(u^{\mu}\de_{\mu}\wh{\psi})^2|0_{\rm R}\rangle=
        \frac{{\rm e}^{-2a\xi}}{2\pi^4a}\int_0^{+\infty}\di \omega \,\omega^2\sinh \left(\frac{\pi \omega}{a}\right)n_{\rm B}\times \\
        &\times \int_{-\infty}^{+\infty}\di^2{\rm k_T}\,{\rm K}_{i\frac{\omega}{a}}\left(\frac{\mT {\rm e}^{a\xi}}{a}\right){\rm K}_{-i\frac{\omega}{a}}\left(\frac{\mT {\rm e}^{a\xi}}{a}\right),
    \end{split}
\end{equation}
where ${\rm K}_{i\frac{\omega}{a}}^*={\rm K}_{-i\frac{\omega}{a}}$ was used.
In the massless limit we have $\mT=|\kT|$, and the integral in the transverse momentum can be carried out analytically
\begin{equation}\label{gteq21}
    \begin{split}
        &\int_{-\infty}^{+\infty}\di^2{\rm k_T}\,{\rm K}_{i\frac{\omega}{a}}\left(\frac{\mT {\rm e}^{a\xi}}{a}\right){\rm K}_{-i\frac{\omega}{a}}\left(\frac{\mT {\rm e}^{a\xi}}{a}\right)\\
        &=\pi a^2{\rm e}^{-2a\xi}\Gamma \left(1+i\frac{\omega}{a}\right)\Gamma \left(1-i\frac{\omega}{a}\right)=
        \pi^2 a{\rm e}^{-2a\xi}\frac{\omega}{\sinh \left(\frac{\pi \omega}{a}\right)}
    \end{split}
\end{equation}
with $\Gamma$ the Gamma function, hence
\begin{equation}
    \langle (u^{\mu}\de_{\mu}\wh{\psi})^2\rangle-\langle 0_{\rm R}|(u^{\mu}\de_{\mu}\wh{\psi})^2|0_{\rm R}\rangle=
    \frac{{\rm e}^{-4a\xi}}{2\pi^2}\int_0^{+\infty}\di \omega \,\frac{\omega^3}{{\rm e}^{\omega/T_0}-1}.
\end{equation}
This integral is carried out by exploiting the standard trick involving the geometric series
\begin{equation}
    \begin{split}
        &\int_0^{+\infty}\di \omega \frac{\omega^3}{{\rm e}^{\omega/T_0}-1}=
        \int_0^{+\infty}\di \omega \,\omega^3\sum_{n=1}^{+\infty}{\rm e}^{-n\frac{\omega}{T_0}}=
        \sum_{n=1}^{+\infty}\int_0^{+\infty}\di \omega \,\omega^3{\rm e}^{-n\frac{\omega}{T_0}}\\
        &=T_0^4\sum_{n=1}^{+\infty}\frac{1}{n^4}\int_0^{+\infty}\di x\,x^3{\rm e}^{-x}=
        6T_0^4\sum_{n=1}^{+\infty}\frac{1}{n^4}=
        6T_0^4\zeta(4)=
        \frac{\pi^4}{15}T_0^4,
    \end{split}
\end{equation}
with $\zeta$ the Riemann zeta function.
The order of the integration and the series could be exchanged thanks to the convergence of the series.
Thus, we obtain the result found in \cite{Becattini:2017ljh}
\begin{equation}
    \langle (u^{\mu}\de_{\mu}\wh{\psi})^2\rangle-\langle 0_{\rm R}|(u^{\mu}\de_{\mu}\wh{\psi})^2|0_{\rm R}\rangle=
    \frac{\pi^2}{30}T_0^4{\rm e}^{-4a\xi}=
    \frac{\pi^2}{30\beta^4}=
    \frac{\pi^2}{30}T^4,
\end{equation}
where we used the expression of the proper temperature in Rindler coordinates
\begin{equation}\label{gteq22}
    \beta^2=\frac{a^2}{T_0^2}\left({z'}^2-t^2\right)=\frac{{\rm e}^{2a\xi}}{T_0^2}.
\end{equation}

As for the second term in \eqref{gteq20}, it is convenient to use the Klein-Gordon equation of motion to rewrite it as
\begin{equation}
    \langle \wh{\cal L}\rangle=\frac{1}{4}\langle \Box \wh{\psi}^2\rangle=\frac{1}{4}\Box \langle \wh{\psi}^2\rangle.
\end{equation}
The D'Alambert operator can be taken out of the thermal expectation value because $\Box \wh{\rho}=0$ at global thermodynamic equilibrium with acceleration.
Using the quantum field expansion \eqref{gteq13} and the equations \eqref{gteq14a}--\eqref{gteq14c}, we readily obtain
\begin{equation}\label{gteq36}
    \langle \wh{\psi}^2\rangle=
    \int_0^{+\infty}\di \omega \int_{-\infty}^{+\infty}\di^2{\rm k_T}\,|u_{\omega,\kT}|^2(2n_{\rm B}+1).
\end{equation}
Once again, the $+1$ term stemming from the commutation relations \eqref{gteq15} gives rise to divergencies and must be renormalized.
In the Rindler vacuum renormalization scheme, we have
\begin{equation}
    \begin{split}
    \langle \wh{\psi}^2\rangle&-\langle 0_{\rm R}|\wh{\psi}^2|0_{\rm R}\rangle=
    \frac{1}{2\pi^4a}\int_0^{+\infty}\di \omega \,\sinh \left(\frac{\pi \omega}{a}\right)n_{\rm B}\times \\
    &\times \int_{-\infty}^{+\infty}\di^2{\rm k_T}\,{\rm K}_{i\frac{\omega}{a}}\left(\frac{\mT {\rm e}^{a\xi}}{a}\right){\rm K}_{-i\frac{\omega}{a}}\left(\frac{\mT {\rm e}^{a\xi}}{a}\right)
    \end{split}
\end{equation}
The integral in the transverse momentum is performed analytically in the massless limit by using \eqref{gteq21}, hence
\begin{equation}
    \langle \wh{\psi}^2\rangle-\langle 0_{\rm R}|\wh{\psi}^2|0_{\rm R}\rangle=
    \frac{{\rm e}^{-2a\xi}}{2\pi^2}\int_0^{+\infty}\di \omega \frac{\omega}{{\rm e}^{\omega/T_0}-1}=
    \frac{T_0^2}{12}{\rm e}^{-2a\xi}.
\end{equation}
In the last step, the integral was carried out by using again the geometric series trick as above, with now $\zeta(2)$ instead of $\zeta(4)$.
The D'Alambert operator in Rindler coordinates reads
\begin{equation}
    \Box={\rm e}^{-2a\xi}\left(\de_{\tau}^2-\de_{\xi}^2\right)-\left(\de_x^2+\de_y^2\right),
\end{equation}
hence the result found in \cite{Becattini:2019poj}
\begin{equation}
    \langle \wh{\cal L}\rangle-\langle 0_{\rm R}|\wh{\cal L}|0_{\rm R}\rangle=
    \frac{1}{4}\Box \left(\langle \wh{\psi}^2\rangle-\langle 0_{\rm R}|\wh{\psi}^2|0_{\rm R}\rangle \right)=
    -\frac{a^2T_0^2}{12}{\rm e}^{-4a\xi}=
    \frac{\alpha^2}{12\beta^4}=
    \frac{\alpha^2}{12}T^4,
\end{equation}
where \eqref{gteq22} and \eqref{gteq23} were used to express the result in terms of $\alpha^2$ and $\beta^2$.
In summary, the energy density for a free real scalar field in the massless limit renormalized with respect to the Rindler vacuum is
\begin{equation}
    \rho_{\rm R}\equiv
    \left(\langle \wh{T}^{\mu \nu}\rangle-\langle 0_{\rm R}|\wh{T}^{\mu \nu}|0_{\rm R}\rangle \right)u_{\mu}u_{\nu}=
    \frac{\pi^2}{30\beta^4}-\frac{\alpha^2}{12\beta^4}=
    \left(\frac{\pi^2}{30}-\frac{\alpha^2}{12}\right)T^4.
\end{equation}
This result was found in \cite{Buzzegoli:2017cqy, Becattini:2015nva} with a perturbative expansion of the density operator \eqref{gteq04} in $\alpha^{\mu}$ at order $\alpha^2$, so we conclude that the perturbative series for the free real scalar field is simply a polynomial in $\alpha^{\mu}$ of order 2.

The pressure $p$ and the anisotropic pressure $k$ can be worked out following similar steps.
By defining their values in the same renormalization scheme as
\begin{equation}
    p_{\rm R}\equiv \frac{1}{2}\left(\langle \wh{T}^{\mu \nu}\rangle-\langle 0_{\rm R}|\wh{T}^{\mu \nu}|0_{\rm R}\rangle \right)\left(\frac{A_{\mu}A_{\nu}}{A^2}-\Delta_{\mu \nu}\right)
\end{equation}
\begin{equation}
    k_{\rm R}\equiv \frac{1}{2A^2}\left(\langle \wh{T}^{\mu \nu}\rangle-\langle 0_{\rm R}|\wh{T}^{\mu \nu}|0_{\rm R}\rangle \right)\left(3\frac{A_{\mu}A_{\nu}}{A^2}-\Delta_{\mu \nu}\right)
\end{equation}
the final expressions read
\begin{equation}\label{gteq31}
    \rho_{\rm R}=
    \frac{\pi^2}{30\beta^4}-\frac{\alpha^2}{12\beta^4}=
    \left(\frac{\pi^2}{30}-\frac{\alpha^2}{12}\right)T^4,
    \qquad (m=0)
\end{equation}
\begin{equation}\label{gteq32}
    p_{\rm R}=\frac{\pi^2}{90\beta^4}+\frac{\alpha^2}{18\beta^4}=
    \left(\frac{\pi^2}{90}+\frac{\alpha^2}{18}\right)T^4,
    \qquad (m=0)
\end{equation}
\begin{equation}\label{gteq33}
    k_{\rm R}=-\frac{\alpha^2}{12\beta^4}=-\frac{\alpha^2}{12}T^4,
    \qquad (m=0).
\end{equation}
This is the thermal expectation value of the energy-momentum tensor of a free real scalar field at global thermodynamic equilibrium in the right Rindler wedge in the massless limit.
As expected, for vanishing acceleration field the anisotropic pressure term vanishes, the ideal form is recovered and the energy density and the isotropic pressure take on their familiar forms.
Let us also emphasize that, by definition of $\alpha^{\mu}$ \eqref{gteq24} and by equation \eqref{gteq25}, the $\alpha^2$ terms are quantum corrections, in the sense that they are proportional to $\hbar^2$ thus vanishing in the limit $\hbar \to 0$.


\subsection{Particles and vacuum: the Unruh effect}
\label{sec:gte_unruh_effect}

As we had occasion to convince ourselves a few times already, thermal expectation values of operators quadratic in the quantum field are divergent owing to the $+1$ term in \eqref{gteq14b} stemming from the commutation relations of creation and annihilation operators \eqref{gteq15}.
In free field theory, they are usually renormalized by subtracting the vacuum expectation value; however,
the vacuum state is defined by the creation and annihilation operators, which in turn depend on the choice of the positive-frequency modes.
As it is known, especially in the context of Quantum Field Theory in curved spacetime, such a choice is not unique in general, therefore we face an ambiguity.
In our case, the quantum field could either be expanded in the whole Minkowski spacetime in the usual plane waves form as in \eqref{gteq27}, or in the two Rindler wedges in terms of the Bessel functions as in \eqref{gteq13} and \eqref{gteq26}.
Consequently, it is found that Rindler creation and annihilation operators are related to the Minkowski ones by a non-trivial Bogolyubov transformation, therefore the Rindler vacuum $|0_{\rm R}\rangle$ and the Minkowski vacuum $|0_{\rm M}\rangle$ are two different states, as first pointed out by Fulling \cite{Fulling:1972md}.
In particular, if we take the standpoint of the Rindler observer, the Minkowski vacuum is seen as a thermal state of particles at the temperature $T_0=\frac{a}{2\pi}$
\begin{equation}\label{gteq34}
    \langle 0_{\rm M}|\wh{a}^{\rm R\dagger}_{\omega,\kT}\wh{a}^{\rm R}_{\omega',\kT'}|0_{\rm M}\rangle=
    \langle 0_{\rm M}|\wh{a}^{\rm L\dagger}_{\omega,\kT}\wh{a}^{\rm L}_{\omega',\kT'}|0_{\rm M}\rangle=
    \frac{1}{{\rm e}^{\frac{2\pi}{a}\omega}-1}\delta(\omega-\omega')\,\delta^2(\kT-\kT').
\end{equation}
This is the famous \textit{Unruh effect}, and $T_0=\frac{a}{2\pi}$ is called the \textit{Unruh temperature} \cite{Unruh:1976db}.
In short, the Unruh effect states that uniformly accelerated observers in Minkowski spacetime, i.e.\ linearly accelerated observers with constant proper acceleration called the Rindler observers, associate a thermal bath of Rindler particles to the no-particle state of inertial observers, i.e.\ the Minkowski vacuum.
Rindler particles are associated with positive-frequency modes as defined by Rindler observers in contrast to Minkowski particles, which are associated with positive-frequency modes as defined by inertial observers.
This is a conceptually subtle Quantum Field Theory result, which has played a crucial role in our understanding that the particle content of a field theory is, in this sense, observer-dependent.

In view of these findings, which vacuum should we choose to renormalize thermal expectation values?
The answer is that, as long as both choices give finite results, it depends on which kind of observer we want to take the standpoint of.
We have seen that taking the point of view of the Rindler observer, namely renormalizing with respect to the Rindler vacuum, means the subtraction of the $T_0=0$ contribution, which is tantamount to neglecting the divergent $+1$ term in equation \eqref{gteq14b}.
In this renormalization scheme, the thermal expectation value of the energy-momentum tensor is given in \eqref{gteq31}--\eqref{gteq33}.
On the other hand, from \eqref{gteq34} it is clear that thermal expectation values of products of Rindler creation and annihilation operators in the Minkowski vacuum take on the same form as \eqref{gteq14a}--\eqref{gteq14c} calculated at $T_0=\frac{a}{2\pi}$.
Therefore, taking the standpoint of the inertial observer, that is renormalizing with respect to the Minkowski vacuum, means the subtraction of the $T_0=\frac{a}{2\pi}$ contribution.

The former eventuality has already been studied, so let us focus here on the latter.
In the Minkowski renormalization scheme, equations \eqref{gteq14a}--\eqref{gteq14c} read
\begin{equation}
    \langle \wh{a}^{\rm R\dagger}_{\omega,\kT}\wh{a}^{\rm R}_{\omega',\kT'}\rangle-\langle 0_{\rm M}|\wh{a}^{\rm R\dagger}_{\omega,\kT}\wh{a}^{\rm R}_{\omega',\kT'}|0_{\rm M}\rangle=
    \left(\frac{1}{{\rm e}^{\omega/T_0}-1}-\frac{1}{{\rm e}^{\frac{2\pi}{a}\omega}-1}\right)\delta(\omega-\omega')\,\delta^2(\kT-\kT')
\end{equation}
\begin{equation}
    \langle \wh{a}^{\rm R}_{\omega,\kT}\wh{a}^{\rm R\dagger}_{\omega',\kT'}\rangle-\langle 0_{\rm M}|\wh{a}^{\rm R}_{\omega,\kT}\wh{a}^{\rm R\dagger}_{\omega',\kT'}|0_{\rm M}\rangle=
    \left(\frac{1}{{\rm e}^{\omega/T_0}-1}-\frac{1}{{\rm e}^{\frac{2\pi}{a}\omega}-1}\right)\delta(\omega-\omega')\,\delta^2(\kT-\kT')
\end{equation}
\begin{equation}
    \langle \wh{a}^{\rm R}_{\omega,\kT}\wh{a}^{\rm R}_{\omega',\kT'}\rangle-\langle 0_{\rm M}|\wh{a}^{\rm R}_{\omega,\kT}\wh{a}^{\rm R}_{\omega',\kT'}|0_{\rm M}\rangle=0=\langle \wh{a}^{\rm R\dagger}_{\omega,\kT}\wh{a}^{\rm R\dagger}_{\omega',\kT'}\rangle-\langle 0_{\rm M}|\wh{a}^{\rm R\dagger}_{\omega,\kT}\wh{a}^{\rm R\dagger}_{\omega',\kT'}|0_{\rm M}\rangle.
\end{equation}
These expressions are suggestive as they make it apparent that the thermal expectation value of any operator quadratic in the quantum field, such as the energy-momentum tensor, vanish when the temperature is as low as the Unruh temperature.
By looking at non-renormalized equations \eqref{gteq35} and \eqref{gteq36}, it is clear that the energy density not only vanishes at $T_0=\frac{a}{2\pi}$, but becomes negative for lower temperatures.
And this is actually not a prerogative of the energy density, but a general feature of operators quadratic in the quantum field.
Thus, in this sense, the Unruh temperature is an absolute lower bound for the temperature, a result found in \cite{Becattini:2017ljh}.
It is important to stress that this conclusion holds locally, for a comoving observer.
From the expression of the proper temperature $T$  \eqref{gteq09} and the acceleration field $A^{\mu}$ \eqref{gteq37}, it is readily seen that in general one can write $T=T_0\sqrt{|A^2|}/a$, so, at the level of the proper temperature, the bound $T_0\ge \frac{a}{2\pi}$ implies
\begin{equation}
    T\ge T_{\rm U}\equiv \frac{\sqrt{|A^2|}}{2\pi},
\end{equation}
where $T_{\rm U}$ is called the \textit{comoving} or \textit{proper Unruh temperature}.
That is, the temperature measured by a comoving thermometer cannot be lower than the magnitude of the acceleration field divided by $2\pi$.
Recall that the magnitude of the acceleration field is constant along a fixed flow line and varies from one flow line to another, therefore so does the proper Unruh temperature, thus behaving like the proper temperature itself.
The feature of the proper Unruh temperature being a lower bound for the proper temperature is apparent in the thermal expectation value of the energy-momentum tensor renormalized with repsect to the Minkowski vacuum.
By defining
\begin{equation}
    \rho_{\rm M}\equiv \left(\langle \wh{T}^{\mu \nu}\rangle-\langle 0_{\rm M}|\wh{T}^{\mu \nu}|0_{\rm M}\rangle \right)u_{\mu}u_{\nu},
\end{equation}
\begin{equation}
    p_{\rm M}\equiv \frac{1}{2}\left(\langle \wh{T}^{\mu \nu}\rangle-\langle 0_{\rm M}|\wh{T}^{\mu \nu}|0_{\rm M}\rangle \right)\left(\frac{A_{\mu}A_{\nu}}{A^2}-\Delta_{\mu \nu}\right),
\end{equation}
\begin{equation}
    k_{\rm M}\equiv \frac{1}{2A^2}\left(\langle \wh{T}^{\mu \nu}\rangle-\langle 0_{\rm M}|\wh{T}^{\mu \nu}|0_{\rm M}\rangle \right)\left(3\frac{A_{\mu}A_{\nu}}{A^2}-\Delta_{\mu \nu}\right),
\end{equation}
one can find the following results in the massless case
\begin{equation}
    \rho_{\rm M}=
    \left(\frac{\pi^2}{30}-\frac{\alpha^2}{12}\right)(T^4-T_{\rm U}^4),
    \qquad (m=0)
\end{equation}
\begin{equation}
    p_{\rm M}=
    \left(\frac{\pi^2}{90}+\frac{\alpha^2}{18}\right)(T^4-T_{\rm U}^4),
    \qquad (m=0)
\end{equation}
\begin{equation}
    k_{\rm M}=
    -\frac{\alpha^2}{12}(T^4-T_{\rm U}^4),
    \qquad (m=0).
\end{equation}

Although the conclusion that scalar thermodynamic functions renormalized with respect to the Minkowski vacuum can be expressed in the above fashion was obtained for a free field theory, it is likely to hold for any interacting field theory as well.
In fact, the Unruh effect was derived for general interacting field theories in \cite{Bisognano:1975ih, Bisognano:1976za} within an axiomatic quantum field theory approach by taking advantage of the KMS feature of the expectation values for the density operator at hand.
For a recent discussion, see also \cite{Gransee:2015aba, Gransee:2016hep}.


\section{Entropy current in the right Rindler wedge}
\label{sec:gte_entropy_current_in_RRW}

With the thermal expectation value of the energy-momentum tensor, we are now in a position to use our method to calculate the entropy current including quantum corrections.
Its divergence will provide information on the entropy production rate, which is expected to vanish at global thermodynamic equilibrium, while its integral will give us the total entropy.
As we shall see, thanks to the factorization property of the density operator, the entropy thereby obtained will be interestingly related to some entanglement entropy.


\subsection{Entropy current calculation}

As a quick recap, in order to apply the method for the calculation of the entropy current put forward in Section \ref{sec:zubarev_entropy_current_method}, we need two ingredients:
\begin{enumerate}
    \item the thermal expectation value the energy-momentum tensor $\langle \wh{T}^{\mu \nu}\rangle$, and
    \item the eigenvector $|0_{\Upsilon}\rangle$ corresponding to the lowest, non-degenerate eigenvalue of $\wh{\Upsilon}$.
\end{enumerate}
Then, the algorithm is the following:
\begin{enumerate}
    \item Take $\langle \wh{T}^{\mu \nu}\rangle$ and subtract $\langle 0_{\Upsilon}|\wh{T}^{\mu \nu}|0_{\Upsilon}\rangle$ by using the $\lambda$-dependent density operator $\wh{\rho}_{\rm R}(\lambda)$ defined according to \eqref{zubeq21}.
    This way, the result is $\lambda$-dependent.
    \item Contract with $\beta_{\nu}$, which is $\lambda$-independent, and integrate in $\lambda$ from $\lambda=1$ to $\lambda=+\infty$ in order to obtain the thermodynamic potential current $\phi^{\mu}$ defined in \eqref{zubeq26}.
    \item Plug the result into \eqref{zubeq27} and obtain the entropy current $s^{\mu}$.
\end{enumerate}

By comparing the definition \eqref{zubeq23} of the operator $\wh{\Upsilon}$ with the expression \eqref{gteq28} of the reduced density operator at global thermodynamic equilibrium with acceleration in the right Rindler wedge, we immediately understand that
\begin{equation}\label{gteq38}
    \wh{\Upsilon}=\frac{\wh{\Pi}_{\rm R}}{T_0}=
    \frac{1}{T_0}\int_0^{+\infty}\di \omega \int_{-\infty}^{+\infty}\di^2{\rm k_T}\,\omega \wh{a}^{\rm R\dagger}_{\omega,\kT}\wh{a}^{\rm R}_{\omega,\kT}.
\end{equation}
Thus, the lowest eigenvector of $\wh{\Upsilon}$ is the Rindler vacuum, $|0_{\Upsilon}\rangle=|0_{\rm R}\rangle$, which is non-degenerate, and the corresponding eigenvalue is $\Upsilon_0=0$.
Hence,
\begin{equation}
    \langle \wh{T}^{\mu \nu}\rangle-\langle 0_{\Upsilon}|\wh{T}^{\mu \nu}|0_{\Upsilon}\rangle=
    \langle \wh{T}^{\mu \nu}\rangle-\langle 0_{\rm R}|\wh{T}^{\mu \nu}|0_{\rm R}\rangle,
\end{equation}
which was already calculated in \eqref{gteq31}--\eqref{gteq33} for a massless field, so we are good to go for what concerns the ingredients.

Now, according to the method, the above quantity should be calculated with the modified density operator $\wh{\rho}_{\rm R}(\lambda)$ in order to be $\lambda$-dependent.
Let us take a closer look at what this operation actually means at a practical level.
By comparing equation \eqref{zubeq21} with \eqref{gteq04} or with \eqref{gteq38}, we can tell that in our case the introduction of $\lambda$ is but a rescaling of $T_0$ as $T_0\mapsto T_0/\lambda$.
It can be easily checked that this transformation gives precisely the desired result.
Thus, we have
\begin{equation}
    \langle \wh{T}^{\mu \nu}\rangle(\lambda)-\langle 0_{\Upsilon}|\wh{T}^{\mu \nu}|0_{\Upsilon}\rangle(\lambda)=
    \rho_{\rm R}(\lambda)u^{\mu}u^{\nu}-p_{\rm R}(\lambda)\Delta^{\mu \nu}+k_{\rm R}(\lambda)A^{\mu}A^{\nu},
\end{equation}
where the rescaling $T_0\mapsto T_0/\lambda$ affects only $\rho_{\rm R}$, $p_{\rm R}$ and $k_{\rm R}$ because $u^{\mu}$, $\Delta^{\mu \nu}$ and $A^{\mu}$ are independent of $T_0$.
This result should now be contracted with $\beta_{\nu}$.
Care must be taken that this $\beta_{\nu}$ does not undergo the temperature rescaling: it concerns only the thermal expectation value $\langle \wh{T}^{\mu \nu}\rangle-\langle 0_{\Upsilon}|\wh{T}^{\mu \nu}|0_{\Upsilon}\rangle$, as can be clearly seen following the steps of the method derivation in Section \ref{sec:zubarev_entropy_current_method}.
Whence
\begin{equation}
    \left(\langle \wh{T}^{\mu \nu}\rangle(\lambda)-\langle 0_{\Upsilon}|\wh{T}^{\mu \nu}|0_{\Upsilon}\rangle(\lambda)\right)\beta_{\nu}=
    \rho_{\rm R}(\lambda)\beta^{\mu}.
\end{equation}
In other words, we do not need the thermal expectation value of the whole energy-momentum tensor, only the energy density contributes to the thermodynamic potential current and the entropy current.
The expression of $\rho_{\rm R}(\lambda)$ is obtained from \eqref{gteq31} by making explicit the $T_0$-dependence using \eqref{gteq09} and \eqref{gteq23}.
To this purpose, it is also convenient to extract the $T_0$-dependence from the four-temperature $\beta^{\mu}$ of \eqref{gteq08} by defining the Killing vector field $\gamma^{\mu}\equiv T_0\beta^{\mu}$, which is $T_0$-independent, hence
\begin{equation}
    \rho_{\rm R}=
    \left(\frac{\pi^2}{30}-\frac{\alpha^2}{12}\right)\frac{1}{\beta^4}=
    \left(\frac{\pi^2}{30}+\frac{a^2}{12T_0^2}\right)\frac{T_0^4}{\gamma^4}=
    \frac{\pi^2}{30\gamma^4}T_0^4+\frac{a^2}{12\gamma^4}T_0^2,
\end{equation}
thus
\begin{equation}
    \rho_{\rm R}(\lambda)=
    \frac{\pi^2}{30\gamma^4}\frac{T_0^4}{\lambda^4}+\frac{a^2}{12\gamma^4}\frac{T_0^2}{\lambda^2}.
\end{equation}
According to \eqref{zubeq26}, the thermodynamic potential current is readily obtained
\begin{equation}
    \begin{split}
        \phi^{\mu}=&\beta^{\mu}\int_1^{+\infty}\di \lambda \,\rho_{\rm R}(\lambda)=\left(\frac{\pi^2T_0^4}{30\gamma^4}\int_1^{+\infty}\frac{\di \lambda}{\lambda^4}+\frac{a^2T_0^2}{12\gamma^4}\int_1^{+\infty}\frac{\di \lambda}{\lambda^2}\right)\beta^{\mu}\\
        =&\left(\frac{\pi^2}{90}\frac{T_0^4}{\gamma^4}+\frac{a^2}{12T_0^2}\frac{T_0^4}{\gamma^4}\right)\beta^{\mu}=
        \left(\frac{\pi^2}{90\beta^4}-\frac{\alpha^2}{12\beta^4}\right)\beta^{\mu}.
    \end{split}
\end{equation}
Plugging this result into \eqref{zubeq27}, we finally have the entropy current
\begin{equation}
    \begin{split}
        s^{\mu}=&
        \phi^{\mu}+\left(\langle \wh{T}^{\mu \nu}\rangle-\langle 0_{\rm R}|\wh{T}^{\mu \nu}|0_{\rm R}\rangle \right)\beta_{\nu}=
        \phi^{\mu}+\rho_{\rm R}\beta^{\mu}\\
        =&\left(\frac{\pi^2}{90\beta^4}-\frac{\alpha^2}{12\beta^4}+\frac{\pi^2}{30\beta^4}-\frac{\alpha^2}{12\beta^4}\right)\beta^{\mu}=
        \left(\frac{2\pi^2}{45\beta^4}-\frac{\alpha^2}{6\beta^4}\right)\beta^{\mu}.
    \end{split}
\end{equation}
In summary,
\begin{equation}
    \phi^{\mu}=
    \left(\frac{\pi^2}{90\beta^4}-\frac{\alpha^2}{12\beta^4}\right)\beta^{\mu}=
    \left(\frac{\pi^2}{90}-\frac{\alpha^2}{12}\right)T^3u^{\mu},\qquad (m=0)
\end{equation}
\begin{equation}\label{gteq39}
    s^{\mu}=
    \left(\frac{2\pi^2}{45\beta^4}-\frac{\alpha^2}{6\beta^4}\right)\beta^{\mu}=
    \left(\frac{2\pi^2}{45}-\frac{\alpha^2}{6}\right)T^3u^{\mu},\qquad (m=0).
\end{equation}
These are the final expressions of the thermodynamic potential current and the entropy current of a free real scalar massless field at global thermodynamic equilibrium with acceleration in the right Rindler wedge \cite{Becattini:2019poj}.
Now some comments are in order.

First of all, we stress that, by definition of $\alpha^{\mu}$ \eqref{gteq24} and by equation \eqref{gteq25}, the $\alpha^2$ terms are quantum corrections, in the sense that they are proportional to $\hbar^2$ thus vanishing in the limit $\hbar \to 0$.

It is also easy to check that the entropy current, and the thermodynamic potential current as well, actually, are divergenceless.
In fact, both of the currents are given by a constant term multiplied by $\beta^{\mu}/\beta^4$, so
\begin{equation}
    \begin{split}
        \de_{\mu}s^{\mu}\propto &\de_{\mu}\phi^{\mu}\propto
        \de_{\mu}\frac{\beta^{\mu}}{\beta^4}=
        \de_t\frac{\beta^t}{\beta^4}+\de_{z'}\frac{\beta^{z'}}{\beta^4}\\
        =&\partial_t\left[\frac{T_0^4}{a^4}\left({z'}^2-t^2\right)^{-2}\frac{a}{T_0}z'\right]+\partial_{z'}\left[\frac{T_0^4}{a^4}\left({z'}^2-t^2\right)^{-2}\frac{a}{T_0}t\right]\\
        =&\frac{T_0^3}{a^3}\left[4\left({z'}^2-t^2\right)^{-3}tz'-4\left({z'}^2-t^2\right)^{-3}z't\right]=0.
    \end{split}
\end{equation}
This was expected because, as a general fact of global thermodynamic equilibrium, the entropy is constant in time, so the entropy production rate is identically null.

As we discussed earlier, the Unruh temperature is an absolute lower bound for the temperature at global thermodynamic equilibrium with acceleration, so the question arises if the entropy current vanishes or not at that temperature.
It turns out that it does not, in particular we have
\begin{equation}\label{gteq40}
    s^{\mu}(T_{\rm U})=\frac{32\pi^2}{45}T_{\rm U}^3u^{\mu},\qquad (m=0).
\end{equation}
The reason for a non-vanishing entropy current in the Minkowski vacuum might be the fact that the field degrees of freedom in the left Rindler wedge were traced out, in the sense that $s^{\mu}$ is the entropy current pertaining the entropy in the right Rindler wedge, thus calculated with the reduced density operator $\wh{\rho}_{\rm R}$.
This partial trace operation often leads to thermal states, thus a non-vanishing entropy current.

Another interesting observation is the following.
From equations \eqref{zubeq26} and \eqref{zubeq27}, it is clear that the thermodynamic potential current and the entropy current depend on the specific energy-momentum tensor quantum operator we decide to choose.
Our choice so far has been the canonical tensor \eqref{gteq19}, namely the minimal coupling one, but this is not the only possibility.
For instance, we could as well consider an ``improved'' tensor which is traceless for a massless field, namely the conformal coupling one,
\begin{equation}
    \wh{T}^{\mu \nu}_{\rm imp}\equiv \wh{T}^{\mu \nu}-\frac{1}{6}\left(\de^{\mu}\de^{\nu}-g^{\mu \nu}\Box \right)\wh{\psi}^2,
\end{equation}
with $\wh{T}^{\mu \nu}$ appearing there being the canonical tensor.
In \cite{Becattini:2011ev} it is argued that, as a general feature of global thermodynamic equilibrium with acceleration or rotation, the energy density pertaining to this improved tensor is different from the one corresponding to the canonical tensor.
Indeed, by using the previously calculated expressions, the additional term with respect to the canonical value in the Rindler renormalization scheme for a massless field is
\begin{equation}
    -\frac{1}{6}\left(u^{\mu}u^{\nu}\de_{\mu}\de_{\nu}-\Box \right)\left(\langle \wh{\psi}^2\rangle-\langle 0_{\rm R}|\wh{\psi}^2|0_{\rm R}\rangle \right)=
    -\frac{1}{6}\left(u^{\mu}u^{\nu}\de_{\mu}\de_{\nu}-\Box \right)\frac{1}{12\beta^2}=
    \frac{\alpha^2}{12\beta^4},
\end{equation}
hence the ``improved'' energy density
\begin{equation}
    \rho^{\rm imp}_{\rm R}=\frac{\pi^2}{30\beta^4}=\frac{\pi^2}{30}T^4,\qquad (m=0).
\end{equation}
That is, the energy density calculated with the ``improved'' energy-momentum tensor for the massless free real scalar field depends only on $\beta^2$ and not on $\alpha^2$, which is a somewhat surprising feature.
Likewise, the entropy current gets modified and one is left with only the first term of equation \eqref{gteq39}
\begin{equation}
    s^{\mu}_{\rm imp}=\frac{2\pi^2}{45\beta^4}\beta^{\mu}=
    \frac{2\pi^2}{45}T^3u^{\mu},\qquad (m=0).
\end{equation}
At the Unruh temperature, we get an expression that, oddly enough, differs from the one obtained from the canonical tensor \eqref{gteq40} by a factor 16
\begin{equation}
    s^{\mu}_{\rm imp}(T_{\rm U})=\frac{2\pi^2}{45}T_{\rm U}^3u^{\mu},\qquad (m=0).
\end{equation}


\subsection{Entropy in the right Rindler wedge and the entanglement entropy}
\label{subsec:gte_entropy}

\begin{figure}
    \begin{center}
	    \includegraphics[width=0.75\textwidth]{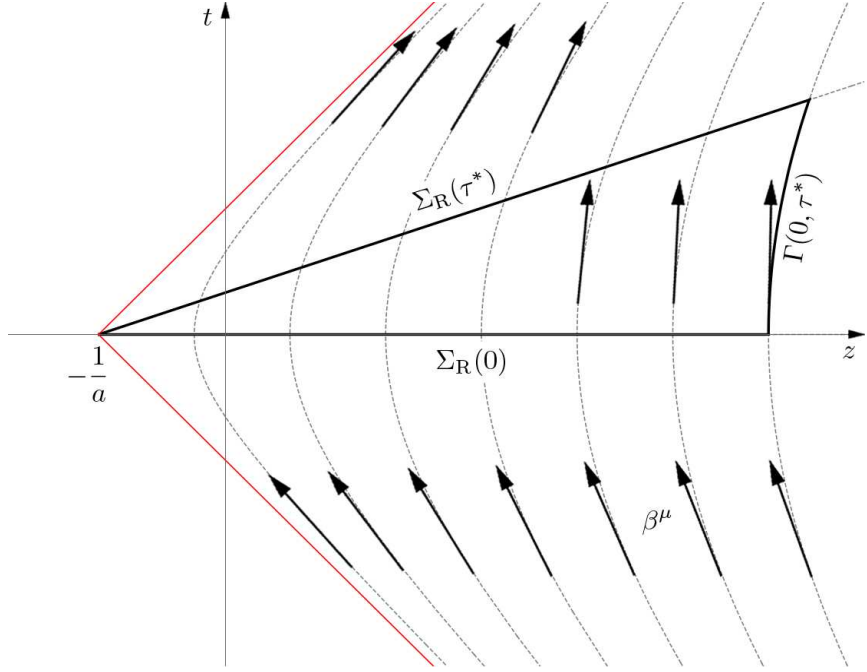}
	    \caption{In this 2-dimensional $(t,z)$-section of the right Rindler wedge, $\Sigma_{\rm R}(0)$ and $\Sigma_{\rm R}(\tau^*)$ are two constant-$\tau$ hypersurfaces of the foliation, corresponding to $\tau=0$ and $\tau=\tau^*$ respectively.
	    $\Gamma(0,\tau^*)$ is the timelike boundary at infinity joining $\Sigma_{\rm R}(0)$ and $\Sigma_{\rm R}(\tau^*)$.
	    The hyperbolae are the flow lines of the four-temperature, which is tangent to $\Gamma(0,\tau^*)$.}
	    \label{fig:rindler_entropy}
	\end{center}
\end{figure}

Equation \eqref{gteq39} is the entropy current in the right Rindler wedge, therefore its integral on a 3-dimensional spacelike hypersurface $\Sigma_{\rm R}$ contained in the right Rindler wedge and whose boundary is the 2-dimensional surface $(t=0,z'=0)$ is the entropy in that subspace
\begin{equation}
    S_{\rm R}=\int_{\Sigma_{\rm R}}\di \Sigma_{\mu}s^{\mu}.
\end{equation}
As mentioned already, the right Rindler wedge is foliated by constant-$\tau$ hypersurfaces.
Let $\Sigma_{\rm R}(0)$ and $\Sigma_{\rm R}(\tau^*)$ be two such hypersurfaces, in particular the $\tau=0$ and $\tau=\tau^*$ ones respectively, as shown in Figure \ref{fig:rindler_entropy}.
In order for the integral to be independent on the choice of the hypersurface of the foliation, two conditions have to be met.
The first one is that the entropy current must be divergenceless, which we already checked.
And the second one is the vanishing of the flux through the timelike boundary at infinity joining the two hypersurfaces of the foliation, indicated as $\Gamma(0,\tau^*)$ in Figure \ref{fig:rindler_entropy}.
This condition is checked just as readily, for the entropy current is parallel to the four-temperature, which is tangent to $\Gamma(0,\tau^*)$, hence $\di \Sigma_{\mu}s^{\mu}=\di \Sigma \,n_{\mu}s^{\mu}=0$.
In summary, we can choose whatever hypersurface of the foliation, as expected from global thermodynamic equilibrium.
A straightforward calculation of the entropy on $\Sigma_{\rm R}(0)$ with the entropy current \eqref{gteq39} pertaining to the canonical energy-momentum tensor yields
\begin{equation}\label{gteq47}
    S_{\rm R}=\left(\int_{-\infty}^{+\infty}\di^2{\rm x_T}\right)\left(\frac{2\pi^2}{45}-\frac{\alpha^2}{6}\right)\frac{T_0^3}{a^3}\lim_{z'\to 0}\frac{1}{2{z'}^2}.
\end{equation}
This expression has two main properties.
The one is the proportionality to the area of the 2-dimensional boundary surface separating the left and right Rindler wedges, that is the so-called \textit{area law}, which is clearly reminiscent of the Bekenstein-Hawking formula for the black hole entropy \cite{Bekenstein:1973ur}.
The other is the divergence as $z'\to 0$, owing to the fact that the proper temperature diverges at the Killing horizon.
Our result is in full agreement with previous literature \cite{Bombelli:1986rw, Srednicki:1993im}.

This entropy we just worked out is the entropy in the right Rindler wedge, namely
the entropy calculated with the reduced density operator pertaining to the right Rindler wedge
\begin{equation}\label{gteq44}
    S_{\rm R}=-\tr_{\rm R}(\wh{\rho}_{\rm R}\log \wh{\rho}_{\rm R}).
\end{equation}
Recall from \eqref{gteq41} and \eqref{gteq42} that the full density operator $\wh{\rho}$ in Minkowski spacetime is factorized as $\wh{\rho}=\wh{\rho}_{\rm R}\otimes \wh{\rho}_{\rm L}$, with $\wh{\rho}_{\rm R}=\tr_{\rm L}(\wh{\rho})$ and $\wh{\rho}_{\rm L}=\tr_{\rm R}(\wh{\rho})$ the reduced density operators of the right and left Rindler wedges respectively.
In Quantum Information Theory, the entropy obtained from the reduced density operator of a bipartite system is called the \textit{entanglement entropy}, which, in some sense, is a measure for the entanglement of a system.
Thus, the above $S_{\rm R}$ is in fact the entanglement entropy of the right Rindler wedge with the left Rindler wedge.
The area law is a property characteristic of the entanglement entropy, and the fact that $S_{\rm R}$ exhibits it too, somehow strengthens our belief in this connection.

The calculation of the entanglement entropy in Quantum Mechanics may not be necessarily hard, depending on the specific system at hand; whereas, as a matter of fact, when it comes to Quantum Field Theory it turns out to be a formidable task which one is not in general able to accomplish.
Only few exact analytical results are known, and most of them heavily rely on the power of conformal symmetry in 2 dimensions \cite{Calabrese:2004eu, Calabrese:2005zw, Calabrese:2009qy, Casini:2005zv, Casini:2005rm}.
The prescription for a holographic dual of the entanglement entropy in the context of the AdS/CFT correspondence has turned this complicated procedure into a conceptually far more intuitive one, which eventually amounts to finding minimal-area surfaces in some geometry, and as such it constitutes a milestone indeed \cite{Ryu:2006bv, Ryu:2006ef}.
However, it is still far from easy to obtain analytical results for non-trivial systems.
In this respect, having been able to perform an exact analytical calculation for a system which is both mathematically non-trivial and of physical concern, makes our result obtained in \cite{Becattini:2019poj} even more interesting.

Note that $S_{\rm R}$ does not vanish in the Minkowski vacuum, that is when the temperature is as low as the Unruh temperature.
This should not be completely surprising as $S_{\rm R}$ is obtained from $\wh{\rho}_{\rm R}$, which in turn is obtained from $\wh{\rho}$ by taking the partial trace over the field degrees of freedom of the left Rindler wedge.
This operation is in general likely to lead to thermal states, as it does indeed in our case too.

Note also that, if we used the ``improved'' energy-momentum tensor instead of the canonical one, the area law and the divergence at the Killing horizon would have stayed the same, so $S_{\rm R}$ would still be infinite overall; however, the constant factor in front of it would have changed to $2\pi^2/45$.
This is somewhat unexpected for the following reason.
As we mentioned, the expression of the entropy current does depend on the specific form of the energy-momentum tensor quantum operator.
On the other hand, the generators of the Poincaré group \eqref{gteq43} should not \cite{Becattini:2011ev}.
Consequently, neither should the entropy, because $S_{\rm R}$ is obtained from $\wh{\rho}_{\rm R}$ as in \eqref{gteq44}, and $\wh{\rho}_{\rm R}$, referring to \eqref{gteq04}, can in turn be written in terms of the Poincaré generators as
\begin{equation}
    \wh{\rho}_{\rm R}=
    \tr_{\rm L}\left(\frac{1}{Z}\exp \left[-\frac{\wh{H}}{T_0}+\frac{a}{T_0}\wh{K}_z\right]\right).
\end{equation}
Nevertheless, $\wh{\Pi}_{\rm R}$ and $\wh{\Pi}_{\rm L}$ might inherit a dependence on the energy-momentum tensor quantum operator because of the truncation at $z'=0$.
This issue will be the subject of further investigation.

We conclude with a final remark referring to a comment we made at the end of Subsection \ref{sec:entr_prod_rate}, which essentially states the following.
Since the entropy current is divergenceless at global thermodynamic equilibrium, if its domain is topologically contractible it can be expressed as the divergence of an antisymmetric rank-2 tensor called the \textit{potential} of $s^{\mu}$
\begin{equation}
    \de_{\mu}s^{\mu}=0\qquad \Rightarrow \qquad s^{\mu}=\de_{\nu}\varsigma^{\mu \nu},\qquad \varsigma^{\mu \nu}=-\varsigma^{\nu \mu}.
\end{equation}
With this potential, the entropy can be cast into a 2-dimensional integral as in \eqref{zubeq31}, i.e.\
\begin{equation}\label{gteq48}
    S=\int_{\Sigma}\di \Sigma_{\mu}s^{\mu}=
    \int_{\Sigma}\di \Sigma_{\mu}\de_{\nu}\varsigma^{\mu \nu}=
    \frac{1}{2}\int_{\de \Sigma}\di \tilde{S}_{\mu \nu}\varsigma^{\mu \nu}=
    -\frac{1}{4}\int_{\de \Sigma}\di S^{\rho \sigma}\sqrt{|g|}\epsilon_{\mu \nu \rho \sigma}\varsigma^{\mu \nu},
\end{equation}
where $\de \Sigma$ is the 2-dimensional boundary surface of $\Sigma$.
The right Rindler wedge is topologically contractible, therefore such a potential for the entropy current \eqref{gteq39} ought to exist.
In order to work it out, we follow the same kind of philosophy as for the derivation of the thermal expectation value of the energy-momentum tensor \eqref{gteq45}.
To form an antisymmetric tensor, we can only use the vector fields $\beta^{\mu}$ and $\alpha^{\mu}$, hence the only possible combination is $\alpha^{\mu}\beta^{\nu}-\alpha^{\nu}\beta^{\mu}$.
According to the decomposition \eqref{gteq18}, this is just proportional to the thermal vorticity, so we have in general
\begin{equation}\label{gteq46}
    \varsigma^{\mu \nu}=G\varpi^{\mu \nu}
\end{equation}
for some thermodynamic scalar function $G=G(\alpha^2,\beta^2)$.
The divergence of this expression must reproduce the entropy current
\begin{equation}
    s^{\mu}=\de_{\nu}(G\varpi^{\mu \nu})=\varpi^{\mu \nu}\de_{\nu}G,
\end{equation}
where we took into account that the thermal vorticity is constant at global thermodynamic equilibrium.
By introducing the entropy density $s=u_{\mu}s^{\mu}$, we get
\begin{equation}
    s=u_{\mu}s^{\mu}=u_{\mu}\varpi^{\mu \nu}\de_{\nu}G=
    -\alpha^{\nu}\de_{\nu}G=
    -\alpha^{\nu}\frac{\de G}{\de \beta^2}\de_{\nu}\beta^2,
\end{equation}
where we used again the decomposition \eqref{gteq18} of the thermal vorticity and, in the last equality, the fact that $\alpha^2$ is constant, so its derivative is identically null.
Using the Killing equation and \eqref{gteq18} once again, we find
\begin{equation}
    \de_{\nu}\beta^2=
    g^{\rho \sigma}\de_{\nu}(\beta_{\rho}\beta_{\sigma})=
    g^{\rho \sigma}(\beta_{\rho}\de_{\nu}\beta_{\sigma}+\beta_{\sigma}\de_{\nu}\beta_{\rho})=
    \beta^{\sigma}\varpi_{\sigma \nu}+\beta^{\rho}\varpi_{\rho \nu}=
    -2\sqrt{\beta^2}\alpha_{\nu},
\end{equation}
hence
\begin{equation}
    s=2\alpha^2\sqrt{\beta^2}\frac{\de G}{\de \beta^2}\qquad \Rightarrow \qquad
    G=\int \di \beta^2\frac{s}{2\alpha^2\sqrt{\beta^2}}\frac{\de G}{\de \beta^2}.
\end{equation}
The entropy density obtained from the entropy current \eqref{gteq39} is
\begin{equation}
    s=u_{\mu}s^{\mu}=\left(\frac{2\pi^2}{45}-\frac{\alpha^2}{6}\right)\frac{1}{(\beta^2)^{3/2}},
\end{equation}
so we can work out the expression of $G$
\begin{equation}
    \begin{split}
        G=&
        \int \di \beta^2 \left(\frac{2\pi^2}{45}-\frac{\alpha^2}{6}\right)\frac{1}{(\beta^2)^{3/2}} \frac{1}{2\alpha^2\sqrt{\beta^2}}\\
        =&\frac{1}{2\alpha^2}\left(\frac{2\pi^2}{45}-\frac{\alpha^2}{6}\right)\int \di \beta^2 \frac{1}{(\beta^2)^2}\\
        =&-\frac{1}{2\alpha^2}\left(\frac{2\pi^2}{45}-\frac{\alpha^2}{6}\right)\frac{1}{\beta^2}=
        -\frac{s}{2\alpha^2}\sqrt{\beta^2}.
    \end{split}
\end{equation}
Plugging this into \eqref{gteq46}, we finally obtain the potential of the entropy current \cite{Becattini:2019poj}
\begin{equation}
    \varsigma^{\mu \nu}=\frac{s}{2\alpha^2}(\beta^{\mu}\alpha^{\nu}-\beta^{\nu}\alpha^{\mu}).
\end{equation}
The boundary of the hypersurface $\Sigma_{\rm R}(0)$ is the $(x,y)$ plane $(t=0,z'=0)$ together with the plane $(t=0,z'=+\infty)$.
In the latter, the argument of the last integral in \eqref{gteq48} vanishes for $s \propto {z'}^{-3}$, $\alpha^2$ is constant
and $\beta^0 \alpha^3 \propto z'$.
We are thus left with the $(x,y)$ plane, and taking into account that the indices $\rho$ and $\sigma$ can only take on values $1$ and $2$ and of the dependence of $\beta^{\mu}$ and $\alpha^{\mu}$ on $(t,z')$, we end up with the same result as \eqref{gteq47}.


\section{Summary and outlook}

In this Chapter, we considered a relativistic quantum fluid at global thermodynamic equilibrium with acceleration, a non-trivial instance of global thermodynamic equilibrium in Minkowski spacetime of both phenomenological and theoretical concern.
For this particular system, the four-temperature has a Killing horizon making it timelike and future-oriented only in the right Rindler wedge, therefore that is the subspace we had to restrict in order to make thermodynamics and hydrodynamics out of it.
Moreover, the density operator is factorized into two reduced density operators each pertaining to a single Rindler wedge, so that thermal expectation values in each wedge can be calculated by taking the partial trace with the corresponding reduced density operator.

In order to obtain the entropy current, we calculated the thermal expectation value of the energy-momentum tensor in the right Rindler wedge.
In particular, we considered the canonical tensor for the simple instance of a free real scalar quantum field, and the calculation was performed analytically in the massless case.
The plain expectation value was expectedly divergent, therefore it needed renormalization.
We discussed both the Rindler and the Minkowski renormalization schemes, and saw how the Unruh effect arises naturally with the Unruh temperature being a lower bound for the temperature.

We then practically tested our method by working out the entropy current for this system including quantum corrections.
The corresponding entropy production rate was identically null, in agreement with the general condition of global thermodynamic equilibrium.
The entropy current did not vanish in the Minkowski vacuum, that is at a temperature as low as the Unruh temperature, which is supposedly due to having traced out the field degrees of freedom in the left Rindler wedge.
As clear from the general method, the expression of the entropy current depends on the form of the energy-momentum tensor quantum operator.
All our calculations were performed with the canonical tensor, i.e.\ the minimal coupling one, but then we compared the result with the one we would have obtained by using an ``improved'' tensor which is traceless for a massless field, i.e.\ the conformal coupling one.
Interestingly enough, the quantum correction due to the acceleration field disappears in the latter case.

Finally, we integrated the above entropy current and obtained the entropy in the right Rindler wedge.
The result has two apparent properties, both in full agreement with known literature: the one is the so-called area law, namely the proportionality to the area of the 2-dimensional surface separating the two Rindler wedges, clearly reminiscent of the Bekenstein-Hawking formula for the black hole entropy; whereas the other is the divergence due to the proper temperature being infinite at the Killing horizon.
This entropy can also be seen as the one obtained by using the von Neumann formula with the reduced density operator of the right Rindler wedge.
By virtue of the factorization of the overall density operator, this is also, by definition, the entanglement entropy of the right Rindler wedge with the left Rindler wedge.
The calculation of the entanglement entropy in Quantum Field Theory is a very hard task and a hot topic in Quantum Information Theory in general.
Having been able to perform it in an exact analytical way by using a method completely orthogonal to the standard ones in the case of a system both mathematically non-trivial and of physical concern adds all the more interest to our result.
We also worked out the potential of the entropy current and recast the entropy in the form of a 2-dimensional surface integral.
Oddly, while maintaining the area law and the divergence at the Killing horizon, the expression of the entropy seemed to change by a constant factor if we considered the ``improved'' energy-momentum tensor instead of the canonical one.
This is a somewhat unexpected feature that deserves further study.



\chapter{Relativistic Quantum Fluid out of Equilibrium with Boost Invariance}
\label{chapter:boost}

In this Chapter, we will consider a relativistic quantum fluid with boost invariance,
also known in literature as the Bjorken model, named after J.\ D.\ Bjorken who first proposed it as a symmetry approximately realized in the central-rapidity region in heavy-ion collisions, and as such phenomenologically interesting.
Modern hydrodynamic calculations can go beyond the Bjorken model by including transverse expansion and, eventually, by breaking boost invariance itself; however, they still follow its general principles.
A boost-invariant fluid is inherently out of thermodynamic equilibrium.
Such fluids are notoriously hard to study, one of the reasons being that the thermodynamic fields are essentially unknown.
However, boost invariance will put special constraints providing us with enough information to at least attempt actual calculations.
In particular, a local thermodynamic equilibrium analysis will be feasible exactly, making this, to the best of our knowledge, the first ever case of an exactly solvable system at local thermodynamic equilibrium, therefore quite interesting on the theoretical side.
This will also give us the opportunity to apply our method for the entropy current at local thermodynamic equilibrium.
A full non-equilibrium analysis will also be carried out in some approximation, making this system a benchmark for the vast and mostly unexplored subject that is non-equilibrium thermal field theory.

We will start by defining boost invariance as a symmetry of the density operator and seeing how it can be realized in the future light-cone, what will make it convenient to introduce the so-called Milne coordinates.
As for the underlying Quantum Field Theory, we will consider again a free real scalar field theory, but this time there will be no analog of the Unruh effect.
Next, we will perform a local thermodynamic equilibrium analysis and calculate exactly the thermal expectation value of the energy-momentum tensor, discussing also the possible renormalization schemes.
With this ingredient, we will be able to calculate the entropy current by using our method at local thermodynamic equilibrium.
Finally, we will present an analysis fully out of thermodynamic equilibrium.
The thermal expectation value of the energy-momentum tensor will be worked out by performing an asymptotic expansion at late times, and renormalization will be discussed.

This Chapter will be based entirely on \cite{Rindori:2021quq} by D.\ R., L.\ Tinti, F.\ Becattini and D.\ H.\ Rischke, which in turn was motivated in the first place by the recent works \cite{Akkelin:2018hpu, Akkelin:2020cfs} by S.\ Akkelin.
For additional information on the Bjorken model and its role in heavy-ion collisions, we refer to Bjorken's original paper \cite{Bjorken:1982qr} as well as the textbook \cite{Florkowski:2010zz}; whereas concerning Quantum Field Theory in Milne coordinates, we suggest the textbook \cite{Birrell:1982ix}.




\section{Thermal field theory with boost invariance}
\label{sec:boost_boost_invariance}

In order to be able to make any thermal field theory, we must first determine the density operator representing the state at hand, which will then enable us to calculate thermal expectation values.
In this Section, we will present the density operator of a relativistic quantum fluid with boost invariance, a non-trivial instance of a non-equilibrium state in the future light-cone particularly interesting in the context of heavy-ion collisions.


\subsection{Boost invariance}

\begin{figure}
    \begin{center}
	    \includegraphics[width=0.8\textwidth]{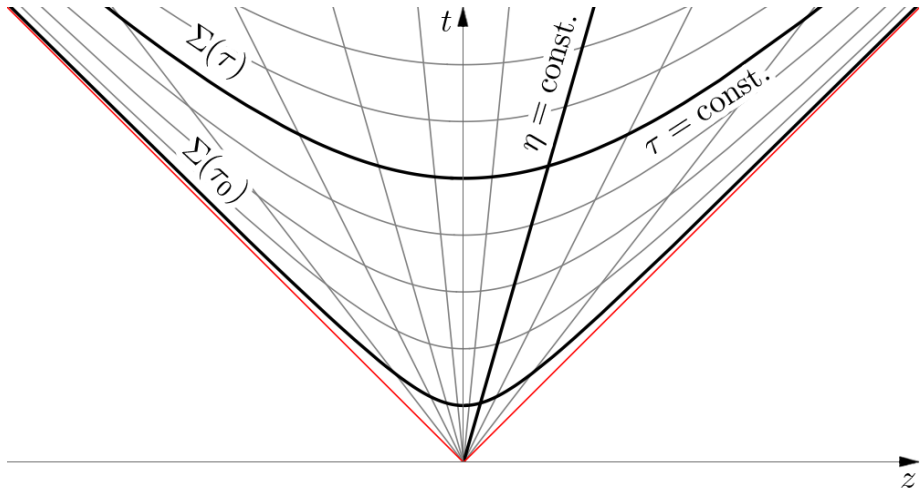}
	    \caption{2-dimensional section of the future light-cone in the $(t,z)$ plane, spanned by the Milne coordinates $(\tau,\eta)$.
	    The hyperbolae are $\tau={\rm const.}$ hypersurfaces, while the straight lines through the origin are $\eta={\rm const.}$ hypersurfaces.}
	    \label{fig:milne_coord}
	\end{center}
\end{figure}

In Chapter \ref{chapter:gteacceleration}, we considered a relativistic quantum fluid at global thermodynamic equilibrium in the right Rindler wedge.
This time, we focus on another patch of Minkowski spacetime, that is the future light-cone defined by $t>|z|$.
Much in the same way as we introduced the Rindler coordinates to span the right Rindler wedge, we now introduce a new set of hyperbolic coordinates in order to span the future light-cone.
These are the so-called \textit{Milne coordinates} $(\tau,\xT,\eta)$, where the transverse coordinates $\xT=(x,y)$ are once again the same as Minkowski's, while $(\tau,\eta)$ are related to $(t,z)$ by
\begin{equation}
    \tau \equiv \sqrt{t^2-z^2},\qquad
    \eta \equiv \frac{1}{2}\log \left(\frac{t+z}{t-z}\right)
\end{equation}
whose inverse read
\begin{equation}
    t\equiv \tau \cosh \eta,\qquad
    z\equiv \tau \sinh \eta.
\end{equation}
The hyperbolic time $\tau$ can be referred to as the \textit{Milne time}, while the coordinate $\eta$ is called the \textit{spacetime rapidity}.
The Milne coordinates span indeed the future light-cone, as shown in Figure \ref{fig:milne_coord},
and the reason for their name is that the line element is but the metrics of the so-called \textit{Milne universe}
\begin{equation}
    \di s^2=
    \di t^2-\di x^2-\di y^2-\di z^2=
    \di \tau^2-\di x^2-\di y^2-\tau^2 \di \eta^2.
\end{equation}
The foliation of the future light-cone with the family of 3-dimensional spacelike hypersurfaces $\{\Sigma(\tau)\}$ given by the constant-$\tau$ hyperboloids is well-defined, for the vorticity-free condition \eqref{zubeq07} is met by the timelike vector field of unit magnitude orthogonal to them,
\begin{equation}
    n^{\mu}=\frac{1}{\sqrt{t^2-z^2}}(t,0,0,z)=
    (\cosh \eta,0,0,\sinh \eta).
\end{equation}
This vector field can be used to choose the hydrodynamic velocity field consistently with the foliation by taking $u^{\mu}=n^{\mu}$, which is essentially the $\beta$-frame, thus
\begin{equation}\label{boosteq07}
    u^{\mu}=(\cosh \eta,0,0,\sinh \eta).
\end{equation}
Then, the tetrad onto which tensors are decomposed is most naturally given by
\begin{subequations}
    \begin{align}
        (\de_{\tau})^{\mu}=&(\cosh \eta,0,0,\sinh \eta)=u^{\mu},\label{boosteq06a}\\
        (\de_x)^{\mu}=&(0,1,0,0)\equiv \hat{i}^{\mu},\label{boosteq06b}\\
        (\de_y)^{\mu}=&(0,0,1,0)\equiv \hat{j}^{\mu},\label{boosteq06c}\\
        (\de_{\eta})^{\mu}=&\tau(\sinh \eta,0,0,\cosh \eta)\equiv \tau \hat{\eta}^{\mu},\label{boosteq06d}
    \end{align}
\end{subequations}
where the hat $\hat{}$ stands for unit magnitude, namely $\hat{i}^2=\hat{j}^2=\hat{\eta}^2=-1$.

For reasons that are going to become clear shortly, let us assume that the proper temperature and the fugacity are scalar functions depending on the Milne time $\tau$ only, that is $T=T(\tau)$ and $\zeta=\zeta(\tau)$, so the four-temperature reads
\begin{equation}\label{boosteq03}
    \beta^{\mu}=\frac{u^{\mu}}{T(\tau)}=\frac{1}{T(\tau)}(\cosh \eta,0,0,\sinh \eta).
\end{equation}
Suppose also that the system reaches local thermodynamic equilibrium at some $\tau_0$, so, according to the discussion in Subsection \ref{sec:LE_to_GE_and_NE}, the stationary state in the Heisenberg picture is the local thermodynamic equilibrium density operator at $\tau_0$
\begin{equation}\label{boosteq01}
    \wh{\rho}=\wh{\rho}_{\rm LE}(\tau_0)=
    \frac{1}{Z_{\rm LE}(\tau_0)}\exp \left[-\int_{\Sigma(\tau_0)}\di \Sigma_{\mu}\left(\wh{T}^{\mu \nu}\beta_{\nu}-\zeta \wh{j}^{\mu}\right)\right],
\end{equation}
where the argument of the integral is intended to be evaluated at $\tau_0$.
The four-temperature is timelike on $\Sigma(\tau_0)$, hence thermodynamically meaningful, but it is not a Killing vector field, as can be checked.
The Milne universe, in fact, is not a stationary spacetime, and as such it does not possess any global timelike Killing vector field.
From the cosmological point of view, it represents instead a linearly expanding universe foliated by hyperboloids with constant negative curvature.
As a consequence, global thermodynamic equilibrium states cannot exist in the Milne universe, therefore the density operator \eqref{boosteq01} represents a non-equilibrium state.

In Subsection \ref{sec:LE_to_GE_and_NE} it was pointed out that if the spacelike hypersurface $\Sigma(\tau_0)$ is invariant under any transformation $g$ of a subgroup ${\rm G}$ of the proper orthochronous Poincar\'e group ${\rm IO}(1,3)^{\uparrow}_+$ and the thermodynamic fields change according to the transformation laws \eqref{zubeq12}, which we hereby report
\begin{equation}\label{boosteq02}
    {D(g^{-1})^{\nu}}_{\sigma}\beta_{\nu}(g^{-1}(y))=\beta_{\sigma}(y),\qquad
    \zeta(g^{-1}(y))=\zeta(y),
\end{equation}
then ${\rm G}$ is a symmetry group for the density operator, in the sense that $\wh{U}(g)\wh{\rho}\wh{U}^{-1}(g)=\wh{\rho}$ with $\wh{U}(g)$ a unitary representation of $g\in {\rm G}$.
The thermodynamic fields appearing in the density operator \eqref{boosteq01}, namely the four-temperature \eqref{boosteq03} at $\tau_0$ and the fugacity $\zeta(\tau_0)$, meet the conditions \eqref{boosteq02} for the following transformations:
\begin{itemize}
    \item boosts with hyperbolic angle $\xi$ along the $z$ direction, indicated as ${\sf L}_3(\xi)$,
    \item translations along the $x$ and $y$ directions, indicated as ${\sf T}_1(a)$ and ${\sf T}_2(a)$ respectively,
    \item rotations of angle $\varphi$ in the $(x,y)$ plane, indicated as ${\sf R}_{12}(\varphi)$,
\end{itemize}
and the hyperboloid $\Sigma(\tau_0)$ is invariant under the same transformations.
This is but the Euclidean group in 2-dimensions times the Lorentz group in 1-dimension, namely ${\rm IO}(2)\otimes {\rm SO}(1,1)$.
Moreover, the density operator is also invariant under:
\begin{itemize}
    \item space reflections mapping $(x,y,z)$ to $(-x,-y,-z)$.
\end{itemize}
Altogether, we refer to this symmetry of the density operator as \textit{boost invariance}.
Accordingly, the description of the system represented by $\wh{\rho}$ must have the same form in all frames boosted in the longitudinal direction.
This is essentially what in literature is often known as the \textit{Bjorken symmetry}, named after J.\ D.\ Bjorken who first proposed it as a symmetry approximately realized in the central-rapidity region in heavy-ion collisions \cite{Bjorken:1982qr}.

As it should be clear by now, the symmetries of the system constrain the form of quantities at the particular thermodynamic configuration at hand.
These constraints are given by the vanishing of Lie derivatives of such quantities along the vector fields associated to the symmetry group, which in the present case can be readily found
\begin{subequations}
    \begin{align}
        \left(\frac{\di {\sf L}_3(\xi)x}{\di \xi}\right)^{\mu}=&(z,0,0,t)=\tau \hat{\eta}^{\mu},\label{boosteq04a}\\
        \left(\frac{\di {\sf T}_1(a)x}{\di a}\right)^{\mu}=&(0,1,0,0)=\hat{i}^{\mu},\label{boosteq04b}\\
        \left(\frac{\di {\sf T}_2(a)y}{\di a}\right)^{\mu}=&(0,0,1,0)=\hat{j}^{\mu},\label{boosteq04c}\\
        \left(\frac{\di {\sf R}_{12}(\varphi)x}{\di \varphi}\right)^{\mu}=&(0,-y,x,0)=r \hat{\varphi}^{\mu}\label{boosteq04d},
    \end{align}
\end{subequations}
where $r\equiv \sqrt{x^2+y^2}$ and $\hat{\varphi}^{\mu}\equiv \frac{1}{r}(0,-y,x,0)$.
Note that  \eqref{boosteq04a}--\eqref{boosteq04c} are just \eqref{boosteq06b}--\eqref{boosteq06d} of the above tetrad, which, by construction, have vanishing Lie derivatives along each other.
Scalars, such as the proper temperature $T$ and the fugacity $\zeta$, can only depend on $\tau$ as a consequence of ${\cal L}_XT=0$ and ${\cal L}_X\zeta=0$ with $X$ either of \eqref{boosteq04a}--\eqref{boosteq04d}
\begin{equation}\label{boosteq05}
    T=T(\tau),\qquad \zeta=\zeta(\tau).
\end{equation}
Let us prove it for $T$ as an example.
With little effort, one can show that
\begin{equation}
    {\cal L}_{\hat{i}}T=\hat{i}^{\mu}\de_{\mu}T=\de_xT=0,
\end{equation}
\begin{equation}
    {\cal L}_{\hat{j}}T=\hat{j}^{\mu}\de_{\mu}T=\de_yT=0,
\end{equation}
\begin{equation}
    {\cal L}_{\tau \hat{\eta}}T=\tau \hat{\eta}^{\mu}\de_{\mu}T=\de_{\eta}T=0,
\end{equation}
\begin{equation}
    {\cal L}_{r\hat{\varphi}}T=r\hat{\varphi}^{\mu}\de_{\mu}T=-y\de_yT+x\de_xT=0.
\end{equation}
The first two equations imply that $T$ cannot depend neither on $x$ nor on $y$, making the last equation trivial, while the third equation forbids the dependence on $\eta$, so the only possible dependence left is in fact on $\tau$.
The same holds for $\zeta$, hence \eqref{boosteq05}, and that is the reason why we previously assumed the temperature and the fugacity to be scalar functions of $\tau$ only.
This philosophy can be further applied to vectors and tensors of any rank.
A vector field $V^{\mu}$ is decomposed, in principle, as
\begin{equation}
    V^{\mu}=A(\tau)u^{\mu}+B(\tau)\hat{i}^{\mu}+C(\tau)\hat{j}^{\mu}+D(\tau)\tau \hat{\eta}^{\mu},
\end{equation}
however, the constrain ${\cal L}_{r\hat{\varphi}}V=0$ implies $B=C=0$, and the invariance under space reflections demands $D=0$ as it maps $\eta \mapsto -\eta$, hence
\begin{equation}
    V^{\mu}=A(\tau)u^{\mu}.
\end{equation}
Concerning vector fields, a well-known feature of the Bjorken model is the existence of a unique timelike vector field of unit magnitude compatible with boost invariance, and this is given precisely by \eqref{boosteq07} \cite{Bjorken:1982qr}.
That is to say that the hydrodynamic velocity field is in fact completely fixed by the symmetry, which is a very powerful statement for the following reason.
At global thermodynamic equilibrium, the thermodynamic fields are determined by the geometric conditions \eqref{zubeq10}, namely the four-temperature must be a Killing vector field and the fugacity a constant.
Out of global thermodynamic equilibrium, these conditions no longer hold and the thermodynamic fields can be whatever, so we have no information on them whatsoever.
This is one of the reasons that make systems out of global thermodynamic equilibrium much harder to study.
However, boost invariance completely fixes the hydrodynamic velocity field and constrains the proper temperature and the fugacity to depend on the Milne time only, therefore, the form of the four temperature cannot be but \eqref{boosteq03}.
The expressions of the temperature and the fugacity as functions of $\tau$ are obtained from the constraint equations of local thermodynamic equilibrium \eqref{zubeq04}, that is
\begin{equation}
	n_{\mu}\tr(\wh{\rho}_{\rm LE}(\tau)\wh{T}^{\mu \nu})_{\rm ren}=n_{\mu}T^{\mu \nu},\qquad
	n_{\mu}\tr(\wh{\rho}_{\rm LE}(\tau)\wh{j}^{\mu})_{\rm ren}=n_{\mu}j^{\mu}.
\end{equation}
Thus, although out of thermodynamic equilibrium, boost invariance actually provides us with information on the thermodynamic fields, enough so to at least attempt calculations.
This constitutes a precious opportunity for us to scratch the surface of the vast and mostly unexplored topic that is non-equilibrium physics.
In particular, we will be interested in the energy-momentum tensor as usual.
Much in the same way as above, the most general form a symmetric rank-2 tensor, such as the thermal expectation value of the energy-momentum tensor, is constrained to be
\begin{equation}\label{boosteq12}
    T^{\mu \nu}=\ed(\tau)u^{\mu}u^{\nu}+\pt(\tau)(\hat{i}^{\mu}\hat{i}^{\nu}+\hat{j}^{\mu}\hat{j}^{\nu})+\pl(\tau)\hat{\eta}^{\mu}\hat{\eta}^{\nu},
\end{equation}
where $\ed$ is the energy density, $\pt$ the transverse pressure and $\pl$ the longitudinal pressure.
For the sake of clarity, the left-hand side stands explicitly for
\begin{equation}\label{boosteq08}
	T^{\mu \nu}=\tr(\wh{\rho}\wh{T}^{\mu \nu})=
	\tr(\wh{\rho}_{\rm LE}(\tau_0)\wh{T}^{\mu \nu}(\tau)),
\end{equation}
namely the non-equilibrium density operator is at fixed $\tau_0$, whereas the energy-momentum tensor quantum operator varies with $\tau$, hence the $\tau$-dependence of $\ed$, $\pt$ and $\pl$.
The pressures in the $x$ and $y$ directions are equal and given by the transverse pressure due to rotational symmetry in the $(x,y)$ plane.
However, as full rotational symmetry in the $(x,y,z)$ space is lacking, the longitudinal pressure will be in general different.
This is reminiscent of the case of global thermodynamic equilibrium with acceleration studied in Chapter \ref{chapter:gteacceleration}, where an anisotropic quantum term was present in the longitudinal direction.
Thus, the anisotropy in the current case can also expected to be a quantum effect.

Provided an underlying Quantum Field Theory is chosen, the three scalar functions $\ed$, $\pt$ and $\pl$ can be determined once we are able to calculate thermal expectation values with the non-equilibrium density operator as in \eqref{boosteq08}.
This is a task one is not able to accomplish in general.
A possible approach to tackle this problem is that of ideal hydrodynamics, where local thermodynamic equilibrium is assumed.
In that case, the dissipations are supposed to be small, so that the non-equilibrium state is in fact close to thermodynamic equilibrium, and by equation \eqref{zubeq09}, the non-equilibrium density operator is approximated by the local thermodynamic equilibrium one.
However, this should not be misleading: unless one uses linear response theory to further approximate the local thermodynamic equilibrium density operator with the global thermodynamic equilibrium one \cite{Becattini:2014yxa}, the calculation of thermal expectation values at local thermodynamic equilibrium is usually far from feasible.
And even if it were, the actual non-equilibrium expressions of $\ed$, $\pt$ and $\pl$, that is the constitutive equations, would still be unknown.
Nevertheless, for our system at hand, boost invariance is so powerful that a non-equilibrium analysis turns out to be possible to some extent, at least for a free Quantum Field Theory.
For the sake of simplicity, in the following we will suppose that there is no charged current present, that is $\zeta \equiv 0$.
Thus, by using the same symbols as in \eqref{gteq10}, the non-equilibrium density operator is
\begin{equation}\label{boosteq16}
    \wh{\rho}=\frac{1}{Z}\exp \left[-\frac{\wh{\Pi}(\tau_0)}{T(\tau_0)}\right],
\end{equation}
where $Z=Z_{\rm LE}(\tau_0)$, $n_{\mu}=u_{\mu}$ and the operator $\wh{\Pi}(\tau_0)$ is given by
\begin{equation}
    \wh{\Pi}(\tau_0)=
    \int_{\Sigma(\tau_0)}\di \Sigma \,\wh{T}^{\mu \nu}(\tau_0)u_{\mu}u_{\nu}=
    \tau_0\int_{-\infty}^{+\infty}\di^2{\rm x_T}\,\di \eta \,\wh{T}^{\mu \nu}(\tau_0)u_{\mu}u_{\nu}
\end{equation}
with $\di \Sigma=\tau_0\,\di^2{\rm x_T}\,\di \eta$ the measure on $\Sigma(\tau_0)$ in Milne coordinates.
All the integration variables are intended to be integrated from $-\infty$ to $+\infty$.
The expression of the energy-momentum tensor quantum operator depends on the specific underlying Quantum Field Theory.
For a free Quantum Field Theory, this density operator represents a relativistic quantum fluid with hydrodynamic velocity field given by \eqref{boosteq07} which is initially interacting, ceases interactions at $\tau_0$ and propagates freely thereafter.
Thus, in the classical limit, the kinetic theory solutions of the free-streaming Boltzmann equation starting from the local thermodynamic 
equilibrium expressions with proper temperature $T(\tau_0)$ and flow velocity $u^{\mu}$ at $\tau_0$, reported in Appendix \ref{appendix:free_sreaming}, are expected to be recovered.

We want to conclude with the following remark.
In Subsection \ref{sec:LE_to_GE_and_NE}, the question arose if a symmetry group for the stationary state $\wh{\rho}=\wh{\rho}_{\rm LE}(\tau_0)$ is also a symmetry group for the local thermodynamic equilibrium density operator $\wh{\rho}_{\rm LE}(\tau)$ at any $\tau$.
Intuitively, if the hypersurface $\Sigma(\tau)$ is mapped into itself and the thermodynamic fields change according to \eqref{boosteq02} at any $\tau$, then the answer is yes.
The constant-$\tau$ hypersurfaces, the four-temperature \eqref{boosteq03} and the fugacity $\zeta(\tau)$ meet indeed those conditions at any $\tau$, therefore boost invariance is also a symmetry of $\wh{\rho}_{\rm LE}(\tau)$.
The expression of the local thermodynamic equilibrium density operator is formally the same as that of the non-equilibrium one at $\tau$ instead of $\tau_0$
\begin{equation}\label{boosteq15}
    \wh{\rho}_{\rm LE}(\tau)=\frac{1}{Z_{\rm LE}(\tau)}\exp \left[-\frac{\wh{\Pi}(\tau)}{T(\tau)}\right],
\end{equation}
where $\wh{\Pi}(\tau)$ is in general different from $\wh{\Pi}(\tau_0)$ and given by
\begin{equation}
    \wh{\Pi}(\tau)=
    \int_{\Sigma(\tau)}\di \Sigma \,\wh{T}^{\mu \nu}(\tau)u_{\mu}u_{\nu}=
    \tau \int_{-\infty}^{+\infty}\di^2{\rm x_T}\,\di \eta \,\wh{T}^{\mu \nu}(\tau)u_{\mu}u_{\nu}.
\end{equation}
In the case of a free Quantum Field Theory, the physical system described by this density operator is a relativistic quantum fluid with hydrodynamic velocity field given by \eqref{boosteq07} who never ceases interactions and does not decouple.
Of course, the forms of scalar, vector and tensor quantities are still those discussed above, as they descend purely from symmetry considerations.
In particular, the expression of the thermal expectation value of the energy-momentum tensor is
\begin{equation}\label{boosteq13}
    \langle \wh{T}^{\mu \nu}\rangle_{\rm LE}=
    \tr(\wh{\rho}_{\rm LE}(\tau)\wh{T}^{\mu \nu}(\tau))=
    \ed_{\rm LE}(\tau)u^{\mu}u^{\nu}+\pt_{\rm LE}(\tau)(\hat{i}^{\mu}\hat{i}^{\nu}+\hat{j}^{\mu}\hat{j}^{\nu})+\pl_{\rm LE}(\tau)\hat{\eta}^{\mu}\hat{\eta}^{\nu}.
\end{equation}
Since the non-equilibrium density operator $\wh{\rho}$ is simply the local thermodynamic equilibrium one $\wh{\rho}_{\rm LE}(\tau)$ evaluated at $\tau_0$, the operator $\wh{\Pi}(\tau_0)$ will just be $\wh{\Pi}(\tau)$ evaluated at $\tau_0$.
However, comparing equations \eqref{boosteq08} and \eqref{boosteq13}, it is clear that the non-equilibrium thermal expectation value \eqref{boosteq12} is not the same as the local thermodynamic equilibrium one \eqref{boosteq13} evaluated at $\tau_0$.
Once again, the energy-momentum tensor quantum operator, and therefore the scalar functions in the above thermal expectation value, depend on the specific underlying Quantum Field Theory.
In analogy with Chapter \ref{chapter:gteacceleration}, in the following we will consider a free real scalar quantum field in the future light-cone.


\subsection{Free scalar field theory in the future light-cone}

The problem of a free real scalar quantum field in the future light-cone is a known one in literature, especially in Quantum Field Theory in curved spacetime and cosmology in general \cite{Birrell:1982ix, Padmanabhan:1990fm, Arcuri:1994kd}, but also in the context of multiparticle production \cite{Berges:2017zws, Berges:2017hne}.
In \cite{Arcuri:1994kd} it was shown that the free real scalar field can be expanded in the future light-cone in a complete set of solutions that do not mix the positive- and negative-frequency modes of the usual plane waves expansion.
Then, S.\ Akkelin showed in \cite{Akkelin:2018hpu} that these modes can be obtained starting from the usual plane waves modes.
Here, we follow his derivation by using a slightly different notation and adding some clues.

The Klein Gordon equation for a free real scalar quantum field $\wh{\psi}$ of mass $m$ in Milne coordinates reads
\begin{equation}
    (\Box+m^2)\wh{\psi}=
    \left[\frac{1}{\tau}(\tau \de_{\tau})-\de_x^2-\de_y^2-\frac{1}{\tau^2}\de_{\eta}^2+m^2\right]\wh{\psi}=0.
\end{equation}
The solution is given by the following quantum field expansion
\begin{equation}\label{boosteq09}
    \wh{\psi}(\tau,\xT,\eta)=
    \int \frac{\di^2{\rm p_T}\,\di \mu}{4\pi \sqrt{2}}
    \left[h({\sf p},\tau){\rm e}^{i(\pT \cdot \xT+\mu \eta)}\wh{b}_{\sf p}+h^*({\sf p},\tau){\rm e}^{-i(\pT \cdot \xT+\mu \eta)}\wh{b}_{\sf p}^{\dagger}\right].
\end{equation}
Here, we indicated the momentum in Milne coordinates as ${\sf p}\equiv(\pT,\mu)$ to distinguish it from the Cartesian vector ${\bf p}=(\pT,p_z)$, where $\pT=(p_x,p_y)$ is the transverse momentum and $\mu$ is the momentum component along the $\eta$-direction.
The integral is intended to run from $-\infty$ to $+\infty$ for each integration variable.
The $\tau$-dependence is stored in the complex functions $h$ and $h^*$ solutions of the equation
\begin{equation}
    \left[\frac{1}{\tau}\de_{\tau}(\tau \de_{\tau})+\mT^2+\frac{\mu^2}{\tau^2}\right]h({\sf p},\tau)=0,
\end{equation}
which is indeed a Bessel equation with the transverse mass given by
\begin{equation}
    \mT=\sqrt{\pT^2+m^2}.
\end{equation}
In fact, $h$ and $h^*$ are defined by means of the Hankel functions ${\rm H}^{(2)}$ and ${\rm H}^{(1)}$ as
\begin{equation}\label{boosteq72}
    \begin{split}
        h({\sf p},\tau)\equiv&
        -i{\rm e}^{\frac{\pi}{2}\mu}{\rm H}^{(2)}_{i\mu}(\mT \tau)\\
        h^*({\sf p},\tau)\equiv&
        i{\rm e}^{-\frac{\pi}{2}\mu}{\rm H}^{(1)}_{i\mu}(\mT \tau),
    \end{split}
\end{equation}
which can be checked to be the complex conjugate of each other and also invariant under momentum reflections ${\sf p}\mapsto -{\sf p}$ by using the integral representation \cite{Gradshteyn:1702455}
\begin{equation}\label{boosteq10}
    \begin{split}
        {\rm H}^{(2)}_{i\mu}(\mT \tau)=&
        -\frac{1}{i\pi}{\rm e}^{-\frac{\pi}{2}\mu}\int_{-\infty}^{+\infty}\di \theta \,{\rm e}^{-i\mT \tau \cosh \theta+i\mu \theta}\\
        {\rm H}^{(1)}_{i\mu}(\mT \tau)=&
        \frac{1}{i\pi}{\rm e}^{\frac{\pi}{2}\mu}\int_{-\infty}^{+\infty}\di \theta \,{\rm e}^{i\mT \tau \cosh \theta-i\mu \theta}.
    \end{split}
\end{equation}
The following quantity is also usually defined in Milne coordinates
\begin{equation}\label{boosteq30}
    \omega^2({\sf p},\tau)\equiv \mT^2+\frac{\mu^2}{\tau^2}={\sf p}^2+m^2,
\end{equation}
which is invariant under momentum reflections ${\sf p}\mapsto -{\sf p}$.
Finally, $\wh{b}_{\sf p}^{\dagger}$ and $\wh{b}_{\sf p}$ are creation and annihilation operators respectively in the future light-cone obeying the usual algebra
\begin{equation}\label{boosteq18}
    \begin{split}
        [\wh{b}_{\sf p},\wh{b}_{{\sf p}'}^{\dagger}]=&\delta^2(\pT-\pT')\,\delta(\mu-\mu'),\\
        [\wh{b}_{\sf p},\wh{b}_{{\sf p}'}]=&0=[\wh{b}_{\sf p}^{\dagger},\wh{b}_{{\sf p}'}^{\dagger}].
    \end{split}
\end{equation}
This allows us to clarify the interpretation of $\mu$, which turns out to be the eigenvalue of the boost operator along the longitudinal direction $\wh{K}_z$, that is
\begin{equation}
    \wh{U}({\sf L}_3(\xi))\wh{b}_{\sf p}^{\dagger}\wh{U}^{-1}({\sf L}_3(\xi))={\rm e}^{-i\xi \wh{K}_z}\wh{b}_{\sf p}^{\dagger}{\rm e}^{i\xi \wh{K}_z}={\rm e}^{-i\xi \mu}\wh{b}_{\sf p}^{\dagger}
\end{equation}
with $\wh{U}$ a unitary representation.
In other words, the operator $\wh{b}_{\sf p}^{\dagger}$ creates a state with eigenvalue $\mu$, much in the same way as for the angular momentum operator in the familiar Quantum Mechanics.

In \cite{Akkelin:2018hpu} it was shown explicitly that the quantum field expansion \eqref{boosteq09} in the future light-cone can also be worked out starting from the standard plane waves expansion in Minkowski spacetime \eqref{gteq27}.
The frequency of the plane wave \eqref{gteq49} is
\begin{equation}
    \omega_{\bf p}^2=\mT^2+p_z^2.
\end{equation}
One can then express $\omega_{\bf p}$ and $p_z$ in terms of a hyperbolic angle ${\rm y}$ called the \textit{momentum rapidity}, or simply the \textit{rapidity}, as
\begin{equation}
    \omega_{\bf p}=\mT \cosh {\rm y},\qquad
    p_z=\mT \sinh {\rm y}.
\end{equation}
Thus, by performing the following ``Fourier-like'' transformation of the plane wave creation and annihilation operators $\wh{a}_{\bf p}^{\dagger}$ and $\wh{a}_{\bf p}$ in the plane waves expansion
\begin{equation}\label{boosteq11}
    \wh{a}_{\bf p}^{\dagger}=
    \frac{1}{\sqrt{2\pi \mT \cosh {\rm y}}}\int_{-\infty}^{+\infty}\di \mu \,{\rm e}^{-i\mu {\rm y}}\wh{b}_{\sf p}^{\dagger},
\end{equation}
changing integration variable from $p_z$ to ${\rm y}$ and then to $\theta={\rm y}-\eta$ and exploiting the integral representation \eqref{boosteq10}, the field expansion \eqref{boosteq09} is finally obtained \cite{Akkelin:2018hpu}.
Equation \eqref{boosteq11} shows that the creation operators in the future light-cone depend only on the creation operators in Minkowski spacetime, and same for the corresponding annihilation operators.
No mixing between creation and annihilation operators is present.
Consequently, the vacuum state of the operators $\wh{b}_{\sf p}$ in the future light-cone coincides with that of $\wh{a}_{\bf p}$ in Minkowski spacetime, that is the Minkowski vacuum $|0_{\rm M}\rangle$.
The ultimate reason for this is that the function $h$ can be expressed can be expressed as a linear combination of plane wave modes with only positive frequencies \cite{Mukhanov:2007zz}.
This is clearly at variance with the case of global thermodynamic equilibrium with acceleration considered in Chapter \ref{chapter:gteacceleration}, where the quantum field was expanded in the Rindler wedges in terms of the Rindler creation and annihilation operators, whose relation with the plane waves creation and annihilation operators was a non-trivial Bogolyubov transformation, therefore the Rindler vacuum and the Minkowski vacuum did not coincide.
The consequences of this fact will be further explored in Subsection \ref{sec:boost_lte_renormalization}.

Now we turn to the density operator.
It is important to stress that the expression of the non-equilibrium density operator \eqref{boosteq16} is exactly the same as the local thermodynamic equilibrium one \eqref{boosteq15} evaluated at $\tau_0$.
Whether one or the other is used makes indeed a difference when it comes to calculating thermal expectation values, as exemplified by equations \eqref{boosteq08} and \eqref{boosteq13}.
However, as far as the very structure of the density operator is concerned, they really are equivalent.
Therefore, without loss of generality, we can let $\tau$ be whatever and focus on the local thermodynamic equilibrium one for the rest of this part.
Then, in Section \ref{sec:boost_lte_analysis} we will perform a local thermodynamic equilibrium analysis, whereas in Section \ref{sec:boost_ne_analysis} a non-equilibrium one, thus distinguishing explicitly the two cases.
Moreover, the ${\sf p}$-dependence will be omitted throughout for ease of notation, restoring it only when necessary.

As for the energy-momentum tensor operator, we consider again the canonical tensor
\begin{equation}\label{boosteq50}
    \wh{T}^{\mu \nu}=\frac{1}{2}\left(\de^{\mu}\wh{\psi}\de^{\nu}\wh{\psi}+\de^{\nu}\wh{\psi}\de^{\mu}\wh{\psi}\right)-g^{\mu \nu}\wh{\cal L},\qquad
    \wh{\cal L}=\frac{1}{2}\left(g^{\mu \nu}\de_{\mu}\wh{\psi}\de_{\nu}\wh{\psi}-m^2\wh{\psi}^2\right).
\end{equation}
In order to work $\wh{\Pi}(\tau)$ out, the above expression must be contracted twice with the hydrodynamic velocity field, yielding
\begin{equation}\label{boosteq34}
    \wh{T}^{\mu \nu}u_{\mu}u_{\nu}=
    \frac{1}{2}\left[(\de_{\tau}\wh{\psi})^2+(\de_x\wh{\psi})^2+(\de_y\wh{\psi})^2+\frac{1}{\tau^2}(\de_{\eta}\wh{\psi})^2+m^2\wh{\psi}^2\right].
\end{equation}
By using the quantum field expansion \eqref{boosteq09} together with the integral representation of the Hankel functions \eqref{boosteq10} and exploiting the invariance of $h(\tau)$ and $h^*(\tau)$ under reflections ${\sf p}\mapsto -{\sf p}$, the following expression is obtained
\begin{equation}
    \begin{split}
        \wh{\Pi}(\tau)=&
        \tau \int \di^2{\rm p_T}\,\di \mu \frac{\pi}{8}\left[\left(|\de_{\tau}h(\tau)|^2+\omega^2(\tau)|h(\tau)|^2\right)\left(\wh{b}_{\sf p}\wh{b}_{\sf p}^{\dagger}+\wh{b}_{\sf p}^{\dagger}\wh{b}_{\sf p}\right)\right.\\
        &\left.+\left((\de_{\tau}h(\tau))^2+\omega^2(\tau)h^2(\tau)\right)\wh{b}_{\sf p}\wh{b}_{-{\sf p}}+
        \left((\de_{\tau}h^*(\tau))^2+\omega^2(\tau)(h^*(\tau))^2\right)\wh{b}_{\sf p}^{\dagger}\wh{b}_{-{\sf p}}^{\dagger}\right].
    \end{split}
\end{equation}
It is convenient to define the positive real function $K(\tau)$ and the complex function $\Lambda(\tau)$ as
\begin{equation}\label{boosteq35}
    K(\tau)\equiv \frac{\pi \tau}{4\omega(\tau)}\left[|\de_{\tau}h(\tau)|^2+\omega^2(\tau)|h(\tau)|^2\right],
\end{equation}
\begin{equation}\label{boosteq36}
    \Lambda(\tau)\equiv \frac{\pi \tau}{4\omega(\tau)}\left[(\de_{\tau}h(\tau))^2+\omega^2(\tau)h^2(\tau)\right].
\end{equation}
Let us stress that the ${\sf p}$-dependence is intended.
Being $h(\tau)$ and $\omega(\tau)$ invariant under momentum reflections ${\sf p}\mapsto -{\sf p}$, so are $K(\tau)$ and $\Lambda(\tau)$, but more interestingly they are related by
\begin{equation}\label{boosteq14}
    K^2(\tau)-|\Lambda(\tau)|^2=1.
\end{equation}
This occurs thanks to the left-hand side being proportional to the Wronskian of the Hankel functions
\begin{equation}
    K^2(\tau)-|\Lambda(\tau)|^2=
    -\left(\frac{\pi \mT \tau}{4}\right)^2
    \left(W[{\rm H}^{(2)}_{i\mu}(\mT \tau),{\rm H}^{(1)}_{i\mu}(\mT \tau)]\right)^2,
\end{equation}
which not only is known, but it turns out to be a very simple function of the argument \cite{Gradshteyn:1702455}
\begin{equation}
    W[{\rm H}^{(2)}_{i\nu}(x),{\rm H}^{(1)}_{i\nu}(x)]=
    {\rm H}^{(2)'}_{i\nu}(x){\rm H}^{(1)}_{i\nu}(x)-{\rm H}^{(1)'}_{i\nu}(x){\rm H}^{(2)}_{i\nu}(x)=
    \frac{4i}{\pi x}.
\end{equation}
This will greatly simplify the calculation of thermal expectation values at local thermodynamic equilibrium in the next Section.
The property \eqref{boosteq14} can also be rephrased by expressing $K(\tau)$ and $\Lambda(\tau)$ in terms of some hyperbolic angle $\Theta({\sf p},\tau)$ and phase $\chi({\sf p},\tau)$
\begin{equation}\label{boosteq22}
    K(\tau)=\cosh(2\Theta(\tau)),\qquad
    \Lambda(\tau)={\rm e}^{i\chi(\tau)}\sinh(2\Theta(\tau)),
\end{equation}
which will be useful later on.
For now, $\wh{\Pi}(\tau)$ is recast in terms of $K(\tau)$ and $\Lambda(\tau)$ as
\begin{equation}\label{boosteq17}
    \wh{\Pi}(\tau)=\int \di^2{\rm p_T}\,\di \mu \frac{\omega(\tau)}{2}
    \left[K(\tau)\left(\wh{b}_{\sf p}\wh{b}_{\sf p}^{\dagger}+\wh{b}_{\sf p}^{\dagger}\wh{b}_{\sf p}\right)+\Lambda(\tau)\wh{b}_{\sf p}\wh{b}_{-{\sf p}}+\Lambda^*(\tau)\wh{b}_{\sf p}^{\dagger}\wh{b}_{-{\sf p}}^{\dagger}\right]
\end{equation}
At variance with equations \eqref{gteq28} and \eqref{gteq29} found in Chapter \ref{chapter:gteacceleration}, this expression is not diagonal in the creation and annihilation operators due to the terms proportional to $\Lambda(\tau)$ and $\Lambda^*(\tau)$.
This is quite unfortunate, for if it were diagonal, thermal expectation values of products of creation and annihilation operators, whence of operators quadratic in the field, could be calculated by using standard methods.
In the following, we look for a suitable Bogolyubov transformation to diagonalize $\wh{\Pi}(\tau)$.


\subsection{Diagonalization of the density operator}

Let us define a new set of creation and annihilation operators by combining $\wh{b}_{\sf p}$ and $\wh{b}_{\sf p}^{\dagger}$ to attempt a diagonalization of $\wh{\Pi}(\tau)$ in \eqref{boosteq17}, whence $\wh{\rho}_{\rm LE}(\tau)$ in \eqref{boosteq15}.
Since the terms we wish to get rid of mix $\wh{b}_{\sf p}$ with $\wh{b}_{-{\sf p}}$ and their Hermitean conjugates, we start with a Bogolyubov transformation of the following form
\begin{equation}\label{boosteq19}
    \begin{split}
        \wh{\xi}_{\sf p}^{\dagger}(\tau)\equiv &A({\sf p},\tau)\wh{b}_{\sf p}^{\dagger}-B({\sf p},\tau)\wh{b}_{-{\sf p}}\\
        \wh{\xi}_{\sf p}(\tau)\equiv &A^*({\sf p},\tau)\wh{b}_{\sf p}-B^*({\sf p},\tau)\wh{b}_{-{\sf p}}^{\dagger}
    \end{split}
\end{equation}
where $A$ and $B$ are complex functions to be determined.
In order for $\wh{\xi}_{\sf p}^{\dagger}(\tau)$ and $\wh{\xi}_{\sf p}(\tau)$ to be creation and annihilation operators respectively, they are demanded to obey the usual algebra
\begin{equation}\label{boosteq27}
    \begin{split}
        [\wh{\xi}_{\sf p}(\tau),\wh{\xi}_{{\sf p}'}^{\dagger}(\tau)]=&
        \delta^2(\pT-\pT')\,\delta(\mu-\mu')\\
        [\wh{\xi}_{\sf p}(\tau),\wh{\xi}_{{\sf p}'}(\tau)]=&0=
        [\wh{\xi}_{\sf p}^{\dagger}(\tau),\wh{\xi}_{{\sf p}'}^{\dagger}(\tau)].
    \end{split}
\end{equation}
By enforcing the commutation relations \eqref{boosteq18}, these in turn imply
\begin{equation}
    \begin{split}
        &\left(|A({\sf p},\tau)|^2-|B({\sf p},\tau)|^2\right)\delta^2(\pT-\pT')\,\delta(\mu-\mu')=\delta^2(\pT-\pT')\,\delta(\mu-\mu')\\
        &\left[A^*(-{\sf p},\tau)B^*({\sf p},\tau)-A^*({\sf p},\tau)B^*(-{\sf p},\tau)\right]\delta^2(\pT+\pT')\,\delta(\mu+\mu')=0\\
        &\left[A({\sf p},\tau)B(-{\sf p},\tau)-A(-{\sf p},\tau)B({\sf p},\tau)\right]\delta^2(\pT+\pT')\,\delta(\mu+\mu')=0
    \end{split}
\end{equation}
whose a possible solution is
\begin{equation}\label{boosteq20}
    A({\sf p},\tau)=A(-{\sf p},\tau),\qquad
    B({\sf p},\tau)=B(-{\sf p},\tau),\qquad
    |A({\sf p},\tau)|^2-|B({\sf p},\tau)|^2=1
\end{equation}
hence the following hyperbolic expressions
\begin{equation}\label{boosteq21}
    A({\sf p},\tau)=\cosh \theta({\sf p},\tau) \,{\rm e}^{i\chi_A({\sf p},\tau)},\qquad
    B({\sf p},\tau)=\sinh \theta({\sf p},\tau) \,{\rm e}^{i\chi_B({\sf p},\tau)}
\end{equation}
for some hyperbolic angle $\theta({\sf p},\tau)$ and phases $\chi_A({\sf p},\tau)$ and $\chi_B({\sf p},\tau)$.
This is not the only possible solution, but we choose it as it makes it easy to invert the Bogolyubov transformation \eqref{boosteq19} as
\begin{equation}\label{boosteq24}
    \begin{split}
        \wh{b}_{\sf p} =& A({\sf p},\tau)\wh{\xi}_{\sf p}(\tau) + B^*({\sf p},\tau)\wh{\xi}_{-{\sf p}}^{\dagger}(\tau)\\
        \wh{b}_{\sf p}^{\dagger} =& A^*({\sf p},\tau)\wh{\xi}_{\sf p}^{\dagger}(\tau) + B({\sf p},\tau)\wh{\xi}_{-{\sf p}}(\tau),
    \end{split}
\end{equation}
plug it into \eqref{boosteq17} and obtain
\begin{equation}\label{boosteq26}
    \begin{split}
        \wh{\Pi}(\tau)=&
        \int{\rm d}^2{\rm p_T}\,{\rm d}\mu\,\frac{\omega}{2}\left\{
        \left[K\left(|A|^2+|B|^2\right)+\Lambda AB^*+\Lambda^*A^*B\right]\left(\wh{\xi}_{\sf p}\wh{\xi}_{\sf p}^{\dagger}+\wh{\xi}_{\sf p}^{\dagger}\wh{\xi}_{\sf p}\right)\right.\\
        &\left.+\left(2KAB+\Lambda A^2+\Lambda^*B^2\right)\wh{\xi}_{\sf p}\wh{\xi}_{-{\sf p}}+\left(2KA^*B^*+\Lambda^*{A^*}^2+\Lambda{B^*}^2\right)\wh{\xi}_{\sf p}^{\dagger}\wh{\xi}_{-{\sf p}}^{\dagger}\right\},
    \end{split}
\end{equation}
where the invariance under momentum reflections ${\sf p}\mapsto -{\sf p}$ was used.
In order to make $\wh{\Pi}(\tau)$ diagonal, the second line must vanish, so we look for $A$ and $B$ such that
\begin{subequations}
    \begin{align}
        2KAB+\Lambda A^2+\Lambda^*B^2=&0\label{boosteq23a}\\
        2KA^*B^*+\Lambda^*{A^*}^2+\Lambda{B^*}^2=&0.
    \end{align}
\end{subequations}
The two equations are the complex conjugate of each other, so they are in fact one complex equation, that is two real equations.
Apparently there are four real unknowns, namely the moduli and phases of $A$ and $B$, but the moduli are related by the last equation in \eqref{boosteq20}.
Moreover, the two equations are invariant under an overall phase shift of both $A$ and $B$, so they actually depend on the difference between the two phases instead of the phases themselves.
In summary, they are two real equations in two real unknowns, one modulus and the phase difference, so we can try to solve the system.
With the hyperbolic expressions \eqref{boosteq21} and \eqref{boosteq22}, equation \eqref{boosteq23a} can be rewritten as
\begin{equation}
    \cosh(2\Theta)\sinh(2\theta)\,{\rm e}^{i(\chi_A+\chi_B)}+\sinh(2\Theta)\left(\cosh^2\theta \,{\rm e}^{i(\chi+2\chi_A)}+\sinh^2\theta \,{\rm e}^{i(2\chi_B-\chi)}\right)=0
\end{equation}
whose solution is
\begin{equation}
    \chi_B({\sf p},\tau)-\chi_A({\sf p},\tau)=\chi({\sf p},\tau),\qquad
    \theta({\sf p},\tau)=-\Theta({\sf p},\tau).
\end{equation}
By setting $\chi_A\equiv 0$ without loss of generality, the functions $A$ and $B$ fulfilling the Bogolyubov relations \eqref{boosteq19} are
\begin{equation}\label{boosteq25}
    A({\sf p},\tau)=\cosh \Theta({\sf p},\tau),\qquad
    B({\sf p},\tau)=-\sinh \Theta({\sf p},\tau) \,{\rm e}^{i\chi({\sf p},\tau)},
\end{equation}
so the inverse transformation \eqref{boosteq24} reads
\begin{equation}\label{boosteq29}
    \begin{split}
        \wh{b}_{\sf p}=&\cosh \Theta({\sf p}, \tau)\wh{\xi}_{\sf p}(\tau)-\sinh \Theta({\sf p},\tau)\,{\rm e}^{-i\chi({\sf p},\tau)}\wh{\xi}_{-{\sf p}}^{\dagger}(\tau)\\
        \wh{b}_{\sf p}^{\dagger}=&\cosh \Theta({\sf p}, \tau) \wh{\xi}_{\sf p}^{\dagger}(\tau)-\sinh \Theta ({\sf p},\tau)\,{\rm e}^{i\chi({\sf p},\tau)}\wh{\xi}_{-{\sf p}}(\tau).
    \end{split}
\end{equation}
Plugging \eqref{boosteq25} and \eqref{boosteq22} into \eqref{boosteq26} and using the commutation relations \eqref{boosteq27} we finally obtain the following result
\begin{equation}
    \begin{split}
        \wh{\Pi}(\tau)=&
        \int \di^2{\rm p_T}\,\di \mu \, \frac{\omega(\tau)}{2}\left(\wh{\xi}_{\sf p}(\tau)\wh{\xi}_{\sf p}^{\dagger}(\tau)+\wh{\xi}_{\sf p}^{\dagger}(\tau)\wh{\xi}_{\sf p}(\tau)\right)\\
        =&\int \di^2{\rm p_T}\,\di \mu \, \omega(\tau) \left(\wh{\xi}_{\sf p}^{\dagger}(\tau)\wh{\xi}_{\sf p}(\tau)+\frac{1}{2}\right).
    \end{split}
\end{equation}
Since this is diagonal, so is the density operator; however, note that the density operator being diagonal in a set of creation and annihilation operators different from that appearing in the quantum field expansion is at variance with the case of global thermodynamic equilibrium with acceleration studied in Chapter \ref{chapter:gteacceleration}, as can be checked by looking at equations \eqref{gteq13} and \eqref{gteq26} together with \eqref{gteq28} and \eqref{gteq29}.
With a diagonal density operator, we are now in a good spot to calculate thermal expectation values of operators quadratic in the quantum field, such as the energy-momentum tensor, by using standard methods.


\section{Local thermodynamic equilibrium analysis}
\label{sec:boost_lte_analysis}

So far, we have been concerned only with the structure of the density operator, therefore we could take either $\tau$ or $\tau_0$ equivalently since $\wh{\rho}_{\rm LE}(\tau)$ and $\wh{\rho}=\wh{\rho}_{\rm LE}(\tau_0)$ have the same properties.
However, when it comes to thermal expectation values it does indeed make a difference whether the density operator is evaluated at $\tau$ or $\tau_0$.
In this Section, we will perform a local thermodynamic equilibrium analysis, therefore the $\tau$-dependence will be highlighted in order to dispel any potential doubt.
On the other hand, for ease of notation, the dependence on the momentum ${\sf p}$ will be omitted throughout whenever unnecessary.
As we shall see, we will be able to obtain analytical results, making this the first example of an exactly solvable system at local thermodynamic equilibrium, to the best of our knowledge.


\subsection{Thermal expectation values}

As a quick recap, the thermal expectation value of the energy-momentum tensor calculated with the local thermodynamic equilibrium density operator is
\begin{equation}\label{boosteq28}
    \langle \wh{T}^{\mu \nu}\rangle_{\rm LE}=\tr(\wh{\rho}_{\rm LE}(\tau)\wh{T}^{\mu \nu}(\tau)),
\end{equation}
and boost invariance constrains its form to be
\begin{equation}\label{boosteq32}
    \langle \wh{T}^{\mu \nu}\rangle_{\rm LE}=
    \ed_{\rm LE}(\tau)u^{\mu}u^{\nu}+\pt_{\rm LE}(\tau)(\hat{i}^{\mu}\hat{i}^{\nu}+\hat{j}^{\mu}\hat{j}^{\nu})+\pl_{\rm LE}(\tau)\hat{\eta}^{\mu}\hat{\eta}^{\nu}.
\end{equation}
The local thermodynamic equilibrium density operator built with the canonical energy-momentum tensor operator of a free real scalar quantum field is not diagonal in the creation and annihilation operators $\wh{b}_{\sf p}^{\dagger}$ and $\wh{b}_{\sf p}$, but it is so in the creation and annihilation operators $\wh{\xi}_{\sf p}^{\dagger}(\tau)$ and $\wh{\xi}_{\sf p}(\tau)$ obtained from the former through the Bogolyubov transformation \eqref{boosteq19}
\begin{equation}
    \wh{\rho}_{\rm LE}(\tau)=\frac{1}{Z_{\rm LE}(\tau)}\exp \left[-\frac{\wh{\Pi}(\tau)}{T(\tau)}\right],\qquad
    \wh{\Pi}(\tau)=\int \di^2{\rm p_T}\,\di \mu \,\omega(\tau) \left(\wh{\xi}_{\sf p}^{\dagger}(\tau)\wh{\xi}_{\sf p}(\tau)+\frac{1}{2}\right).
\end{equation}
This diagonal expression allows for thermal expectation values of products of $\wh{\xi}_{\sf p}^{\dagger}(\tau)$ and $\wh{\xi}_{\sf p}(\tau)$ to be obtained by using the standard method shown through \eqref{gteq50}--\eqref{gteq51} in Chapter \ref{chapter:gteacceleration}, hence the result
\begin{subequations}
    \begin{align}
        \langle \wh{\xi}_{\sf p}^{\dagger}(\tau)\wh{\xi}_{{\sf p}'}(\tau)\rangle_{\rm LE}=&n_{\rm B}(\tau)\,\delta^2(\pT-\pT')\,\delta(\mu-\mu')\label{boosteq31a}\\
        \langle \wh{\xi}_{\sf p}(\tau)\wh{\xi}_{{\sf p}'}^{\dagger}(\tau)\rangle_{\rm LE}=&\left(n_{\rm B}(\tau)+1\right)\delta^2(\pT-\pT')\,\delta(\mu-\mu')\label{boosteq31b}\\
        \langle \wh{\xi}_{\sf p}(\tau)\wh{\xi}_{{\sf p}'}(\tau)\rangle_{\rm LE}=&0=\langle \wh{\xi}_{\sf p}^{\dagger}(\tau)\wh{\xi}_{{\sf p}'}^{\dagger}(\tau)\rangle_{\rm LE}\label{boosteq31c},
    \end{align}
\end{subequations}
where $n_{\rm B}(\tau)$ is the Bose-Einstein distribution
\begin{equation}
    n_{\rm B}(\tau)=\frac{1}{{\rm e}^{\omega(\tau)/T(\tau)}-1}.
\end{equation}
with $\omega(\tau)$ given by \eqref{boosteq30}.
This is indeed the average number of excitations of the operator $\wh{\xi}_{\sf p}^{\dagger}(\tau)$, but by no means a density of particles.
Since the quantum field is expanded in terms of the creation and annihilation operators $\wh{b}_{\sf p}^{\dagger}$ and $\wh{b}_{\sf p}$, the particle density is given by $\langle \wh{b}_{\sf p}^{\dagger}\wh{b}_{{\sf p}'}\rangle_{\rm LE}$.
Another reason to be interested in thermal expectation values of products of $\wh{b}_{\sf p}^{\dagger}$ and $\wh{b}_{\sf p}$ is the fact that the energy-momentum tensor operator is quadratic in the quantum field, therefore it contains such terms.
The thermal expectation values of these products are most readily obtained from \eqref{boosteq31a}--\eqref{boosteq31c} by means of the inverse Bogolyubov transformation \eqref{boosteq29}
\begin{subequations}
    \begin{align}
        \langle \wh{b}_{\sf p}\wh{b}_{{\sf p}'}\rangle_{\rm LE}=&
        -\frac{1}{2}\sinh \left(2\Theta(\tau)\right){\rm e}^{-i\chi(\tau)}\left(2n_{\rm B}(\tau)+1\right)\delta^2(\pT-\pT')\,\delta(\mu-\mu')\label{boosteq33a}\\
        \langle \wh{b}_{\sf p}^{\dagger}\wh{b}_{{\sf p}'}^{\dagger}\rangle_{\rm LE}=&
        -\frac{1}{2}\sinh \left(2\Theta(\tau)\right){\rm e}^{i\chi(\tau)}\left(2n_{\rm B}(\tau)+1\right)\delta^2(\pT-\pT')\,\delta(\mu-\mu')\label{boosteq33b}\\
        \langle \wh{b}_{\sf p}\wh{b}_{{\sf p}'}^{\dagger}\rangle_{\rm LE}=&
        \left[\cosh \left(2\Theta(\tau)\right)n_{\rm B}(\tau)+\cosh^2\Theta(\tau)\right]\delta^2(\pT-\pT')\,\delta(\mu-\mu')\label{boosteq33c}\\
        \langle \wh{b}_{\sf p}^{\dagger}\wh{b}_{{\sf p}'}\rangle_{\rm LE}=&
        \left[\cosh \left(2\Theta(\tau)\right)n_{\rm B}(\tau)+\sinh^2\Theta(\tau)\right]\delta^2(\pT-\pT')\,\delta(\mu-\mu').\label{boosteq33d}
    \end{align}
\end{subequations}
These equations make it clear that the hyperbolic angle $\Theta(\tau)$, which can be determined from \eqref{boosteq22}, accounts for field vacuum effects.
More concerning the vacuum will be said shortly in Subsection \ref{sec:boost_lte_renormalization}.

Note that the $+1$ term in \eqref{boosteq31b}, stemming from the commutation relations \eqref{boosteq27}, gives rise to divergencies.
The details of renormalization will be shortly discussed in Subsection \ref{sec:boost_lte_renormalization}.
For the time being, we simply point out that renormalizing with respect to the vacuum of the operators $\wh{\xi}_{\sf p}(\tau)$, whence of $\wh{\Pi}(\tau)$, is but the subtraction of the $T(\tau)=0$ contribution, which eventually amount to canceling the $+1$ in \eqref{boosteq31b}.

With equations \eqref{boosteq33a}--\eqref{boosteq33d}, we are in a good position to calculate the thermal expectation value of the energy-momentum tensor \eqref{boosteq32}, that is the energy density, transverse pressure and longitudinal pressure at local thermodynamic equilibrium.
First of all, these scalar functions are obtained from the following contractions
\begin{equation}
    \ed_{\rm LE}(\tau)=\langle \wh{T}^{\mu \nu}\rangle_{\rm LE}u_{\mu}u_{\nu},
\end{equation}
\begin{equation}
    \pt_{\rm LE}(\tau)=\langle \wh{T}^{\mu \nu}\rangle_{\rm LE}\hat{i}_{\mu}\hat{i}_{\nu}=
    \langle \wh{T}^{\mu \nu}\rangle_{\rm LE}\hat{j}_{\mu}\hat{j}_{\nu}=
    \frac{1}{2}\langle \wh{T}^{\mu \nu}\rangle_{\rm LE}\left(\hat{i}_{\mu}\hat{i}_{\nu}+\hat{j}_{\mu}\hat{j}_{\nu}\right),
\end{equation}
\begin{equation}
    \pl_{\rm LE}(\tau)=\langle \wh{T}^{\mu \nu}\rangle_{\rm LE}\hat{\eta}_{\mu}\hat{\eta}_{\nu}.
\end{equation}
We take the energy density as an example, the pressures are worked out through the same procedure.
By plugging the quantum field expansion \eqref{boosteq09} into the contraction \eqref{boosteq34}, we have the following expression of the energy density in terms of thermal expectation values of products of creation and annihilation operators $\wh{b}_{\sf p}^{\dagger}$ and $\wh{b}_{\sf p}$
\begin{equation}\label{boosteq52}
    \begin{split}
        \ed_{\rm LE}(\tau)=&\langle \wh{T}^{\mu \nu}\rangle_{\rm LE}u_{\mu}u_{\nu}=
        \int\frac{{\rm d^2p_T}\,{\rm d}\mu\,{\rm d^2p_T'}\,{\rm d}\mu'}{4(4\pi)^2}\times \\
        &\times \left\{\left[(\partial_{\tau}h({\sf p},\tau))(\partial_{\tau}h({\sf p}',\tau))-\left(p_xp_x'+p_yp_y'+\frac{1}{\tau^2}\mu\mu'-m^2\right)h({\sf p},\tau)h({\sf p}',\tau)\right]\times \right.\\
        &\times {\rm e}^{i[(\pT+\pT')\cdot {\bf x}_{\rm T}+(\mu+\mu')\eta]}\langle \wh{b}_{\sf p}\wh{b}_{{\sf p}'}\rangle_{\rm LE}+\\
        &+\left[(\partial_{\tau}h({\sf p},\tau))(\partial_{\tau}h^*({\sf p}',\tau))+\left(p_xp_x'+p_yp_y'+\frac{1}{\tau^2}\mu\mu'+m^2\right)h({\sf p},\tau)h^*({\sf p}',\tau)\right]\times \\
        &\times {\rm e}^{i[(\pT-\pT')\cdot {\bf x}_{\rm T}+(\mu-\mu')\eta]}\langle \wh{b}_{\sf p}\wh{b}_{{\sf p}'}^{\dagger}\rangle_{\rm LE}+\\
        &+\left[(\partial_{\tau}h^*({\sf p},\tau))(\partial_{\tau}h({\sf p}',\tau))+\left(p_xp_x'+p_yp_y'+\frac{1}{\tau^2}\mu\mu'+m^2\right)h^*({\sf p},\tau)h({\sf p}',\tau)\right]\times \\
        &\times {\rm e}^{-i[(\pT-\pT')\cdot {\bf x}_{\rm T}+(\mu-\mu')\eta]}\langle \wh{b}_{\sf p}^{\dagger}\wh{b}_{{\sf p}'}\rangle_{\rm LE}+\\
        &+\left[(\partial_{\tau}h^*({\sf p},\tau))(\partial_{\tau}h^*({\sf p}',\tau))-\left(p_xp_x'+p_yp_y'+\frac{1}{\tau^2}\mu\mu'-m^2\right)h^*({\sf p},\tau)h^*({\sf p}',\tau)\right]\times \\
        &\left.\times {\rm e}^{-i[(\pT+\pT')\cdot {\bf x}_{\rm T}+(\mu+\mu')\eta]}\langle \wh{b}_{\sf p}^{\dagger}\wh{b}_{{\sf p}'}^{\dagger}\rangle_{\rm LE}\right\}.
    \end{split}
\end{equation}
Then, by using equations \eqref{boosteq33a}--\eqref{boosteq33d}, we obtain
\begin{equation}
    \begin{split}
        \ed_{\rm LE}(\tau)=&
        \int \frac{{\rm d^2p_T}\,{\rm d}\mu}{4(4\pi)^2}
        \bigg[
        \left[|\partial_{\tau}h(\tau)|^2+\omega^2(\tau)|h(\tau)|^2\right]\left(2n_{\rm B}(\tau)+1\right)\cosh(2\Theta(\tau))
        \\
        &-\frac{1}{2}\left[(\partial_{\tau}h(\tau))^2+\omega^2(\tau)h^2(\tau)\right]\left(2n_{\rm B}(\tau)+1\right)\sinh(2\Theta(\tau))\,{\rm e}^{-i\chi(\tau)}\\
        &\left.
        -\frac{1}{2}\left[(\partial_{\tau}h^*(\tau))^2+\omega^2(\tau)(h^*(\tau))^2\right]\left(2n_{\rm B}(\tau)+1\right)\sinh(2\Theta(\tau))\,{\rm e}^{i\chi(\tau)}.
        \right]
    \end{split}
\end{equation}
In square brackets we recognize the functions $K$ and $\Lambda$ defined in \eqref{boosteq35} and \eqref{boosteq36}.
These are expressed in terms of the hyperbolic angle $\Theta(\tau)$ and the phase $\chi(\tau)$ as in \eqref{boosteq22}, hence
\begin{equation}
    \begin{split}
        \ed_{\rm LE}(\tau)=&
        \int \di^2{\rm p_T}\,\di \mu \frac{\omega(\tau)}{16\pi^3\tau}
        \left[\cosh^2(2\Theta(\tau))-\sinh^2(2\Theta(\tau))\right]\left(2n_{\rm B}(\tau)+1\right)\\
        =&\int \frac{\di^2{\rm p_T}\,\di \mu}{(2\pi)^3\tau}\omega(\tau)\left(n_{\rm B}(\tau)+\frac{1}{2}\right).
    \end{split}
\end{equation}
The transverse and longitudinal pressures are worked out following the same steps, and the final expressions read
\begin{equation}\label{boosteq37}
    \ed_{\rm LE}(\tau)=\int \frac{\di^2{\rm p_T}\,\di \mu}{(2\pi)^3\tau \omega(\tau)}\omega^2(\tau)\left(n_{\rm B}(\tau)+\frac{1}{2}\right),
\end{equation}
\begin{equation}\label{boosteq38}
    \pt_{\rm LE}(\tau)=\int \frac{\di^2{\rm p_T}\,\di \mu}{(2\pi)^3\tau \omega(\tau)}\frac{|\pT|^2}{2}\left(n_{\rm B}(\tau)+\frac{1}{2}\right),
\end{equation}
\begin{equation}\label{boosteq39}
    \pl_{\rm LE}(\tau)=\int \frac{\di^2{\rm p_T}\,\di \mu}{(2\pi)^3\tau \omega(\tau)}\frac{\mu^2}{\tau^2}\left(n_{\rm B}(\tau)+\frac{1}{2}\right).
\end{equation}
These are the integral expressions of the thermal expectation value of the canonical energy-momentum tensor of a free real scalar quantum field at local thermodynamic equilibrium in the future light-cone.
To the best of our knowledge, this was first obtained in \cite{Rindori:2021quq}.

From the mathematical point of view, there is a key fact that greatly eases the calculations to the point that simple analytical results such as \eqref{boosteq37}--\eqref{boosteq39} can be obtained, and that is the following.
The functions $A(\tau)$ and $B(\tau)$ of the Bogolyubov transformation, appearing above in the thermal expectation values of products of $\wh{b}_{\sf p}$ and $\wh{b}_{\sf p}^{\dagger}$, are expressed in terms of the functions $K(\tau)$ and $\Lambda(\tau)$ through equation \eqref{boosteq25}.
The functions $K(\tau)$ and $\Lambda(\tau)$, in turn, can be written in terms of the hyperbolic angle $\Theta(\tau)$ and the phase $\chi(\tau)$ thanks to the relation \eqref{boosteq14}, which eventually is a consequence of the Wronskian of the Hankel functions being reconstructed.
In summary, the ultimate reason for the above analytical results is the reconstruction of the Wronskian of the Hankel functions, occurring because the density operator and the energy-momentum tensor operator in \eqref{boosteq28} are evaluated at the same $\tau$.
In the non-equilibrium analysis to be carried out in Section \ref{sec:boost_ne_analysis}, this will no longer hold true because of the mixing of $\tau$ and $\tau_0$, as exemplified in \eqref{boosteq08}.
The difficulties we will encounter there will make us appreciate the above exact analytical results even more.

Equations \eqref{boosteq37}--\eqref{boosteq39} can be recast in a more compact form.
Let us define the following real function taking on different expressions for either the energy density, the transverse pressure or the longitudinal pressure
\begin{equation}
    \gamma(\tau)\equiv \left\{
    \begin{aligned}
        \omega^2(\tau),\qquad \qquad &\ed_{\rm LE}(\tau)\\
        -\frac{\mu^2}{\tau^2}-m^2,\qquad \qquad &\pt_{\rm LE}(\tau)\\
        -\mT^2+\frac{\mu^2}{\tau^2},\qquad \qquad &\pl_{\rm LE}(\tau)
    \end{aligned}
    \right.
\end{equation}
and a modification of the functions $K(\tau)$ and $\Lambda(\tau)$ defined in \eqref{boosteq35} and \eqref{boosteq36} respectively
\begin{equation}\label{boosteq41}
    K_{\gamma}(\tau)\equiv \frac{\pi \tau}{4\omega(\tau)}\left[|\de_{\tau}h(\tau)|^2+\gamma(\tau)|h(\tau)|^2\right],
\end{equation}
\begin{equation}\label{boosteq42}
    \Lambda_{\gamma}(\tau)\equiv \frac{\pi \tau}{4\omega(\tau)}\left[(\de_{\tau}h(\tau))^2+\gamma(\tau)h^2(\tau)\right].
\end{equation}
Once again, the ${\sf p}$-dependence is omitted in $\gamma(\tau)$, $K_{\gamma}(\tau)$ and $\Lambda_{\gamma}(\tau)$ both to ease the notation and to highlight the $\tau$-dependence, which is the conceptually important one.
Much in the same way as $K(\tau)$ and $\Lambda(\tau)$ satisfy \eqref{boosteq14} thanks to the Wrosnkian of the Hankel functions, $K_{\gamma}(\tau)$ and $\Lambda_{\gamma}(\tau)$ are related by
\begin{equation}\label{boosteq40}
    K^2_{\gamma}(\tau)-|\Lambda_{\gamma}(\tau)|^2=
    \frac{\gamma(\tau)}{\omega^2(\tau)}.
\end{equation}
In the case of the energy density, $K_{\gamma}(\tau)$ and $\Lambda_{\gamma}(\tau)$ coincide with $K(\tau)$ and $\Lambda(\tau)$ respectively, thus \eqref{boosteq40} reduces correctly to \eqref{boosteq14}.
With these new functions, we can further define $\Gamma^{\gamma}_{\rm LE}\equiv \{\ed_{\rm LE},\pt_{\rm LE},\pl_{\rm LE}\}$ as the set of functions that make up the thermal expectation value of the energy-momentum tensor distinguished by the value of the function $\gamma(\tau)$, thus the \eqref{boosteq37}--\eqref{boosteq39} can be compactly rearranged as
\begin{equation}\label{boosteq45}
    \Gamma^{\gamma}_{\rm LE}(\tau)=
    \int \frac{\di^2{\rm p_T}\,\di \mu}{(2\pi)^3\tau \omega(\tau)}\omega^2(\tau)
    \left[K_{\gamma}(\tau)K(\tau)-{\rm Re}\left(\Lambda_{\gamma}(\tau)\Lambda^*(\tau)\right)\right]
    \left(n_{\rm B}(\tau)+\frac{1}{2}\right).
\end{equation}
That is the same structure we will recover in the non-equilibrium analysis too.
However, in the present local thermodynamic equilibrium analysis we can take it a step further by recasting the expression in square brackets in the following form by using the definitions \eqref{boosteq35}, \eqref{boosteq36}, \eqref{boosteq41} and \eqref{boosteq42} and exploiting again the Wronskian of the Hankel functions
\begin{equation}
    K_{\gamma}(\tau)K(\tau)-{\rm Re}\left(\Lambda_{\gamma}(\tau)\Lambda^*(\tau)\right)=
    \frac{\omega^2(\tau)+\gamma(\tau)}{2\omega^2(\tau)},
\end{equation}
hence
\begin{equation}\label{boosteq46}
    \Gamma^{\gamma}_{\rm LE}(\tau)=
    \int \frac{\di^2{\rm p_T}\,\di \mu}{(2\pi)^3\tau \omega(\tau)}
    \frac{\omega^2(\tau)+\gamma(\tau)}{2}\left(n_{\rm B}(\tau)+\frac{1}{2}\right).
\end{equation}

This can be further recast in a more familiar form.
By changing the integration variable $\mu$ to $p_z\equiv \mu/\tau$, the measure becomes
\begin{equation}\label{boosteq43}
    \di^2{\rm p_T}\frac{\di \mu}{\tau}=\di^2{\rm p_T}\,\di p_z=\di^3{\rm p},
\end{equation}
and $\omega(\tau)$ the on-shell energy
\begin{equation}\label{boosteq44}
    \omega(\tau)=\sqrt{\mT^2+\frac{\mu^2}{\tau^2}}=\sqrt{\mT^2+p_z^2}=\sqrt{|\pT|^2+p_z^2+m^2}\equiv \varepsilon.
\end{equation}
In turn, $n_{\rm B}(\tau)$ becomes the energy-dependent Bose-Einstein distribution $n_{\rm B}(\varepsilon,T(\tau))$ in the phase space, hence the the energy density \eqref{boosteq37} as well as the transverse and longitudinal pressures \eqref{boosteq38} and \eqref{boosteq39} can be written as the familiar momentum integrals of the relativistic neutral Bose gas.
Altogether, the non-renormalized thermal expectation value of the energy-momentum tensor at local equilibrium reads
\begin{equation}\label{boosteq47}
    \langle \wh{T}^{\mu \nu}\rangle_{\rm LE}=\int \frac{\di^3{\rm p}}{\varepsilon}P^{\mu}P^{\nu}\left(\frac{1}{{\rm e}^{\beta_{\lambda}P^{\lambda}}-1}+\frac{1}{2}\right),
\end{equation}
where $\beta^{\mu}$ is the four-temperature \eqref{boosteq03}.
Therefore, the thermodynamic functions $\Gamma^{\gamma}_{\rm LE}(\tau)$ are just the familiar functions of $T(\tau)$ as the usual ideal relativistic gas.
In particular, the transverse and the longitudinal pressures are in fact equal, namely
\begin{equation}
    \pt_{\rm LE}(\tau)=\pl_{\rm LE}(\tau)\equiv p_{\rm LE}(\tau).
\end{equation}
We can convince ourselves of this equality simply by noticing that the measure \eqref{boosteq43}, the energy \eqref{boosteq44} and the Bose-Einstein distribution $n_{\rm B}(\tau)$ are all invariant under rotations of the momentum ${\bf p}$, so it makes no difference between $p_x$, $p_y$ and $p_z$.
Thus, the transverse pressure \eqref{boosteq38} and the longitudinal pressure \eqref{boosteq39} are equal.
In other words, the thermal expectation value of the energy-momentum tensor at local thermodynamic equilibrium has the ideal form
\begin{equation}
    \langle \wh{T}^{\mu \nu}\rangle_{\rm LE}=
    \ed_{\rm LE}(\tau)u^{\mu}u^{\nu}-p_{\rm LE}(\tau)\Delta^{\mu \nu}.
\end{equation}
This is a somewhat unexpected feature of local thermodynamic equilibrium, totally at variance with the case of global thermodynamic equilibrium with acceleration studied in Chapter \ref{chapter:gteacceleration}.

Of course, the plain thermal expectation values \eqref{boosteq45}, or equivalently \eqref{boosteq46} or \eqref{boosteq47}, are divergent owing to the $1/2$ terms stemming from the commutation relations \eqref{boosteq27}, therefore they must be suitably renormalized.


\subsection{Renormalization}
\label{sec:boost_lte_renormalization}

Renormalization in free Quantum Field Theory is most readily established by the subtraction of some vacuum expectation value.
In the present case of a relativistic quantum fluid with boost invariance, we face an ambiguity in the choice of the vacuum which is somehow different from the one discussed in Subsection \ref{sec:gte_unruh_effect} for a relativistic quantum fluid at global thermodynamic equilibrium with acceleration.
There, the quantum field was expanded in the Rindler wedges in terms of the Rindler creation and annihilation operators, and the density operator was diagonal with respect to them.
Their vacuum was then the Rindler vacuum.
Those operators were related to the usual plane wave ones by a non-trivial Bogolyubov transformation, therefore the Rindler vacuum did not coincide with the Minkowski vacuum.
Thus, the ambiguity back then was choosing between the two natural vacua given by the Rindler and the Minkowski vacuum.
By taking the standpoint of the Rindler observer, the Minkowski vacuum appeared not as empty of particles but populated by a thermal distribution of them at a temperature proportional to the acceleration.
Together with the vacuum choice ambiguity, the key ingredient that let the Unruh effect emerge was the fact that the four-temperature possessed a Killing horizon.

Here, the quantum field is expanded in the creation and annihilation operators $\wh{b}_{\sf p}^{\dagger}$ and $\wh{b}_{\sf p}$, but the density operator is diagonal in a different set of operators, the $\wh{\xi}_{\sf p}^{\dagger}(\tau)$ and $\wh{\xi}_{\sf p}(\tau)$, related to the previous ones by a non-trivial Bogolyubov transformation.
Thus, the vacua of the two set of operators do not coincide.
The $\wh{b}_{\sf p}^{\dagger}$ and $\wh{b}_{\sf p}$ can be obtained from the usual plane wave ones $\wh{a}_{\bf p}^{\dagger}$ and $\wh{a}_{\bf p}$ by means of the ``Fourier-like'' transformation \eqref{boosteq11}, which does not involve any mixing of creation and annihilation operators, therefore the vacuum of $\wh{b}_{\sf p}$ is in fact the Minkowski vacuum $|0_{\rm M}\rangle$.
The Minkowski vacuum, of course, is a static state.
On the other hand, the vacuum of $\wh{\xi}_{\sf p}(\tau)$ is a $\tau$-dependent state, and we indicated it as $|0_{\tau}\rangle$ to emphasize this fact
\begin{equation}\label{boosteq70}
    \wh{\xi}_{\sf p}(\tau)|0_{\tau}\rangle=0.
\end{equation}
By using standard methods \cite{Mukhanov:2007zz}, from the coefficients of the Bogolyubov transformation \eqref{boosteq29}, it can be shown that $|0_{\tau}\rangle$ is related to $|0_{\rm M}\rangle$ by \cite{Rindori:2021quq}
\begin{equation}
    |0_{\tau}\rangle = \prod_{{\sf p}} \frac{1}{|\cosh \Theta({\sf p},\tau)|^{1/2}}\exp \left[ -\frac{1}{2}\tanh \Theta({\sf p},\tau) {\rm e}^{-i\chi({\sf p},\tau)} \wh{b}_{\sf p}^{\dagger}\wh{b}_{-{\sf p}}^{\dagger}\right] |0_{\rm M}\rangle
\end{equation}
In summary, the ambiguity here lies in the choice of either $|0_{\tau}\rangle$ or $|0_{\rm M}\rangle$, leading to two different renormalization schemes
\begin{equation}\label{boosteq48}
    {\langle \wh{T}^{\mu \nu}\rangle_{\rm LE}}_{\tau}\equiv
    \langle \wh{T}^{\mu \nu}\rangle_{\rm LE}-\langle 0_{\tau}|\wh{T}^{\mu \nu}|0_{\tau}\rangle,
\end{equation}
\begin{equation}\label{boosteq49}
    {\langle \wh{T}^{\mu \nu}\rangle_{\rm LE}}_{\rm M}\equiv
    \langle \wh{T}^{\mu \nu}\rangle_{\rm LE}-\langle 0_{\rm M}|\wh{T}^{\mu \nu}|0_{\rm M}\rangle.
\end{equation}

This situation is reminiscent of the Unruh effect, and indeed from the velocity field \eqref{boosteq07} a non-vanishing acceleration field can be derived; however, the relation \eqref{boosteq11} between the plane wave operators $\wh{a}_{\bf p}^{\dagger}$ and $\wh{a}_{\bf p}$ and the Milne ones $\wh{b}_{\sf p}^{\dagger}$ and $\wh{b}_{\sf p}$ is linear, making the vacua of the two sets of operators coincide.
Therefore, Milne observers count the same particles as inertial observers, given by
\begin{equation}
    \langle \wh{a}_{\bf p}^{\dagger}\wh{a}_{\bf p}\rangle_{\rm LE}=
    \frac{1}{2\pi \mT \sqrt{\cosh {\rm y}\cosh {\rm y}'}}\int_{-\infty}^{+\infty}\di \mu \,{\rm e}^{-i\mu({\rm y}-{\rm y}')}\langle \wh{b}_{\sf p}^{\dagger}\wh{b}_{\sf p}\rangle_{\rm LE}.
\end{equation}
In other words, no analog of the Unruh effect is present between inertial and Milne observers.
The ultimate reason for this seems also to be the fact that the future light-cone is not a stationary spacetime, and as such it does not possess any global timelike Killing vector field.
In fact, the four-temperature \eqref{boosteq03}, which is the only possible one compatible with boost invariance, is not a Killing vector field, so it has no Killing horizon whatsoever.
Without a Killing horizon, no Unruh effect seems to emerge.

The next step is the comparison between the two renornalization schemes \eqref{boosteq48} and \eqref{boosteq49}.
Let us start from the former, which is also the most straightforward one.
In the limit of vanishing proper temperature, the local thermodynamic equilibrium density operator reduces to the vacuum $|0_{\tau}\rangle$, that is
\begin{equation}\label{boosteq71}
    \lim_{T(\tau)\to 0}\wh{\rho}_{\rm LE}(\tau)=
    \lim_{T(\tau)\to 0}\frac{1}{Z_{\rm LE}(\tau)}\exp \left[-\frac{\wh{\Pi}(\tau)}{T(\tau)}\right]=
    |0_{\tau}\rangle \langle 0_{\tau}|.
\end{equation}
The first thing we can deduce from this fact is that the vacuum $|0_{\tau}\rangle \langle 0_{\tau}|$ has the same symmetries as the local thermodynamic equilibrium density operator $\wh{\rho}_{\rm LE}(\tau)$ which, however, is less symmetric than the supposedly Poincar\'e-invariant Minkowski vacuum.
This does not imply that $|0_{\tau}\rangle$ is degenerate, but that Poincar\'e transformations will give rise to non-vanishing components of excited states.
The second one is that renormalization with respect to the vacuum $|0_{\tau}\rangle$ is essentially the subtraction of the $T(\tau)=0$ contribution.
This could just as well be realized by looking at the thermal expectation values of products of $\wh{\xi}_{\sf p}(\tau)$ and $\wh{\xi}_{\sf p}^{\dagger}(\tau)$ \eqref{boosteq31a}--\eqref{boosteq31c}.
There, the subtraction of the $T(\tau)=0$ contribution eventually amounts to canceling the $+1$ term in \eqref{boosteq31b}, which in turn stems from the commutation relations \eqref{boosteq27}.
In this scheme, the energy density and the pressure at local thermodynamic equilibrium become
\begin{equation}\label{boosteq56}
    {\ed_{\rm LE}}_{\tau}(\tau)\equiv
    \ed_{\rm LE}(\tau)-\langle 0_{\tau}|\wh{T}^{\mu \nu}|0_{\tau}\rangle u_{\mu}u_{\nu}=
    \int \frac{\di^2{\rm p_T}\,\di \mu}{(2\pi)^3\tau \omega(\tau)}\omega^2(\tau)n_{\rm B}(\tau)
\end{equation}
\begin{equation}\label{boosteq57}
    {p_{\rm LE}}_{\tau}(\tau)\equiv
    p_{\rm LE}(\tau)-\langle 0_{\tau}|\wh{T}^{\mu \nu}|0_{\tau}\rangle \hat{e}_{\mu}\hat{e}_{\nu}=
    \int \frac{\di^2{\rm p_T}\,\di \mu}{(2\pi)^3\tau \omega(\tau)}p_i^2(\tau)n_{\rm B}(\tau)
\end{equation}
with $\hat{e}_{\mu}$ either $\hat{i}_{\mu}$, $\hat{j}_{\mu}$ or $\hat{\eta}_{\mu}$ and $p_i$ either $p_x$, $p_y$ or $\mu/\tau$.
Equivalently, these can be recast in the more compact forms \eqref{boosteq45} and \eqref{boosteq46} as
\begin{equation}\label{boosteq58}
    \begin{split}
        {\Gamma^{\gamma}_{\rm LE}}_{\tau}(\tau)=&
        \int \frac{\di^2{\rm p_T}\,\di \mu}{(2\pi)^3\tau \omega(\tau)}\omega^2(\tau)\left[K_{\gamma}(\tau)K(\tau)-{\rm Re}\left(\Lambda_{\gamma}(\tau)\Lambda^*(\tau)\right)\right]n_{\rm B}(\tau)\\
        =&\int \frac{\di^2{\rm p_T}\,\di \mu}{(2\pi)^3\tau \omega(\tau)}
        \frac{\omega^2(\tau)+\gamma(\tau)}{2}n_{\rm B}(\tau).
    \end{split}
\end{equation}
These integrals are convergent due to the exponential suppression for large values of the momentum ${\sf p}$ provided by the Bose-Einstein distribution.
In particular, in the massless limit they can be carried out analytically to yield the familiar expressions of the energy density and pressure of a neutral Bose gas at temperature $T(\tau)$
\begin{equation}\label{boosteq59}
    {\ed_{\rm LE}}_{\tau}(\tau)=\frac{\pi^2}{30}T^4(\tau),\qquad (m=0),
\end{equation}
\begin{equation}\label{boosteq60}
    {p_{\rm LE}}_{\tau}(\tau)=\frac{1}{3}{\ed_{\rm LE}}_{\tau}(\tau)=
    \frac{\pi^2}{90}T^4(\tau),\qquad (m=0).
\end{equation}

Now we turn to the renormalization scheme \eqref{boosteq49}, namely we take the standpoint either of the inertial or Milne observer and we subtract the Minkowski vacuum contribution.
In order to calculate the expectation value of the energy-momentum tensor operator in the Minkowski vacuum, we exploit the fact that $\wh{T}^{\mu \nu}$ is built with the quantum field as in \eqref{boosteq50}, and the quantum field is in turn expanded in Milne coordinates as in \eqref{boosteq09} in terms of $\wh{b}_{\sf p}$ and $b_{\sf p}^{\dagger}$, whose vacuum is $|0_{\rm M}\rangle$.
Much in the same way as previously done, we take the energy density as a prototype, with the pressure to be worked out following analogous steps.
The energy density in the Minkowski vacuum is given by \eqref{boosteq52} where the thermal expectation values of products of $\wh{b}_{\sf p}$ and $\wh{b}_{\sf p}^{\dagger}$ are supposed to be calculated on $|0_{\rm M}\rangle$ instead of using $\wh{\rho}_{\rm LE}(\tau)$.
Of course, the products $\wh{b}_{\sf p}\wh{b}_{{\sf p}'}$, $\wh{b}_{\sf p}^{\dagger}\wh{b}_{{\sf p}'}^{\dagger}$ and $\wh{b}_{\sf p}^{\dagger}\wh{b}_{{\sf p}'}$ have vanishing expectation value on the Minkowski vacuum as $\wh{b}_{\sf p}|0_{\rm M}\rangle=0$.
On the other hand, the expectation value of $\wh{b}_{\sf p}\wh{b}^{\dagger}_{{\sf p}'}$ is non-vanishing and worked out with the commutation relations \eqref{boosteq18}
\begin{equation}\label{boosteq51}
	\langle 0_{\rm M}|\wh{b}_{\sf p}\wh{b}^{\dagger}_{{\sf p}'}|0_{\rm M}\rangle=
	\langle 0_{\rm M}|\wh{b}^{\dagger}_{{\sf p}'}\wh{b}_{\sf p}|0_{\rm M}\rangle+\langle 0_{\rm M}|[\wh{b}_{\sf p},\wh{b}^{\dagger}_{{\sf p}'}]|0_{\rm M}\rangle=
	\delta^2(\pT-\pT')\,\delta(\mu-\mu').
\end{equation}
Thus, the energy density in the Minkowski vacuum is readily obtained from \eqref{boosteq52}
\begin{equation}
    \begin{split}
        \langle 0_{\rm M}|\wh{T}^{\mu \nu}|0_{\rm M}\rangle u_{\mu}u_{\nu}=&
        \int \frac{{\rm d^2p_T}\,{\rm d}\mu}{4(4\pi)^2}\left(|\partial_{\tau}h(\tau)|^2+\omega^2(\tau)|h(\tau)|^2\right)\\
	    =&\int \frac{{\rm d^2p_T}\,{\rm d}\mu}{(2\pi)^3\tau \omega(\tau)}\omega^2(\tau) \frac{K(\tau)}{2}.
	\end{split}
\end{equation}
Repeating the same steps for the pressure, the same result is found with $K(\tau)$ replaced by $K_{\gamma}(\tau)$, the two of them coinciding in the case of the energy density.
We can therefore conclude that the thermal expectation value of the energy-momentum tensor renormalized with respect to the Minkowski vacuum reads
\begin{equation}
    \begin{split}
        {\Gamma^{\gamma}_{\rm LE}}_{\rm M}(\tau)=&
        \int \frac{\di^2{\rm p_T}\,\di \mu}{(2\pi)^3\tau \omega(\tau)}\omega^2(\tau)\times \\
        &\times \left\{
        \left[K_{\gamma}(\tau)K(\tau)-{\rm Re}\left(\Lambda_{\gamma}(\tau)\Lambda^*(\tau)\right)\right]\left(n_{\rm B}(\tau)+\frac{1}{2}\right)-\frac{K_{\gamma}(\tau)}{2}
        \right\}.
    \end{split}
\end{equation}
This expression, however, subtly hides some problem.
Let us take once again the energy density as an example and split it into two terms as
\begin{equation}\label{boosteq53}
    \begin{split}
        {\ed_{\rm LE}}_{\rm M}(\tau)=&
        \int \frac{\di^2{\rm p_T}\,\di \mu}{(2\pi)^3\tau}\omega(\tau)n_{\rm B}(\tau)-
        \int \frac{\di^2{\rm p_T}\,\di \mu}{(2\pi)^3\tau}\omega(\tau)\frac{K(\tau)-1}{2}.
    \end{split}
\end{equation}
The first term coincides with \eqref{boosteq56} found in the former renormalization scheme and is clearly convergent.
The second term depends on the transverse momentum $\pT$ only through its modulus ${\rm p_T}\equiv |\pT|$, therefore we can change to polar variables in the transverse momentum and factorize the angular part
\begin{equation}
    \int \frac{\di^2{\rm p_T}\,\di \mu}{(2\pi)^3\tau}\omega(\tau)\frac{K(\tau)-1}{2}=
    \frac{1}{(2\pi)^2\tau}\int_{-\infty}^{+\infty}\di \mu \int_0^{+\infty}\di {\rm p_T}\,{\rm p_T}\,\omega(\tau)\frac{K(\tau)-1}{2}.
\end{equation}
Changing variable from ${\rm p_T}$ to the transverse mass $\mT$ and then to $v\equiv \mT \tau$ we have
\begin{equation}
    \frac{1}{(2\pi)^2\tau^2}\int_{-\infty}^{+\infty}\di \mu \int_{m\tau}^{+\infty}\di v\,v \sqrt{v^2+\mu^2}\frac{K(v)-1}{2}.
\end{equation}
In \cite{Rindori:2021quq}, an asymptotic analysis was performed for $m\tau \gg 1$, or, restoring the natural constants, $m\tau c^2\gg \hbar$.
As the details of this analysis are quite involved, they will be addressed in a dedicated space in Subsection \ref{sec:boost_ne_asymptotic_analysis}.
For the time being, we simply mention that the trend of $K(\tau)-1$ in this limit is
\begin{equation}\label{boosteq54}
    K(\tau)-1=\cosh \Theta(\tau)-1\simeq \frac{\Theta^2(\tau)}{2}\simeq \frac{\mT^4}{\tau^2\omega^6(\tau)}.
\end{equation}
In this limit we also have $v\gg 1$, and using \eqref{boosteq54} the integral in $v$ turns out to diverge
\begin{equation}
    \int_{m\tau}^{+\infty}\di v\,v \sqrt{v^2+\mu^2}\frac{v^4}{\left(\sqrt{v^4+\mu^2}\right)^6}\sim
    \int_{m\tau}^{+\infty}\di v\,v\frac{v^4}{v^5}\sim \infty.
\end{equation}
Of course this is not a formal proof, but a strong indication nonetheless that renormalizing with respect to the Minkowski vacuum provides us with thermal expectation values that are still divergent.
This fact seems to automatically single out $|0_{\tau}\rangle$ as a preferred choice, however, it must be emphasized that $|0_{\tau}\rangle$ is a $\tau$-dependent vacuum, therefore thermal expectation values renormalized accordingly might inherit some eventually undesired $\tau$-dependence.
These issues will be further discussed in Subsection \ref{sec:boost_ne_renormalization} in the context of the non-equilibrium analysis.
For the time being, we would like the takeaway to be the fact that the Minkowski vacuum renormalization scheme should be discarded as it appears to provide us with thermal expectation values that are still divergent.
Moreover, the next Subsection will make it clear why at some point we would have needed to study the effects of the $|0_{\tau}\rangle$ contributions subtraction anyway.


\subsection{Entropy current in the future light-cone}
\label{sec:boost_lte_entropy_current}

With the plain thermal expectation value of the energy-momentum tensor, we are now in a good spot to calculate the entropy current at local thermodynamic equilibrium by using the method put forward Section \ref{sec:zubarev_entropy_current_method}.
That is in fact well-suited for systems at local thermodynamic equilibrium, which is the underlying hypothesis of relativistic hydrodynamics, and global thermodynamic equilibrium is included as a particular case.
However, no known expression for the entropy current nor for the entropy production exists for a Quantum Field Theory out of thermodynamic equilibrium, therefore in the non-equilibrium analysis to be carried out in Section \ref{sec:boost_ne_analysis} we will not be able to re-apply our method.

As a quick recap, the method demands to be provided with two ingredients:
\begin{enumerate}
    \item the thermal expectation value the energy-momentum tensor $\langle \wh{T}^{\mu \nu}\rangle_{\rm LE}$, and
    \item the eigenvector $|0_{\Upsilon}\rangle$ corresponding to the lowest, non-degenerate eigenvalue of $\wh{\Upsilon}(\tau)$.
\end{enumerate}
Then, the algorithm is the following:
\begin{enumerate}
    \item Take $\langle \wh{T}^{\mu \nu}\rangle_{\rm LE}$ and subtract $\langle 0_{\Upsilon}|\wh{T}^{\mu \nu}|0_{\Upsilon}\rangle$ by using the $\lambda$-dependent density operator $\wh{\rho}_{\rm LE}(\lambda)$ defined according to \eqref{zubeq21}.
    This way, the result is $\lambda$-dependent.
    \item Contract with $\beta_{\nu}$, which is $\lambda$-independent, and integrate in $\lambda$ from $\lambda=1$ to $\lambda=+\infty$ in order to obtain the thermodynamic potential current $\phi^{\mu}$ defined in \eqref{zubeq26}.
    \item Plug the result into \eqref{zubeq27} and obtain the entropy current $s^{\mu}$.
\end{enumerate}

By comparing the definition \eqref{zubeq23} of the operator $\wh{\Upsilon}(\tau)$ with the expression \eqref{boosteq15} of the local thermodynamic equilibrium density operator with boost invariance in the future light-cone, we immediately understand that
\begin{equation}\label{boosteq55}
    \wh{\Upsilon}(\tau)=\frac{\wh{\Pi}(\tau)}{T(\tau)}=
    \frac{1}{T(\tau)}\int \di^2{\rm p_T}\,\di \mu \,\omega(\tau)\left(\wh{\xi}_{\sf p}^{\dagger}(\tau)\wh{\xi}_{\sf p}(\tau)+\frac{1}{2}\right).
\end{equation}
Thus, the lowest eigenvector of $\wh{\Upsilon}(\tau)$ is the vacuum of $\wh{\Pi}(\tau)$, that is $|0_{\tau}\rangle$.
As we discussed already in Subsection \ref{sec:boost_lte_renormalization}, $|0_{\tau}\rangle$ is a $\tau$-dependent and non-degenerate state which does not coincide with the Minkowski vacuum and possesses the same symmetries as the local thermodynamic equilibrium density operator.
Hence
\begin{equation}
    \langle \wh{T}^{\mu \nu}\rangle_{\rm LE}-\langle 0_{\Upsilon}|\wh{T}^{\mu \nu}|0_{\Upsilon}\rangle=
    \langle \wh{T}^{\mu \nu}\rangle_{\rm LE}-\langle 0_{\tau}|\wh{T}^{\mu \nu}|0_{\tau}\rangle,
\end{equation}
which was calculated in \eqref{boosteq56} and \eqref{boosteq57} or equivalently \eqref{boosteq58}, with analytical results given by \eqref{boosteq59} and \eqref{boosteq60} for a massless field.
So, we are good to go for what concerns the ingredients.

According to the method, the above quantity should be calculated with the modified density operator $\wh{\rho}_{\rm LE}(\tau,\lambda)$ defined in \eqref{zubeq21} in order to be $\lambda$-dependent.
Once again, by comparing \eqref{zubeq21} with \eqref{boosteq55}, we can tell that this is but a rescaling of the proper temperature as $T(\tau)\mapsto T(\tau)/\lambda$, in fact this transformation gives precisely the sought result.
Thus, we have
\begin{equation}
    \langle \wh{T}^{\mu \nu}\rangle_{\rm LE}(\lambda)-\langle 0_{\Upsilon}|\wh{T}^{\mu \nu}|0_{\Upsilon}\rangle(\lambda)=
    {\ed_{\rm LE}}_{\tau}(\tau,\lambda)u^{\mu}u^{\nu}-
    {p_{\rm LE}}_{\tau}(\tau,\lambda)\Delta^{\mu \nu},
\end{equation}
where the rescaling $T(\tau)\mapsto T(\tau)/\lambda$ affects only the energy density and the pressure because $u^{\mu}$ and $\Delta^{\mu \nu}$ are independent of $T(\tau)$.
This result should be contracted with the four-temperature while taking care that $\beta_{\nu}$ does not undergo the temperature rescaling, for it concerns only the thermal expectation value $\langle \wh{T}^{\mu \nu}\rangle_{\rm LE}-\langle 0_{\Upsilon}|\wh{T}^{\mu \nu}|0_{\Upsilon}\rangle$, as can be clearly seen in the derivation in Section \ref{sec:zubarev_entropy_current_method}.
Whence
\begin{equation}
    \left(\langle \wh{T}^{\mu \nu}\rangle_{\rm LE}(\lambda)-\langle 0_{\Upsilon}|\wh{T}^{\mu \nu}|0_{\Upsilon}\rangle(\lambda)\right)\beta_{\nu}=
    {\ed_{\rm LE}}_{\tau}(\tau,\lambda)\beta^{\mu},
\end{equation}
therefore we do not need to know the thermal expectation value of the whole energy-momentum tensor, only the energy density contributes to the thermodynamic potential current and the entropy current.
By looking at equation \eqref{boosteq56}, we see that the only dependence of the energy density on the proper temperature is in the Bose-Einstein distribution, thus
\begin{equation}
    {\ed_{\rm LE}}_{\tau}(\tau,\lambda)=
    \int \frac{\di^2{\rm p_T}\,\di \mu}{(2\pi)^3\tau \omega(\tau)}\omega^2(\tau)\frac{1}{{\rm e}^{\lambda \omega(\tau)/T(\tau)}-1}.
\end{equation}
For a massless field, the $\lambda$-integration can be carried out analytically, yielding
\begin{equation}
    \int_1^{+\infty}\di \lambda \,{\ed_{\rm LE}}_{\tau}(\tau,\lambda)=
    \int_1^{+\infty}\di \lambda \,\frac{\pi^2}{30}\frac{T^4(\tau)}{\lambda^4}=
    \frac{\pi^2}{90}T^4(\tau),\qquad (m=0)
\end{equation}
and the following expression for the thermodynamic potential current is readily obtained
\begin{equation}
    \phi^{\mu}=\frac{\pi^2}{90\beta^4}\beta^{\mu}=
    \frac{\pi^2}{90}T^3(\tau)u^{\mu},\qquad (m=0).
\end{equation}
Plugging this result into \eqref{zubeq27}, we finally obtain the entropy current 
\begin{equation}
    s^{\mu}=\frac{2\pi^2}{45\beta^4}\beta^{\mu}=
    \frac{2\pi^2}{45}T^3(\tau)u^{\mu},\qquad (m=0).
\end{equation}
These are the expressions of the thermodynamic potential current and the entropy current of a free real massless scalar field at local thermodynamic equilibrium with boost invariance in the future light-cone \cite{Rindori:2021quq}.
The entropy current coincides with the classical expression of the usual relativistic kinetic theory, therefore, no quantum corrections are apparently present, which is somewhat unexpected.


\section{Non-equilibrium analysis}
\label{sec:boost_ne_analysis}

So far, we have been concerned with an analysis at local thermodynamic equilibrium, which is a non-equilibrium configurations but still close to global thermodynamic equilibrium, in the sense that the volume term in \eqref{zubeq09} containing the derivatives of the thermodynamic fields is small enough to be neglected.
In this Section, we perform a full non-equilibrium analysis instead, therefore we consider the full non-equilibrium density operator $\wh{\rho}=\wh{\rho}_{\rm LE}(\tau_0)$.
The calculation of the thermal expectation value of the energy-momentum tensor goes on much in the same way as we saw in the local thermodynamic equilibrium analysis.
However, there are now complications due to the mixing of terms calculated at $\tau$ and others at $\tau_0$.
It is therefore mandatory to pay attention to these dependencies, so the dependence on ${\sf p}$ will still be omitted whenever possible.


\subsection{Thermal expectation values}
\label{subsec:ne_tev}

As previously mentioned, the non-equilibrium density operator and the local thermodynamic equilibrium one have the same symmetries, so the former will simply be the latter calculated at $\tau_0$, namely
\begin{equation}\label{boosteq69}
    \wh{\rho}=\wh{\rho}_{\rm LE}(\tau_0)=
    \frac{1}{Z_{\rm LE}(\tau_0)}\exp \left[-\frac{\wh{\Pi}(\tau_0)}{T(\tau_0)}\right]
\end{equation}
with $\wh{\Pi}(\tau_0)$ given by
\begin{equation}
    \wh{\Pi}(\tau_0)=\int \di^2{\rm p_T}\,\di \mu \,\omega(\tau_0)\left(\wh{\xi}_{\sf p}^{\dagger}(\tau_0)\wh{\xi}_{\sf p}(\tau_0)+\frac{1}{2}\right).
\end{equation}
The operators $\wh{\xi}_{\sf p}(\tau_0)$ and $\wh{\xi}_{\sf p}^{\dagger}(\tau_0)$ are those that diagonalize $\wh{\Pi}(\tau_0)$, which is but $\wh{\Pi}(\tau)$ at $\tau_0$, so they are just $\wh{\xi}_{\sf p}(\tau)$ and $\wh{\xi}_{\sf p}^{\dagger}(\tau)$ respectively at $\tau_0$.
In terms of $\wh{b}_{\sf p}$ and $\wh{b}_{\sf p}^{\dagger}$, they are obtained from \eqref{boosteq29} at $\tau_0$, namely
\begin{equation}
    \begin{split}
        \wh{b}_{\sf p}=&\cosh \Theta({\sf p}, \tau_0)\wh{\xi}_{\sf p}(\tau_0)-\sinh \Theta({\sf p},\tau_0)\,{\rm e}^{-i\chi({\sf p},\tau_0)}\wh{\xi}_{-{\sf p}}^{\dagger}(\tau_0),\\
        \wh{b}_{\sf p}^{\dagger}=&\cosh \Theta({\sf p}, \tau_0) \wh{\xi}_{\sf p}^{\dagger}(\tau_0)-\sinh \Theta ({\sf p},\tau_0)\,{\rm e}^{i\chi({\sf p},\tau_0)}\wh{\xi}_{-{\sf p}}(\tau_0).
    \end{split}
\end{equation}

Thermal expectation values out of thermodynamic equilibrium are calculated with this density operator as anticipated in \eqref{boosteq08}, which we hereby report.
The non-equilibrium thermal expectation value of the energy-momentum tensor is
\begin{equation}
    T^{\mu \nu}=\tr(\wh{\rho}\wh{T}^{\mu \nu})=\tr(\wh{\rho}_{\rm LE}(\tau_0)\wh{T}^{\mu \nu}(\tau)).
\end{equation}
This is constrained by boost invariance to be of the form \eqref{boosteq12}, that is
\begin{equation}
    T^{\mu \nu}=\ed(\tau)u^{\mu}u^{\nu}+\pt(\tau)(\hat{i}^{\mu}\hat{i}^{\nu}+\hat{j}^{\mu}\hat{j}^{\nu})+\pl(\tau)\hat{\eta}^{\mu}\hat{\eta}^{\nu}.
\end{equation}
In order to calculate the energy density and the two pressures, we need to know the non-equilibrium thermal expectation values of products of $\wh{b}_{\sf p}$ and $\wh{b}_{\sf p}^{\dagger}$ with $\wh{\rho}$, that is
\begin{subequations}
    \begin{align}
        \langle \wh{b}_{\sf p}\wh{b}_{{\sf p}'}\rangle=&
        \tr(\wh{\rho}\,\wh{b}_{\sf p}\wh{b}_{{\sf p}'})=
        \tr(\wh{\rho}_{\rm LE}(\tau_0)\wh{b}_{\sf p}\wh{b}_{{\sf p}'})\\
        \langle \wh{b}_{\sf p}^{\dagger}\wh{b}_{{\sf p}'}^{\dagger}\rangle=&
        \tr(\wh{\rho}\,\wh{b}_{\sf p}^{\dagger}\wh{b}_{{\sf p}'}^{\dagger})=
        \tr(\wh{\rho}_{\rm LE}(\tau_0)\wh{b}_{\sf p}^{\dagger}\wh{b}_{{\sf p}'}^{\dagger})\\
        \langle \wh{b}_{\sf p}\wh{b}_{{\sf p}'}^{\dagger}\rangle=&
        \tr(\wh{\rho}\,\wh{b}_{\sf p}\wh{b}_{{\sf p}'}^{\dagger})=
        \tr(\wh{\rho}_{\rm LE}(\tau_0)\wh{b}_{\sf p}\wh{b}_{{\sf p}'}^{\dagger})\\
        \langle \wh{b}_{\sf p}^{\dagger}\wh{b}_{{\sf p}'}\rangle=&
        \tr(\wh{\rho}\,\wh{b}_{\sf p}^{\dagger}\wh{b}_{{\sf p}'})=
        \tr(\wh{\rho}_{\rm LE}(\tau_0)\wh{b}_{\sf p}^{\dagger}\wh{b}_{{\sf p}'}).
    \end{align}
\end{subequations}
These are most readily given by evaluating \eqref{boosteq33a}--\eqref{boosteq33d} at $\tau_0$
\begin{subequations}
    \begin{align}
        \langle \wh{b}_{\sf p}\wh{b}_{{\sf p}'}\rangle=&
        -\frac{1}{2}\sinh \left(2\Theta(\tau_0)\right){\rm e}^{-i\chi(\tau_0)}\left(2n_{\rm B}(\tau_0)+1\right)\delta^2(\pT-\pT')\,\delta(\mu-\mu')\label{boosteq61a}\\
        \langle \wh{b}_{\sf p}^{\dagger}\wh{b}_{{\sf p}'}^{\dagger}\rangle=&
        -\frac{1}{2}\sinh \left(2\Theta(\tau_0)\right){\rm e}^{i\chi(\tau_0)}\left(2n_{\rm B}(\tau_0)+1\right)\delta^2(\pT-\pT')\,\delta(\mu-\mu')\label{boosteq61b}\\
        \langle \wh{b}_{\sf p}\wh{b}_{{\sf p}'}^{\dagger}\rangle=&
        \left[\cosh \left(2\Theta(\tau_0)\right)n_{\rm B}(\tau_0)+\cosh^2\Theta(\tau_0)\right]\delta^2(\pT-\pT')\,\delta(\mu-\mu')\label{boosteq61c}\\
        \langle \wh{b}_{\sf p}^{\dagger}\wh{b}_{{\sf p}'}\rangle=&
        \left[\cosh \left(2\Theta(\tau_0)\right)n_{\rm B}(\tau_0)+\sinh^2\Theta(\tau_0)\right]\delta^2(\pT-\pT')\,\delta(\mu-\mu').\label{boosteq61d}
    \end{align}
\end{subequations}
Taking once again the energy density as a prototype, its non-equilibrium thermal expectation value is obtained from \eqref{boosteq52} by replacing \eqref{boosteq33a}--\eqref{boosteq33d} with \eqref{boosteq61a}--\eqref{boosteq61d}
\begin{equation}
    \begin{split}
        \ed(\tau)=&
        \tr(\wh{\rho}\wh{T}^{\mu \nu})u_{\mu}u_{\nu}=
        \tr(\wh{\rho}_{\rm LE}(\tau_0)\wh{T}^{\mu \nu}(\tau))u_{\mu}u_{\nu}\\
        =&\int \frac{\di^2{\rm p_T}\,\di \mu}{(2\pi)^3\tau}\omega(\tau)
        \left[K(\tau)K(\tau_0)-{\rm Re}\left(\Lambda(\tau)\Lambda^*(\tau_0)\right)\right]
        \left(n_{\rm B}(\tau_0)+\frac{1}{2}\right).
    \end{split}
\end{equation}
As in the local thermodynamic equilibrium analysis, the pressures are worked out following the same steps.
Thus, the non-equilibrium analog of equation \eqref{boosteq45} reads
\begin{equation}\label{boosteq65}
    \Gamma^{\gamma}(\tau)=
    \int \frac{\di^2{\rm p_T}\,\di \mu}{(2\pi)^3\tau \omega(\tau)}\omega^2(\tau)
    \left[K_{\gamma}(\tau)K(\tau_0)-{\rm Re}\left(\Lambda_{\gamma}(\tau)\Lambda^*(\tau_0)\right)\right]
    \left(n_{\rm B}(\tau_0)+\frac{1}{2}\right).
\end{equation}

Note that at $\tau=\tau_0$, namely at local thermodynamic equilibrium, this equation reduces to \eqref{boosteq45}, and so the results of the local thermodynamic equilibrium analysis are recovered as expected.
However, at later times $\tau>\tau_0$, because of the mixing of some functions at $\tau$ and others at $\tau_0$, the Wronskian of the Hankel functions is, unfortunately, not reconstructed.
When written explicitly, the above equation is given by a non-trivial combination of Hankel fuctions of the same order but with different arguments which cannot be recast in any simple form.
We can therefore tell that for $\tau >\tau_0$ the thermal expectation value of the energy-momentum tensor differs from its local thermodynamic equilibrium form, but we are not able to perform the integration of this combination of Hankel functions in general.
Indeed, since we are dealing with a free Quantum Field Theory, one expects to find the same expression as for the free-streaming solution of Boltzmann equation in Milne coordinates, which is reported in Appendix \ref{appendix:free_sreaming}.
However, there will also be quantum corrections due to vacuum subtraction.

In the next Subsection, we will consider a suitable approximation in which the calculation of the non-equilibrium thermal expectation value of the energy-momentum tensor becomes feasible.


\subsection{Asymptotic analysis}
\label{sec:boost_ne_asymptotic_analysis}

In \cite{Rindori:2021quq}, the behaviour of the stress-energy tensor and related quantities for late times $\tau$ was studied.
The starting point is the function $h(\tau)$ defined in \eqref{boosteq72}, that is
\begin{equation}\label{hankelint}
    h(\tau)=-i{\rm e}^{\frac{\pi}{2}\mu}{\rm H}^{(2)}_{i\mu}(\mT \tau).
\end{equation}
The asymptotic expansion of the Hankel functions for large arguments is \cite{Gradshteyn:1702455}
\begin{equation}\label{large_x}
    {\rm H}^{(2)}_{\nu}(x)\sim\sqrt{\frac{2}{\pi x}}{\rm e}^{-i\left( x-\frac{\pi}{2}\nu -\frac{\pi}{4} \right)}\sum_n
    \frac{1}{(2ix)^n} \frac{\Gamma(\nu+1/2 +n)}{n!\Gamma(\nu+1/2-n)}
\end{equation}
which is valid for ${\rm Re}(\nu)>-1/2$ and $|{\rm arg}(x)|<\pi$.
By making use of the property $z\Gamma(z) = \Gamma(z+1)$, substituting $x=m_{\rm T}\tau$ and $\nu=i\mu$, and plugging into \eqref{hankelint}, we get
\begin{equation}\label{h_large_arg}
    h(\tau) \sim \sqrt{\frac{-2i}{\pi m_{\rm T}\tau}}{\rm e}^{-im_{\rm T}\tau}\sum_n
    \frac{1}{(2im_{\rm T}\tau)^n} \frac{\left(i\mu +\frac{1}{2} -n \right)^{(2n)}}{n!},
\end{equation}
valid for large $m_{\rm T}\tau$.
Similarly, using the exact relation \cite{Gradshteyn:1702455}
\begin{equation}
    z \partial_z {\rm H}^{(2)}_{\nu}(z)
    =\nu {\rm H}^{(2)}_{\nu}(z) -z {\rm H}^{(2)}_{\nu+1}(z),
\end{equation}
along with the expansion \eqref{large_x}, one obtains the expansion for the derivative $\partial_\tau h$
\begin{equation}\label{h_dot_large_arg}
    \begin{split}
    \partial_\tau h(\tau) \sim&
    -im_{\rm T}\sqrt{\frac{-2i}{\pi m_{\rm T}\tau}}{\rm e}^{-im_{\rm T}\tau}\times \\
    &\times \left[1+  \sum_{n>0} \frac{1}{(2im_{\rm T}\tau)^{n}}\left(-2i\mu \frac{\left(i\mu +\frac{3}{2} -n \right)^{(2n-2)}}{(n-1)!} + \frac{\left(i\mu +\frac{3}{2} -n \right)^{(2n)}}{n!}\right) \right].
    \end{split}
\end{equation}
Particularly, retaining the terms up to first order (next to leading) in $m_{\rm T}\tau$ we get
\begin{equation}
    \begin{split}
        h(\tau) \simeq&
        \sqrt{\frac{-2i}{\pi m_{\rm T}\tau}}{\rm e}^{-im_{\rm T}\tau}\left[ 1 -\frac{i}{2m_{\rm T}\tau} \left( i\mu -\frac{1}{2}\right)\left( i\mu +\frac{1}{2}\right)\right]\\
        =&\sqrt{\frac{-2i}{\pi m_{\rm T}\tau}}{\rm e}^{-im_{\rm T}\tau}\left[1 +i\frac{1+4\mu^2}{8m_{\rm T}\tau} \right],
    \end{split}
\end{equation}
\begin{equation}
    \partial_{\tau} h(\tau) \simeq-im_{\rm T}\sqrt{\frac{-2i}{\pi m_{\rm T}\tau}}e^{-im_{\rm T}\tau}\left[1-i\frac{3-4\mu^2}{8m_{\rm T}\tau}\right].
\end{equation}
Feeding the above expansions in the definitions \eqref{boosteq41} and \eqref{boosteq42}, we obtain
\begin{equation}\label{kappa3}
    K_{\gamma}(\tau)\simeq
    \frac{1}{2 m_{\rm T}\omega(\tau)} \left[m_{\rm T}^2
    +\gamma(\tau)\right],
\end{equation}
\begin{equation}\label{lambda3}
    \Lambda_{\gamma}(\tau)\simeq
    \frac{1}{2m_{\rm T}\omega(\tau)}
    {\rm e}^{-2im_{\rm T}\tau}
    \left[-m_{\rm T}^2\left(1-i\frac{3-4\mu^2}{4m_{\rm T}\tau} \right) +\gamma(\tau)\left(1+i\frac{1+4\mu^2}{4m_{\rm T}\tau} \right) \right].
\end{equation}
In the case of $K_{\gamma}(\tau)$, the remainder is of the order $1/[m_{\rm T}\omega(\tau)(m_{\rm T}\tau)^2]$, while for $\Lambda_{\gamma}$ is of order ${\rm e}^{-2im_{\rm T}\tau}/[m_{\rm T}\omega(\tau)(m_{\rm T}\tau)^2]$.

The last formulae are indicative of the behaviour both at late times and for large values of the transverse momentum ${\rm p_T}$, whence the transverse mass $\mT$, necessary for the convergence check of the renormalized thermal expectation values.
In both cases, $m_{T}\tau\gg1$ and the previous asymptotic expansion is appropriate.
Concerning the large transverse momentum behaviour, equation \eqref{kappa3} tells us that $K(\tau)\to 1$, therefore the leading order of $K(\tau)-1$ is zero, at least at first order.
However, from equation \eqref{lambda3} and the exact relation \eqref{boosteq22}, one can obtain the value of the second order expansion.
Indeed, for $\gamma(\tau)=\omega(\tau)$, in the large $m_{\rm T}$ limit
\begin{equation}
    \Lambda(\tau)=
    \frac{1}{2m_{\rm T} \tau}{\rm e}^{-2im_{\rm T}\tau}
    +{\rm ord}\left( \frac{{\rm e}^{-2im_{\rm T}\tau}}{m_{\rm T}^2} \right),
\end{equation}
therefore
\begin{equation}
    \frac{1}{2m_{\rm T}\tau}\simeq |\Lambda(\tau)|=
    \sinh \Theta(\tau)\simeq \Theta(\tau),
\end{equation}
then, still in the limit of large $m_{\rm T}$,
\begin{equation}
    K(\tau)-1\simeq
    \frac{1}{2}\Theta^2(\tau)\simeq
    \frac{1}{8m_{\rm T}^2\tau^2}.
\end{equation}

A very similar approach can be used for the late time behaviour $\tau\to\infty$.
Namely, using the asymptotic formulae \eqref{kappa3} and \eqref{lambda3}, making an expansion of $\omega(\tau)$ and $\gamma(\tau)$ for large $\tau$ keeping only the first orders, that is up to the order $1/\tau$ for $K_{\gamma}(\tau)$ and to the order ${\rm e}^{-2im_{\rm T}\tau}/\tau$ for $\Lambda_{\gamma}(\tau)$.
This difference between the order of $K_{\gamma}(\tau)$ and $\Lambda_{\gamma}(\tau)$ is because of the different gauge functions in the two cases from the large argument expansion of the Hankel functions.
Therefore, one has
\begin{equation}
    K_{\gamma}(\tau)\simeq
    \frac{m_{\rm T}^2 +\tilde{\gamma}}{2 m_{\rm T}^2},
\end{equation}
\begin{equation}
    \Lambda_{\gamma}(\tau)\simeq
    \frac{1}{2m_{\rm T}^2 }{\rm e}^{-2im_{\rm T}\tau}
    \left[\tilde{\gamma} - m_{\rm T}^2 +i\frac{m_{\rm T}^2(3-4\mu^2)+\tilde{\gamma}(1+4\mu^2)}{4m_{\rm T}\tau}\right],
\end{equation}
with $\tilde{\gamma}$ being
\begin{equation}
    \tilde{\gamma}\equiv
    \lim_{\tau \to \infty}\gamma(\tau)=
    \begin{cases}
        m_{\rm T}^2, & \mbox{for }\ed(\tau)\\
        -m^2, & \mbox{for } \pt(\tau)\\
        -m_{\rm T}^2, &\mbox{for } \pl(\tau).
    \end{cases}
\end{equation}
It is important to note that, beside the case related to the energy density and only at the leading order, $\Lambda_{\gamma}(\tau)$ has a rapidly oscillating phase preventing a proper limit in the function domain.
However, it converges in the distribution domain, which is fine since it has to be integrated.
In fact the limits
\begin{equation}
    \lim_{\tau \to \infty} \sin(2m_{\rm T}\tau), \qquad
    \lim_{\tau \to \infty} \cos(2m_{\rm T}\tau)
\end{equation}
are proportional to Dirac delta functions.
Making use of the formula for the delta families
\begin{equation}
    \delta(x)=\lim_{\epsilon\to 0} \frac{1}{\epsilon}f\left(\frac{x}{\epsilon}\right),
\end{equation}
for a generic function $f$ normalized to $1$, by indicating $\epsilon \equiv 1/\tau$ in the first case and $\epsilon \equiv 1/\sqrt{\tau}$ in the second one, we have
\begin{align}
        \lim_{\epsilon \to 0} \left(\frac{1}{\pi}\frac{\sin(x/\epsilon)}{x}\right)=\delta(x)
    \qquad &\Rightarrow \qquad
    \sin(2m_{\rm T}\tau)\;\stackrel{\tau\to \infty}{\longrightarrow}\;2\pi m_{\rm T}\,\delta(2m_{\rm T}),\\
    \lim_{\epsilon \to 0}
    \left(\frac{1}{\epsilon \sqrt{\pi}}\cos\left(\frac{x^2}{\epsilon^2}\right)\right)
    =\delta(x) \qquad &\Rightarrow \qquad
    \cos(2m_{\rm T}\tau)\;\stackrel{\tau \to \infty}{\longrightarrow}\;
    \sqrt{\frac{\pi}{\tau}}\,\delta(\sqrt{2m_{\rm T}}).
\end{align}
In both cases, the Dirac delta function is outside of the domain of integration, and all these integrals are vanishing in the long proper time limit.


These expressions can now be used to calculate the asymptotic value of the non-equilibrium thermal expectation value \eqref{boosteq65} at late times.
In particular, we want to focus on the energy density, in which case $K_{\gamma}(\tau)$ and $\Lambda_{\gamma}(\tau)$ coincide with $K(\tau)$ and $\Lambda(\tau)$ respectively.
As $\tau$ grows to infinity, $K(\tau)$ 
tends to the constant value $1$, whereas $\Lambda(\tau)$ 
vanishes, thus
\begin{equation}\label{boosteq68}
    \begin{split}
        \ed(\tau)=&
        \int \frac{\di^2{\rm p_T}\,\di \mu}{(2\pi)^3\tau}\omega(\tau)
        \left[K(\tau)K(\tau_0)-{\rm Re}\left(\Lambda(\tau)\Lambda^*(\tau_0)\right)\right]
        \left(n_{\rm B}(\tau_0)+\frac{1}{2}\right)\\
        \simeq&
        \int \frac{\di^2{\rm p_T}\,\di \mu}{(2\pi)^3\tau}\omega(\tau)
        K(\tau_0)\left(n_{\rm B}(\tau_0)+\frac{1}{2}\right)\\
        =&\int \frac{\di^2{\rm p_T}\,\di \mu}{(2\pi)^3\tau}\omega(\tau)
        \left(n_{\rm B}(\tau_0)+\frac{1}{2}\right)+
        \int \frac{\di^2{\rm p_T}\,\di \mu}{(2\pi)^3\tau}\omega(\tau)
        \left(K(\tau_0)-1\right)\left(n_{\rm B}(\tau_0)+\frac{1}{2}\right)\\
        =&\int \frac{\di^2{\rm p_T}\,\di \mu}{(2\pi)^3\tau}\omega(\tau)
        \left(n_{\rm B}(\tau_0)+\frac{1}{2}\right)+
        \int \frac{\di^2{\rm p_T}\,\di \mu}{(2\pi)^3\tau}\omega(\tau)
        2\sinh^2 \Theta(\tau_0)\left(n_{\rm B}(\tau_0)+\frac{1}{2}\right)
    \end{split}
\end{equation}
where equation \eqref{boosteq22} was used in the last step.
Renormalization is about to be discussed in Subsection \ref{sec:boost_ne_renormalization}.
As we shall shortly see there, once suitably renormalized, the first integral will correspond to the classical free-streaming solution in Milne coordinates, reported in Appendix \ref{appendix:free_sreaming}.
On the other hand, the second integral will be a pure quantum correction due to vacuum effects, for it will vanish only for $\Theta(\tau_0)=0$.


\subsection{Renormalization}
\label{sec:boost_ne_renormalization}

The non-equilibrium thermal expectation values \eqref{boosteq65} as well as the asymptotic energy density \eqref{boosteq68} are divergent due to the $1/2$ terms stemming from the commutation relations between $\wh{\xi}_{\sf p}(\tau_0)$ and $\wh{\xi}_{\sf p}^{\dagger}(\tau_0)$ that must be suitably renormalized.
What we are facing here is essentially the same ambiguity of which vacuum contribution should be subtracted as the one discussed in Subsection \ref{sec:boost_lte_renormalization} in the context of the local thermodynamic equilibrium analysis.
The choice there was between the two vacua $|0_{\tau}\rangle$ and $|0_{\rm M}\rangle$, the former being the $\tau$-dependent vacuum of the local thermodynamic equilibrium density operator whereas the latter being the static Minkowski vacuum.
In particular, it was shown that by taking the standpoint of the inertial or the Milne observer and renormalizing with respect to the Minkowski vacuum, we ended up with thermal expectation values that were still divergent.
This persuaded us that the Minkowski vacuum renormalization scheme should be discarded.
That conclusion still holds in the present non-equilibrium analysis, but now we have to take a closer look at what the subtraction of the contribution of the $\tau$-dependent vacuum $|0_{\tau}\rangle$ implies.

The renormalized thermal expectation value of the energy-momentum tensor ought to fulfill the hydrodynamic equations $\de_{\mu}T^{\mu \nu}=0$.
However, in the $|0_{\tau}\rangle$ scheme, the renormalized tensor
\begin{equation}
    T^{\mu \nu}_{\tau}\equiv
    \tr(\wh{\rho}\,\wh{T}^{\mu \nu})-\langle 0_{\tau}|\wh{T}^{\mu \nu}|0_{\tau}\rangle=
    \tr \left[(\wh{\rho}-|0_{\tau}\rangle \langle 0_{\tau}|)\wh{T}^{\mu \nu}\right]
\end{equation}
has a non-vanishing divergence due to the $\tau$-dependence of $|0_{\tau}\rangle$
\begin{equation}
    \begin{split}
        \de_{\mu}T^{\mu \nu}_{\tau}=&
        \de_{\mu}\tr \left[(\wh{\rho}-|0_{\tau}\rangle \langle 0_{\tau}|)\wh{T}^{\mu \nu}\right]\\
        =&\tr \left[(\wh{\rho}-|0_{\tau}\rangle \langle 0_{\tau}|)\de_{\mu}\wh{T}^{\mu \nu}\right]-
        \tr \left[(\de_{\mu}|0_{\tau}\rangle \langle 0_{\tau}|)\wh{T}^{\mu \nu}\right]\\
        =&-\tr \left[u_{\mu}\left(\de_{\tau}|0_{\tau}\rangle \langle 0_{\tau}|\right)\wh{T}^{\mu \nu}\right]\ne 0,
    \end{split}
\end{equation}
where the conservation law $\de_{\mu}\wh{T}^{\mu \nu}=0$ and the stationarity of $\wh{\rho}$ were used.
Therefore, although the energy-momentum tensor quantum operator is conserved, its renormalized thermal expectation value does not fulfill the hydrodynamic equations in this scheme.
We conclude that, in order to have a thermal expectation which is both finite and conserved, the vacuum to be subtracted must be stationary.
Since the Minkowski vacuum can neither be chosen, our most reasonable alternative is the vacuum of the non-equilibrium density operator \eqref{boosteq69}, namely the vacuum of $\wh{\Pi}(\tau_0)$.
That is but the state annihilated by the operators $\wh{\xi}_{\sf p}(\tau_0)$, so, in analogy with \eqref{boosteq70}, we indicate it as $|0_{\tau_0}\rangle$
\begin{equation}
    \wh{\xi}_{\sf p}(\tau_0)|0_{\tau_0}\rangle \equiv 0.
\end{equation}
As mentioned already, the operators $\wh{\xi}_{\sf p}(\tau_0)$ are just $\wh{\xi}_{\sf p}(\tau)$ evaluated at a fixed $\tau_0$, hence $|0_{\tau_0}\rangle$ is $|0_{\tau}\rangle$ at $\tau_0$, and as such it is a stationary state suitable for a proper renormalization.

Much in the same way as $|0_{\tau}\rangle$ is obtained from $\wh{\rho}_{\rm LE}(\tau)$ in the $T(\tau)\to 0$ limit as in \eqref{boosteq71}, $|0_{\tau_0}\rangle$ is obtained from $\wh{\rho}$ in the $T(\tau_0)\to 0$ limit
\begin{equation}
    \lim_{T(\tau_0)\to 0}\wh{\rho}=
    \lim_{T(\tau_0)\to 0}\wh{\rho}_{\rm LE}(\tau_0)=
    \lim_{T(\tau_0)\to 0}\frac{1}{Z_{\rm LE}(\tau_0)}\exp \left[-\frac{\wh{\Pi}(\tau_0)}{T(\tau_0)}\right]=
    |0_{\tau_0}\rangle \langle 0_{\tau_0}|.
\end{equation}
Thus, renormalization with respect to $|0_{\tau_0}\rangle$ is but the subtraction of the $T(\tau_0)=0$ contribution.
The non-equilibrium thermal expectation value of the energy-momentum tensor thereby renormalized is
\begin{equation}
    T^{\mu \nu}_{\tau_0}\equiv
    \tr(\wh{\rho}\wh{T}^{\mu \nu})-\langle 0_{\tau_0}|\wh{T}^{\mu \nu}|0_{\tau_0}\rangle=
    \tr(\wh{\rho}_{\rm LE}(\tau_0)\wh{T}^{\mu \nu})-\langle 0_{\tau_0}|\wh{T}^{\mu \nu}|0_{\tau_0}\rangle,
\end{equation}
in particular, equation \eqref{boosteq65} becomes
\begin{equation}
    \Gamma^{\gamma}_{\tau_0}(\tau)=
    \int \frac{\di^2{\rm p_T}\,\di \mu}{(2\pi)^3\tau \omega(\tau)}\omega^2(\tau)
    \left[K_{\gamma}(\tau)K(\tau_0)-{\rm Re}\left(\Lambda_{\gamma}(\tau)\Lambda^*(\tau_0)\right)\right]
    n_{\rm B}(\tau_0),
\end{equation}
whereas the asymptotic expression of the energy-density at late times \eqref{boosteq68} reads
\begin{equation}
    \ed_{\tau_0}(\tau)=
    \int \frac{\di^2{\rm p_T}\,\di \mu}{(2\pi)^3\tau}\omega(\tau)
    n_{\rm B}(\tau_0)+
    \int \frac{\di^2{\rm p_T}\,\di \mu}{(2\pi)^3\tau}\omega(\tau)
    2\sinh^2 \Theta(\tau_0)n_{\rm B}(\tau_0).
\end{equation}
The first term corresponds to the classical free-streaming solution in Milne coordinates, as shown in Appendix \ref{appendix:free_sreaming}.
On the other hand, the second one is a pure quantum correction due to vacuum effects, interpretation supported by the fact that it vanishes only if $\Theta(\tau_0)=0$.
Interestingly enough, the latter does not vanish at later times, in fact it can as well be comparable with the former if the main argument of $\Theta(\tau_0)$, namely $\mT \tau_0$, is of order ${\cal O}(1)$, that is for an early decoupling of the system.


\section{Summary and outlook}

In this Chapter, we considered a relativistic quantum fluid with boost invariance.
This is of particular concern for the physics of the quark-gluon plasma as boost invariance is approximately realized in the central-rapidity region in heavy-ion collisions, although modern hydrodynamic calculations can go beyond this model by including transverse expansion and, eventually, by breaking boost invariance itself.
A boost-invariant fluid is inherently out of thermodynamic equilibrium, so the thermodynamic fields are in general unknown.
However, geometrical constraints descended from this particular symmetry that provided us with enough information on the thermodynamic fields and other quantities to at least attempt actual calculations.

After thoroughly presenting boost invariance and its properties, we considered as a Quantum Field Theory underlying the hydrodynamic theory a free real scalar field theory in the future light-cone.
This had quite some differences with respect to the case considered in Chapter \ref{chapter:gteacceleration} in the right Rindler wedge.
In the Rindler wedges, the quantum field was expanded in terms of Rindler creation and annihilation operators, which were different from the usual operators appearing in the plane waves expansion, and the density operator was diagonal with respect to them.
In particular, these two sets of operators were related by a non-trivial Bogolyubov transformation, which, together with the four-temperature having a Killing horizon, ultimately gave rise to the Unruh effect.
On the other hand, in the future light-cone the quantum field was expanded in terms of creation and annihilation operators linearly related to the plane waves ones, but the density operator was diagonalized by a new set of operators related to the former by a non-trivial Bogolyubov transformation.
Moreover, the four-temperature had no Killing horizon for it was not a Killing vector field.
In fact, the future-light-cone is not a stationary spacetime, therefore it does not possess any global timelike Killing vector field.
In view of these facts, there appeared to be no analog of the Unruh effect in the future light-cone.

With a diagonal density operator, we were in a good position to calculate thermal expectation values.
First, we performed a local thermodynamic equilibrium analysis.
At local thermodynamic equilibrium, in fact, due to the reconstruction of the Wronskian of the Hankel functions, thermal expectation values of operators of physical interest, such as the energy-momentum tensor, could be calculated in an exact analytical way.
Somehow unexpectedly, the ideal form of the energy-momentum tensor was obtained.
Renormalization was also discussed, pointing out peculiar vacuum effects: the subtraction of the Minkowski vacuum contribution ended up in thermal expectation values which were still infinite, at variance, for instance, with the case studied in Chapter \ref{chapter:gteacceleration}.
The entropy current thereby obtained seemed not to contain any apparent quantum correction, which was equally unexpected.

Finally, we performed a full non-equilibrium analysis.
This time, however, because of the mixing of terms calculated at different times, thermal expectation values became non-trivial combinations of Hankel functions evaluated at different arguments.
Neither the Wronskian could be reconstructed, nor we were able to carry out exact integration of such combinations.
In order to go on, we considered an asymptotic expansion which simplified the expressions and eventually allowed us to obtain results.
Renormalization was also discussed in this limit, highlighting again the interesting vacuum effects.
The subtraction of the Minkowski vacuum contribution still ended up in divergent thermal expectation values, therefore we had to subtract the $|0_{\tau_0}\rangle$ contribution, with $|0_{\tau_0}\rangle$ being the stationary vacuum of $\wh{\xi}_{\sf p}(\tau_0)$ at fixed $\tau_0$.
Interestingly, in this renormalization scheme the energy density turned out to be the sum of two terms: the former being the classical free-streaming solution in Milne coordinates, while the second one being a pure quantum correction due to a vacuum term.
No entropy current or entropy production rate could be calculated since no formula is currently known for such quantities in the case of a Quantum Field Theory fully out of thermodynamic equilibrium.



\chapter{Conclusions}
\label{chapter:conclusions}

The leitmotif of this work is the determination of the entropy current of relativistic quantum fluids at local thermodynamic equilibrium.
The entropy current is, in fact, a fundamental ingredient of relativistic hydrodynamics which makes sure that the second law of thermodynamics is enforced in a covariant language, and whose expression entails the constitutive equations of the conserved currents in terms of the derivatives of the intensive thermodynamic parameters.
However, in the Israel-Stewart theory the structure of the entropy current is essentially postulated rather than derived.

In this work, we proved that, from the point of view of the quantum statistical foundations of relativistic hydrodynamics, a rigorous definition of the entropy current can indeed be provided at local thermodynamic equilibrium and a method to calculate it can be put forward.
This was done by using the Zubarev formalism of relativistic quantum statistical mechanics, which fully embeds the quantum field theoretical nature of the system underlying its hydrodynamic picture.
As such, it is particularly suitable for the description of strongly interacting fluids such as the quark-gluon plasma produced in heavy-ion collisions, for which the relativistic kinetic theoretical approach breaks down.
It should be emphasized that our method holds at local thermodynamic equilibrium at most, since no equation is currently known for the entropy current or the entropy production of relativistic quantum fluids fully out of thermodynamic equilibrium.
In this framework, we showed that a sufficient condition for the existence of the entropy current at local thermodynamic equilibrium is that the quantum operator appearing in the argument of the exponential function in the density operator should be bounded from below with non-degenerate lowest eigenvalue.
In this case, the logarithm of the partition function was proven to be extensive, meaning that it is expressed as the integral of a vector field, called the thermodynamic potential current, on a 3-dimensional spacelike hypersurface, a usually tacitly understood hypothesis.
The expression of the entropy current was thereby obtained in terms of the thermal expectation values at local thermodynamic equilibrium of the conserved currents, namely the energy-momentum tensor and some possible charged current quantum operators.

The next step was the actual calculation of the entropy current for two different kinds of fluids.
This, however, should not be mistaken for a mere test of our method, but rather it should be regarded as an opportunity for us to study two systems interesting in their own right, especially in the context of heavy-ion collisions, and use the framework of relativistic quantum statistical mechanics to work out exact analytical results.
Thus, this ought to be considered as an equally important objective of this work.

To begin with, we explored the case of a relativistic quantum fluid at global thermodynamic equilibrium with acceleration field of constant magnitude, a non-trivial instance of global thermodynamic equilibrium in Minkowski spacetime.
We managed to work out the thermal expectation value of the energy-momentum tensor of a free real scalar quantum field in the right Rindler wedge in an exact analytical way, at least in the massless case.
This revealed to possess a longitudinal anisotropy in the pressure due to the acceleration field, which was indeed a quantum term.
The Unruh effect also emerged, not surprisingly, thanks to the combination of two major points.
The first one was the density operator being diagonal in the same set of creation and annihilation operators appearing in the quantum field expansion, namely the Rindler ones, and these were related to the familiar plane waves set by a non-trivial Bogolyubov transformation.
The second one being the fact that the four-temperature possessed a Killing horizon.
Renormalization was then discussed by pointing out the differences between the two natural schemes at hand, namely the subtraction of either the Minkowski vacuum contribution or the Rindler vacuum one.
After that, we calculated the entropy current by employing our method: the result revealed to be divergenceless, in agreement with the general condition of global thermodynamic equilibrium, and also to possess a quantum term proportional to the acceleration field.
A non-vanishing entropy current was found in the Minkowski vacuum, probably owing to having traced out the field degrees of freedom in the left Rindler wedge.
The entropy thereby obtained by integration was in full agreement with previously known literature and it turned out to coincide with the entanglement entropy of the right Rindler wedge with the left Rindler wedge by virtue of the factorization of the density operator.
The calculation of the entanglement entropy in Quantum Field Theory is a notoriously hard task to accomplish: only few exact calculations are present in literature and most of them heavily rely on either the 2-dimensional conformal symmetry or the AdS/CFT correspondence.
Having been able to perform it in an exact analytical way by using a method completely different from the standard ones for a system of physical concern adds all the more interest to our result.
The entropy current certainly depends on the specific form of the energy-momentum tensor quantum operator, whereas the entropy should not.
The above findings were worked out by using the canonical tensor, corresponding to minimal coupling, but we also compared them with those we would have obtained with an ``improved'' tensor, which is traceless for a massless field and corresponding to conformal coupling.
The entropy current did indeed change, in particular the quantum term proportional to the acceleration field disappeared.
However, the entropy seemed to change as well by a constant factor, which was somewhat unexpected.
This will be subject for further studies.

Finally, we examined the case of a relativistic quantum fluid with boost invariance in the future light-cone, which is inherently out of thermodynamic equilibrium.
Very little information, if any, is usually known about the thermodynamic fields for non-equilibrium systems, making them remarkably hard to study.
Nevertheless, we showed that boost invariance implied special constraints providing us with enough information to at least attempt actual calculations.
After a thorough presentation of boost invariance, we considered free real scalar field theory in the future light-cone.
Interestingly, the quantum field was expanded in a set of creation and annihilation operators related to the familiar ones of the plane waves expansion by a linear transformation, therefore their vacuum did coincide with the Minkowski vacuum.
On the other hand, the density operator was diagonalized in a different set of creation and annihilation operators which were related to the previous ones by a non-trivial and time-dependent Bogolyubov transformation, so its time-dependent vacuum was different from the Minkowski vacuum.
Because of these differences with respect to the previously analyzed system and because of the four-temperature not being a Killing vector field, whence not possessing any Killing horizon, no analog of the Unruh effect emerged in the future light-cone.
With a diagonalized density operator, we worked out the thermal expectation value of the canonical energy-momentum tensor at local thermodynamic equilibrium in an exact analytical way.
The result proved to possess the well-known ideal form with the transverse and longitudinal pressures being equal to each other, which was somewhat unexpected.
Renormalization was then discussed.
Curiously, the subtraction of the Minkowski vacuum contribution ended up giving thermal expectation values which were still divergent, therefore it had to be discarded.
The remaining natural choice was the subtraction of the contribution corresponding to the time-dependent vacuum of the density operator, which gave indeed finite quantities.
This gave us the opportunity to use our method to calculate the entropy current at local thermodynamic equilibrium.
Surprisingly, the result coincided with the classical free-streaming one in Milne coordinates, without any apparent quantum correction.
Finally, we also performed an analysis fully out of thermodynamic equilibrium.
However, because of non-trivial combinations of Hankel functions evaluated at different arguments, we were forced to introduce some approximation to eventually carry out calculations.
In particular, we considered the asymptotic behaviour at late times.
Renormalization was the discussed in this limit highlighting interesting vacuum effects.
The subtraction of the Minkowski vacuum contribution still gave us divergent quantities, so we subtracted the contribution corresponding to the vacuum of the non-equilibrium density operator, which was stationary by definition.
Interestingly, the energy density thereby obtained was given by the sum of two terms, the one being the classical free-streaming one in Milne coordinates, while the other one being a pure quantum correction due to the vacuum.
Neither the entropy current nor the entropy production rate could be calculated since no formula is currently known for a Quantum Field Theory fully out of thermodynamic equilibrium.


\appendix

\chapter{Free-streaming in Milne coordinates}
\label{appendix:free_sreaming}
The free-streaming equation in classical relativistic kinetic theory reads
\begin{equation}\label{appAeq01}
    p^{\mu}\de_{\mu}f(x,{\bf p})=0,
\end{equation}
where $p^{\mu}$ is the four-momentum and $f$ the distribution function.
Its explicit solution in Cartesian coordinates is
\begin{equation}
    f(x,{\bf p}) = f_0\left({\bf x}-\frac{t-t_0}{\varepsilon}{\bf p},{\bf p}\right),
\end{equation}
wherein $f_0({\bf x},{\bf p}) = f(t_0,{\bf x};{\bf p})$ is the initial condition in a 
generic inertial reference frame, and $\varepsilon=\sqrt{m^2+p^2}$ is the (on-shell) energy.

In the the longitudinal boost-invariant symmetry, the initial condition is given at some Milne time $\tau_0$ rather than the time $t_0$ in Cartesian coordinates.
Nevertheless, there is a very simple solution in this case too.
Since the distribution function is a scalar, it must be invariant under the symmetry transformations at stake, that are longitudinal boosts and rotations and translations in the transverse plane.
Hence, it depends only on the independent scalars that may be formed with combinations of the space and time coordinates and the momentum vector which are invariant under the group of transformations ${\rm IO}(2) \otimes {\rm SO}(1,1)$.
These scalars are
\begin{equation}
    \tau=\sqrt{t^2-z^2},\qquad
    {\rm p_T}=\sqrt{p_x^2+p_y^2},\qquad
    w=z\varepsilon-tp_z.
\end{equation}
The last variable can be shown to be equivalent to the covariant component $p_\eta$ of the the four-momentum vector in Milne coordinates.
Indeed, there is a fourth invariant scalar
\begin{equation}
    v=t\varepsilon-zp_z=
    \tau \sqrt{m^2+{\rm p}_{\rm T}^2+\frac{w^2}{\tau^2}},
\end{equation}
but it is redundant because of the on-shell condition (and positivity) of the energy and because $t>|z|$ in the future light cone.
The invariance under space reflections $(x,y,z)\mapsto (-x,-y,-z)$ makes  $f$ depending on the square of $w$ rather than $w$ itself.
By plugging these arguments into equation \eqref{appAeq01}, we obtain
\begin{equation}
    \frac{v}{\tau}\frac{\de}{\de \tau}f(\tau,{\rm p_T},w^2)=0, 
\end{equation}
since the contribution in the partial derivatives with respect to $w$ cancels out.
The free-streaming solution is then very simple, a constant in $\tau$: 
\begin{equation}
    f(\tau,{\rm p_T},w^2)=
    f(\tau_0,{\rm p_T},w)\equiv
    f_0({\rm p_T},w^2).
\end{equation}

We are now in a position to calculate the free-streaming solution for the energy-momentum tensor from its classical kinetic definition
\begin{equation}
    T^{\mu \nu}=\int \frac{\di^3{\rm p}}{(2\pi)^3\varepsilon}\,p^{\mu}p^{\nu}\,f,
\end{equation}
where the energy density, the transverse pressure and the longitudinal pressure read
\begin{subequations}
    \begin{align}
        \ed=&T^{\mu \nu}u_{\mu}u_{\nu}=
        \int \frac{\di^3{\rm p}}{(2\pi)^3\varepsilon}\frac{v^2}{\tau^2}f,\\
        \pt=&T^{\mu \nu}\frac{1}{2}\left(\hat{i}_{\mu}\hat{i}_{\nu}+\hat{j}_{\mu}\hat{j}_{\nu}\right)=
        \int \frac{\di^3{\rm p}}{(2\pi)^3\varepsilon}\frac{{\rm p}_{\rm T}^2}{2}f,\\
        \pl=&T^{\mu \nu}\hat{\eta}_{\mu}\hat{\eta}_{\nu}=
        \int \frac{\di^3{\rm p}}{(2\pi)^3\varepsilon}\frac{w^2}{\tau^2}f.
    \end{align}
\end{subequations}
By changing the integration variable as
\begin{equation}
    w=z\varepsilon-tp_z
    \qquad \Rightarrow \qquad
    \di w=\left|-\frac{v}{\varepsilon}\right|\di p_z
    \qquad \Rightarrow \qquad
    \frac{\di p_z}{\varepsilon}=\frac{\di w}{v},
\end{equation}
we obtain
\begin{equation}\label{appAeq02}
    \ed=
    \int \frac{\di^2{\rm p}_{\rm T}}{(2\pi)^3}\frac{\di w}{v}\,\frac{v^2}{\tau^2}\,f_0({\rm p_T},w^2)=
    \int \frac{\di^2{\rm p}_{\rm T}\,\di w}{(2\pi)^3\tau}\,\sqrt{m_{\rm T}^2+\frac{w^2}{\tau^2}}\,f_0({\rm p_T},w^2),
\end{equation}
\begin{equation}\label{appAeq03}
    \pt=
    \int \frac{\di^2{\rm p}_{\rm T}}{(2\pi)^3}\frac{\di w}{v}\,\frac{{\rm p}_{\rm T}^2}{2}\,f_0({\rm p_T},w^2)
    =\int \frac{\di^2{\rm p}_{\rm T}\,\di w}{(2\pi)^3\tau \sqrt{m_{\rm T}^2+\frac{w^2}{\tau^2}}}\,\frac{{\rm p}_{\rm T}^2}{2}\,f_0({\rm p_T},w^2),
\end{equation}
\begin{equation}\label{appAeq04}
    \pl=
    \int \frac{\di^2{\rm p}_{\rm T}}{(2\pi)^3}\frac{\di w}{v}\, \frac{w^2}{\tau^2}\,f_0({\rm p_T},w^2)
    =\int \frac{\di^2{\rm p}_{\rm T}\,\di w}{(2\pi)^3 \tau \sqrt{m_{\rm T}^2+\frac{w^2}{\tau^2}}}\,\frac{w^2}{\tau^2}\,f_0({\rm p_T},w^2).
\end{equation}
The change of variable introduces in the integral an explicit dependence on the proper time.
Equations \eqref{appAeq02}--\eqref{appAeq04} are the classical relativistic expressions of the energy density and pressures of a free-streaming gas and coincide with the leading terms obtained in Subsection \ref{subsec:ne_tev} with the substitution $w\mapsto \mu$ and with the initial distribution equal to the Bose-Einstein local equilibrium distribution function $f_0=n_{\rm B}^0$.



\clearpage
\bibliography{bibliography.bib}
\bibliographystyle{unsrt}
\addcontentsline{toc}{chapter}{Bibliography}


\end{document}